%% file: thesis.tex
\documentclass[12pt]{myucthesis}

\usepackage[letterpaper,includehead,margin=1in,headheight=15pt]{geometry}
\usepackage{fancyhdr}
\usepackage{multirow}
\usepackage[T1]{fontenc}
\usepackage{longtable}
\usepackage{booktabs}
\usepackage{xstring}

\makeatletter
\renewcommand{\@chapapp}{}% Not necessary...
\newenvironment{chapquote}[2][2em]
  {\setlength{\@tempdima}{#1}%
   \def\chapquote@author{#2}%
   \parshape 1 \@tempdima \dimexpr\textwidth-2\@tempdima\relax%
   \itshape}
  {\par\normalfont\hfill--\ \chapquote@author\hspace*{\@tempdima}\par\bigskip}
\makeatother

\usepackage{natbib}
\usepackage[T1]{fontenc} % see http://tinyurl.com/67zdxwf
\usepackage[colorlinks,urlcolor=blue,citecolor=blue,linkcolor=blue,pdfusetitle]{hyperref}
\usepackage{pdflscape} % allows landscape-oriented figures with PDF page rotation
\usepackage{aasmacros,amsmath,amssymb,graphicx}
\usepackage{mydeluxetable} % deluxetable customized to play well with ukasesZ

\usepackage{ccaption}
\usepackage{xspace}
\usepackage{topcapt}
\usepackage[super]{nth}
\usepackage[xspace]{ellipsis}

\begin{document}
\ssp % single spacing
\hypersetup{pageanchor=false}
\include{setup}
\maketitle
\copyrightpage
\input{nc}

\begin{abstract}
In this thesis, I explore two topics in exoplanet science. The first is the prevalence of Earth-size planets in the Milky Way Galaxy. To determine the occurrence of planets having different sizes, orbital periods, and other properties, I conducted a survey of extrasolar planets using data collected by NASA's {\em Kepler Space Telescope}. This project involved writing new algorithms to analyze \Kepler data, finding planets, and conducting follow-up work using ground-based telescopes. I found that most stars have at least one planet at or within Earth's orbit and that 26\% of Sun-like stars have an Earth-size planet with an orbital period of 100~days or less. 

The second topic is the connection between the properties of planets and their host stars. The precise characterization of exoplanet hosts helps to bring planet properties like mass, size, and equilibrium temperature into sharper focus and probes the physical processes that form planets. I studied the abundance of carbon and oxygen in over 1000 nearby stars using optical spectra taken by the California Planet Search. I found a large range in the relative abundance of carbon and oxygen in this sample, including a handful of carbon-rich stars. I also developed a new technique called \SpecMatch for extracting fundamental stellar parameters from optical spectra. \SpecMatch is particularly applicable to the relatively faint planet-hosting stars discovered by \Kepler.
\end{abstract}

\hypersetup{pageanchor=true}
\begin{frontmatter}

\begin{dedication}
\null\vfil
{\large
\begin{center}
For Alana.
\end{center}}
\null\vfil
\end{dedication}

\tableofcontents
\listoffigures % optional
\listoftables % optional

% If using code.sty, can also add:
%% \listofcodes
%% \addcontentsline{toc}{chapter}{List of Code Examples}

\begin{acknowledgements}
As my stint as an Berkeley Astronomy Grad comes to a close, I've enjoyed the chance to reflect on the last five years. This thesis would not have been possible without a vast network of people who helped me along the way. Thank you for helping me to pursue my dreams.

I'd like to acknowledge the support I've received from the US taxpayers. I've enjoyed financial support from the National Science Foundation and the University of California, Berkeley. My thesis depended heavily on data from the Keck Observatory, {\em Kepler Space Telescope}, and the NERSC supercomputing center. All of these institutions receive public support, and I'm grateful to live in a society that values the pursuit of knowledge for its own sake.

Geoff Marcy, my PhD adviser, deserves special recognition. Geoff: as a young kid, I dreamed of being a scientist, but I never thought I'd get to work with one of the most brilliant, creative, and inspiring scientists on the planet Earth. I am so grateful for the extraordinary effort you took to guide me as a student and launch my career, even if it meant Skype meetings at midnight. You showed me what it takes to be a good scientist. As I step out of the nest and try to strike out my own, I know I'll depend on the many pearls of wisdom you gave me during our time together. ``Data is precious and doesn't grow on trees.'' ``If a project isn't exciting and fun, don't do it!'' I feel so lucky to have you as a mentor, colleague, and friend.

I've also had the good fortune to have a number of other mentors. Particle physicist Yury Kolomensky took me under his wing when I was a freshman at Cal and gave me my first taste of real scientific research. Andrew Howard generously hosted me for a year at the University of Hawaii, which proved to be one of the most exciting years of my life. Eugene Chiang poured his heart into improving the graduate student experience and into his classes, each of which was a masterpiece. Eugene: thanks for holding me to a high standard, and inspiring me to do better. I'll miss our lively discussions on the BART. Peter Nugent taught me nearly everything I know about high-performance computing and always made time to help me with my software. Eliot Quataert and Michael Manga served on my qual and thesis committees  (along with Geoff and Eugene) and helped guide me toward finishing my dissertation.

I had the good fortune of starting my PhD as {\em Kepler Space Telescope} began beaming its precious photometry back to Earth. The {\em Kepler Mission} was a monumental and heroic accomplishment that began before I was born. There are hundreds of scientists and engineers at NASA and Ball Aerospace who poured years of their lives into making \Kepler a reality. A few who I'd like to single out are Bill Borucki, David Koch, Natalie Batalha, Jon Jenkins, and Doug Caldwell. Thank you.

Nearly every chapter in this thesis benefited directly from data collected at the Keck Observatory. Thinking back to my high school years, when I would squint to see Saturn's rings with my 80~mm refractor, it's remarkable that only a few years later, I'd observe at one of the largest telescopes in the world. I'm grateful for the hard work of the telescope operators, support astronomers, and other staff that keep the mighty Keck telescope running. While in Hawaii, I had the privilege of meeting many people of Hawaiian descent. In our conversations, I was able to glimpse the powerful connection many feel to the land of their ancestors. The summit of Mauna Kea has profound cultural significance and deserves respect. I am deeply grateful that we astronomers have been treated as guests on the mountain.

It's been a joy to work in a department with so many wonderful colleagues. Thanks to all the staff that kept the department humming along and helped me focus on my research. Thanks in particular to Bill Boyd, Nina Ruymaker, Rayna Helgens, Barb Hoversten, and especially Dexter Stewart. Dexter: every time I see you, I leave with a smile on my face. Through my nine years at Berkeley, it meant so much to know that you cared about my well-being.

A big shout out to my fellow BADGrads. Graduate school is filled with hard work and uncertainty, and having you all there made navigating the uncertain waters easier. Thanks for all your support, and thanks also for making grad school so fun. The rest of my cohort deserves special recognition: Casey Stark, Garrett ``Karto'' Keating, and Francesca Fornasini. When I first met you five years ago in Eugene's radiation class, I couldn't have guessed that you would soon turn into some of my closest friends. I'll especially cherish all the delicious home-cooked meals we shared.

I've also had the privilege of working alongside the other fantastic astronomers in the Marcy group: Howard Isaacson, Lauren Weiss, Lea Hirsch, and Rea Kolbl. Thanks for your willingness to help me with my research---you all feel more like family than co-workers. Some of my fondest memories of grad school are the group pizza lunches in the courtyard of HFA.

Finally, I'm grateful to have an amazing family which is a limitless fountain of love, support, and encouragement. Mom and Dad, thanks for renting those Cosmos laserdiscs all those years ago. I think that got the ball rolling. Thanks for supporting my dreams. Ryan, I'm lucky to have a brother who I know is rooting for me to succeed. I really appreciate your quirky humor, your sense of adventure, and your reminders to come out and play basketball. And, of course, there's Alana. Getting a PhD involved its fair share of frustration and you bore the brunt of my emotional pain. You helped soften the pain of setbacks and amplified the joy of my successes, all while wrestling with the demands of medical school. You're an amazing woman, and I couldn't have done it with you.

% Feel free to modify or remove this acknowledgment:
\let\thefootnote\relax\footnote{This dissertation was typeset using the
\href{https://github.com/pkgw/ucastrothesis}{\textsf{ucastrothesis}}
\LaTeX\ template. Thanks, Peter!}

\end{acknowledgements}
\end{frontmatter}

% Including in chapters

\include{intro/intro}

\include{modes/modes}

\include{terra50d/terra50d}

\include{terra1yr/terra1yr}

\include{terra1yr/terra1yr_si}

\include{co/co}
\include{sm/smcat}

\bibliography{thesis}
\end{document}

%% file: setup.tex
\title{Prevalence of Earth-size Planets Orbiting Sun-like Stars}
\author{Erik Ardeshir Petigura} % must match BearFacts!
\degreesemester{Spring}
\degreeyear{2015}
\degree{Doctor of Philosophy}
\numberofmembers{4}
\chair{Professor Geoffrey Marcy}
\othermembers{
Professor Eugene Chiang \\
Professor Michael Manga \\
Professor Eliot Quataert 
}
\field{Astrophysics}
\campus{Berkeley}

%% file: nc.tex
\newcommand{\Mjup}{\ensuremath{ M_{J} }\xspace}
\newcommand{\Msini}{\ensuremath{ M \sin i }\xspace}
\newcommand{\Mp}{\ensuremath{ M_{P}}\xspace}
\newcommand{\Me}{\ensuremath{M_{E}}\xspace }

\newcommand{\nc}{\newcommand}
\newcommand{\logepso}{\log \epsilon_O}
\newcommand{\logepsni}{\log \epsilon_{Ni}}
\newcommand{\logepsc}{\log \epsilon_{C}}
\newcommand{\teff}{$T_{eff}$}
\newcommand{\monh}{[M/H]}
\newcommand{\feh}{[Fe/H]}
\newcommand{\euh}{[Eu/H]}
\newcommand{\kms}{km/s}
\newcommand{\nd}{\nodata}
\newcommand{\vsini}{$v \sin i$}
\newcommand{\um}{$\mu$m}
\newcommand{\chiexC}{8.537} % excitation energy for carbon. eV
\nc{\fTellFailO}{53}
\nc{\fTellFailC}{43}
\nc{\errFloor}{0.03}

\newcommand{\nThreeD}{1025}
\newcommand{\nSampStars}{1070} %Total number of sample stars
\newcommand{\nSpecTot}{15,000}

\newcommand{\hlf}{\frac{1}{2}}
\newcommand{\twd}{\Delta T}
\newcommand{\ep}{t_{0}}
\newcommand{\df}{\delta F}
\newcommand{\mdf}{\overline{\df}}
\newcommand{\mtx}[1]{\ensuremath{\mathbf{#1}}}

% \xspace guesses whether a space is needed after the command
\newcommand{\Kepler}{\textit{Kepler}\xspace} 
\newcommand{\Kmag}{\textit{Kp}\xspace}

% New commands for terra50d

\renewcommand{\hlf}{\tfrac{1}{2}}
\newcommand{\tdur}{\ensuremath{\Delta T}\xspace}
\renewcommand{\ep}{\ensuremath{t_{0}}\xspace}
\newcommand{\Per}{\ensuremath{P}\xspace}

\renewcommand{\df}{\delta F}
\renewcommand{\mdf}{\overline{\df}}
\renewcommand{\mtx}[1]{\ensuremath{\mathbf{#1}}}
\renewcommand{\teff}{\ensuremath{ {\rm{T}_{\rm eff}} }\xspace}
\newcommand{\logg}{\ensuremath{ {\log g }}\xspace}
\newcommand{\sm}{\ensuremath{ {\sim}}\xspace}
\newcommand{\Pcad}{\ensuremath{P_{\text{cad}}}\xspace}

\newcommand{\SNRC}{\ensuremath{\text{SNR}_\text{CDPP}}\xspace}
\newcommand{\NSNRnoteKOI}{609\xspace}
\newcommand{\NSNR}{738\xspace}
\newcommand{\NSNRP}{391\xspace} \newcommand{\NSNRnotP}{347\xspace} \newcommand{\NnotSolarSubset}{10\xspace} 
\newcommand{\Np}{\ensuremath{n_{\text{pl,cell}}}\xspace}
\newcommand{\NpAug}{\ensuremath{n_{\text{pl,aug,cell}}}}
\newcommand{\fcell}{\ensuremath{f_{\text{cell}}}\xspace }
\newcommand{\fcellBa}{\ensuremath{f_{\text{cell,Batalha}}}\xspace}
\newcommand{\flogA}{\ensuremath{d^{2}\fcell / d\log{P} / d\log{R_P}}\xspace }
\newcommand{\NstarAmen}{ n_{\star,\text{amen}} }

\newcommand{\Rsun}{\ensuremath{R_{\odot}}\xspace }
\newcommand{\fonetwo}{\ensuremath{15.1^{+1.8}_{-2.7}\%}\xspace}
\renewcommand{\Kepler}{\textit{Kepler}\xspace} 
\renewcommand{\Kmag}{\textit{Kp}\xspace} 
\newcommand{\SpecMatch}{SpecMatch\xspace}
\newcommand{\TERRA}{{\tt TERRA}\xspace}

\renewcommand{\Re}{\ensuremath{ R_{E} }\xspace} 
\newcommand{\Rp}{\ensuremath{ R_{P} }\xspace}
\newcommand{\Rstar}{\ensuremath{R_{\star}}\xspace} 
\newcommand{\rrat}{\ensuremath{\Rp / \Rstar}\xspace}  
\newcommand{\rratfrac}{\ensuremath{\frac{\Rp}{\Rstar}}\xspace}  

\renewcommand{\nc}[2]{\newcommand{#1}{\ensuremath{#2}\xspace}}

\nc{\Mstar}{M_\star} 
\nc{\dfsec}{\delta F_{\text{sec}}}
\nc{\dfpri}{\delta F_{\text{pri}}}
\nc{\teffpri}{\text{T}_{\text{eff,1}}}
\nc{\teffsec}{\text{T}_{\text{eff,2}}}
\nc{\teq}{\text{T}_{\text{eq}} } 

\renewcommand{\Re}{\ensuremath{ R_{\oplus} }\xspace} 
\nc{\ncell}{n_{\text{cell}}}
\nc{\nstar}{n_{\star}}
\nc{\PerMax}{P_{\text{max}}}
\renewcommand{\deg}{\ensuremath{^{\circ}\xspace}}

\nc{\FE}{F_{\oplus}}
\nc{\Fp}{F_{P}}
\nc{\PT}{P_{\text{T}}}
\nc{\Lstar}{L_{\star}}

%% file: intro/intro.tex
% This sample file is dedicated to the public domain.

\chapter{Introduction}

\begin{chapquote}{Carl Sagan, \textit{Pale Blue Dot}}
``There are 400 billion stars in the Milky Way Galaxy. Of this immense multitude, could it be that our humdrum Sun is the only one with an inhabited planet? Maybe\ldots Or, here and there, peppered across space, orbiting other suns, maybe there are worlds something like our own, on which other beings gaze up and wonder as we do about who else lives in the dark.''
\end{chapquote}

\label{c.intro}

Our species has, for a long time, looked up at the night sky and wondered if there are other worlds like our own out there among the stars. Until recently, this question was relegated to the realm of philosophy and metaphysics given the limitations of our eyes and telescopes. Today, we live in a privileged time. Over the past twenty years, we have discovered thousands of planets around other stars. Their existence proves that the processes that form planets are not unique to the Solar System and that the real estate for life may be plentiful throughout the Universe. 

Over the past five years, I have been fortunate to play a part in the rapidly growing field of extrasolar planet science. My thesis consists of two major themes: (i) the prevalence of planets in the Milky Way and (ii) the precise characterization of stars with planets. In Chapter~\ref{c.intro}, I give a brief historical review of exoplanet discoveries and provide some background for the rest of my thesis. I measured the occurrence of planets by analyzing data collected by NASA's {\em Kepler Space Telescope}. This survey involved constructing my own photometric pipeline to remove instrumental systematics from \Kepler photometry (Chapter~\ref{c.modes}). I analyzed \Kepler data to measure the occurrence of planets out to Mercury's orbit (Chapter~\ref{c.terra50d}), which I later extended to include Earth-like orbits (Chapters~\ref{c.terra1yr} and \ref{c.terra1yr_si}). I also worked on two projects involving the precision characterization of planet-hosting stars. I studied carbon and oxygen in nearby stars with an eye toward finding carbon-rich stars (Chapter~\ref{c.co}). In Chapter~\ref{c.sm}, I present a new tool called \SpecMatch for extracting fundamental stellar properties from high-resolution optical spectra with an emphasis on faint stars, where existing techniques are challenged.

\section{Exoplanets: A Brief History}
The study of extrasolar planets touches on some of the core questions that define us as a species: ``Are there other Earths?'' ``Is there life among the stars?'' and ``Are there other beings that revel at the miracle of life and ponder their own origins?'' These questions were debated by the philosophers of ancient Greece. Some, like Democritus (c.~460 -- c.~370~BC) and Epicurus (341 -- 270~BC), thought that there were other planets like Earth. Epicurus wrote, ``There are infinite worlds both like and unlike this world of ours \ldots\ We must believe in all worlds there are living creatures and plants and other things we see in this world'' \citep{Seager10}. The idea that there might be a plurality of worlds was also considered, but rejected, by Plato (c.~420 -- c.~348~BC) and Aristotle (384 -- 332~BC). Platonic and Aristotelian philosophy had a profound influence on Western thought, while only fragments of Democritus' and Epicurus' work survive today \citep{Billings13}.

Astronomical techniques capable of detecting extrasolar planets emerged in the second half of the twentieth century. In 1952, Otto Struve wrote a two-page paper outlining two techniques that could detect planets around other stars \citep{Struve52}. The first technique involves looking for the slight wobble of a star as it is tugged by an unseen planetary companion. The wobble of a star can be detected by Doppler shifts in the star's spectral lines over a planet's orbital period. This technique is called the radial velocity (RV) or Doppler technique. For circular orbits seen edge on, the line of sight velocity of a planet hosting star is:
\begin{equation*}
v_{\star} = \frac{M_{p}}{M_{\star}} v_{p},
\end{equation*}
where $M_{p}$ and $M_{\star}$ are the masses of the planet and host star and $v_p$ is the orbital velocity of the planet. For inclined orbits, we replace $M_p$ by $M_p \sin i$ where $i$ is the orbital inclination.\footnote{$i = 90^{\circ}$ for edge on orbits and $0^{\circ}$ for face on orbits.}

Struve also noted that planets that happen to pass in front of their host stars as seen from Earth would dim their stellar hosts once per orbit. A transiting planet dims its host star by an amount equal to the fraction of the stellar disk blocked:
\begin{equation*}
\frac{\Delta F}{F} = \left(\frac{\Rp}{\Rstar}\right)^{2},
\end{equation*}
where \Rp and \Rstar are the planet and star radii, respectively. As a point of reference, Jupiter dims the Sun by 1\%, while the Earth blocks only 0.01\% of the Sun's light. The radial velocity and transit techniques have enabled the vast majority of exoplanet discoveries to date.

Doppler searches for extrasolar planets began in earnest the early 1980s \citep{Campbell83,Marcy83,Mayor85,Latham85}. \cite{Latham89} reported a sub-stellar object orbiting HD~114762 with \Msini = 11 \Mjup, on the dividing line between brown dwarf and giant planet.\footnote{Objects more massive than 13 \Mjup are considered are considered brown dwarfs because they fuse Deuterium} \cite{Wolszczan92} made the first 
unambiguous detection of planet-mass objects outside the Solar System. They found two planets weighing 2.8 and 3.4 Earth-masses orbiting the millisecond pulsar PSR~B1257+12.\footnote{The PSR~B1257+12 discoveries were a strange twist of fate. PSR~B1257+12 was the third millisecond pulsar discovered and its associated planets suggested that planets around pulsars might be common. However, today, we know of roughly 300 millisecond pulsars, and only one other pulsar planet: PSR B1620-26b. Given current pulsar timing precision, we could easily detect Earth-analog planets around all of them. (Scott Ransom, priv. communication 2015).} In 1995, \citeauthor{Mayor95} announced 51~Pegasi~b, a Jupiter-mass planet orbiting the star 51~Pegasi with a 4.2 day orbital period. This discovery ushered in a torrent of planets discovered with the radial velocity technique. Within months of the 51~Pegasi~b discovery, two more giant planets were announced around 70~Virginis \citep{Marcy96} and 47~Ursae~Majoris \citep{Butler96}. By 2000, RVs had uncovered roughly a dozen planets and by 2010, that number had grown to around 300 \citep{Han14}.

Detecting extrasolar planets by the transit method also began to receive serious consideration in 1980s. \cite{Borucki84} concluded that, while it was possible to detect the transits of giant planets from the ground, Earth-size planets required a dedicated space-born mission. In 1984, Borucki began a thirty-year effort to design, fund, build, and fly a telescope called \Kepler, which today has discovered roughly 4000 of the 5000 exoplanets known.

In late 1999, eleven extrasolar planets had been discovered with RVs, but none had been observed transiting their host stars \citep{Charbonneau00}. Shortly after the RV discovery of HD~209458b, \cite{Henry00} and \cite{Charbonneau00} observed the periodic dimming of HD~209458 due to the occultation by a Jupiter-sized companion. The discovery of a transiting planet was a significant milestone for the exoplanet field. The observations helped dispel doubts that the radial velocity detections were face on binary stars or coherent stellar pulsations. HD~209458b was the first planet with a measured radius and the fact that the orbit was edge on meant the $\Msini \approx M$. Knowing a planet's mass and radius together helps constrain its bulk composition and structure. 

HD~209458b invigorated efforts to detect planets by the transit technique. Starting in the early 2000s, many groups started ground-based transit surveys.%
\footnote{A few of the more prolific ground-based surveys are OGLE \citep{Udalski02}, TrES \citep{Alonso04}, XO \citep{McCullough05}, HAT \citep{Bakos07}, and SuperWASP \citep{Pollacco06}. 
}
These surveys are strongly biased toward detecting close-in planets. Assuming random orbital inclinations, the probability that an extrasolar planet will transit, as seen from earth is $P_T = \Rstar / a$, where \Rstar is the stellar radius and $a$ is the planet-star orbital separation. For the Earth and the Sun, $P_T = \Rsun /\mathrm{1~AU} = 1/200$. Ground-based transit surveys must also contend with photometric noise due to differential extinction, scintillation, and flat-fielding errors \citep{Winn10}. Photometric precision improves dramatically when using space-based facilities. As an example, the original ground-based \cite{Charbonneau00} observations of HD~209458b achieved photometric precision of $\approx$ 3~ppt per minute. \cite{Brown01} observed the same planet using the {\em Hubble Space Telescope} and achieved photometric precision of $\approx$0.11~ppt per minute, sufficient to detect Earth-sized planets. These observations strengthened the case for a dedicated space-based mission to search for Earth-size extrasolar planets. After four attempts beginning in 1992, the {\em Kepler Space Telescope} was approved in December 2001 \citep{Borucki03}.

\begin{figure*}
\centering
\includegraphics[width=\textwidth]{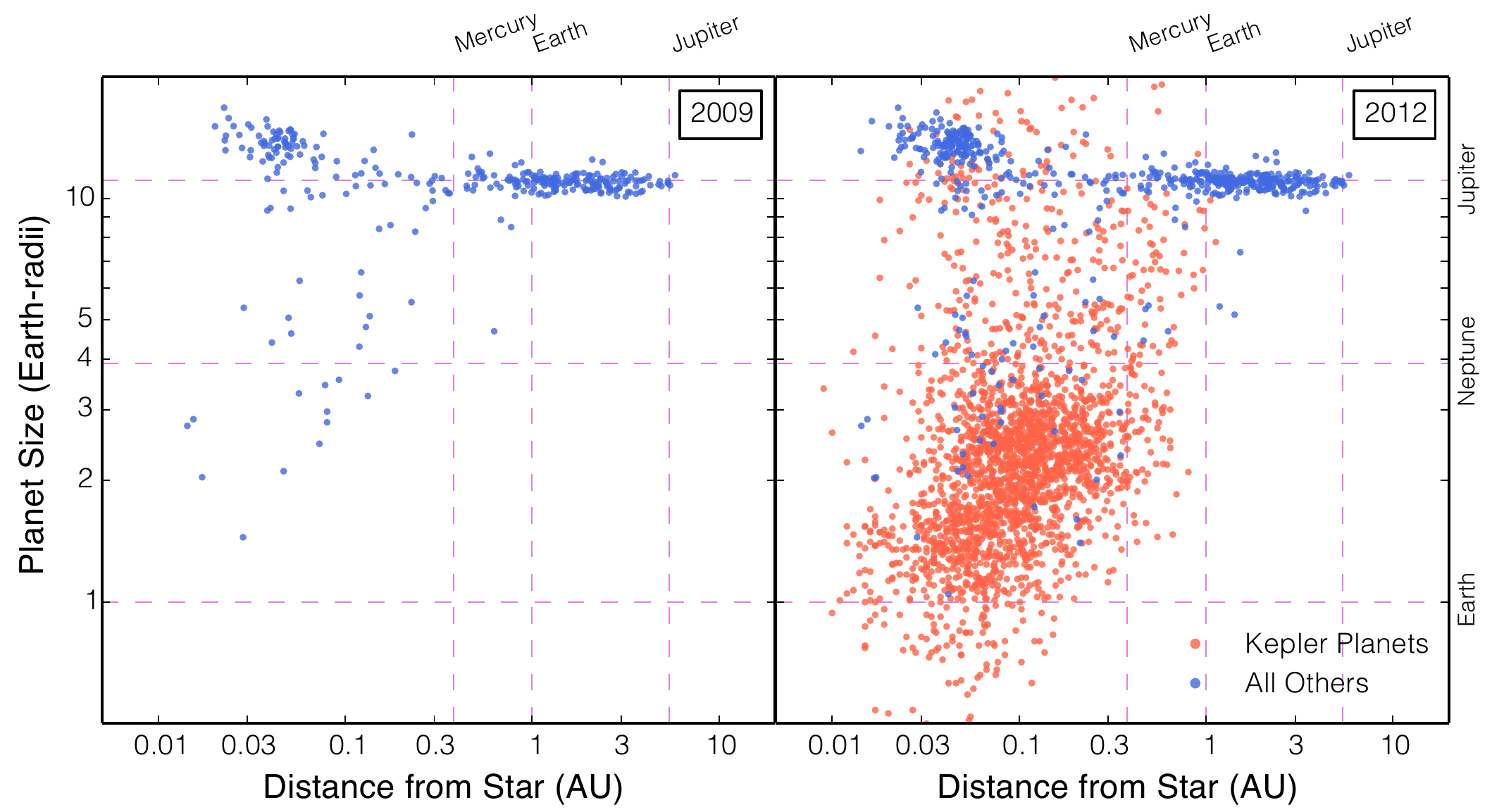}
\caption[Known extrasolar planets before and after the launch of \Kepler]{\Kepler has ushered in a paradigm shift in our understanding of extrasolar planets. The left panel shows the sizes and orbital distances of planets known prior to the launch of \Kepler in 2009. Most planets in this figure were discovered using the Doppler technique. When only planet mass is known, planet size is estimated according to the \cite{Weiss13} mass-radius relationship. Two distinct populations of planets are visible. Long-period giant planets (many of which are on eccentric orbits) and close-in ``hot Jupiters'' (which have inflated radii due to poorly understood processes). By 2012, this picture had changed dramatically with the analysis of just over one year of \Kepler data \citep{Batalha13}. Planets closer than 1~AU between the size of Earth and Neptune are a common outcome of planet formation, yet are absent in our own solar system.}
\label{fig:intro-2009}
\end{figure*}

On March 7, 2009, \Kepler lifted off from Cape Canaveral, Florida into an Earth-trailing orbit. From December 2009 to May 2013, \Kepler took a picture of a $10\deg\times10\deg$ region of the sky in the constellation Cygnus every thirty minutes. On the ground, these images were converted into brightness measurements of $\sm$150,000 stars. \Kepler has completely transformed our knowledge of planets outside the Solar System. The left panel of Figure~\ref{fig:intro-2009} shows known planets when \Kepler was launched. We see a large population of giant planets with orbital periods of a year and longer, a number of hot Jupiters, which are favored by selection effects, and a small number of close-in planets the size of Neptune (4~\Re) and smaller. The right panel of Figure~\ref{fig:intro-2009} includes planets found in the first year of \Kepler photometry \citep{Batalha13}. Planets with sizes between Earth and Neptune are common around other stars, while absent in our own Solar System.

\section{Exoplanet Demographics}
Once it became clear that other stars had planets, the natural next question is ``what fraction of stars have planets?'' The emerging field of exoplanet demographics parallels the work of Hertzsprung, Russell, and others a century ago. Just as the HR diagram shaped our understanding of stellar physics, the prevalence of planets with different sizes, orbital distances, and other properties is shaping our understanding of planet formation and evolution.

The key measured quantity in exoplanet demography is planet occurrence, $f_p$, which is simply 
\[
f_p = \frac{N_p}{N_{\star}},
\]
the number of planets divided by the number of stars. However, all planet surveys contain biases and selection effects, which require careful attention.

Prior to \Kepler, RV surveys probed giant planet occurrence out to $\sm$10-year orbits as well the occurrence of smaller planets on close-in orbits. \cite{Cumming08} found that 10.5\% of Sun-like stars have a giant planet with an orbital period less than 5.5 years (Jupiter has a 12-year orbit). While these ``exo-Jupiters'' are reminiscent of the giant planets in our own solar system, they are often found on highly elliptical orbits in contrast to the nearly circular orbits of the Solar System planets. Hot Jupiters, while easy to detect, are relatively rare. \cite{Marcy05} found an occurrence of $1.2 \pm ± 0.1\%$ for hot Jupiters ($P~\lesssim~$12~d). Through intense monitoring of restricted stellar samples, \cite{Howard10} and \cite{Mayor11} found that close-in ($P < 50$~d) planet occurrence rises steeply toward small planet sizes.  Roughly 15\% of GK stars have a 3-10 \Me planet with $P~<~50$~d. 

One way in which planet occurrence ties in with planet formation is through planet population synthesis modeling. These models attempt to capture the important aspects of planet formation physics to produce synthetic populations of planets that can be compared with observations. One such study is that of \cite{Ida08} who predicted that planets between 1 and 30 \Me and $P~<~1$~year should be extremely rare. However, the RV surveys of \cite{Howard10}, \cite{Mayor11} and the \Kepler-based surveys of \cite{Howard12}, \cite{Fressin13}, and \cite{Petigura13b} showed that sub-Neptune-size planets are very common. The migration models have since been modified to more closely match the observed distribution of planets \citep{Mordasini12}.

%Key results in this chapter are that 26\% of Sun-like stars have a 1-2 Re planet with P = 5-100 days and that roughly 1 in 5 stars like the Sun have an Earth-size planet that receives the same level of stellar irradiation as the Earth, to a factor of 4. 

%I found a rapid rise in the occurrence of planets at three Earth-radii and that roughly 1 in 6 stars like the Sun have a planet between one and two times the size of Earth residing within the orbit of Mercury. 

\section{Precision Characterization of Stars}
Another component of my thesis involves the precise characterization of stars with extrasolar planets. Planet formation is believed to occur early in a star's lifetime. Millimeter observations of young stars show that gas disks dissipate within roughly 10~Myr \citep{Haisch01}. Gas giant planets are thought to form during this time frame. In order to understand planet formation, we would like to observe planets during the process of formation. However, detecting planets around young stars is challenging due to high activity levels which hamper transit and RV methods. However, the composition of the host star offers a window into the protoplanetary disk from which planets form. 

Stellar metallicity is thought to be a good tracer of the amount of solids available in a protoplanetary disk. There is a well-established correlation between giant planet occurrence and host star metallicity \citep{Gonzalez97,Santos04,Fischer05}. However, smaller planets are found around stars having a wide range of metallicities. This new trend was observed among the RV-detected planets \citep{Mayor11} and \citep{Buchhave12nat}. The spectra taken during the course of RV planet detection surveys are valuable in probing the connection between planet and host star. 

In Chapter~\ref{c.co}, I study the carbon and oxygen abundances of over 1000 stars from the California Planet Search \citep{Marcy08}. I find that planet-hosting stars have a wide range of carbon and oxygen abundances. Of particular interest are planets around stars with carbon to oxygen ratios approaching unity. Terrestrial planets in carbon-rich environments are thought to have radically different compositions compared to planets that form in environments with C/O~<~1 \citep{Kuchner05,Bond10}. 

Another reason for studying host star properties is that in many cases, our understanding of planet properties is limited by our imperfect knowledge of the host star. In fitting a transit profile, we measure the planet-star radius ratio, \Rp/\Rstar, not on the physical size of the planet itself. Most stars in the \Kepler field have photometrically determined stellar radii which are uncertain at the 35\% level \citep{Brown11,Batalha13,Burke14}. Large uncertainties in planet size can hide features in the properties of large ensembles of planets. For example, any sharp features in the planet radius distribution that would indicate an important size scale for planet formation are smeared by planet radius errors.

Extracting stellar parameters for \Kepler host stars presented new challenges compared to similar efforts for nearby stars. Stars in the \Kepler field are typically $\sm$1~kpc from Earth. A Sun-like star at 1~kpc has $V~=~14.7$. Traditional spectroscopic methods often utilize high SNR spectra. For example, \cite{Valenti05} analyzed Keck HIRES spectra with SNR/pixel~>~200. However, obtaining a SNR/pixel~=~100 spectrum of a $V = 14.7$ star would take 2.5 hours with HIRES and is impractical for large samples of stars. In order to work with fainter stars, I developed a new tool called SpecMatch that fits large swaths of spectra containing thousands of lines. SpecMatch is able to accurately constrain fundamental stellar properties even for low SNR spectra. SpecMatch will enable measurements of stellar radii good to 5\% for large samples of \Kepler planet hosts.

%% file: modes/modes.tex
\chapter{Identification and Removal of Noise Modes in \Kepler Photometry}
\label{c.modes}

\noindent A version of this chapter was previously published in the {\em Publications of the Astronomical Society of the Pacific}
(Erik~A.~Petigura \& Geoffrey~W.~Marcy, 2012, PASP 124, 1073).\\

We present the Transiting Exoearth Robust Reduction Algorithm (TERRA) --- a novel framework for identifying and removing instrumental noise in \Kepler photometry.  We identify instrumental noise modes by finding common trends in a large ensemble of light curves drawn from the entire \Kepler field of view.  Strategically, these noise modes can be optimized to reveal transits having a specified range of timescales.  For \Kepler target stars of low photometric noise, TERRA produces ensemble-calibrated photometry having 33~ppm RMS scatter in 12-hour bins, rendering individual transits of earth-size planets around sun-like stars detectable as $\sim 3 \sigma$ signals.

\section{Introduction}
The \Kepler Mission is ushering in a new era of exoplanet science.
Landmark discoveries include \Kepler-10b, a rocky planet
\citep{Batalha:2011}; the \Kepler-11 system of six transiting planets
\citep{Lissauer:2011el}; earth-sized \Kepler-20e and 20f
\citep{Fressin12}; KOI-961b, c, and d -- all smaller than earth
\citep{Muirhead12}; and \Kepler-16b a circumbinary planet
\citep{Doyle:2011ev}.  While \Kepler has revealed exciting individual
systems, the mission's legacy will be the first statistical sample of
planets extending down to earth size and out to 1 AU.  \Kepler is the
first instrument capable of answering ``How common are earths?'' --- A
question that dates to antiquity.

Planet candidates are detected by a sophisticated pipeline developed
by the \Kepler team Science Operations Center.  In brief, systematic
effects in the photometry are suppressed by the Pre-search Data
Conditioning (PDC) module, the output of which is fed into the
Transiting Planet Search (TPS) module.  For further information, see
\cite{Jenkins10}.

The \Kepler mission was designed to study astrophysical phenomena with
a wide range of timescales, which include 1-hour transits of hot
Jupiters, 10-hour transits of planets at 1 AU, and weeklong spot
modulation patterns.  The PDC module is charged with removing
instrumental noise while preserving signals with a vast range of
timescales.  We review sources of instrumental errors in
\S~\ref{sec:InstNoise}, highlighting the effects that are most
relevant to transit detection.

The \Kepler team has released candidate planets based on the first 4
and 16 months of data \citep{Borucki11,Batalha12}.  Many of the
candidates have additional followup observations from the ground and
space aimed at ruling out false positive scenarios.  In addition,
statistical arguments suggest that 90-95\% of all candidates and that
$\sim 98$\% of candidates in multi-candidate systems are bonafide
planets \citep{Morton11,Lissauer:2012}.

While \textit{Kepler's} false positive rate is low, its completeness
is largely uncharacterized.  If the completeness decreases
substantially with smaller planet size or longer orbital periods, the
interpretations regarding occurrence drawn from the \cite{Borucki11}
and \cite{Batalha12} catalogs will be incorrect.  Hunting for the
smallest planets, including earth-sized planets in the habitable zone,
will require exquisite suppression of systematic effects.  Without
optimal detrending, systematic noise will prevent the detection of the
smallest planets, possibly the habitable-zone earth-sized planets,
which is the main goal of the \Kepler mission.  Therefore, it is
essential for independent groups to develop pipelines that compliment
both PDC and TPS.  An early example of an outside group successfully
identifying new planet candidates is the Planet Hunters project
\citep{Fischer:2011bb,Lintott:2012ut}, which uses citizen scientists
to visually inspect light curves.  In addition, existing pipelines
from the HAT ground-based search \citep{Huang:2012uj} and the
\textit{CoRoT} space mission \citep{Ofir:2012va} have been brought to
bear on the \Kepler dataset yielding $\sim100$ new planet candidates.

We present the Transiting Exoearth Robust Reduction Algorithm (TERRA)
--- a framework for identifying and removing systematic noise.  We
identify systematic noise terms by searching for photometric trends
common to a large ensemble of stars.  Our implementation is tuned
toward finding trends with transit-length timescales.

\section{Instrumental Noise in \Kepler Photometry}
\label{sec:InstNoise}
\begin{figure}[htbp]
\begin{center}
\includegraphics[width=6.5in]{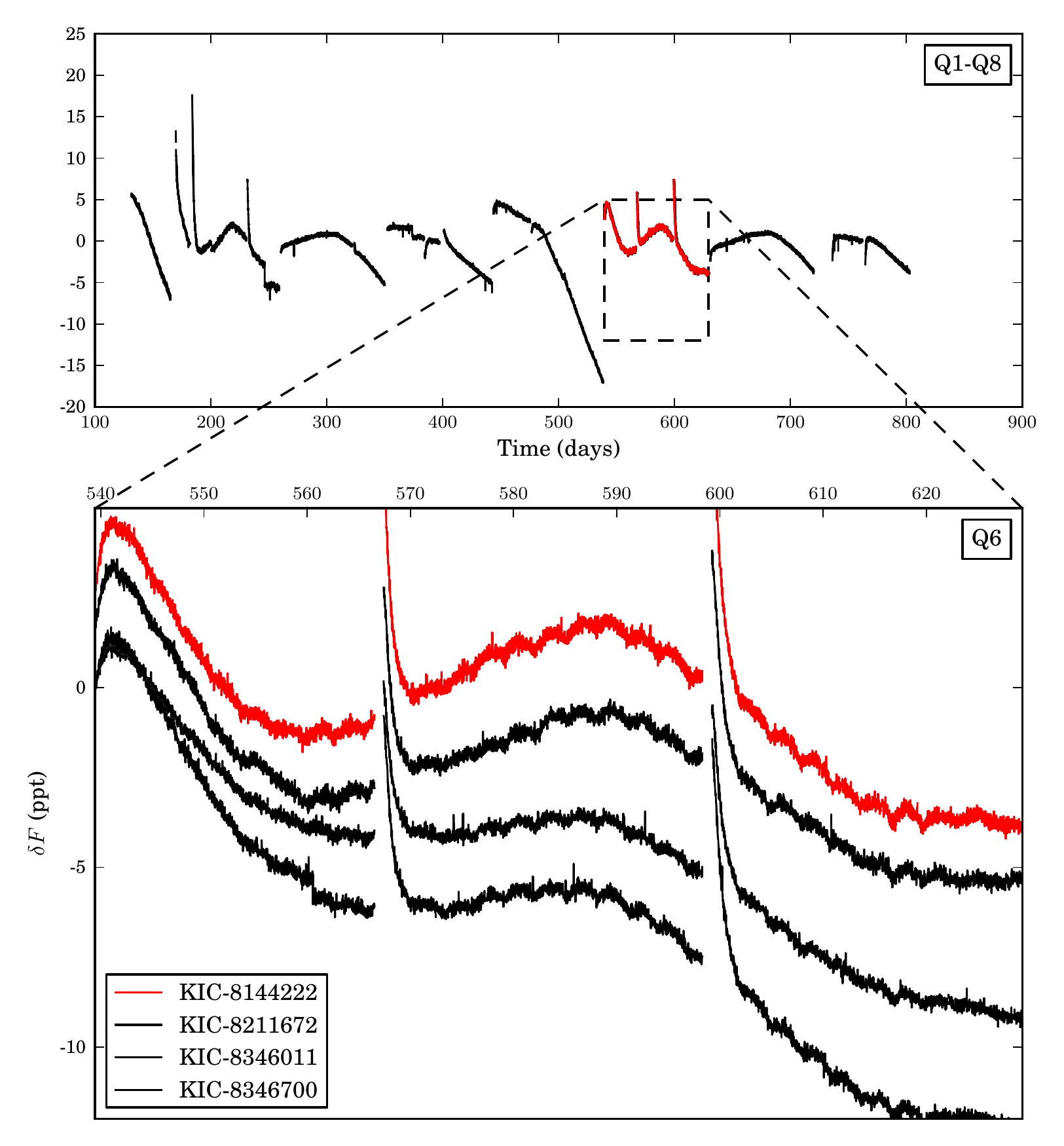}
\caption[Representative \Kepler photometry]{Top: Normalized flux from KIC-8144222 (\Kmag=12.4,
  CDPP12=35.4 ppm) from Quarter 1 through 8 (Q1-Q8).  Bottom: Detail
  of Q6 photometry showing KIC-8144222 along with three stars of
  similar brightness, noise level, and location on the FOV (12.0 $<$
  \Kmag $<$ 13.0, CDPP12 $<$ 40 ppm, mod.out = 16.1).  Much of the
  variability is common to the 4 stars and therefore instrumental in
  origin.  The two spikes are due to thermal settling events, and the
  three-day ripples are due to onboard momentum management.}
\label{fig:SampLC}
\end{center}
\end{figure}

The \Kepler spacecraft makes photometric observations of $\sim$156,000
targets.  Long cadence photometry is computed by summing all the
photoelectrons within a predefined target aperture during a
29.4~minute integration.  The \Kepler team makes this ``Simple
Aperture Photometry'' available to the scientific
community~\citep{KeplerArchiveManual}.  Simple aperture photometry
contains many sources of noise other than Poisson shot noise.  We
illustrate several noise sources in Figure~\ref{fig:SampLC}, where we
show the normalized photometry ($\df$) of KIC-8144222 (\Kmag = 12.4).
$\df = (F-\overline{F})/\overline{F}$ where $F$ is the simple aperture
photometry.

The dominant systematic effect on multi-quarter timescales is
``differential velocity aberration'' \citep{VanCleve:2009}.  As
\Kepler orbits the sun, its velocity relative to the \Kepler field
changes.  When the spacecraft approaches the \Kepler field, stars on
the extremities of the field move toward the center.  Stellar PSFs
move over \Kepler apertures by $\sim$ 1 arcsecond resulting in a
$\sim$ 1 \% effect over 1-year timescales.

We show a detailed view of KIC-8144222 photometry from Quarter 6 (Q6)
in Figure~\ref{fig:SampLC}.  The decaying exponential shapes are
caused by thermal settling after data downlinks.  Each month, \Kepler
rotates to orient its antenna toward earth.  Since \Kepler is not a
uniformly colored sphere, changing the spacecraft orientation with
respect to the sun changes its overall temperature.  After data
downlink, \Kepler takes several days to return to its equilibrium
temperature (Jeffrey Smith, private communication, 2012).  KIC-8144222
photometry also shows a $\sim$0.1\% effect with a 3-day period due to
thermal coupling of telescope optics to the reaction wheels.  We
explore this 3-day cycle in depth in \S~\ref{sec:interpretation}.

Since all of the previously mentioned noise sources are coherent on
timescales longer than one cadence (29.4~minutes), the RMS of binned
photometry does not decrease as $1/\sqrt{N}$, where N is the number of
measurements per bin.  In order to describe the noise on different
timescales, the \Kepler team computes quantities called CDPP3, CDPP6,
and CDPP12 which are measures of the photometric scatter in 3, 6, and
12-hour bins.  KIC-8144222 has CDPP12 35.4~ppm and is a low-noise star
(bottom 10 percentile).  For a more complete description of noise in
\Kepler data see \cite{Christiansen:2011}.

As a comparison, we selected stars which were similar to KIC-8144222
in position on the Field of View (FOV), noise level, and brightness
(mod.out = 16.1, CDPP12 $<$ 40 ppm, 12.0 $<$ \Kmag $<$ 13.0).  From
this 13-star sample, we randomly selected 3 stars and show their light
curves in Figure~\ref{fig:SampLC}.  The photometry from the comparison
stars is strikingly similar to the KIC-8144222 photometry.  Since much
of the variability is correlated, it must be due to the state of the
\Kepler spacecraft.  Common trends among stars can be identified and
removed.  The \Kepler team calls this ``cotrending,'' a term we adopt.

\begin{figure}[htbp]
\begin{center}
\includegraphics[height=6in]{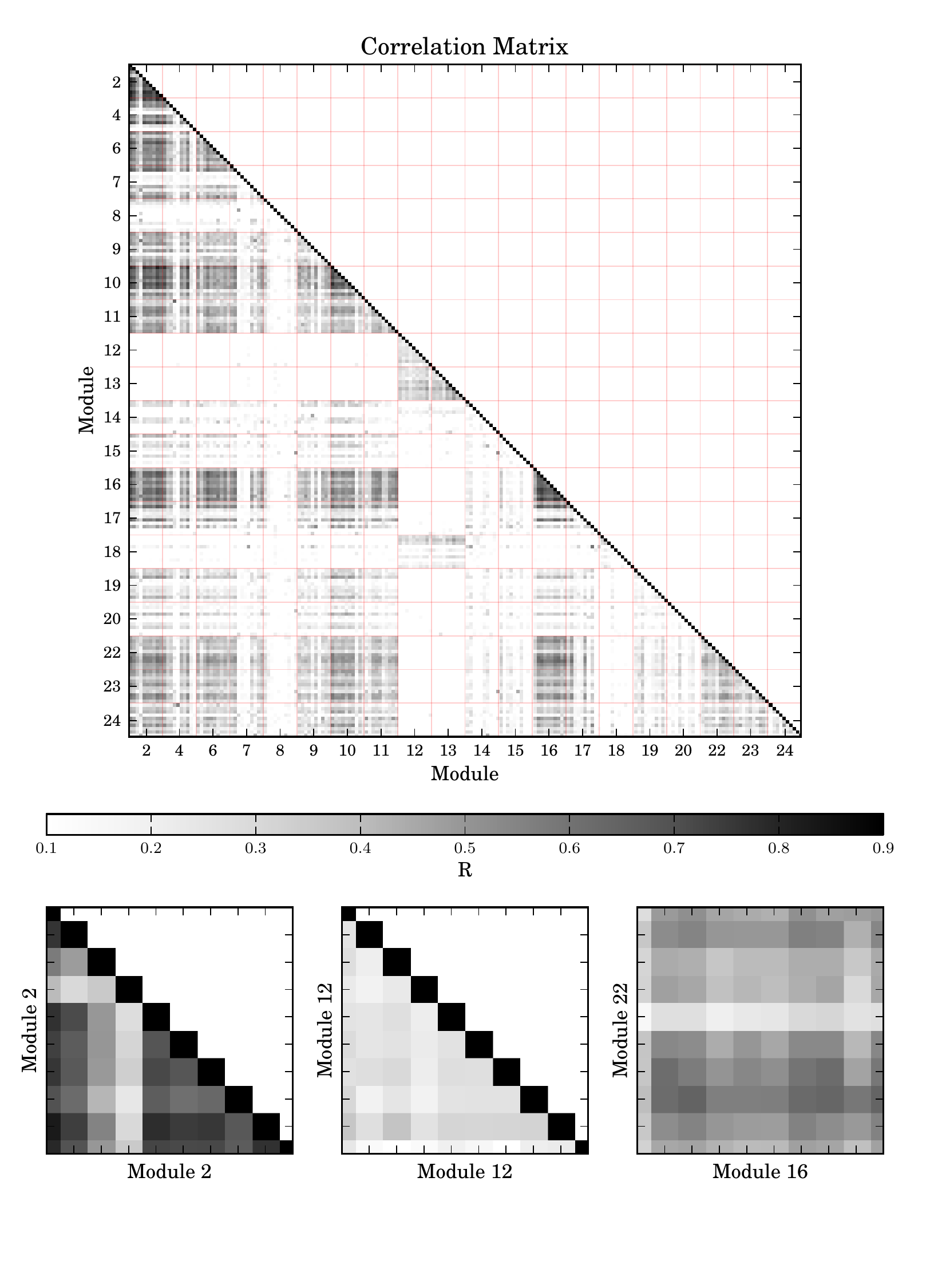}
\end{center}
\caption[Correlation across \Kepler light curves]{Top: Correlation matrix constructed from 200 Q6 light curves.
  The correlation (R-value) between two stars is represented by the
  gray scale, which ranges from 0.1 to 0.9.  The diagonal elements
  have R = 1.  The stars are ordered according to module and the red
  lines delineate one module from another.  We enlarge several 10x10
  regions in the lower panels.  Stars in some modules (such as module
  2) are highly correlated, while other modules (such as module 12)
  show little correlation.  The module 22 - module 16 correlation
  matrix is an example of significant inter-module correlation.  We
  observed the same patterns in a correlation matrix constructed from
  $\sim 1200$ stars, but we show the 200-star correlation matrix so
  that individual elements are discernible as pixels.  }
\label{fig:corrmat}
\end{figure}

Correlated noise with timescales between 1 and 10 hours can mimic
planetary transits and requires careful treatment.  To illustrate the
transit-scale correlations among a large sample of stars, we show a
correlation matrix constructed from 200 Q6 light curves in
Figure~\ref{fig:corrmat}.  The \Kepler photometer is an array of 42
CCDs arranged in 21 modules~\citep{KeplerArchiveManual}.  We organized
the rows and columns of the correlation matrix by module.  We
constructed the correlation matrix using the following steps:
\begin{enumerate}
\item
We randomly selected 10 light curves from each of the 20 total
modules\footnote{Module 3 failed during Q4~\citep{Christiansen:2011}.}
from stars with the following properties: $12.5 < $ \Kmag $ < 13.5$
and CDPP12 $< 40$~ppm.
\item
To highlight transit-scale correlations, we subtracted a best fit
spline from the photometry.  The knots of the spline are fixed at
10-day intervals so that we remove trends $\gtrsim$ 10~days.
\item
We normalized each light curve so that its median absolution deviation
(MAD) is unity.
\item
We evaluated the pairwise correlation (Pearson-R) between all 200
stars.
\end{enumerate}
The correlation matrix shows that stars in some modules (e.g. module
2) correlate strongly with other stars in the same module.  However,
other modules (e.g. module 12) shows little inter-module correlation.
Finally, the large off-diagonal correlations show that stars in some
modules correlate strongly with stars in different modules.

\section{Identification of Photometric Modes}
We have shown that there is significant high-frequency ($\lesssim$
10~days) systematic noise in \Kepler photometry.  In order to recover
the smallest planets, this noise must be carefully characterized and
removed.  We isolate systematic noise by finding common trends in a
large ensemble of stars.  This is an extension of differential
photometry, widely used by ground-based transit surveys to calibrate
out the time-variable effects of the earth's atmosphere.  We find
these trends using Principle Component Analysis (PCA).  This is
similar to the Sys-Rem, TFA, and PDC algorithms
\citep{Tamuz:2005,Kovacs:2005,Twicken10}, but our implementation is
different.  We briefly review PCA in the context of cotrending a large
ensemble of light curves.

\subsection{PCA on Ensemble Photometry}
Consider an ensemble of N light curves each with M photometric
measurements.  We can think of the ensemble as a collection of N
vectors in an M-dimensional space.  Each light curve $\df$ can be
written as a linear combination of M basis vectors that span the
space,
\begin{eqnarray}
\label{eqn:BV}
\df_{1} & = & a_{1,1} V_1 + \hdots  +  a_{1,M} V_M \notag\\
        &\vdots&\\
\df_{N} & = & a_{N,1} V_1 + \hdots  +  a_{N,M} V_M \notag
\end{eqnarray}
where each of the $V_{j}$ basis vectors is the same length as the
original photometric time series.  Equation~\ref{eqn:BV} can be
written more compactly as
\[
\mtx{D} = \mtx{A} \mtx{V}
\]
where
\[
\mtx{D} =  
\begin{pmatrix}
  \df_{1} \\
  \vdots  \\
  \df_{N} \\
\end{pmatrix},
\mtx{A} = 
\begin{pmatrix}
  a_{1,1} & \hdots & a_{1,M} \\
  \vdots  & \ddots & \vdots  \\
  a_{N,1} & \hdots & a_{N,M} \\
\end{pmatrix},
\mtx{V} = 
\begin{pmatrix}
  V_{1} \\
  \vdots       \\
  V_{M} \\
\end{pmatrix}
\]
Singular Value Decomposition (SVD) simultaneously solves for the basis
vectors $\mtx{V}$ and the coefficient matrix $\mtx{A}$ because it
decomposes any matrix $\mtx{D}$ into
\[
\mtx{D} = \mtx{USV}^{\mathsf{T}}.
\]
$\mtx{V}$ is an M x M matrix where the columns are the eigenvectors of
$\mtx{D}^{\mathsf{T}}\mtx{D}$ or ``principle components,'' and the
diagonal elements of S are the corresponding eigenvalues.  The
eigenvalues $\{s_{1,1}, \dots, s_{M,M}\}$ describe the extent to which
each of the principle components capture variability in the ensemble
and are ordered from high to low.  The columns of $\mtx{U}$ are the
eigenvectors of $\mtx{D}\mtx{D}^{\mathsf{T}}$.  Both $\mtx{U}$ and
$\mtx{V}$ are unitary matrices, i.e. $\mtx{U}\mtx{U}^{\mathsf{T}} =
\mtx{I}$ and $\mtx{V}\mtx{V}^{\mathsf{T}} = \mtx{I}$.

As we saw in \S~\ref{sec:InstNoise}, stars show common photometric
trends due to changes in the state of the \Kepler spacecraft.  The
most significant principle components will correspond to these common
trends.  If we identify the first $N_{Mode}$ principle components as
instrumental noise modes, we can remove them via
\begin{equation}
\df_{i,cal} = \df_{i} - \sum_{j=1}^{N_{Mode}} a_{i,j} V_{j}
\label{eqn:fit}
\end{equation}
where $\df_{cal}$ is an ensemble-calibrated light curve.  However
since the collection of $\{V_{i}, \hdots V_{M}\}$ spans the space, the
higher principle components describe astrophysical variability, shot
noise, and exoplanet transits.  We must be careful not to remove too
many components because we would be removing the signals of interest.

\subsection{PCA implementation}
\begin{figure}
\begin{center}
\includegraphics{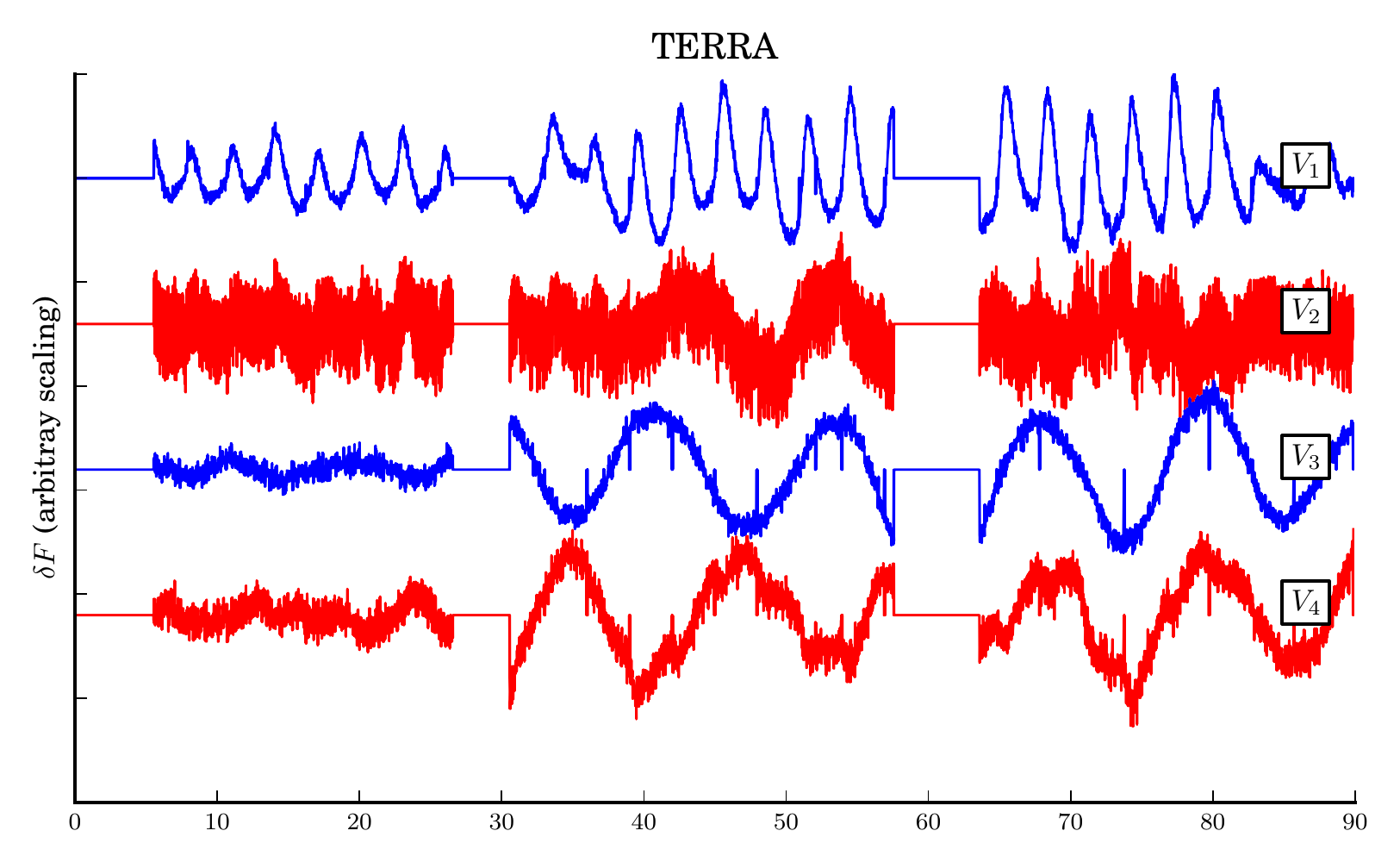}
\end{center}
\caption[TERRA principle components]{ Top: The first four TERRA principle components in our
  1000-light curve ensemble plotted in order of significance.  $V_{1}$
  has a 3-day periodicity and is due to changes in the thermal state
  of the spacecraft caused by a 3-day momentum management cycle.
  $V_{2}$ has a high frequency component (P = 1.68~hours) that could
  be due to a 20~minute thermal cycle from an onboard heater aliased
  with the 29.4~minute observing cadence alias of a 20-minute thermal
  cycle driven by an onboard heater.  }
\label{fig:CompCBV}
\end{figure}

We construct a large reference ensemble of light curves $\{\df_{1},
\dots, \df_{N}\}$ of 1000 stars ($12.5 < $ \Kmag $ < 13.5$, CDPP12 $<
40$~ppm) drawn randomly from the entire FOV.  Before performing SVD,
we remove thermal settling events and trends $\gtrsim$ 10~days as
described in \S~\ref{sec:InstNoise}.  Since SVD finds the eigenvectors
of $\mtx{D}^{\mathsf{T}}\mtx{D}$ it is susceptible to outliers as is
any least squares estimator.  We perform a robust SVD that relies on
iterative outlier rejection following these steps:
\begin{enumerate}
\item
Find principle components and weights for light curve ensemble.
\item
The $i^{\text{th}}$ light curve is considered an outlier if any of the
mode weights ($a_{i,1},\hdots,a_{i,4}$) differ significantly from the
typical mode weight in the ensemble.  We consider $a_{i,j}$ to be
significantly different from the ensemble if
\[
\frac{|a_{i,j} - \text{med}(a_{j})|}{\text{MAD}(a_{j})} > 10
\]
where med$(a_{j})$ and MAD$(a_{j})$ are the median value and the
median absolute deviation of all the $a_{j}$ mode weights.
\item
Remove outlier light curves from the ensemble.
\item
Repeat until no outliers remain.
\end{enumerate}
For our 1000-star sample we identified and removed 51 stars from our
ensemble.  These stars tended to have high amplitude intrinsic
astrophysical variability, i.e. due to spots and flares.  We plot the
four most significant TERRA principle components in
Figure~\ref{fig:CompCBV} and offer some physical interpretations of
the mechanisms behind these modes in the following section.

\subsection{Interpretation of Photometric Modes}
\label{sec:interpretation}
In this section, we associate the variability captured in the
principle components to changes in the state of the \Kepler spacecraft
that couple to photometry.  The three-day cycle isolated in our first
principle component is due to a well-known, three-day momentum
management cycle on the spacecraft \citep{Christiansen:2011}.  To keep
a fixed position angle, \Kepler must counteract external torques by
spinning up reaction wheels.  These reaction wheels have frictional
losses which leak a small amount of heat into the spacecraft, which
changes the PSF width and shape of the stars.

\begin{figure}
\begin{center}
\includegraphics{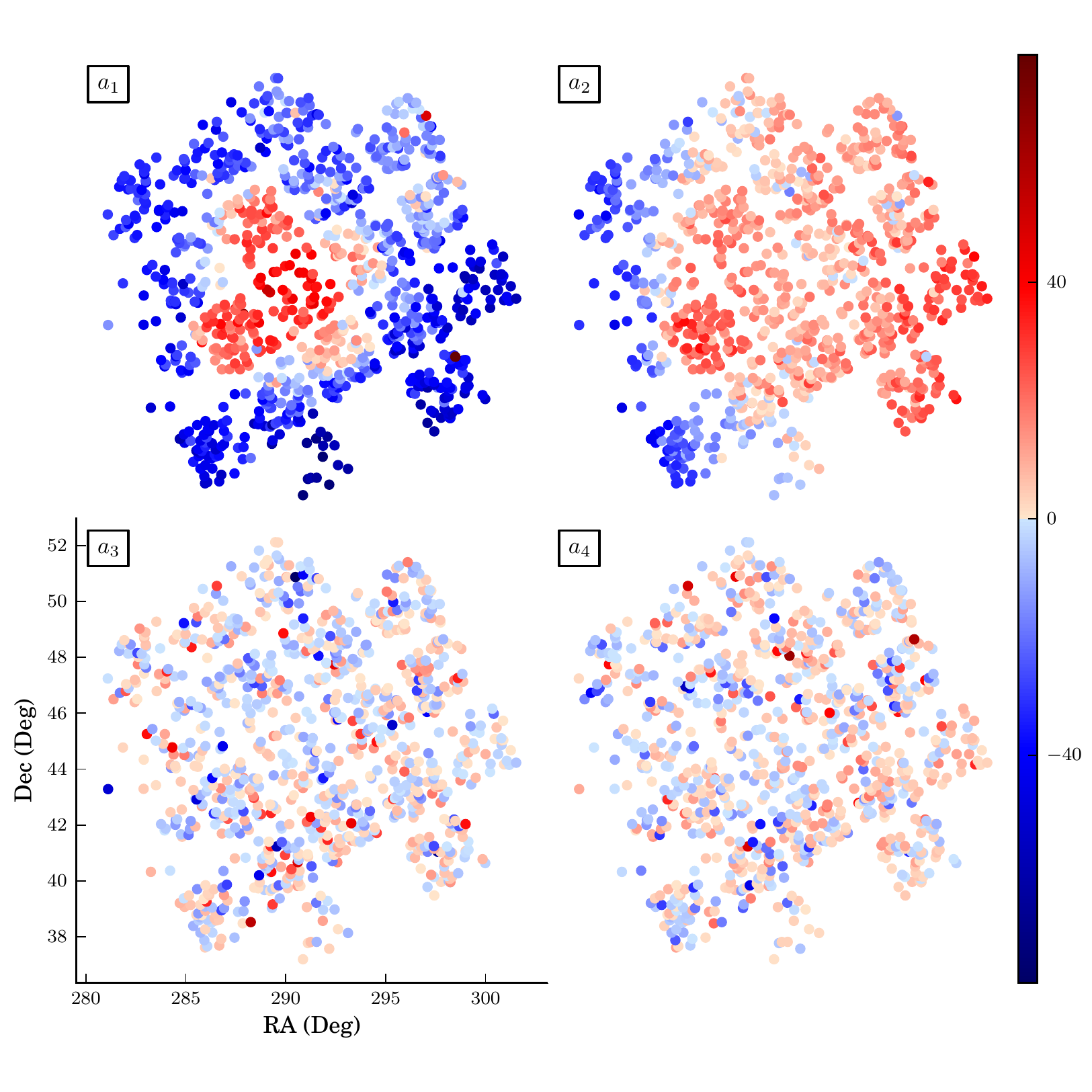}
\end{center}
\caption[Light curve systematics across the \Kepler field of view]{The RA and Dec positions of our 1000-star ensemble.  The
  points are color-coded by $a_{i}$, the weights for mode $V_{i}$.
  Negative values are shown in blue and positive values are shown in
  red.  The fact that the sign and magnitude of $a_{1}$ depends on
  distance from the center of the FOV supports the idea that the
  variability captured by $V_{1}$ is due to PSF breathing of the
  telescope which is driven by the three-day momentum management
  cycle.  The gradient in $a_{2}$ could be due to the thermal coupling
  of an onboard heater to the optics in a tip/tilt sense.  Mode
  weights $a_{3}$ and $a_{4}$ show no spatial correlation and do not
  seem to depend on changes in the PSF width.  }
\label{fig:RADecCoeff}
\end{figure}

We can gain a more detailed understanding of this effect, by examining
how the mode weights for each reference star corresponding to $V_{1}$,
i.e. $\{a_{1,1}, \dots, a_{N,1} \}$, vary across the FOV.  We display
the RA and Dec positions of our 1000-star sample in
Figure~\ref{fig:RADecCoeff} and color-code the points with the value
of $a_{i}$.  The $a_{1}$ and $a_{2}$ mode weights show remarkable
spatial correlation across the FOV.  That $a_{1}$ is positive in the
center of the FOV and negative at the edges of the FOV means the
systematic photometric errors in these two regions respond to the
momentum cycle in an anticorrelated sense.  The telescope is focused
such that the PSF is sharpest at intermediate distances from the
center of the FOV.  Since stars in the center and on the extreme edges
have the blurriest PSFs \citep{VanCleve:2009}, they respond most
strongly to the momentum cycle.

The mechanism behind the variability seen in $V_{2}$ is less clear.
$V_{2}$ includes a high frequency component with a period of 1.68
hours.  The \Kepler team has also noticed this periodicity in the
pixel scale (Douglas Caldwell, private communication, 2012).  A
possible explanation is thermal coupling of the telescope optics to a
heater that turns off and on with a $\sim$20~minute period. The
1.68~hour variability would be an alias of this higher frequency with
the observing cadence of 29.4~minutes.  The gradient in $a_{2}$ across
the FOV suggests the heater is coupled to the telescope optics in a
tip/tilt rather than piston sense.

The higher-order components $a_{3}$ and $a_{4}$ do not show
significant spatial correlation, which suggests that $V_{3}$ and
$V_{4}$ are not due to changes in the local PSF. Since $V_{3}$ and
$V_{4}$ have a $\sim$10-day timescale, they could be the high
frequency component of the differential velocity aberration trend that
was not removed by our 10-day spline.

\section{Calibrated Photometry}
\subsection{Removal of Modes}
\begin{figure}
\begin{center}
\plotone{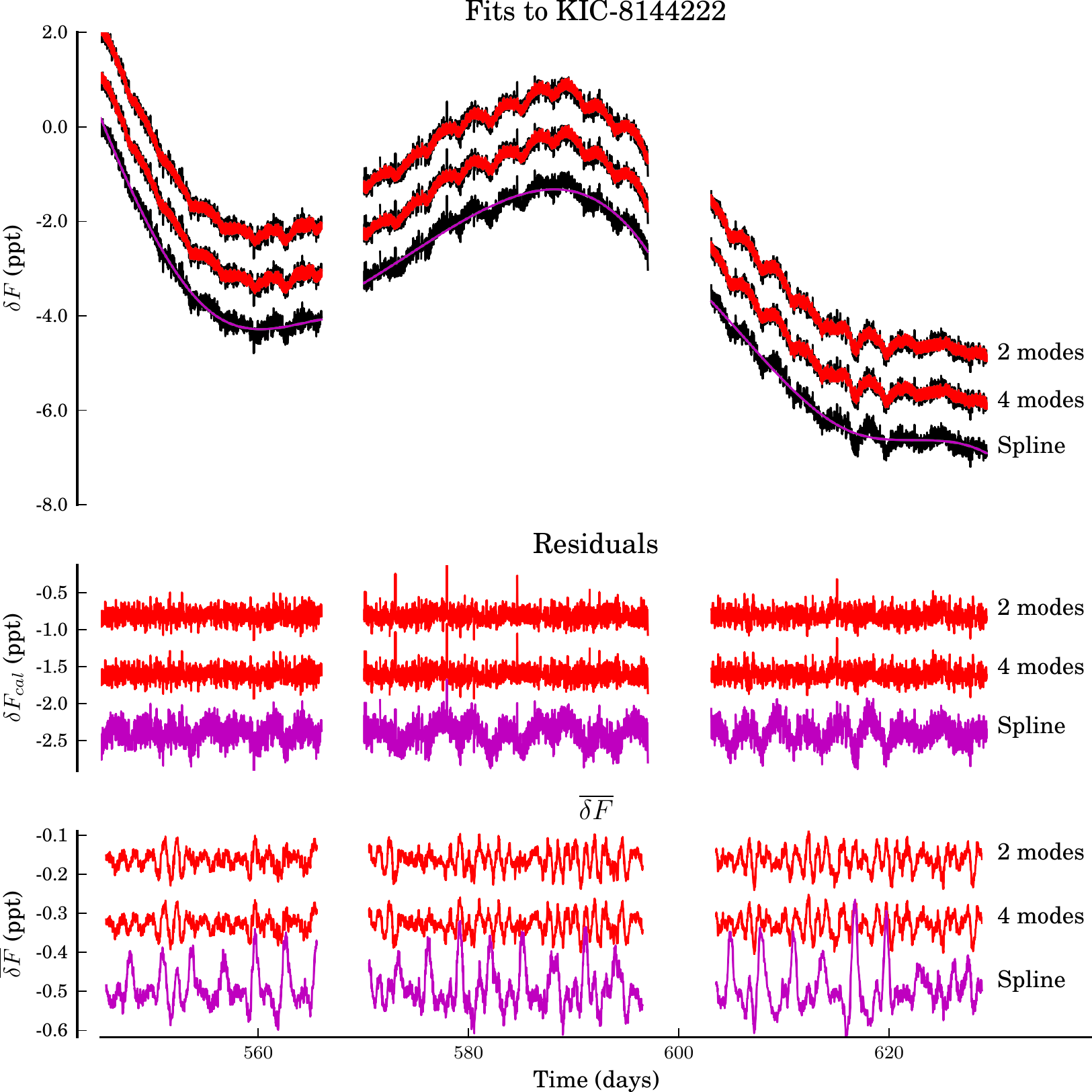}
\end{center}
\caption[TERRA calibrated photometry]{ Least squares fits using TERRA principle components to
  KIC-8144222 Q6 photometry.  The bottom panel shows 12-hour $\mdf$
  where smaller scatter implies greater sensitivity to small transits.
  We show the spline fit (magenta) as a baseline since it incorporates
  no ensemble-based cotrending information.  The $\mdf$ using spline
  detrending shows large spikes at the momentum cycle cusps, which are
  suppressed in the TERRA cotrending.  Using our robust modes, we are
  able to produce a clean, calibrated light curve using only two
  modes.  Decreased model complexity helps guard against overfitting.
}
\label{fig:fits}
\end{figure}

After determining which of the $N_{Mode}$ principle components
correspond to noise modes, we can remove them according to
Equation~\ref{eqn:fit}.  In Figure~\ref{fig:fits}, we show fits to
KIC-8144222 Q6 photometry using different combinations of TERRA
principle components.  We achieve uniform residuals using only 2 of
our modes as we show quantitatively below.  The simplicity of our
model buys some insurance against overfitting.

\subsection{Performance}
For each of the residuals in Figure~\ref{fig:fits}, we computed the
mean depth $\mdf(t_{i})$ of a putative 12-hour transit centered at
$t_{i}$ for every cadence in Q6.  The distribution of $\mdf$ due to
noise determines the minimum transit depth that can be detected by a
transit search algorithm.  $\mdf$ is computed by
\[
\mdf(t_{i}) = [\df * g](t_{i})
\]
where `$*$' denotes convolution and $g$ is the following kernel
\[
g(t_{i}) = \frac{1}{N_{T}}
	\left[
		\underbrace{\hlf, \dots, \hlf}_{\text{length = }N_{T} },
		\underbrace{-1, \dots, -1}_{\text{length = }N_{T}}, 
		\underbrace{\hlf, \dots, \hlf}_{\text{length = }N_{T} }
	\right].
\]
where $N_{T}$ is 24.  For each of the cotrending schemes, we computed
the following statistics describing the distribution of $\mdf$:
standard deviation ($\sigma$), 90 percentile (90 \%), and 99
percentile (99 \%).  The standard deviation is roughly equivalent to
CDPP12.  Since transit search algorithms key off on peaks in $\mdf$,
the percentile statistics are more appropriate figures of merit.  We
list these statistics for KIC-8144222 in Table~\ref{tab:fits}.
Ensemble-calibrated photometry produced tighter distributions in
$\mdf$ than the spline baseline.

\begin{deluxetable}{l c c c }
\tablewidth{0pc}
\tablecaption{Comparison of fits to KIC-8144222 photometry.}
\tablehead{
\colhead{Cotrending}&
\colhead{$\sigma$} &
\colhead{90 \%} & 
\colhead{99 \%} \\
}
\startdata
2 PMs  & 24  & 28 & 53 \\
4 PMs  & 24  & 28 & 53 \\
Spline & 53  & 66 & 146 \\
\enddata
\tablecomments{
Standard deviation, 90 percentile, and 99 percentile (in ppm) of the
$\mdf$ distributions for KIC-8144222 using different cotrending
schemes.  The spline fit is included as a baseline since it
incorporates no ensemble-based cotrending information.  In computing
$\mdf$, we have assumed a 12-hour transit duration.  All cotrending
approaches yield tighter $\mdf$ distributions than the spline
baseline.  }
\label{tab:fits}
\end{deluxetable}

\subsection{Comparison to PDC}
In this section, we offer some simple comparisons between TERRA and
the PDC implementation of~\cite{Twicken10}.  This paper represents
our efforts to improve upon that algorithm.  The \Kepler PDC pipeline
has evolved beyond that presented in \cite{Twicken10} culminating
with PDC-MAP \citep{Stumpe12,Smith12}.  We feel that the
\cite{Twicken10} algorithm is an important touchstone for
comparison given that the most recent release of planets
\citep{Batalha12} was based on photometry that was largely processed
with the \cite{Twicken10} algorithm.

We assess cotrending performance in the context of transit
detectability.  We note that PDC outputs are not directly used in
transit detection.  PDC light curves are subject to additional
detrending (mostly of low frequency content) before the transiting
planet search is run \citep{Tenenbaum:2010}.

\begin{figure}
\begin{center}
\plotone{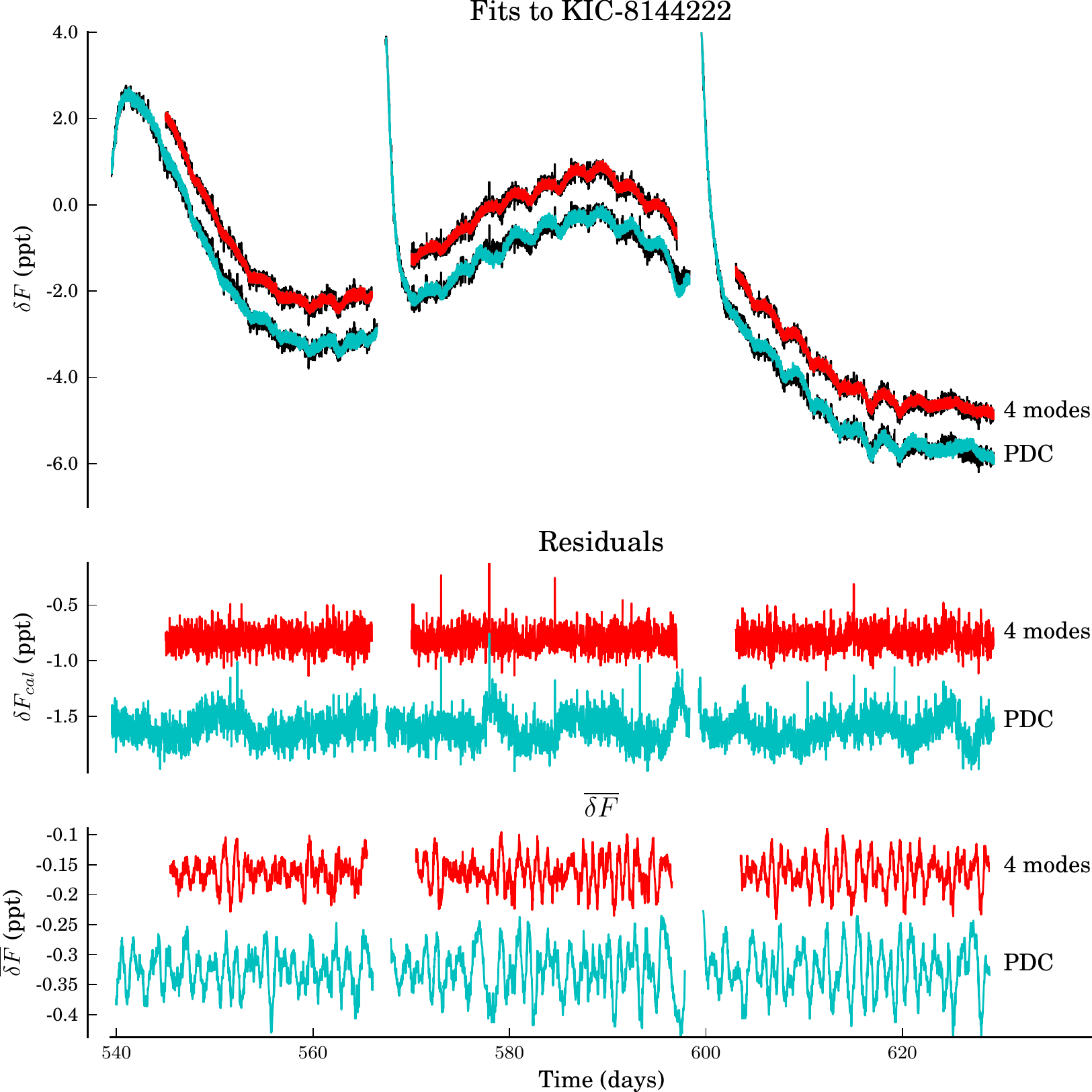}
\end{center}
\caption[PDC calibrated photometry]{ Same as Figure~\ref{fig:fits} except we compare fits using
  the 4 TERRA modes, with the PDC processed photometry.  The bottom
  panel shows 12-hour $\mdf$ where smaller scatter implies greater
  sensitivity to small transits.  The RMS scatter in the 12-hour
  $\mdf$ distribution is 24~ppm for TERRA processed photometry and
  36~ppm for PDC processed photometry.  }
\label{fig:fits2}
\end{figure}

In Figure~\ref{fig:fits2}, we show fits to the KIC-8144222 photometry
using 4 TERRA modes and the PDC algorithm.  While PDC flattens
photometry collected during the thermal transients, it injects high
frequency noise into regions that are featureless in the
TERRA-calibrated photometry.  For KIC-8144222, the RMS scatter in the
12-hour $\mdf$ distribution is 24~ppm for TERRA processed photometry
and 36~ppm for PDC processed photometry.

\begin{figure}
\begin{center}
\includegraphics{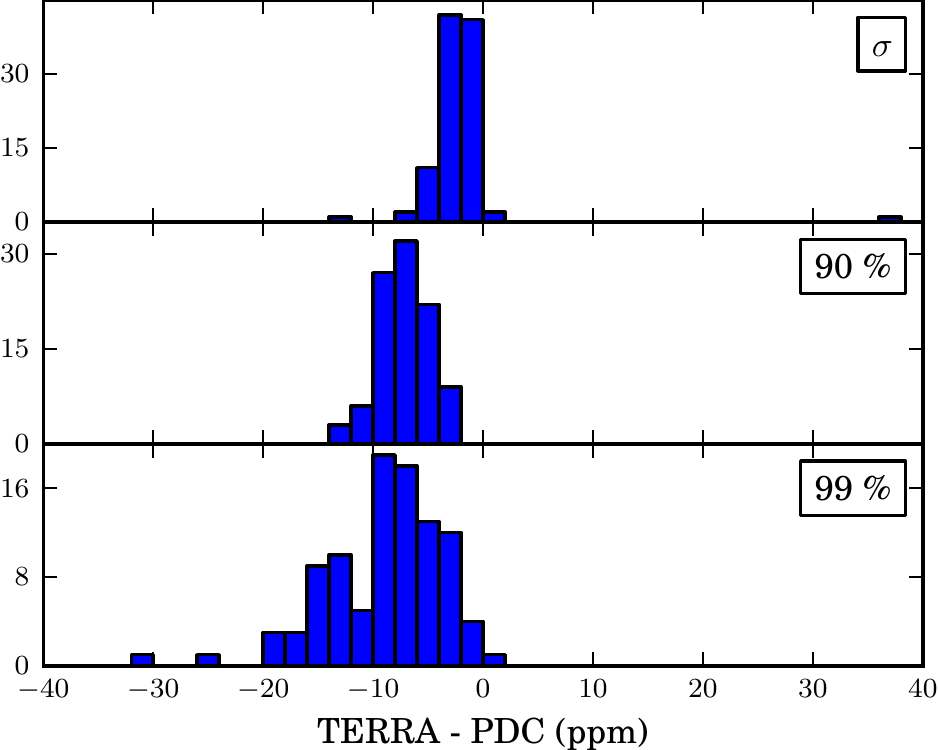}
\end{center}
\caption[Comparison of TERRA and PDC]{ We computed the standard deviation, 90 percentile, and 99
  percentile (in ppm) of 12-hour $\mdf$ for 100 light curves using
  TERRA and PDC cotrending.  The histograms show the difference of the
  TERRA and PDC statistics.  Negative values mean a tighter $\mdf$
  distribution using our cotrending and hence a lower noise floor in a
  transit search.}
\label{fig:stat100}
\end{figure}

Using 4 TERRA modes, we cotrend 100 stars selected at random from our
1000-star reference ensemble.  We then compute 3, 6, and 12-hour
$\mdf$ from TERRA and PDC calibrated light curves.  We then calculate
the difference between the $\sigma$, 90\%, and 99\% statistics for
TERRA and PDC cotrending.  We show the distribution of these
differences for the 12-hour $\mdf$ in Figure~\ref{fig:stat100}.  The
median improvement in $\sigma$, 90\%, and 99\% using TERRA cotrending
is 2.8, 6.6, and 8.7~ppm.  We tabulate the median values of the
$\sigma$, 90\%, and 99\% statistics in Table~\ref{tab:stat100}.

We believe that these comparisons are representative of the stars from
which we constructed our reference ensemble ($12.5 < $ \Kmag $ < 13.5$
and CDPP12 $< 40$~ppm).  These bright, low-noise stars are the most
amenable to exoearth detection. Our comparisons do not pertain to
stars with different brightness or noise level.

\begin{deluxetable}{l c c c c c c}
\tablewidth{0pc}
\tablecaption{Comparison of TERRA and PDC cotrending performance for 100 stars.}
\tablehead{
\colhead{Transit Width}&
\colhead{$\sigma$}&
\colhead{$\sigma$}&
\colhead{90\%}&
\colhead{90\% }&
\colhead{99\%}&
\colhead{99\% }\\
\colhead{(hours)}&
\colhead{TERRA}&
\colhead{PDC}&
\colhead{TERRA}&
\colhead{PDC}&
\colhead{TERRA}&
\colhead{PDC}\\
}
\startdata
3 & 58 & 60 & 68 & 76 & 129 & 141 \\
6 & 43 & 45 & 50 & 57 & 97 & 105 \\
12 & 33 & 37 & 39 & 47 & 76 & 88 \\
\enddata
\tablecomments{
A comparison of the $\mdf$ distributions using TERRA and PDC
cotrending of 100 stars drawn randomly from our 1000-star sample.  We
have assumed a range of transit widths.  We show the median values of
the standard deviation, 90 percentile, and 99 percentile (in ppm) of
the $\mdf$ distributions.  For these 100 stars, TERRA yields tighter
distributions of $\mdf$.  The improvement ranges from 8 to 12~ppm in
the 99 \% statistic.  
}
\label{tab:stat100}
\end{deluxetable}

\section{Conclusions}
\label{sec:conclusions}
TERRA is a new technique for using ensemble photometry to
self-calibrate instrumental systematics in \Kepler light curves.  We
construct a simple noise model by running a high-pass filter and
removing thermal settling events before computing principle
components.  For a typical $12.5 < $ \Kmag $ < 13.5$ and CDPP12 $<
40$~ppm star, TERRA produces ensemble-calibrated photometry with
33~ppm RMS scatter in 12~hour bins.  With this noise level, a 100~ppm
transit from an exoearth will be detected at $\sim 3\sigma$ per
transit.

A potential drawback of removing thermal settling events is discarding
photometry that contains a transit.  Thermal settling events amounted
to 14\% of the valid cadences in Q1-Q8 photometry.  Since signal to
noise grows as the square root of the number of transits, removing
14\% of the photometry results in a 7\% reduction in the signal to
noise of a given transit.  The completeness of the survey may decrease
slightly, since some borderline transits will remain below
threshold. One further complication arises due to the fact that gaps
due to thermal settling occur on regular (monthly) intervals.  Given
the right epoch, a planet with a period that is a multiple of $\sim
30$ days may repeatedly transit during a gap.  Thus, removing gaps
amounts to a more significant reduction of survey completeness for
specific regions in period-epoch space.

Ensemble-based cotrending is most effective when the timescales in the
ensemble are matched to the signal of interest. We are skeptical that
a ``one size fits all'' approach exists and we encourage those who
wish to get the most out of \Kepler data to tune their cotrending to
the timescale of their signals of interest.

%% file: terra50d/terra50d.tex
\chapter{A Plateau in the Planet Population Below Twice the Size of Earth}
\label{c.terra50d}

\noindent A version of this chapter was previously published in the {\em Astrophysical Journal} \\
\noindent (Erik~A.~Petigura, Geoffrey~W.~Marcy, \& Andrew~W.~Howard, 2013, ApJ 770, 69).\\

We carry out an independent search of \Kepler photometry for small transiting planets with sizes 0.5--8.0 times that of Earth and orbital periods between 5 and 
50 days, with the goal of measuring the fraction of stars harboring such planets. We use a new transit search algorithm, \TERRA, optimized to detect small planets around photometrically quiet stars. We restrict our stellar sample to include the 12,000 stars having the lowest photometric noise in the \Kepler survey, thereby maximizing the detectability of Earth-size planets. We report 129 planet candidates having radii less than 6 \Re found in 3 years of \Kepler photometry (quarters 1--12). Forty-seven of these candidates are not in \cite{Batalha12}, which only analyzed photometry from quarters 1--6. We gather Keck HIRES spectra for the majority of these targets leading to precise stellar radii and hence precise planet radii. We make a detailed measurement of the completeness of our planet search. We inject synthetic dimmings from mock transiting planets into the actual \Kepler photometry. We then analyze that injected photometry with our \TERRA pipeline to assess our detection completeness for planets of different sizes and orbital periods. We compute the occurrence of planets as a function of planet radius and period, correcting for the detection completeness as well as the geometric probability of transit, $\Rstar/a$. The resulting distribution of planet sizes exhibits a power law rise in occurrence from 5.7 \Re down to 2 \Re, as found in \cite{Howard12}. That rise clearly ends at 2 \Re. The occurrence of planets is consistent with constant from 2 \Re toward 1 \Re. This unexpected plateau in planet occurrence at 2 \Re suggests distinct planet formation processes for planets above and below 2 \Re. We find that $15.1^{+1.8}_{-2.7}$\% of solar type stars---roughly one in six---has a 1--2~\Re planet with \Per~=~5--50~days.

\section{Introduction}

The \Kepler Mission has discovered an extraordinary sample of more
than 2300 planets with radii ranging from larger than Jupiter to
smaller than Earth \citep{Borucki11,Batalha12}. Cleanly measuring and
debiasing this distribution will be one of {\em Kepler's} great
legacies. \cite{Howard12}, H12 hereafter, took a key step, showing
that the planet radius distribution increases substantially with
decreasing planet size down to at least 2 \Re. While the distribution
of planets of all periods and radii contains a wealth of information,
we choose to focus on the smallest planets.  Currently, only \Kepler
is able to make quantitative statements about the occurrence of
planets down to 1 \Re.

The occurrence distributions in H12 were based on planet
candidates\footnote{The term ``planet candidate'' is used because a
  handful of astrophysical phenomena can mimic a transiting
  planet. However, \cite{Morton11}, \cite{Morton12}, and
  \cite{Fressin13} have shown that the false positive rate among
  \Kepler candidates is low, generally between 5\% and 15\%.}
detected in the first four months of \Kepler photometry
\citep{Borucki11}. These planet candidates were detected by a
sophisticated pipeline developed by the \Kepler team Science
Operations Center \citep{Twicken10,Jenkins10}.\footnote{Since H12,
  \cite{Batalha12} added many candidates, bringing the number of
  public KOIs to >~2300. In addition, the \Kepler team planet search
  pipeline has continued to evolve \citep{Smith12,Stumpe12}.}
Understanding pipeline completeness, the fraction of planets missed by
the pipeline as a function of size and period, is a key component to
measuring planet occurrence. Pipeline completeness can be assessed by
injecting mock dimmings into photometry and measuring the rate at
which injected signals are found. The completeness of the official
\Kepler pipeline has yet to be measured in this manner. This was the
key reason why H12 were cautious interpreting planet occurrence under
2 \Re.

In this work, we focus on determining the occurrence of small
planets. To maximize our sensitivity to small planets, we restrict our
stellar sample to include only the 12,000 stars having the lowest
photometric noise in the \Kepler survey. We comb through quarters
1--12 (Q1--Q12) --- 3 years of \Kepler photometry --- with a new
algorithm, \TERRA, optimized to detect low signal-to-noise transit
events. We determine \TERRA's sensitivity to planets of different
periods and radii by injecting synthetic transits into \Kepler
photometry and measuring the recovery rate as a function of planet
period and radius.

We describe our selection of 12,000 low-noise targets in
Section~\ref{sec:Sample}. We comb their photometry for exoplanet
transits with \TERRA, introduced in Section~\ref{sec:TERRA}. We report
candidates found with \TERRA (Section~\ref{sec:TERRAplanetYield}),
which we combine with our measurement of pipeline completeness
(Section~\ref{sec:MC}) to produce debiased measurements of planet
occurrence (Section~\ref{sec:occurrenceme}). We offer some comparisons
between \TERRA planet candidates and those from \cite{Batalha12} in
Section~\ref{sec:compare} as well as occurrence measured using both
catalogs in Section~\ref{sec:OccurCOMB}. We offer some interpretations
of the constant occurrence rate for planets smaller than 2 \Re in
Section~\ref{sec:Discussion}.

\section{The Best12k Stellar Sample}
\label{sec:Sample}

We restrict our study to the best 12,000 solar type stars from the
perspective of detecting transits by Earth-size planets, hereafter,
the ``Best12k'' sample. For the smallest planets, uncertainty in the
occurrence distribution stems largely from pipeline incompleteness due
to the low signal-to-noise ratio (SNR) of an Earth-size transit.

\begin{figure}
\centering
\includegraphics[width=0.8\textwidth]{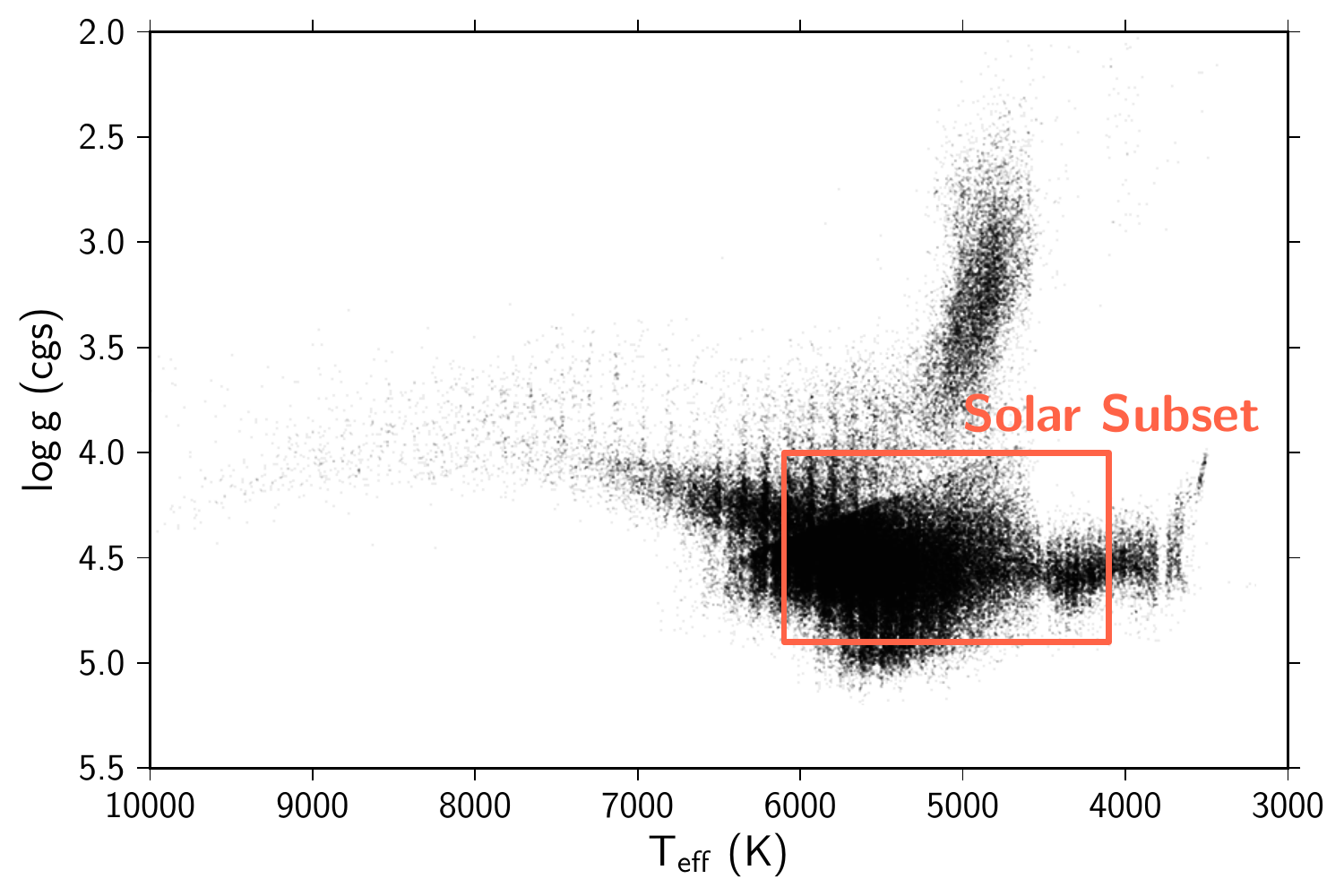}
\caption[Effective temperatures and surface gravities of stellar sample]{\Kepler target stars observed every quarter from Q1--Q9.  The
  rectangle marks the ``solar subset'' of stars with
  \teff~=~4100--6100~K and \logg~=~4.0--4.9 (cgs).}
\label{fig:solarsubset}
\end{figure}

Our initial sample begins with the 102,835 stars that were observed
during every quarter from Q1--Q9.\footnote{We ran \TERRA on Q1-Q12
  photometry, but we selected the Best12k sample before Q10-Q12 were
  available.}  From this sample, following H12, we select 73,757
``solar subset'' stars that are solar-type G and K having \teff =
4100--6100~K and \logg = 4.0--4.9~(cgs). \teff and \logg values are
present in the \Kepler Input Catalog (KIC; \citealp{Brown11}) which is
available online.\footnote{http://archive.stsci.edu/Kepler/kic.html}
Figure~\ref{fig:solarsubset} shows the KIC-based \teff and \logg
values as well as the solar subset. KIC stellar parameters have large
uncertainties: $\sigma(\logg) \sim$ 0.4~dex and $\sigma(\teff) \sim$
200~K \citep{Brown11}. As we will discuss in
Section~\ref{sec:TERRAplanetYield}, we determine stellar parameters
for the majority of \TERRA planet candidates spectroscopically. For
the remaining cases, we use stellar parameters that were determined
photometrically, but incorporated a main sequence prior
\citep{Batalha12}. After refining the stellar parameters, we find that
\NnotSolarSubset of the 129 \TERRA planet candidates fall outside of
the \teff = 4100--6100~K and \logg = 4.0--4.9~(cgs) solar subset.

\begin{figure}[htbp]
\centering
\includegraphics[width=0.8\columnwidth]{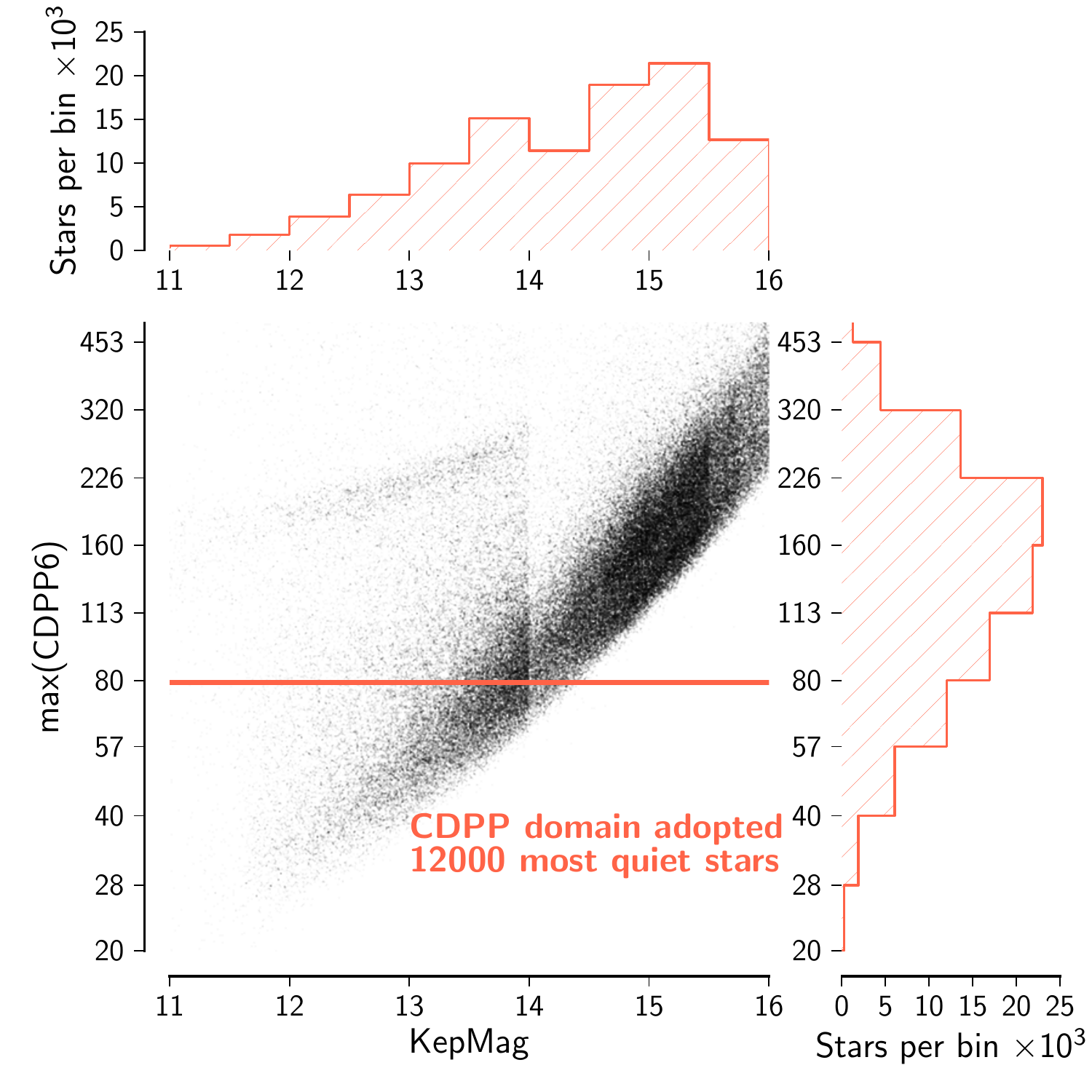}
\caption[Brightness and photometric variability of stellar sample]{Stellar photometric noise level plotted against \Kepler
  magnitude. Noise level is the maximum value of CDPP6 over Q1--Q9. Of
  the 73,757 stars that pass our cuts on \teff and \logg, we select
  the 12,000 most quiet stars. The line shows max(CDPP6) = 79.2~ppm,
  corresponding to the noisiest star in the Best12k sample, well below
  the median value of 143~ppm.}
\label{fig:noise}
\end{figure}

From the 73,757 stars that pass our cuts on \logg and \teff, we choose
the 12,000 lowest noise stars. \Kepler target stars have a wide range
of noise properties, and there are several ways of quantifying
photometric noise. The \Kepler team computes quantities called CDPP3,
CDPP6, and CDPP12, which are measures of the photometric scatter in 3,
6, and 12 hour bins \citep{Jenkins10}. Since CDPP varies by quarter,
we adopt the maximum 6-hour CDPP over Q1--Q9 as our nominal noise
metric. We use the maximum noise level (as opposed to median or mean)
because a single quarter of noisy photometry can set a high noise
floor for planet detection. One may circumvent this problem by
removing noisy regions of photometry, which is a planned upgrade to
\TERRA. Figure~\ref{fig:noise} shows the distribution of max(CDPP6)
among the 73,757 stars considered for our sample.

In choosing our sample, we wanted to include stars amenable to the
detection of planets as small as 1 \Re. We picked the 12,000 quietest
stars based on preliminary completeness estimates. The noisiest star
in the Best12k sample has max(CDPP6) of 79.2~ppm. We estimated that
the $\sim$ 100~ppm transit of an Earth-size planet would be detected
at $\SNRC \sim 1.25$.\footnote{\SNRC, the expected SNR using the
  max(CDPP6) metric, is different from the SNR introduced in
  Section~\ref{sec:TERRAgrid}. \SNRC is more similar to the SNR
  computed by the \Kepler team, which adopts \SNRC > 7.1 as their
  detection threshold.}  Given that Q1-Q12 contains roughly 1000 days
of photometry, we expected to detect a 5-day planet at $ \SNRC \sim
1.25 \times \sqrt{1000/5} \sim 18$ (a strong detection) and to detect
a 50-day planet at $\SNRC \sim 1.25 \times \sqrt{1000/50} \sim 5.6$ (a
marginal detection). In our detailed study of completeness, described
in Section~\ref{sec:MC}, we find that \TERRA recovers most planets
down to 1 \Re having \Per~=~5--50~days.

We draw stars from the H12 solar subset for two reasons. First, we may
compare our planet occurrence to that of H12 without the complication
of varying occurrence with different stellar types. We recognize that
subtle differences may exist between the H12 and Best12k stellar
sample. One such difference is that the Best12k is noise-limited,
while the H12 sample is magnitude-limited. H12 included bright stars
with high photometric variability, which are presumably young and/or
active stars. Planet formation efficiency could depend on stellar
age. Planets may be less common around older stars that formed before
the metallicity of the Galaxy was enriched to current levels. This
work assesses planet occurrence for a set of stars that are
systematically selected to be 3-10 Gyr old by virtue of their reduced
magnetic activity.

The second reason for adopting the H12 solar subset is a practical
consideration of our completeness study. As shown in
Section~\ref{sec:MC}, we parameterize pipeline efficiency as a
function of \Per and \Rp. Because M-dwarfs have smaller radii than
G-dwarfs, an Earth-size planet dims an M-dwarf more substantially and
should be easier for \TERRA to detect. Thus, measuring completeness as
a function of \Per, \Rp, {\em and} \Rstar (or perhaps \Per and
\Rp/\Rstar) is appropriate when analyzing stars of significantly
different sizes. Such extensions are beyond the scope of this paper,
and we consider stars with $\Rstar \sim \Rsun$.

\section{Planet Search Pipeline}
\label{sec:TERRA}
Identifying the smallest transiting planets in \Kepler photometry
requires a sophisticated automated pipeline. Our pipeline is called
``\TERRA'' and consists of three major components. First, \TERRA
calibrates photometry in the time domain. Then, \TERRA combs the
calibrated photometry for periodic, box-shaped signals by evaluating
the signal-to-noise ratio (SNR) over a finely-spaced grid in transit
period (\Per), epoch (\ep) and duration (\tdur). Finally, \TERRA fits
promising signals with a \cite{Mandel02} transit model and rejects
signals that are not consistent with an exoplanet transit. We review
the calibration component in Section~\ref{sec:TERRAcal}, but refer the
reader to \cite{Petigura12} for a detailed description. We present,
for the first time, the grid-search and light curve fitting components
in Sections~\ref{sec:TERRAgrid} and \ref{sec:TERRAdv}.

\subsection{Photometric Calibration}
\label{sec:TERRAcal}
We briefly review the major time domain components of \TERRA; for a
more complete description, please refer to \cite{Petigura12}.  We
begin with \Kepler ``simple aperture long cadence photometry,'' which
we downloaded from the Mikulski Archive for Space Telescopes
(MAST). This photometry is the total photoelectrons accumulated within
a predefined target aperture over a 29.4 minute interval
\citep{KeplerArchiveManual}. We remove thermal settling events
manually and cosmic rays using a median filter.  Next, we remove
photometric trends longer than 10 days with a high-pass filter.
Finally, we identify photometric modes shared by a large ensemble of
stars with using a robust principal components analysis. The optimum
linear combination of the four most significant modes is removed from
each light curve individually.

\subsection{Grid-based Transit Search}
\label{sec:TERRAgrid}

We then search for periodic, box-shaped signals in ensemble-calibrated
photometry. Such a search involves evaluating the SNR over a finely
sampled grid in period (\Per), epoch (\ep), and duration (\tdur), i.e.
 \begin{equation}
\text{SNR} = \text{SNR}(\Per,\ep,\tdur).
\end{equation}
Our approach is similar to the widely-used {\tt BLS} algorithm of
\cite{Kovacs02} as well as to the {\tt TPS} component of the \Kepler
pipeline \citep{Jenkins10TPS}. {\tt BLS}, {\tt TPS}, and \TERRA are
all variants of a ``matched filter'' (North 1943). The way in which
such an algorithm searches through \Per, \ep, and \tdur is up to the
programmer. We choose to search first through \tdur (outer loop), then
\Per, and finally, \ep (inner loop).

For computational simplicity, we consider transit durations that are
integer numbers of long cadence measurements. Since we search for
transits with \Per~=~5--50 days, we try $\tdur = [3,5,7,10,14,18]$
long cadence measurements, which span the range of expected transit
durations, 1.5 to 8.8 hours, for G and K dwarf stars.

After choosing \tdur, we compute the mean depth, $\mdf(t_{i})$, of a
putative transit with duration = \tdur centered at $t_{i}$ for each
cadence. $\mdf$ is computed via

\begin{equation}
\mdf(t_{i}) = \sum_{j} F(t_{i-j}) G_{j}
\end{equation}
where $F(t_{i})$ is the median-normalized stellar flux at time $t_{i}$
and $G_{j}$ is the $j$th element of the following kernel
\begin{equation}
\mathbf{G} = \frac{1}{\tdur}
     \left[
          \underbrace{\hlf, \dots, \hlf}_{\tdur},
          \underbrace{-1, \dots, -1}_{\tdur},
          \underbrace{\hlf, \dots, \hlf}_{\tdur}
     \right].
\end{equation}
As an example, if $\tdur = 3$, 

\begin{equation}
\mathbf{G} = \frac{1}{3}
     \left[
          \hlf, \hlf,\hlf ,-1,-1, -1,\hlf, \hlf,\hlf
     \right].
\end{equation}

We search over a finely sampled grid of trial periods from 5--50 days
and epochs ranging from $t_{\text{start}}$ to $t_{\text{start}}+\Per$,
where $t_{\text{start}}$ is the time of the first photometric
observation. For a given (\Per,\ep,\tdur) there are $N_T$ putative
transits with depths $\mdf_{i}$, for $i = 0, 1, \hdots, N_T-1$. For
each (\Per,\ep,\tdur) triple, we compute SNR from
\begin{equation}
\text{SNR} = \frac{\sqrt{N_T}}{\sigma} \text{mean}(\mdf_i),
\label{eqn:SNR}
\end{equation}
where $\sigma$ is a robust estimate (median absolute deviation) of the noise in bins of length \tdur.

\newcommand{\PcadO}{\ensuremath{ P_{\text{cad,0}}}\xspace}

For computational efficiency, we employ the ``Fast Folding Algorithm''
(FFA) of \cite{Staelin69} as implemented in Petigura \& Marcy (2013;
in prep.). Let \PcadO be a trial period that is an integer number of
long cadence measurements, e.g. \PcadO = 1000 implies $\Per =
1000\times 29.4~\text{min}$ = 20.43~days. Let $N_{\text{cad}} = 51413$
be the length of the Q1-Q12 time series measured in long
cadences. Leveraging the FFA, we compute SNR at the following periods:
\begin{equation}
P_{\text{cad},i} = \PcadO + \frac{i}{M-1}; \; i=0, 1, \hdots, M-1
\end{equation}
where $M = N_{\text{cad}} / \PcadO $ rounded up to the nearest power
of two. In our search from 5--50~days, \PcadO ranges from 245--2445,
and we evaluate SNR at $\sim10^5$ different periods. At each
$P_{\text{cad,i}}$ we evaluate SNR for \PcadO different starting
epochs. All told, for each star, we evaluate SNR at $\sim 10^9$
different combinations of \Per, \ep, and \tdur.

\begin{figure*}
\includegraphics[width=\textwidth]{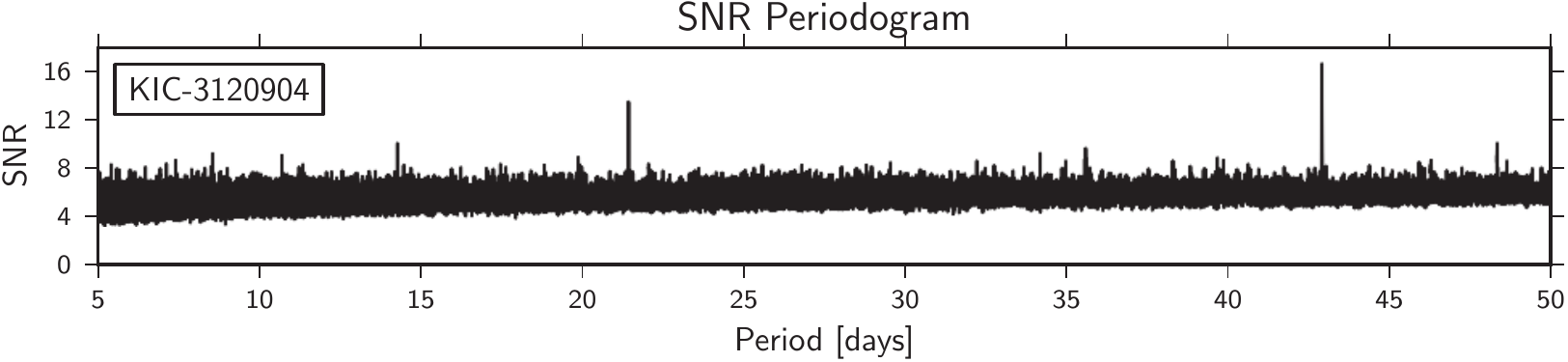}
\caption[SNR periodogram of KIC-3120904 photometry]{SNR periodogram of KIC-3120904 photometry. We evaluate SNR
  over a finely-spaced, three-dimensional grid of \Per, \ep, and
  \tdur. We store the maximum SNR for each trial period, resulting in
  a one-dimensional distribution of SNR. A planet candidate (not in
  \citealt{Batalha12}) produces a SNR peak of 16.6 at
  \Per~=~42.9~days, which rises clearly above the detection floor of
  SNR $\sim$ 8.}
\label{fig:SNR-Periodogram}
\end{figure*}

Due to runtime and memory constraints, we store only one SNR value for
each of the trial periods. \TERRA stores the maximum SNR at that
period for all \tdur and \ep. We refer to this one-dimensional
distribution of SNR as the ``SNR periodogram,'' and we show the
KIC-3120904 SNR periodogram in Figure~\ref{fig:SNR-Periodogram} as an
example. Because we search over many \tdur and \ep at each trial
period, fluctuations often give rise SNR $\sim$ 8 events and set the
detectability floor in the SNR periodogram. For KIC-3120904, a star
not listed in the \cite{Batalha12} planet catalog, we see a SNR peak
of 16.6, which rises clearly above stochastic background.

If the maximum SNR in the SNR periodogram exceeds 12, we pass that
particular (\Per,\ep,\tdur) on to the ``data validation'' (DV) step,
described in the following section, for additional vetting. We chose
12 as our SNR threshold by trial and error. Note that the median
absolute deviation of many samples drawn from a Gaussian distribution
is 0.67 times the standard deviation, i.e. $\sigma_{\rm{MAD}} = 0.67
\sigma_{\rm{STD}}$. Therefore, \TERRA SNR = 12 corresponds roughly to
SNR = 8 in a {\tt BLS} or {\tt TPS} search.

\begin{figure*}[htbp]
\includegraphics[width=\textwidth]{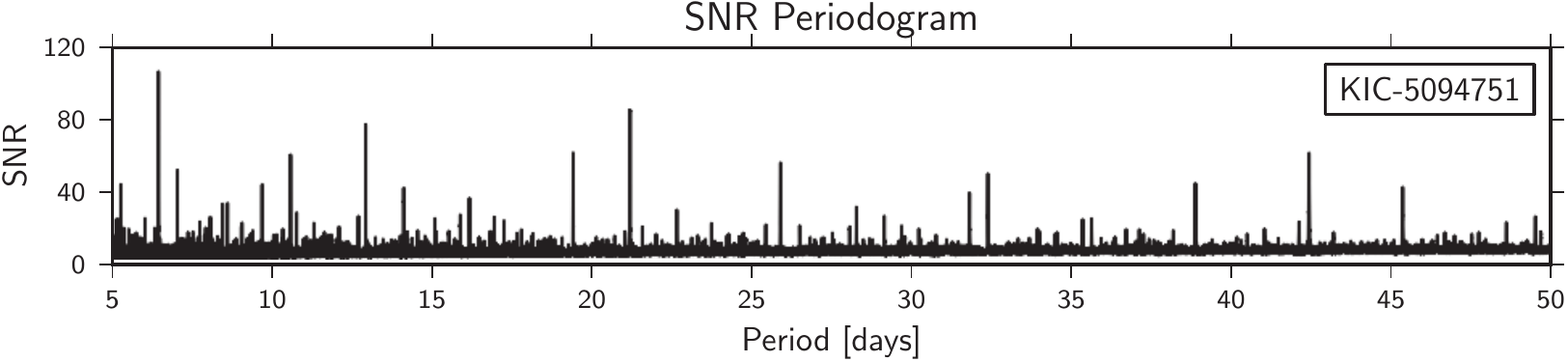}
\caption[SNR periodogram of KIC-5094751 photometry]{SNR periodogram of KIC-5094751 photometry, demonstrating
  \TERRA's insensitivity to lower SNR candidates in multi-candidate
  systems. \cite{Batalha12} lists two planets belonging to
  KIC-5094751, KOI-123.01 and KOI-123.02 with \Per = 6.48 and 21.22
  days, respectively. \TERRA detected KOI-123.01 with a period of 6.48
  days (highest SNR peak). Sub-harmonics belonging to KOI-123.01 are
  visible at $[2, 3, \hdots] \times \Per = [13.0, 19.4,
    \hdots]$~days. A second set of SNR peaks due to KOI-123.02 (\Per =
  21.2 days) is visible at $[0.5, 2, \hdots] \times \Per = [10.6,
    42.4, \hdots]$~days. Had we removed the transit due to KOI-123.01,
  KOI-123.02 would be easily detectible due its high SNR of
  $\sim80$. \TERRA does not yet include multi-candidate logic and is
  thus blind to lower SNR candidates in multi-candidate systems.}
\label{fig:SNRmulti}
\end{figure*}

Since \TERRA only passes the (\Per,\ep,\tdur) triple with the highest
SNR on to DV, \TERRA does not detect additional planets with lower SNR
due to either smaller size or longer orbital period. As an example of
\TERRA's insensitivity to small candidates in multi-candidate systems,
we show the \TERRA SNR periodogram for KIC-5094751 in
Figure~\ref{fig:SNRmulti}. \cite{Batalha12} lists two candidates
belonging to KIC-5094751: KOI-123.01 and KOI-123.02 with P = 6.48 and
21.22 days, respectively. Although the SNR periodogram shows two sets
of peaks coming from two distinct candidates, \TERRA only identifies
the first peak.  Automated identification of multi-candidate systems
is a planned upgrade for \TERRA. Another caveat is that \TERRA assumes
strict periodicity and struggles to detect low SNR transits with
significant transit timing variations, i.e. variations longer than the
transit duration.

\subsection{Data Validation}
\label{sec:TERRAdv}
If the SNR periodogram has a maximum SNR peak > 12, we flag the
corresponding (\Per,\ep,\tdur) for additional vetting. Following the
language of the official \Kepler pipeline, we refer to these triples
as ``threshold crossing events'' (TCEs), since they have high
photometric SNR, but are not necessarily consistent with an exoplanet
transit. \TERRA vets the TCEs in a step called ``data validation,''
again following the nomenclature of the official \Kepler
pipeline. Data validation (DV), as implemented in the official \Kepler
pipeline, is described in \cite{Jenkins10}. We emphasize that \TERRA
DV does not depend on the DV component of the \Kepler team pipeline.

\begin{figure}
\begin{center}
\includegraphics[width=0.8\textwidth]{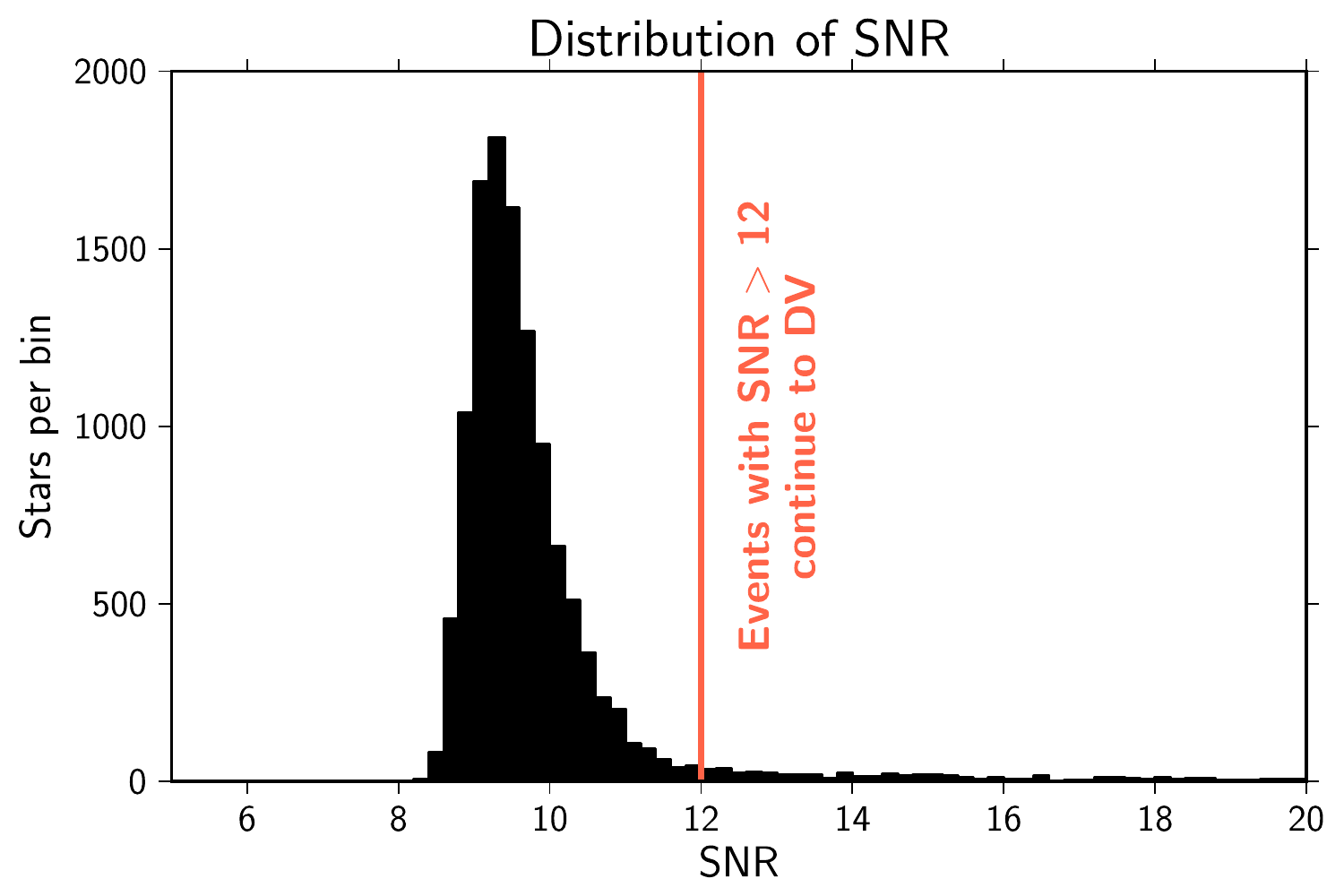}
\caption[Distribution of the highest SNR peak for each star in the
  Best12k sample]{Distribution of the highest SNR peak for each star in the
  Best12k sample. We show SNR = 5--20 to highlight the distribution of
  low SNR events. The \NSNR stars with SNR > 12 are labeled
  ``threshold crossing events'' (TCEs) and are subjected to additional
  scrutiny in the ``data validation'' component of \TERRA.}
\label{fig:SNR-hist}
\end{center}
\end{figure}

We show the distribution of maximum SNR for each Best12k star in
Figure~\ref{fig:SNR-hist}. Among the Best12k stars, \NSNR have a
maximum SNR peak exceeding 12. Adopting SNR = 12 as our threshold
balances two competing needs: the desire to recover small planets (low
SNR) and the desire to remove as many non-transit events as possible
before DV (high SNR). As discussed below, only 129 out of all 738
events with SNR > 12 are consistent with an exoplanet transit, with
noise being responsible for the remaining \NSNRnoteKOI. As shown in
Figure~\ref{fig:SNR-hist}, that number grows rapidly as we lower the
SNR threshold. For example, the number of TCEs grows to 3055 with a
SNR threshold of 10, dramatically increasing the burden on the DV
component.

A substantial number (\NSNRnotP) of TCEs are due to harmonics or
subharmonics of TCEs outside of the \Per= 5--50 day range and are
discarded. In order to pass DV, a TCE must also pass a suite of four
diagnostic metrics. The metrics are designed to test whether a light
curve is consistent with an exoplanet transit. We describe the four
metrics in Table~\ref{tab:DV} along with the criteria the TCE must
satisfy in order pass DV. The metrics and cuts were determined by
trial and error. We recognize that the \TERRA DV metrics and cuts are
not optimal and discard a small number of compelling exoplanet
candidates, as discussed in Section~\ref{sec:catonly}. However, since
we measure \TERRA's completeness by injection and recovery of
synthetic transits, the sub-optimal nature of our metrics and cuts is
incorporated into our completeness corrections.

\begin{figure*}
\begin{center}
\includegraphics[width=\textwidth]{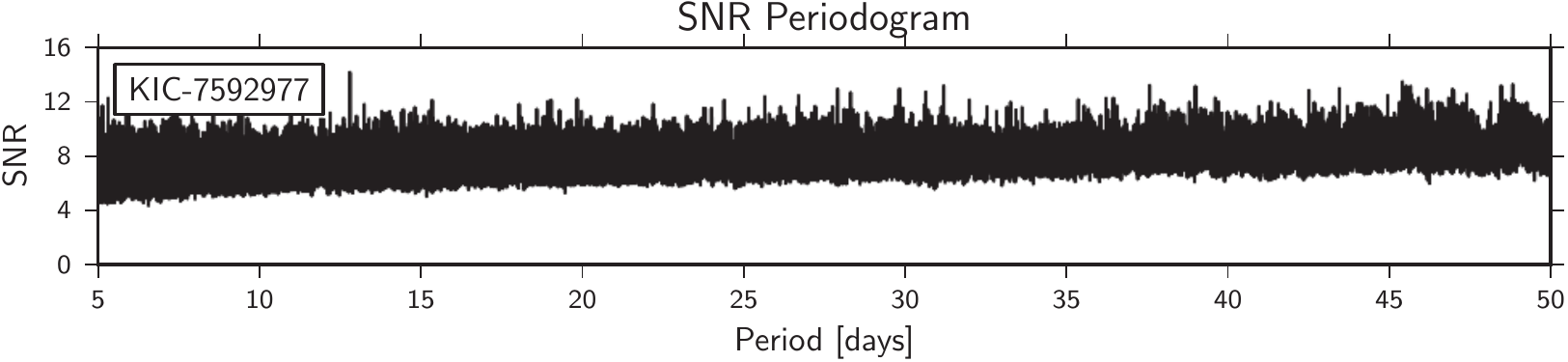}
\caption[SNR periodogram of KIC-7592977 photometry]{SNR periodogram of KIC-7592977, which passed the automated DV
  cuts, but was removed manually. KIC-7592977 photometry exhibited
  short bursts of high photometric scatter, which raised the noise
  floor to SNR $\sim$ 12, up from SNR $\sim$ 8 as in most stars.}
\label{fig:SNR-bad}
\end{center}
\end{figure*}

Our suite of automated cuts removes all but 145 TCEs. We perform a
final round of manual vetting and remove 16 additional TCEs, leaving
129 planet candidates. Most TCEs that we remove manually come from
stars with highly non-stationary photometric noise properties. Some
stars have small regions of photometry that exceed typical noise
levels by a factor of 3. We show the SNR periodogram for one such
star, KIC-7592977, in Figure~\ref{fig:SNR-bad}. Our definition of SNR
(Equation~\ref{eqn:SNR}) incorporates a single measure of photometric
scatter based on the median absolute deviation, which is insensitive
to short bursts of high photometric variability. In such stars,
fluctuations readily produce SNR $\sim$ 12 events and raise the
detectability floor to SNR $\sim$ 12, up from SNR $\sim$ 8 in most
stars. We also visually inspect phase-folded light curves for coherent
out-of-transit variability, not caught by our automated cuts, and for
evidence of a secondary eclipse.

\begin{deluxetable}{  l p{4.5in} r }
\tablecaption{Cuts used during data validation}
\tablewidth{0pt} 
\tablehead{\colhead{name} &
\colhead{description} & 
\colhead{value}}
 \startdata {\tt
  s2n\_out\_on\_in } & Compelling transits have flat out-of-transit
light curves. For a TCE with (\Per,\ep,\tdur), we remove the transit
region from the light curve and evaluate the SNR of all other
($P',\ep',\tdur'$) triples where $P=P'$ and $\tdur=\tdur'$. {\tt
  s2n\_out\_on\_in} is the ratio of the two highest SNR events.  & <
0.7 \\ {\tt med\_on\_mean } & Since the our definition of SNR
(Equation~\ref{eqn:SNR}) depends on the arithmetic mean of individual
transit depths, outliers occasionally produce high SNR TCEs. For each
TCE, we compute a robust SNR,
    \[
	\text{medSNR} = \frac{\sqrt{N_{T}}}{\sigma} \text{median}(\mdf_{i}).
    \]
     {\tt med\_on\_mean} is medSNR divided by SNR as defined in Equation~\ref{eqn:SNR}.        &
     > 1.0
     \\
     {\tt autor }
     &
     We compute the circular autocorrelation of the phase-folded light curve. {\tt autor} is the ratio of the highest autocorrelation peak (at 0 lag) to the second highest peak and is sensitive to out-of-transit variability.    
     &
     > 1.6
	\\
	{\tt taur}
	& 
	We fit the phase-folded light curve with a \cite{Mandel02} model. {\tt taur} is the ratio of the best fit transit duration to the maximum duration given the KIC stellar parameters and assuming a circular orbit.
	&
	< 2.0
	\\
\enddata
\label{tab:DV}
\end{deluxetable}

\section{Small Planets Found by \TERRA}
\label{sec:TERRAplanetYield}
Out of the 12,000 stars in the Best12k sample, \TERRA detected 129
planet candidates achieving SNR > 12 that passed our suite of DV cuts
as well as visual inspection. Table~\ref{tab:planetsTERRA} lists the
129 planet candidates. We derive planet radii using \Rp/\Rstar (from
Mandel-Agol model fits) and \Rstar from spectroscopy (when available)
or broadband photometry.

We obtained spectra for 100 of the 129 stars using HIRES \citep{Vogt94} at the Keck I telescope with standard configuration of the
California Planet Survey \citep{Marcy08}. These spectra have
resolution of $\sim$ 50,000, at a signal-to-noise of 45 per pixel at
5500~\AA. We determine stellar parameters using a routine called {\tt
  \SpecMatch} (Howard et al. 2013, in prep). In brief, {\tt
  \SpecMatch} compares a stellar spectrum to a library of $\sim800$
spectra with \teff = 3500--7500~K and \logg = 2.0--5.0 (determined
from LTE spectral modeling). Once the target spectrum and library
spectrum are placed on the same wavelength scale, we compute $\chi^2$,
the sum of the squares of the pixel-by-pixel differences in normalized
intensity. The weighted mean of the ten spectra with the lowest
$\chi^2$ values is taken as the final value for the effective
temperature, stellar surface gravity, and metallicity. We estimate
{\tt \SpecMatch}-derived stellar radii are uncertain to 10\% RMS,
based on tests of stars having known radii from high resolution
spectroscopy and asteroseismology.

For 27 stars where spectra are not available, we adopt the
photometrically-derived stellar parameters of \cite{Batalha12}. These
parameters are taken from the KIC \citep{Brown11}, but then modified
so that they lie on the Yonsei-Yale stellar evolution models of
\cite{Demarque04}. The resulting stellar radii have uncertainties of
35\% (rms), but can be incorrect by a factor of 2 or more. As an
extreme example, the interpretations of the three planets in the
KOI-961 system \citep{Muirhead12} changed dramatically when HIRES
spectra showed the star to be an M5 dwarf (0.2~\Rsun as opposed to
0.6~\Rsun listed in the KIC). We could not obtain spectra for two
stars, KIC-7345248 and KIC-8429668, which were not present in
\cite{Batalha12}. We determine stellar parameters for these stars by
fitting the KIC photometry to Yonsei-Yale stellar models. We adopt
35\% fractional errors on photometrically-derived stellar radii.

Once we determine \Per and \ep, we fit a \cite{Mandel02} model to the
phase-folded photometry. Such a model has three free parameters:
\rrat, the planet to stellar radius ratio; $\tau$, the time for the
planet to travel a distance \Rstar during transit; and $b$, the impact
parameter. In this work, \rrat is the parameter of interest. However,
$b$ and \rrat are covariant, i.e. a transit with $b$ approaching unity
only traverses the limb of the star, and thus produces a shallower
transit depth. In order to account for this covariance, best fit
parameters were computed via Markov Chain Monte Carlo. We find that
the fractional uncertainty on \rrat, $\frac{\sigma(\rrat)}{\rrat}$ can
be as high as 10\%, but is generally less than 5\%. Therefore, the
error on \Rp due to covariance with $b$ is secondary to the
uncertainty on \Rstar.

\begin{figure*}
\centering
\includegraphics[width=0.8\columnwidth]{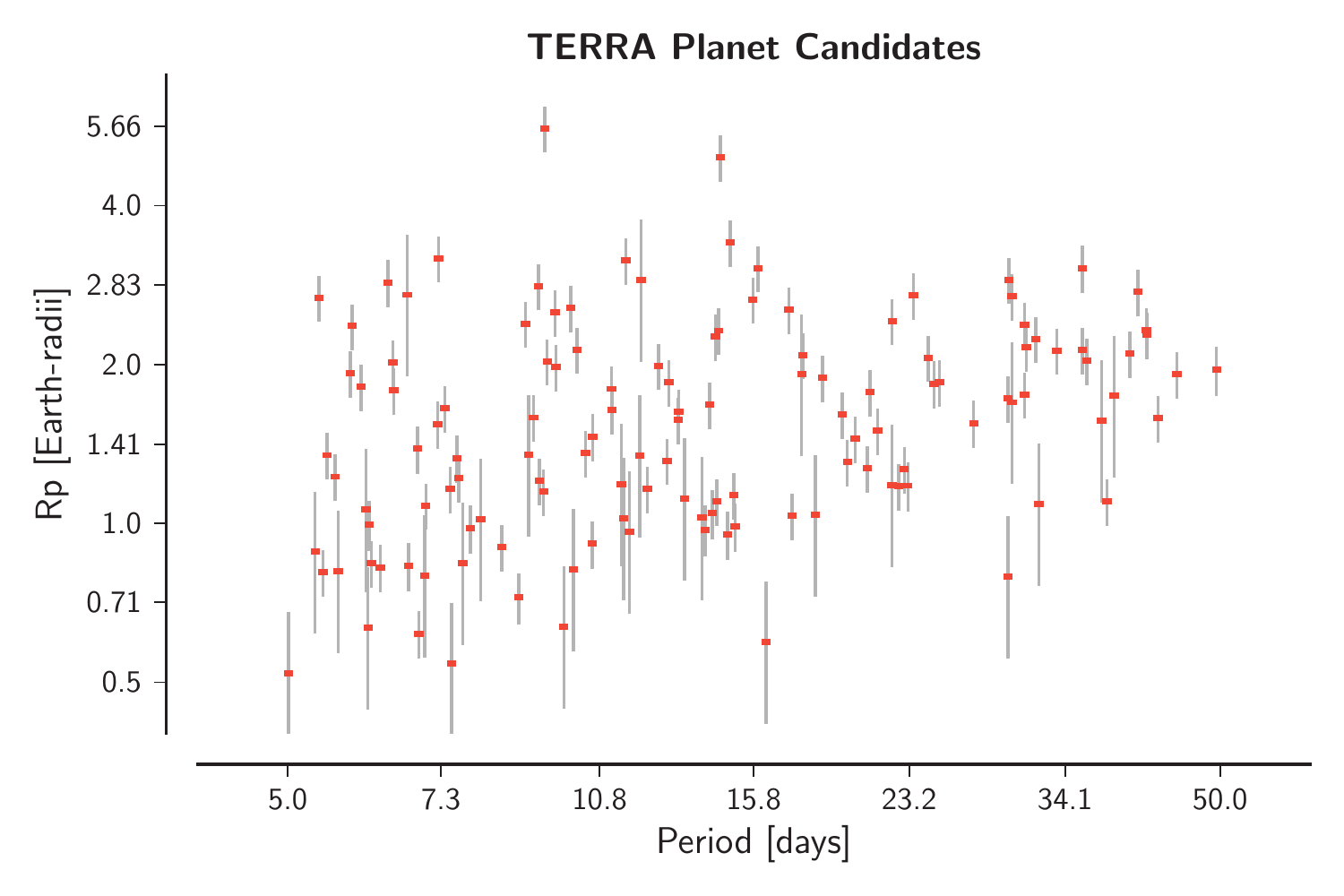}
\caption[TERRA planet candidates]{Periods and radii of 129 planet candidates detected by \TERRA. Errors on \Rp are computed via 
$\frac{ \sigma(\Rp) }{\Rp} =  
\sqrt{
\left( \frac{\sigma(\Rstar)}{\Rstar} \right)^2 +
\left( \frac{\sigma(\rrat)}{\rrat}\right)^2
}$, where \rrat is the radius ratio. The error in \Rp stems largely from the uncertainty in stellar radii. We adopt $\frac{\sigma(\Rstar)}{\Rstar}$ = 10\% for the 100 stars with spectroscopically determined \Rstar and $\frac{\sigma(\Rstar)}{\Rstar}$ = 35\% for the remaining stars with \Rstar determined from photometry. Using MCMC, we find the uncertainty in \rrat  is generally < 5\% and thus a minor component of the overall error budget.}
\label{fig:TERRAraw}
\end{figure*}

\begin{figure*}
\centering
\includegraphics[width=0.8\columnwidth]{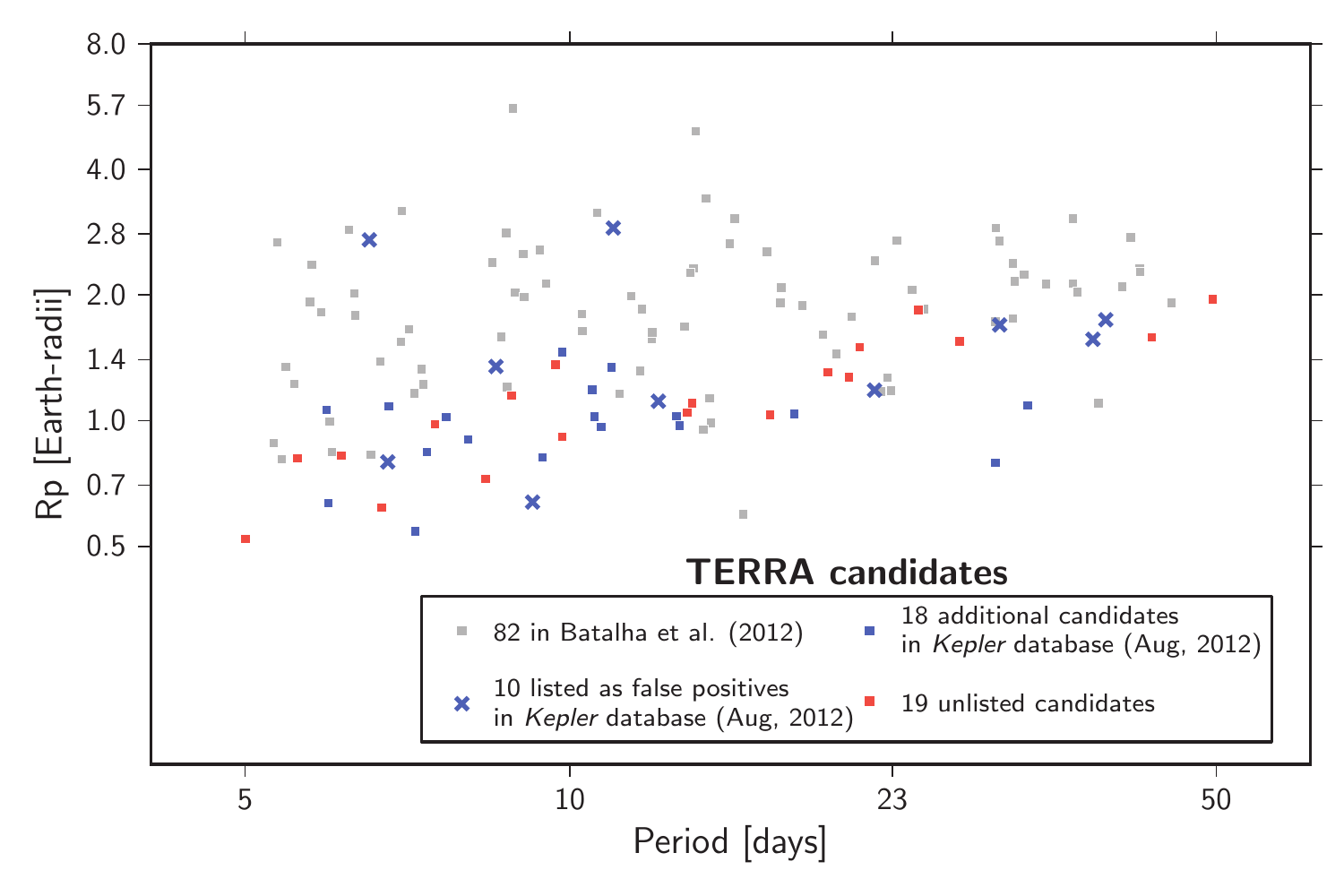}
\caption[TERRA planet candidates listed in other catalogs]{Periods and radii of all 129 \TERRA planet candidates. The
  gray points show candidates that were listed in
  \cite{Batalha12}. The blue crosses represent candidates deemed false
  positives by the \Kepler team as of August 8, 2012 (Jason Rowe,
  private communication 2012). These false positives are removed from
  our sample prior to computing occurrence. Eighteen additional
  candidates were listed in the same \Kepler team database. Red points
  show 19 unlisted \TERRA candidates.}
\label{fig:TERRAcommon}
\end{figure*}

We show the distribution of \TERRA candidates in
Figure~\ref{fig:TERRAraw} over the two-dimensional domain of planet
radius and orbital period. Our 129 candidates range in size from 6.83
\Re to 0.48 \Re (smaller than Mars). The median \TERRA candidate size
is 1.58 \Re. In Figure~\ref{fig:TERRAcommon}, we show the substantial
overlap between the \TERRA planet sample and those produced by the
\Kepler team. \TERRA recovers 82 candidates listed in
\cite{Batalha12}. We discuss the significant overlap between the two
works in detail in Section~\ref{sec:compare}. As of August 8th, 2012,
10 of our \TERRA candidates were listed as false positives in an
internal database of \Kepler planet candidates maintained by Jason
Rowe (Jason Rowe, 2012, private communication) and are shown as blue
crosses in Figure~\ref{fig:TERRAcommon}. We do not include these 10
candidates in our subsequent calculation of
occurrence. Table~\ref{tab:planetsTERRA} lists the KIC identifier,
best fit transit parameters, stellar parameters, planet radius, and
\Kepler team false positive designation of all 129 candidates revealed
by the \TERRA algorithm. The best fit transit parameters include
orbital period, \Per; time of transit center, \ep; planet to star
radius ratio, \rrat; time for planet to cross \Rstar during transit,
$\tau$; and impact parameter, $b$. We list the following stellar
properties: effective temperature, \teff; surface gravity, \logg; and
stellar radius, \Rstar.

\section{Completeness of Planet Catalog}
\label{sec:MC}
When measuring the distribution of planets as a function of \Per and
\Rp, understanding the number of missed planets is as important as
finding planets themselves. H12 accounted for completeness in a rough
sense based on signal-to-noise considerations. For each star in their
sample, they estimated the SNR over a range of \Per and \Rp using CDPP
as an estimate of the photometric noise on transit-length
timescales. H12 chose to accept only planets with SNR > 10 in a single
quarter of photometry for stars brighter than Kp = 15.  This metric
used CDPP and was a reasonable pass on the data, particularly when the
pipeline completeness was unknown.  Determining expected SNR from CDPP
does not incorporate the real noise characteristics of the photometry,
but instead approximates noise on transit timescales as stationary
(CDDP assumed to be constant over a quarter) and Gaussian
distributed. Moreover, identifying small transiting planets with
transit depths comparable to the noise requires a complex, multistage
pipeline. Even if the integrated SNR is above some nominal threshold,
the possibility of missed planets remains a concern.

We characterize the completeness of our pipeline by performing an
extensive suite of injection and recovery experiments. We inject mock
transits into raw photometry, run this photometry though the same
pipeline used to detect planets, and measure the recovery rate. This
simple, albeit brute force, technique captures the idiosyncrasies of
the \TERRA pipeline that are missed by simple signal-to-noise
considerations.

We perform 10,000 injection and recovery experiments using the
following steps:

\begin{enumerate}
\item We select a star randomly from the Best12k sample. 
\item We draw (\Per,\Rp) randomly from log-uniform distributions over 5--50~days and 0.5--16.0~\Re.
\item We draw impact parameter and orbital phase randomly from uniform distributions ranging from 0 to 1.
\item We generate a \cite{Mandel02} model.
\item We inject it into the ``simple aperture photometry'' of the selected star.
\end{enumerate}
We then run the calibration, grid-based search, and data validation
components of \TERRA (Sections~\ref{sec:TERRAcal},
\ref{sec:TERRAgrid}, and \ref{sec:TERRAdv}) on this photometry and
calculate the planet recovery rate. We do not, however, perform the
visual inspection described in Section~\ref{sec:TERRAdv}. An injected
transit is considered recovered if the following two criteria are met:
(1) The highest SNR peak passes all DV cuts and (2) the output period
and epoch are consistent with the injected period and epoch to within
0.01 and 0.1 days, respectively.

\begin{figure*}[htbp]
\includegraphics[clip=True,width=\textwidth]{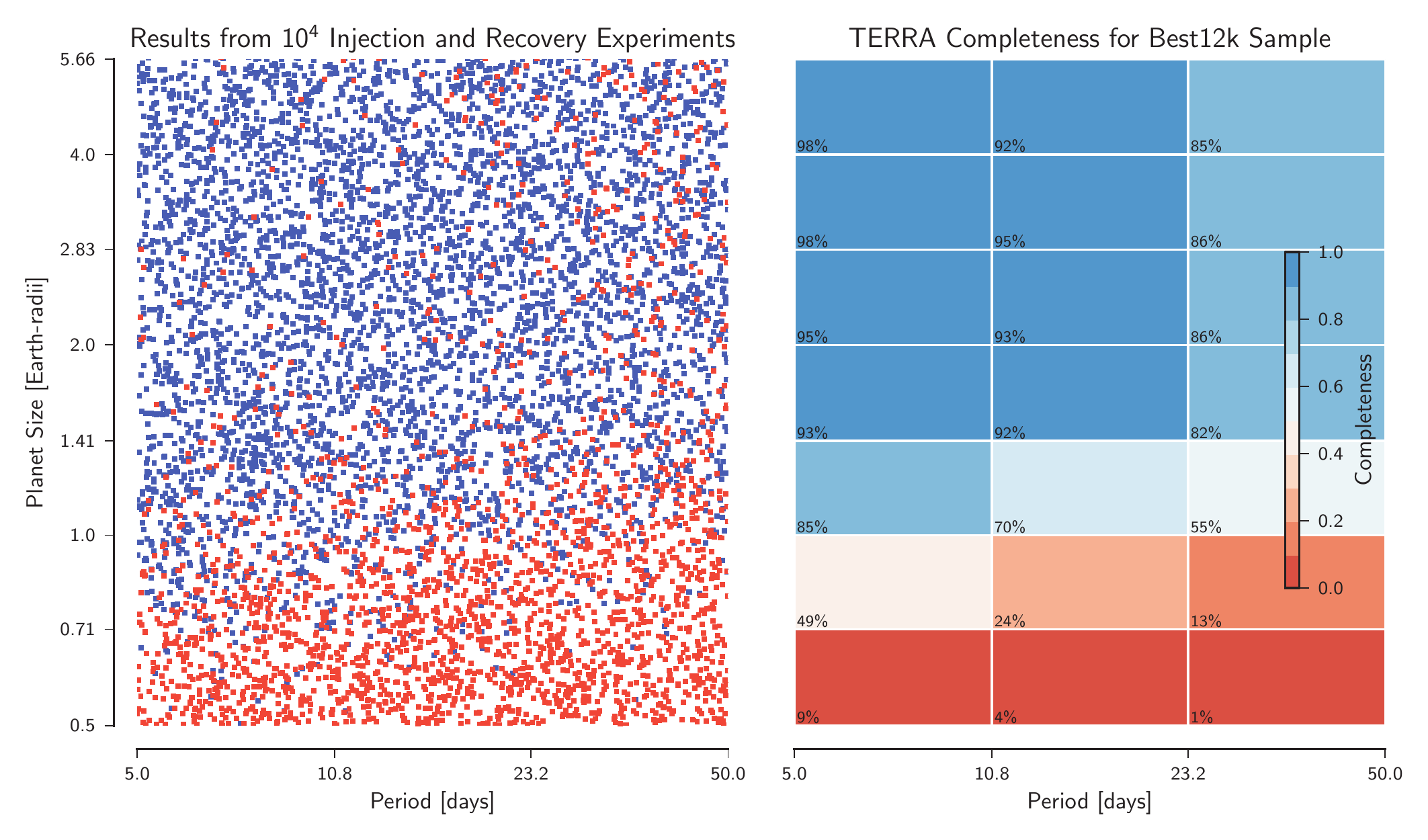}
\caption[Completeness from injection and recovery experiments]{Results from the injection and recovery of 10,000 synthetic
  transit signals into actual photometry of randomly selected stars
  from our Best12k stellar sample. Each point represents the planet
  radius and orbital period of a mock transiting planet. The blue
  points represent signals that passed the DV post-analysis and where
  \TERRA recovers the correct period and epoch. Signals that did not
  pass DV and/or were not successfully recovered, are shown as red
  points. Pipeline completeness is simply the number of blue points
  divided by the total number of points in each bin. The figure shows
  that for planet sizes above 1.0~\Re, our pipeline discovers over
  50\% of the injected planets, and presumably accomplishes a similar
  success rate for actual transiting planets. The completeness for
  planets larger than 1~\Re is thus high enough to compute planet
  occurrence for such small planets, with only moderate completeness
  corrections needed (less than a factor of 2). Note that we are
  measuring the recovery rate of putative planets in the Best12k
  sample with \TERRA. Had we selected a lower noise stellar sample,
  for example the ``Best6k,'' the region of high completeness would
  extend to even small radii.}
\label{fig:comp}
\end{figure*}

Figure~\ref{fig:comp} shows the distribution of recovered simulations
as a function of period and radius. Nearly all simulated planets with
\Rp > 1.4 \Re are recovered, compared to almost none with \Rp < 0.7
\Re. Pipeline completeness is determined in small bins in
(\Per,\Rp)-space by dividing the number of successfully recovered
transits by the total number of injected transits in a bin-by-bin
basis. This ratio is \TERRA's recovery rate of putative planets within
the Best12k sample. Thus, our quoted completeness estimates only
pertain to the low photometric noise Best12k sample. Had we selected
an even more rarified sample, e.g. the ``Best6k,'' the region of high
completeness would extend down toward smaller planets.

\section{Occurrence of Small Planets}
\label{sec:occurrenceme}

Following H12, we define planet occurrence, $f$, as the fraction of a
defined population of stars having planets within a domain of planet
radius and period, including all orbital inclinations. \TERRA,
however, is only sensitive to one candidate (highest SNR) per system,
so we report occurrence as the fraction of stars with {\em one or
  more} planets with \Per = 5--50 days. Our occurrence measurements
apply to the Best12k sample of low-noise, solar-type stars described
in Section~\ref{sec:Sample}.

In computing planet occurrence in the Best12k sample, we follow the
prescription in H12 with minor modifications. Notably, we have
accurate measures of detection completeness described in the previous
section. In contrast, H12 estimated completeness based on the presumed
signal-to-noise of the transit signal, suffering both from approximate
characterization of photometric noise using CDPP and from poor
knowledge of the efficiency of the planet-finding algorithm for all
periods and sizes.

For each \Per-\Rp bin, we count the number of planet candidates,
\Np. Each planet that transits represents many that do not transit
given the orientation of their orbital planes with respect to
\Kepler's line of sight. Assuming random orbital alignment, each
observed planet represents a/\Rstar total planets when non-transiting
geometries are considered. For each cell, we compute the number of
augmented planets, $\NpAug = \sum_i a_i/R_{\star,i}$, which accounts
for planets with non-transiting geometries. We then use Kepler's
\nth{3} law together with \Per and $M_\star$ to compute a/\Rstar
assuming a circular orbit.\footnote{H12 determined a/\Rstar directly
  from light curve fits, but found little change when computing
  occurrence from a/\Rstar using Kepler's third law.}

\begin{figure*}
\centering
\includegraphics[width=\textwidth]{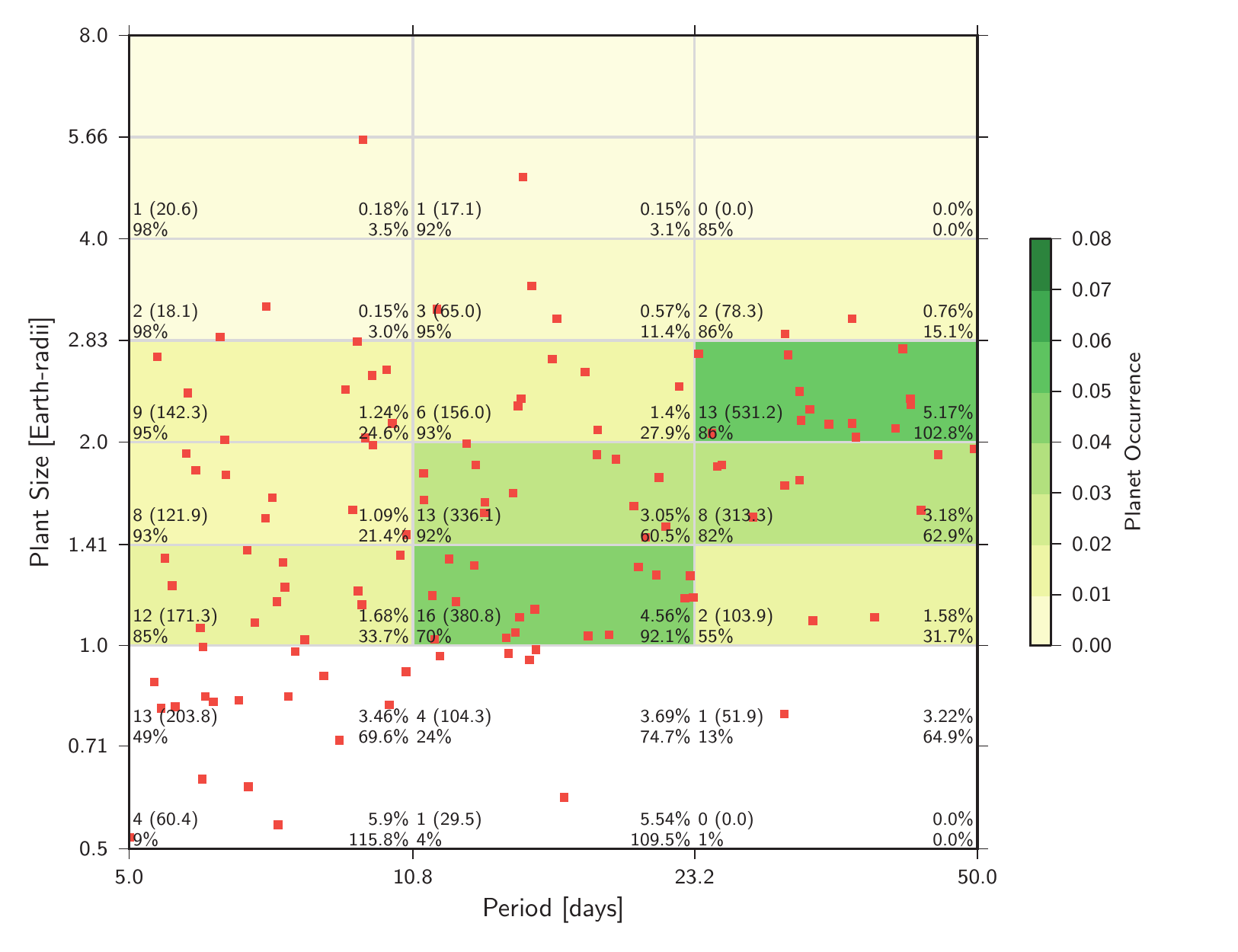}
\caption[Planet occurrence using TERRA planet candidates]{Planet occurrence as a function of orbital period and planet
  radius for \Per = 5--50 days and \Rp = 0.5--8 \Re. \TERRA planet
  candidates are shown as red points. Cell occurrence, \fcell, is
  given by the color scale. We quote the following information for
  each cell: Top left--number of planets (number of augmented
  planets); lower left--completeness; top right--fractional planet
  occurrence, \fcell; bottom right--normalized planet occurrence,
  \flogA. We do not color cells where the completeness is less than
  50\% (i.e. the completeness correction is larger than a factor of
  2).}
\label{fig:TERRAoccur2D}
\end{figure*}

To compute occurrence, we divide the number of stars with planets in a
particular cell by the number of stars amenable to the detection of a
planet in a given cell, $\NstarAmen$. This number is just $N_\star$ =
12,000 times the completeness, computed in our Monte Carlo study. The
debiased fraction of stars with planets per \Per-\Rp bin, $\fcell$, is
given by $\fcell=\NpAug/\NstarAmen$. We show \fcell on the \Per-\Rp
plane in Figure~\ref{fig:TERRAoccur2D} as a color scale. We also
compute \flogA, i.e. planet occurrence divided by the logarithmic area
of each cell, which is a measure of occurrence which does not depend
on bin size. We annotate each \Per-\Rp bin of
Figure~\ref{fig:TERRAoccur2D} with the corresponding value of \Np,
\NpAug, \fcell, and \flogA.

Due to the small number of planets in each cell, errors due to
counting statistics alone are significant. We compute Poisson errors
on \Np for each cell. Errors on \NpAug, \fcell, and \flogA include
only the Poisson errors from \Np. There is also shot noise associated
with the Monte Carlo completeness correction due to the finite number
of simulated planets in each \Per-\Rp cell, but such errors are small
compared to errors on \Np. The orbital alignment correction,
$a/\Rstar$, is also uncertain due to imperfect knowledge of stellar
radii and orbital separations. We do not include such errors in our
occurrence estimates.

\begin{figure}
\centering
\includegraphics[width=0.8\columnwidth]{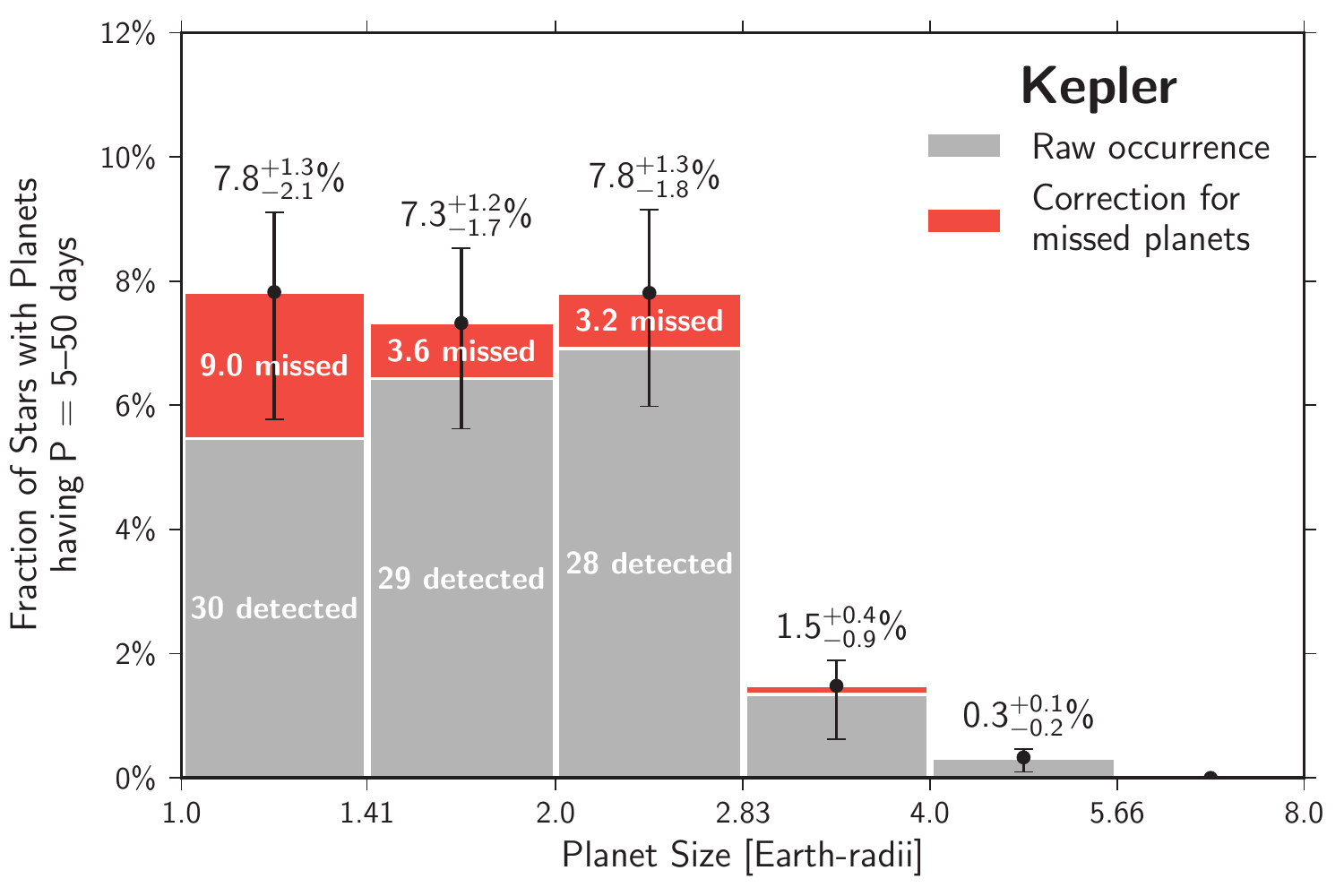}
\caption[Planet size distribution]{Distribution of planet occurrence for \Rp ranging from 1.0 to
  8.0~\Re. We quote the fraction of Sun-like stars harboring a planet
  with \Per~=~5--50~days for each \Rp bin. We observe a rapid rise in
  planet occurrence from 8.0 down to 2.8~\Re, as seen in H12. Below
  2.8 \Re, the occurrence distribution is consistent with flat. This
  result rules out a power law increase in planet occurrence toward
  smaller radii. Adding up the two smallest radius bins, we find
  \fonetwo of Sun-like stars harbor a 1.0--2.0~\Re planet within
  $\sim0.25$~AU. To compute occurrence as a function of \Rp, we simply
  sum occurrence rates for all period bins shown in
  Figure~\ref{fig:TERRAoccur2D}.  Errors due to counting statistics
  are computed by adding errors from each of the three period bins in
  quadrature. The gray portion of the histogram shows occurrence
  values before correcting for missed planets due to pipeline
  incompleteness. Our correction to account for missed planets is
  shown in red, and is determined by the injection and recovery of
  synthetic transits described in Section~\ref{sec:MC}. We do not show
  occurrence values where the completeness is < 50\%.}
\label{fig:TERRAoccur1D}
\end{figure}

Of particular interest is the distribution of planet occurrence with
\Rp for all periods. We marginalize over \Per by summing occurrence
over all period bins from 5 to 50~days. The distribution of radii
shown in Figure~\ref{fig:TERRAoccur1D} shows a rapid rise in
occurrence from 8.0 to 2.8~\Re. H12 also observed a rising occurrence
of planets down to 2.0~\Re, which they modeled as a power law. {\em
  Planet occurrence is consistent with a flat distribution from 2.8 to
  1.0~\Re, ruling out a continuation of a power law increase in
  occurrence for planets smaller than 2.0~\Re. We find \fonetwo of
  Sun-like stars harbor a 1.0--2.0~\Re planet with \Per~=~5--50~days.}
Including larger planets, we find that $24.8^{+2.1}_{-3.4}\%$ of stars
harbor a planet larger than Earth with \Per~=~5--50~days. Occurrence
values assuming a 100\% efficient pipeline are shown as gray bars in
Figure~\ref{fig:TERRAoccur1D}. The red bars show the magnitude of our
completeness correction. Even though \TERRA detects many planets
smaller than 1.0~\Re, we do not report occurrence for planets smaller
than Earth since pipeline completeness drops abruptly below 50\%.

\begin{figure}
\centering
\includegraphics[clip=True,width=0.8\columnwidth]{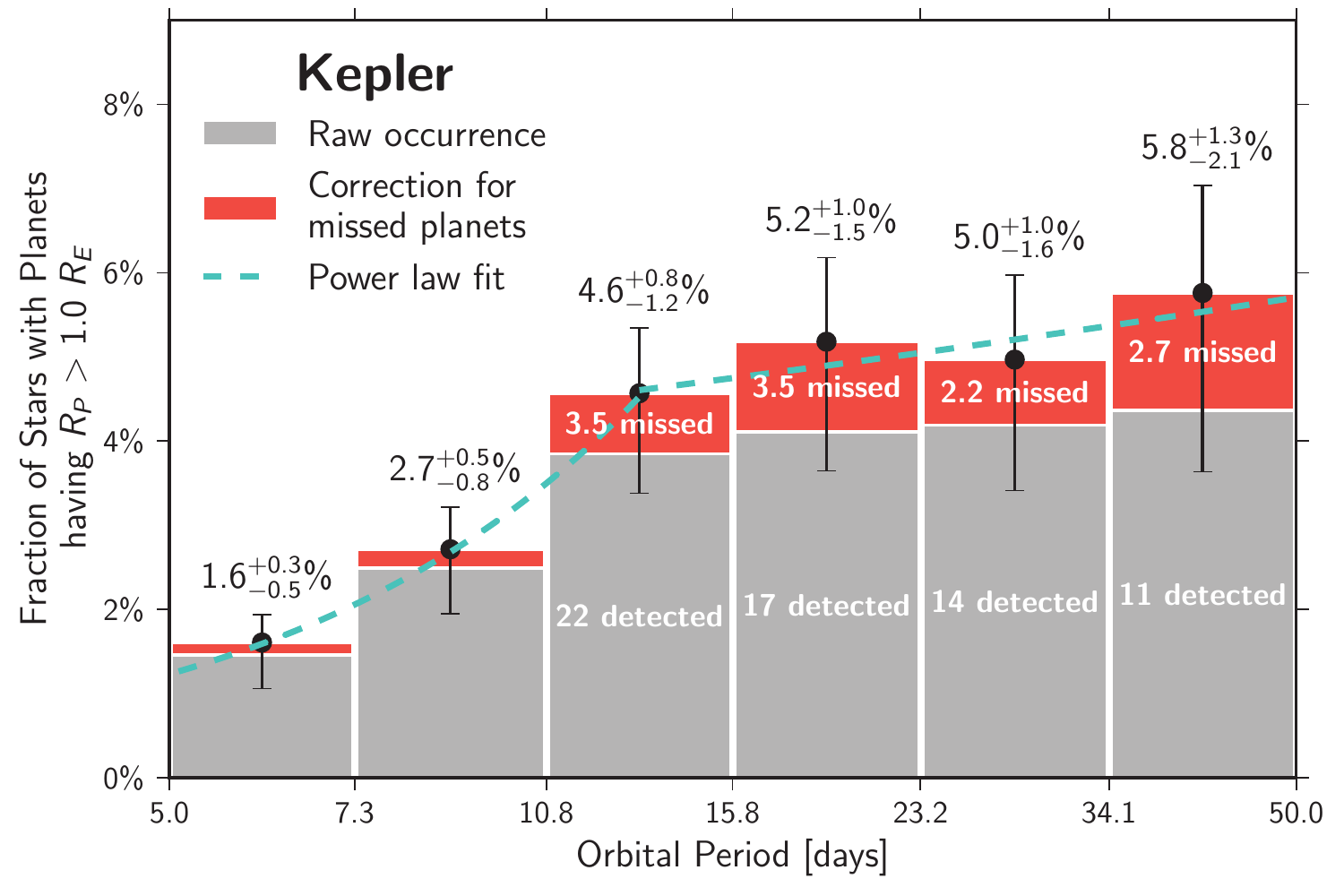}
\caption[Planet orbital period distribution]{Distribution of planet occurrence for different orbital
  periods ranging from 5 to 50~days. We quote the fraction of Sun-like
  stars with a planet Earth-size or larger as a function of orbital
  period. We observe a gradual rise in occurrence from 5.0 to 10.8~\Re
  followed by a leveling off for longer orbital periods. H12 observed
  a similar leveling off in their analysis which included planets
  larger than 2~\Re. We fit the domains above and below 10.8~\Re
  separately with power laws, $d f / d \log \Per = k_{P}
  \Per^{\alpha}$. We find best fit values of $k_{P} =
  0.185^{+0.043}_{-0.035},\ \alpha =0.16 \pm 0.07$ for
  \Per~=~5--10.8~days and $k_{P} = 8.4^{+0.9}_{-0.8} \times
  10^{-3},\ \alpha =1.35 \pm 0.05$ for
  \Per~=~10.8--50~days. Speculatively, we extrapolate the latter power
  law fit another decade in period and estimate $41.7^{+6.8}_{-5.9}$\%
  of Sun-like stars harbor a planet Earth-size or larger with
  \Per~=~50--500~days. As in Figure~\ref{fig:TERRAoccur1D}, the gray
  portion of the histogram shows uncorrected occurrence while the red
  region shows our correction for pipeline incompleteness. Note that
  the number of detected planets decreases as \Per increases from 10.8
  to 50~days, while occurrence remains nearly constant. At longer
  periods, the geometric transit probability is lower, and each
  detected planet counts more toward $d f / d \log \Per$.}
\label{fig:TERRAoccurP}
\end{figure}

We show planet occurrence as a function of orbital period in
Figure~\ref{fig:TERRAoccurP}. In computing this second marginal
distribution, we include radii larger than 1~\Re so that corrections
due to incompleteness are small. Again, as in
Figure~\ref{fig:TERRAoccur1D}, gray bars represent uncorrected
occurrence values while red bars show our correction to account for
planets that \TERRA missed. Planet occurrence rises as orbital period
increases from 5.0 to 10.8~days. Above 10.8~days, planet occurrence is
nearly constant per logarithmic period bin with a slight indication of
a continued rise. This leveling off of the distribution was noted by
H12, who considered \Rp > 2.0 \Re. We fit the distribution of orbital
periods for \Rp > 1.0 \Re with two power laws of the form
\begin{equation}
\frac{d f }{ d \log \Per} = k_{P} \Per^{\alpha}, 
\end{equation}
where $\alpha$ and $k_{P}$ are free parameters. We find best fit
values of $k_{P} = 0.185^{+0.043}_{-0.035},\ \alpha =0.16 \pm 0.07$
for \Per~=~5--10.8~days and $k_{P} = 8.4^{+0.9}_{-0.8} \times
10^{-3},\ \alpha =1.35 \pm 0.05$ for \Per~=~10.8--50~days. We note
that $k_{P}$ and $\alpha$ are strongly covariant. Extrapolating the
latter fit speculatively to \Per > 50~days, we find
$41.7^{+6.8}_{-5.9}$\% of Sun-like stars host a planet 1 \Re or larger
with \Per~=~50--500~days.

\section{Comparison of \TERRA and Batalha et al. (2012) Planet Catalogs}
\label{sec:compare}
Here, we compare our candidates to those of
\cite{Batalha12}. Candidates were deemed in common if their periods
agree to within 0.01 days. We list the union of the \TERRA and
\cite{Batalha12} catalogs in Table~\ref{tab:planetsJOIN}. Eighty-two
candidates appear in both catalogs (Section~\ref{ssec:common}), 47
appear in this work only (Section~\ref{ssec:TERRAOnly}), and 33 appear
in \cite{Batalha12} only (Section~\ref{sec:catonly}). We discuss the
significant overlap between the two catalogs and explain why some
candidates were detected by one pipeline but not the other.

\subsection{Candidates in Common}
\label{ssec:common}
Eighty-two of our candidates appear in the \cite{Batalha12}
catalog. We show these candidates in \Per-\Rp space in
Figure~\ref{fig:TERRAcommon} as grey points. \TERRA detected no new
candidates with \Rp > 2 \Re. This agreement in detected planets having
\Rp > 2 \Re demonstrates high completeness for such planets in both
pipelines for this sample of quiet stars. This is not very surprising
since candidates with \Rp > 2 \Re have high SNR, e.g. min, median, and
max SNR = 19.3, 71.5, and 435 respectively.

Radii for the 82 planets in common were fairly consistent between
\cite{Batalha12} and this work. The two exceptions were KIC-8242434
and KIC-8631504. Using {\tt SpecMatch}, we find stellar radii of 0.68
and 0.72~\Rsun, respectively, down from 1.86 and 1.80 \Rsun in
\cite{Batalha12}. The revised planet radii are smaller by over a
factor of two. Radii for the other planets in common were consistent
to $\sim$ 20\%.

\subsection{\TERRA Candidates Not in \cite{Batalha12} Catalog}
\label{ssec:TERRAOnly}

\TERRA revealed 47 planet candidates that did not appear in
\cite{Batalha12}. Such candidates are colored blue and red in
Figure~\ref{fig:TERRAcommon}. Many of these new detections likely stem
from the fact that we use twice the photometry that was available to
Batalha et al. (2012). To get a sense of how additional photometry
improves the planet yield of the \Kepler pipeline beyond
\cite{Batalha12}, we compared the \TERRA candidates to the \Kepler
team KOI list dated August 8, 2012 (Jason Rowe, private
communication). The 28 candidates in common between the August 8, 2012
\Kepler team sample and this work are colored blue in
Figure~\ref{fig:TERRAcommon}. Of these 28 candidates, 10 are listed as
false positives and denoted as crosses in
Figure~\ref{fig:TERRAcommon}.

We announce 37 new planet candidates with respect to \cite{Batalha12}
that were not listed as false positives in the \Kepler team
sample. These 37 candidates, all with \Rp~$\lessapprox$ 2~\Re, are a
subset of those listed in Table~\ref{tab:planetsTERRA}. As a
convenience, we show this subset in
Table~\ref{tab:planetsTERRAnew}. We remind the reader that all
photometry used in this work is publicly available. We hope that
interested readers will fold the photometry on the ephemeris in
Table~\ref{tab:planetsTERRA} and assess critically whether a planet
interpretation is correct. As a quick reference, we have included
plots of the transits of the 37 new candidates from
Table~\ref{tab:planetsTERRAnew} in the appendix
(Figures~\ref{fig:binned1} and \ref{fig:binned2}). We do not claim
that our additional candidates bring pipeline completeness to unity
for planets with \Rp~$\lessapprox$ 2~\Re. As shown in
Section~\ref{sec:MC}, our planet sample suffers from significant
incompleteness in the same \Per-\Rp space where most of the new
candidates emerged.

\subsection{\cite{Batalha12} Candidates Not in \TERRA catalog}
\label{sec:catonly}
\begin{figure}[htbp]
\centering
\includegraphics[width=0.8\columnwidth]{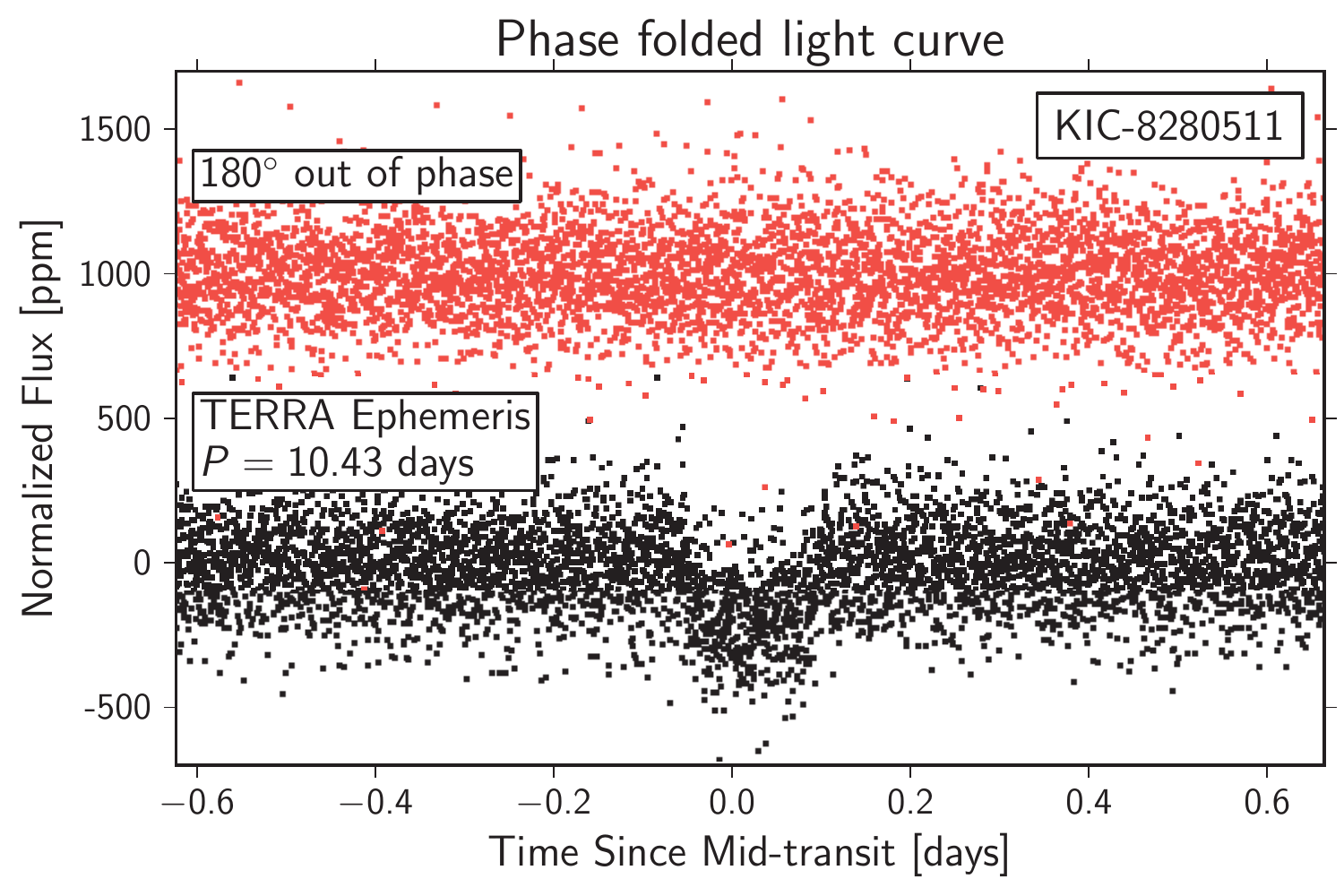}
\caption[Phase-folded photometry of KIC-8280511]{Phase-folded photometry of KIC-8280511 folded on the correct
  10.43~day period found by \TERRA. KOI-1151.01 is listed with \Per =
  5.22~days in Batalha et al. (2012). If the transit was truly on the
  5.22~day period, we should see a transit of equal depth 180 degrees
  out of phase. KOI-1151.01 is listed in \cite{Batalha12} with half
  its true period.}
\label{fig:KIC-8280511}
\end{figure}

There are 33 planet candidates in the \cite{Batalha12} catalog from
Best12k stars that \TERRA missed. Of these, 28 are multi-candidate
systems where one component was identified by \TERRA. \TERRA is
currently insensitive to multiple planet systems (as described in
Section~\ref{sec:TERRAgrid}). \TERRA missed the remaining 5
\cite{Batalha12} candidates for the following reasons:
\begin{itemize}
\item 2581.01 : A bug in the pipeline prevented successful photometric calibration (Section~\ref{sec:TERRAcal}). This bug affected 19 out of   12,000 stars in the Best12k sample.
\item 70.01, 111.01, 119.01 : Failed one of the automated DV cuts
  ({\tt taur}, {\tt med\_on\_mean}, and {\tt taur}, respectively). We
  examined these three light curves in the fashion described in
  Section~\ref{sec:TERRAdv}, and we determined these light curves were
  consistent with an exoplanet transit. The fact that DV is discarding
  compelling transit signals decreases \TERRA's overall
  completeness. Computing DV metrics and choosing the optimum cuts is
  an art. There is room for improvement here.
\item KOI-1151.01 : Period misidentified in \cite{Batalha12}. In \cite{Batalha12} KOI-1151.01 is listed with with a \Per~=~5.22~days. \TERRA found a candidate with \Per~=~10.43~days. Figure~\ref{fig:KIC-8280511} shows phase-folded photometry with the \TERRA ephemeris. A period of 5.22 days would imply dimmings in regions where the light curve is flat.
\end{itemize}

\begin{figure*}[htbp]
\centering
\includegraphics[width=0.8\columnwidth]{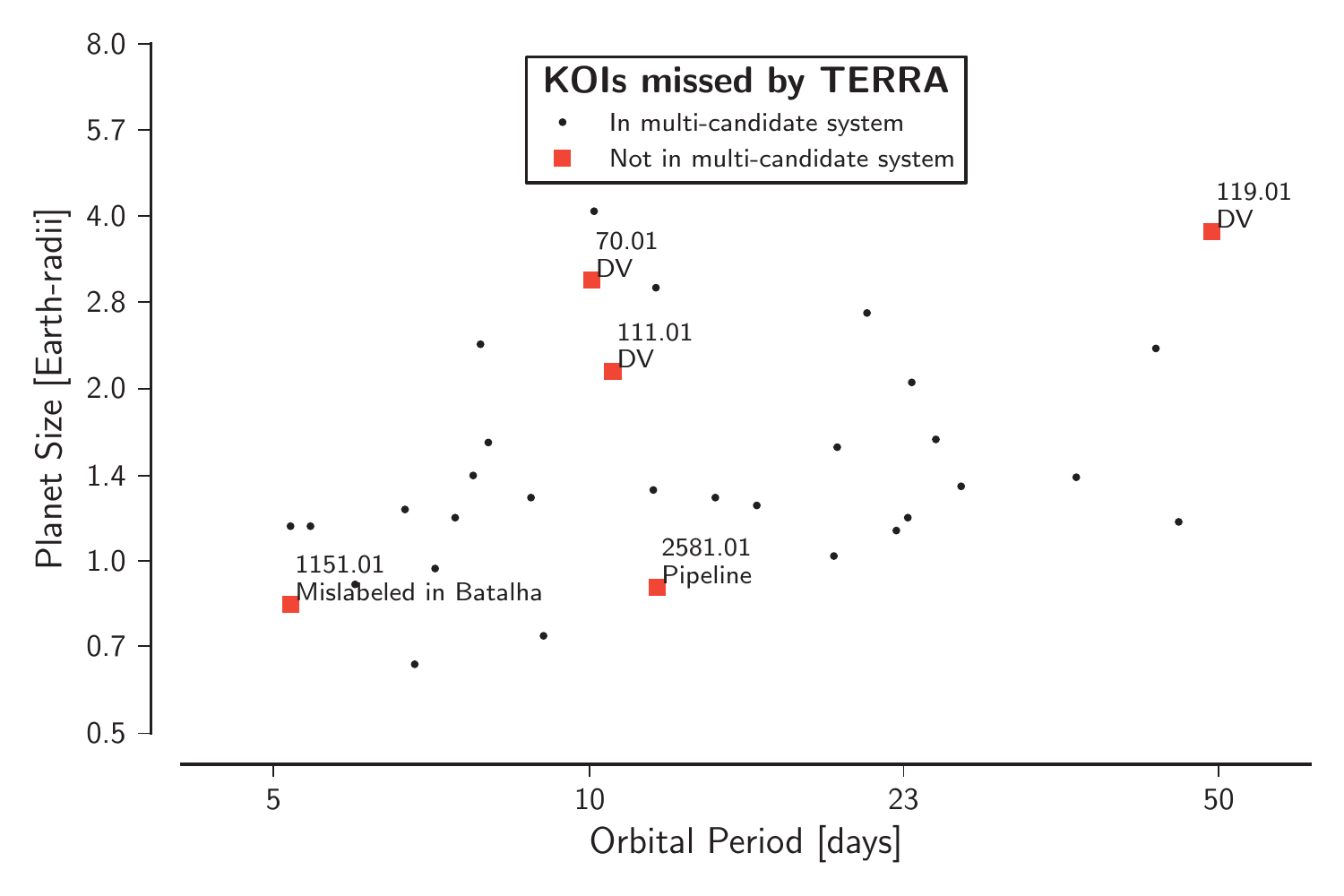}
\caption[KOIs missed by TERRA]{\Per and \Rp for the 33 candidates present in
  \cite{Batalha12} but not found by \TERRA. The small symbols show the
  candidates in mulit-planet systems. \TERRA is blind to such
  candidates. The 5 larger symbols show the other failure modes of
  \TERRA: 2581.01 failed due to a pipeline bug; 70.01, 111.01, and
  119.01 did not pass DV; and \TERRA missed KOI-1151.01 because it is
  listed in \cite{Batalha12} with the incorrect period. Most of the
  missed planets have \Rp < 1.4 \Re.}
\label{fig:TERRAmissed}
\end{figure*}

We plot the 33 total candidates listed in \cite{Batalha12}, but not
found by \TERRA in Figure~\ref{fig:TERRAmissed}. We highlight the 5
missed candidates that cannot be explained by the fact that they are a
lower SNR candidate in a multi-candidate system. \TERRA is blind to
planets in systems with another planet with higher
SNR. Figure~\ref{fig:TERRAmissed} shows that most of these missed
planets occur at \Rp < 1.4 \Re.

\section{Occurrence with Planet Multiplicity Included}
\label{sec:OccurCOMB}
While the \TERRA planet occurrence measurement benefits from
well-characterized completeness, it does not include the contribution
of multis to overall planet occurrence. As discussed in
Sections~\ref{sec:TERRAgrid} and \ref{sec:catonly}, \TERRA only
detects the highest SNR candidate for a given star. Here, we present
planet occurrence including multis from \cite{Batalha12}. Thus, the
occurrence within a bin, $f$, in this section should be interpreted as
the {\em average number of planets} per star with \Per = 5--50
days. The additional planets from \cite{Batalha12} raise the
occurrence values somewhat over those of the previous
section. However, the rise and plateau structure remains the same.

We compute $\fcell$ from the 32 candidates present in
\cite{Batalha12}, but not found by \TERRA (mislabeled KOI-1151.01 was
not included). For clarity, we refer to this separate occurrence
calculation as \fcellBa. Because the completeness of the \Kepler
pipeline is unknown, we apply no completeness correction. This
assumption of 100\% completeness is certainly an overestimate, but we
believe that the sensitivity of the \Kepler pipeline to multis is
nearly complete for \Rp > 1.4 \Re. \TERRA has > 80\% completeness for
\Rp > 1.4 \Re because planets in that size range with \Per = 5--50
days around Best12k stars have high SNR. The \Kepler pipeline should
also be detecting these high SNR candidates. Also, once a KOI is
found, the \Kepler team reprocesses the light curve for additional
transits (Jason Rowe, private communication). Due to this additional
scrutiny, we believe that the \Kepler completeness for multis is
higher than for singles, all else being equal.

\begin{figure*}
\includegraphics[width=\textwidth]{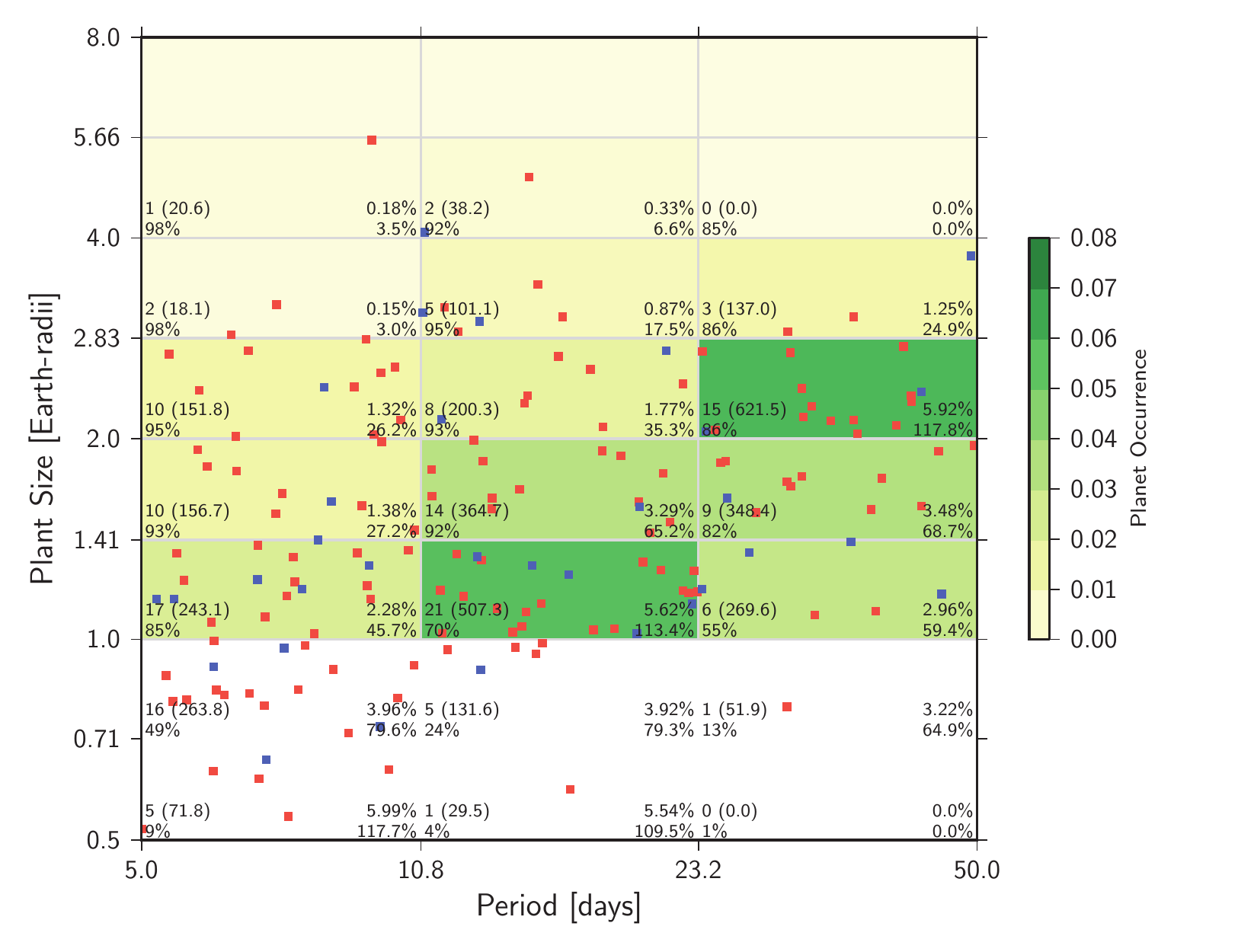}
\caption[Planet occurrence including Batalha et al. (2012) candidates]{As in Figure~\ref{fig:TERRAoccur2D}, red points show 119
  \TERRA-detected planets. Blue points represent additional planets
  from \cite{Batalha12}. Most (28 out of 32) of these new candidates
  are planets in multi-candidate systems where \TERRA successfully
  identifies the higher SNR candidate. We apply no completeness
  correction to these new planets, and we believe this is appropriate
  for \Rp > 1.4 \Re. We quote the following occurrence information for
  each cell: Top left--number of planets (number of augmented
  planets), lower left--completeness, top right--fractional planet
  occurrence \fcell, bottom right--normalized planet occurrence
  \flogA. We do not color cells where the completeness is less than
  50\% (i.e. the completeness correction is larger than a factor of
  2). The planet counts and occurrence values are for the combined
  \TERRA and \cite{Batalha12} sample. The completeness values are the
  same as in Figure~\ref{fig:TERRAoccur2D}.}
\label{fig:COMBoccur2D}
\end{figure*}

\begin{figure}
\centering
\includegraphics[width=0.8\columnwidth]{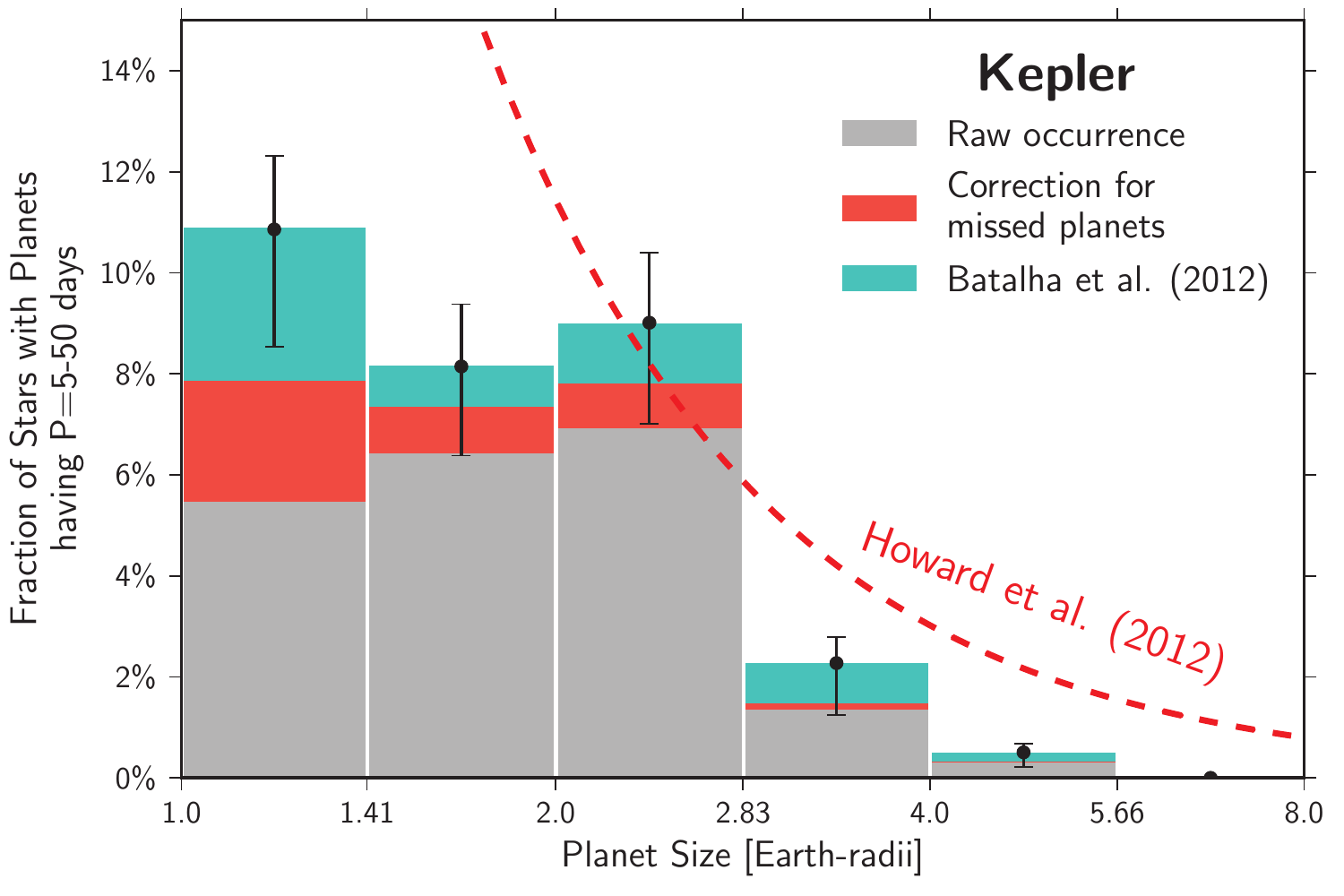}
\caption[Planet radius distribution including Batalha et al. (2012) candidates]{Same as Figure~\ref{fig:TERRAoccur1D} with inclusion of
  planets in multi planet systems. The blue regions represent the
  additional contribution to planet occurrence from the
  \cite{Batalha12} planets. The addition of these new planets does not
  change the overall shape of the distribution. The dashed line is the
  power law fit to the planet size distribution in H12. The fit agrees
  qualitatively for \Rp > 2 \Re, but not within errors. We expect the
  H12 fit to be $\sim25\%$ higher than our occurrence measurements
  since H12 included planets with \Per~<~50~days (not
  \Per~=~5--50~days). Additional discrepancies between occurrence in
  H12 and this work could stem from different characterizations of
  completeness, reliance on photometric versus spectroscopic
  measurements of \Rstar, and magnitude-limited, rather than
  noise-limited, samples.}
\label{fig:COMBoccur1D}
\end{figure}

We then add \fcellBa to $\fcell$ computed in the previous section. We
show occurrence computed using \TERRA and \cite{Batalha12} planets as
a function of \Per and \Rp in Figure~\ref{fig:COMBoccur2D} and as a
function of only \Rp in Figure~\ref{fig:COMBoccur1D}. The 32
additional planets from \cite{Batalha12} do not change the overall
shape of the occurrence distribution: rising from 4.0 to 2.8~\Re and
consistent with flat from 2.8 down to 1.0~\Re.

H12 fit occurrence for \Rp > 2~\Re with a power law, 
\begin{equation}
\frac{d f }{ d \log \Rp} = k_{R} \Rp^{\alpha}, 
\end{equation}
finding $\alpha = -1.92\pm0.11$ and $k_{R} = 2.9^{+0.5}_{-0.4}$
(Section 3.1 of H12). As a point of comparison, we plot the H12 power
law over our combined occurrence distribution in
Figure~\ref{fig:COMBoccur1D}. The fit agrees qualitatively for \Rp > 2
\Re, but not within errors. We expect the H12 fit to be $\sim25\%$
higher than our occurrence measurements since H12 included planets
with \Per~<~50~days (not \Per~=~5--50~days). Additional discrepancies
could stem from different characterizations of completeness, reliance
on photometric versus spectroscopic measurements of \Rstar, and
magnitude-limited, rather than noise-limited, samples.

\section{Discussion}
\label{sec:Discussion}
\subsection{TERRA}

We implement in this work a new pipeline for the detection of
transiting planets in \Kepler photometry and apply it to a sample of
12,000 G and K-type dwarfs stars chosen to be among the most
photometrically quiet of the \Kepler target stars.  These low noise
stars offer the best chance for the detection of small, Earth-size
planets in the \Kepler field and will one day be among the stars from
which $\eta_{\oplus}$---the fraction of Sun-like stars bearing
Earth-size planets in habitable zone orbits---is estimated. In this
work, we focus on the close-in planets having orbital periods of 5--50
days and semi-major axes $\lessapprox$~0.25 AU.  Earth-size planets
with these characteristics are statistically at the margins of
detectability with the current $\sim$ 3 years of photometry in \Kepler
quarters Q1--Q12.

Our \TERRA pipeline has two key features that enable confident
measurement of the occurrence of close-in planets approaching Earth
size. First, \TERRA calibrates the \Kepler photometry and searches for
transit signals independent of the results from the \Kepler Mission's
official pipeline.  In some cases, \TERRA calibration achieves
superior noise suppression compared to the Pre-search Data
Conditioning (PDC) module of the official \Kepler pipeline
\citep{Petigura12}. The transit search algorithm in \TERRA is
efficient at detecting low SNR transits in the calibrated light
curves.  This algorithm successfully rediscovers 82 of 86 stars
bearing planets in \cite{Batalha12}. Recall that the current version
of \TERRA only detects the highest SNR transit signal in each system.
Thus, additional planets orbiting known hosts are not reported here.
We report the occurrence of stars having one or more planets, not the
mean number of planets per star as in H12 and elsewhere. Our pipeline
also detects 37 planets not found in \cite{Batalha12} (19 of which
were not in the catalog of the \Kepler team as of August \nth{8},
2012), albeit with the benefit of 6 quarters of additional photometry
for \TERRA to search.

The second crucial feature of \TERRA is that we have characterized its
detection completeness via the injection and recovery of synthetic
transits in real \Kepler light curves from the Best12k
sample. \textit{This completeness study is crucial to our occurrence
  calculations because it allows us to statistically correct for
  incompleteness variations across the \Per--\Rp plane.}  While the
\Kepler Project has initiated a completeness study of the official
pipeline \citep{Christiansen12}, \TERRA is the only pipeline for
\Kepler photometry whose detection completeness has been calibrated by
injection and recovery tests.  Prior to \TERRA, occurrence
calculations required one to \textit{assume} that the \Kepler planet
detections were complete down to some SNR limit, or to estimate
completeness based on SNR alone without empirical tests of the
performance of the algorithms in the pipeline. For example, H12 made
cuts in stellar brightness ($Kp$ < 15) and transit SNR ($>$ 10 in a
single quarter of photometry) and restricted their search to planets
larger than 2 \Re with orbital periods shorter than 50 days. These
conservative cuts on the planet and star catalogs were driven by the
unknown completeness of the official \Kepler pipeline at low SNR.  H12
applied two statistical corrections to convert their distribution of
detected planets into an occurrence distribution. They corrected for
non-transiting planets with a geometric a/\Rstar correction.  They
also computed the number of stars amenable to the detection (at
SNR~>~10 in a single quarter) of each planet and considered only that
number of stars in the occurrence calculation.  H12 had no empirical
way to determine the actual detection efficiency of the algorithms in
the pipeline. Here, we apply the geometric a/\Rstar correction and
correct for pipeline completeness across the \Per--\Rp plane by
explicit tests of the \TERRA pipeline efficiency, which naturally
incorporates an SNR threshold correction as in H12.

\subsection{Planet Occurrence}
H12 found that for close-in planets, the planet radius function rises
steeply from Jupiter size to 2 \Re.  For smaller planets of
$\sim$~1--2 \Re, occurrence was approximately constant in logarithmic
\Rp bins, but H12 were skeptical of the result below 2 \Re because of
unknown pipeline completeness and the small number statistics near 1
\Re in the \cite{Borucki11} planet catalog. In this work, we strongly
confirm the power law rise in occurrence from 4 to 2 \Re using a
superior assessment of completeness and nine times more photometry
than in H12.  Using \TERRA, we can empirically and confidently compute
occurrence down to 1~\Re. Our key result is the plateau of planet
occurrence for the size range 1--2.8 \Re for planets having orbital
periods 5--50~days around Sun-like stars.  In that size range of
1--2.8~\Re, 23\% of stars have a planet orbiting with periods between
5 and 50 days.  Including the multiple planets within each system, we
find 0.28 planets per star within the size range 1--2.8~\Re and with
periods between 5--50~days. These results apply, of course, to the
\Kepler field, with its still unknown distribution of masses, ages,
and metallicities in the Galactic disk.

As shown in Figure~\ref{fig:TERRAoccur2D}, \TERRA detects many
sub-Earth size planets (<~1.0~\Re). These sub-Earths appear in regions
of low completeness, and, provocatively, may represent just the tip of
the iceberg. A rich population of sub-Earths may await discovery given
more photometry and continued pipeline improvements. With 8~years of
total photometry in an extended \Kepler mission (compared to 3 years
here), the computational machinery of \TERRA---including its light
curve calibration, transit search, and completeness calibration---will
enable a measurement of $\eta_{\oplus}$ for habitable zone orbits.

\subsection{Interpretation}
We are not the first to note the huge population of close-in planets
smaller or less massive than Neptune.  Using Doppler surveys,
\cite{Howard10} and \cite{Mayor11} showed that the planet mass
function rises steeply with decreasing mass, at least for close-in
planets. In \Kepler data, the excess of close-in, small planets was
obvious in the initial planet catalogs released by the \Kepler Project
\citep{Borucki11a,Borucki11}.  H12 characterized the occurrence
distribution of these small planets as a function of their size,
orbital period, and host star temperature.  These occurrence
measurements, based on official \Kepler planet catalogs, were refined
and extended by \cite{Youdin11}, \cite{Traub12}, \cite{Dong12},
\cite{Beauge13}, and others.  Our contribution here shows a clear
plateau in occurrence in the 1--2.8 \Re size range and certified by an
independent search of Kepler photometry using a pipeline calibrated by
injection and recovery tests.  The onset of the plateau at $\sim$ 2.8
\Re suggests that there is a preferred size scale for the formation of
close-in planets.

H12 and \cite{Youdin11} noted falling planet occurrence for periods
shorter than $\sim$ 7 days. We also observe declining planet
occurrence for short orbital periods, but find that the transition
occurs closer to $\sim$ 10 days. We consider our period distribution
to be in qualitative agreement with those of H12 and
\cite{Youdin11}. Planet formation and/or migration seems to discourage
very close-in planets (\Per $\lessapprox$ 10 days).

Close-in, small planets are now the most abundant planets detected by
current transit and Doppler searches, yet they are absent from the
solar system. The solar system is devoid of planets between 1 and
3.88~\Re (Earth and Neptune) and planets with periods less than
Mercury's (\Per < 88.0 days). The formation mechanisms and possible
subsequent migration of such planets are hotly debated.  The
population synthesis models of \cite{Ida10} and \cite{Mordasini12}
suggest that they form near or beyond the ice line and then migrate
quiescently in the protoplanetary disk.  These models follow the
growth and migration of planets over a wide range of parameters (from
Jupiter mass down to Earth mass orbiting at distances out to $\sim10$
AU) and they predict ``deserts'' of planet occurrence that are not
detected.

More recently, \cite{Hansen12} and \cite{Chiang12} have argued for the
{\em in situ} formation of close-in planets of Neptune size and
smaller.  In these models, close-in rocky planets of a few Earth
masses form from protoplanetary disks more massive than the minimum
mass solar nebula.  Multiple planets per disk form commonly in these
models and accretion is fast ($\sim 10^5$ years) and efficient due to
the short dynamical timescales of close-in orbits.  The rocky cores
form before the protoplanetary disk has dissipated, accreting nebular
gas that adds typically $\sim$ 3\% to the mass of the planet
\citep{Chiang12}.  But the small amounts of gas can significantly
swell the radii of these otherwise rocky planets.  For example,
\cite{Adams08} found that adding a H/He gas envelope equivalent to
0.2--20\% of the mass of a solid 5 $M_E$ planet increases the radius
8--110\% above the gas-free value.

We find the {\em in situ} model plausible because it naturally
explains the large number of close, sub-Neptune-size planets, the high
rate of planet multiplicity and nearly co-planar and circular orbits
\citep{Lissauer11, Fang12}, and does not require tuning of planet
migration models. Our result of a plateau in the planet size
distribution for 1--2.8~\Re with a sharp falloff in occurrence for
larger planets along with decreasing occurrence for \Per $\lessapprox$
10~days are two significant observed properties of planets around
Sun-like stars that must be reproduced by models that form planets
{\em in situ} or otherwise and by associated population synthesis
models.

The {\em in situ} model seems supported by the sheer large occurrence
of sub-Neptune-size planets within 0.25~AU. It seems unlikely that all
such planets form beyond the snow line at $\sim$ 2~AU, which would
require inward migration to within 0.25~AU, but not all the way into
the star. Such models of formation beyond the snow line seem to
require fine tuning of migration and parking mechanisms, as well as
the tuning of available water or gas beyond 2~AU, while avoiding
runaway gas accretion toward Jupiter masses. Still, {\em in situ}
formation seems to require higher densities than those normally
assumed in a minimum mass solar nebula \citep{Chiang12} in order to
form the sub-Neptune planets before removal of the gas. If this {\em
  in situ} model is correct, we expect these sub-Neptune-size planets
to be composed of rock plus H and He, rather than rock plus water
\citep{Chiang12}. Thus, a test of the {\em in situ} mode of formation
involves spectroscopic measurements of the chemical composition of the
close-in sub-Neptunes.

\begin{landscape}
\begin{deluxetable}{@{\extracolsep{-6pt}}l*{14}{r}ccc}
\tabletypesize{\footnotesize}
\tablecaption{Planet candidates identified with \TERRA}
\tablewidth{0pt}
\tablehead{
  \colhead{} & 
  \multicolumn{8}{c}{Light Curve Fit}   & 
  \colhead{} &
  \multicolumn{3}{c}{Stellar Parameters} &
  \colhead{}\\ 
  \cline{2-9}\cline{11-13}\\[-1.5ex]
  \colhead{KIC}	& 
  \colhead{\Per} &
  \colhead{$\ep\tablenotemark{a}$} &  
  \colhead{$\rratfrac$ } &
  \colhead{$\sigma(\rratfrac)$ } &   
  \colhead{$\tau$} &   
  \colhead{$\sigma(\tau)$} &   
  \colhead{ $b$ } &   
  \colhead{ $\sigma(b)$ } &
  \colhead{} &
  \colhead{\teff} &
  \colhead{\logg} &
  \colhead{\Rstar} &
  \colhead{\Rp} & 
  \colhead{$\sigma(\Rp)$} & 
  \colhead{source\tablenotemark{b}} &
  \colhead{FP} &
  \colhead{B12} \\
  \colhead{} & 
  \colhead{(d)} &  
  \colhead{(d)}	&  
  \colhead{ (\%) } &
  \colhead{} &   
  \colhead{(hrs)} &   
  \colhead{} &   
  \colhead{} &  
  \colhead{} &
  \colhead{} &
  \colhead{(K)} &
  \colhead{(cgs)} &
  \colhead{(\Rsun)} &
  \colhead{(\Re)} &
  \colhead{} & 
  \colhead{} & 
  \colhead{} &
  \colhead{} \\[-2.5ex]
}
\startdata
\input{terra50d/tab1.tex}
\enddata
\tablecomments{Orbital period, \Per; time of transit center, \ep;
  planet-to-star radius ratio, \rrat; the time for the planet to
  travel \Rstar during transit, $\tau$; and transit impact parameter,
  $b$ are all determined from the \cite{Mandel02} light curve fit. By
  default, stellar parameters \Rstar, \teff, and \logg come from
  \SpecMatch. If \SpecMatch parameters do not exist, parameters are
  taken from the corrected KIC values, described in
  Section~\ref{sec:TERRAplanetYield}. The FP column lists whether a
  candidate was designated a false positive by the \Kepler team
  (`Y'--yes, `N'--no, ` '--no designation). The B12 column lists
  whether a candidate was present in \cite{Batalha12}.}
\tablenotetext{a}{Time of transit center (BJD-2454900).}
\tablenotetext{b}{Source of stellar parameters: `S'--{\tt
    SpecMatch}-derived parameters using Keck HIRES spectra,
  `P1'--photometrically-derived parameters from \cite{Batalha12},
  `P2'--photometrically-derived parameters computed by the
  authors. See Section~\ref{sec:TERRAplanetYield} for more details. }
\label{tab:planetsTERRA}
\end{deluxetable}

\begin{deluxetable}{@{\extracolsep{-6pt}}l *{14}{r} c }
  \tabletypesize{\footnotesize} \tablecaption{New candidates
    identified with \TERRA} \tablewidth{0pt} \tablehead{ \colhead{} &
    \multicolumn{8}{c}{Light Curve Fit} & \colhead{} &
    \multicolumn{3}{c}{Stellar Parameters} & \colhead{} \\
    \cline{2-9}\cline{11-13} \\[-1.5ex]
    \colhead{KIC} & 
    \colhead{\Per} &  
    \colhead{$\ep\tablenotemark{a}$} &  
    \colhead{$\rratfrac$ } &
    \colhead{$\sigma(\rratfrac)$ } &   
    \colhead{$\tau$} &   
    \colhead{$\sigma(\tau)$} &   
    \colhead{$b$ } &   
    \colhead{$\sigma(b)$ } &
    \colhead{} &
    \colhead{\teff} &
    \colhead{\logg} &
    \colhead{\Rstar} &
    \colhead{\Rp} & 
    \colhead{$\sigma(\Rp)$} & 
    \colhead{source\tablenotemark{b}}  \\
    \colhead{} & 
    \colhead{(d)} &  
    \colhead{(d)} &  
    \colhead{ (\%) } &
    \colhead{} &   
    \colhead{ (hrs)} &   
    \colhead{} &   
    \colhead{} &  
    \colhead{} &
    \colhead{} &
    \colhead{(K)} &
    \colhead{(cgs)} &
    \colhead{(\Rsun)} &
    \colhead{(\Re)} &
    \colhead{} & 
    \colhead{} \\[-2.5ex]
}
\startdata
\input{terra50d/tab3.tex}
\enddata

\tablecomments{The 37 \TERRA candidates not in \cite{Batalha12} and
  not listed as false positives by the \Kepler team. The column
  definitions are the same as in Table~\ref{tab:planetsTERRA}}

\label{tab:planetsTERRAnew}
\end{deluxetable}

\end{landscape}

\begin{deluxetable}{l*{7}{r}}
\tabletypesize{\footnotesize}
\setlength{\tabcolsep}{0.15in}
\tablecaption{Union of \cite{Batalha12} and \TERRA planet candidate catalogs}
\tablewidth{0pt}
\tablehead{
	\colhead{} &
	\multicolumn{3}{c}{Batalha} &
	\colhead{}&
	\multicolumn{2}{c}{\TERRA} \\
	\cline{2-4}\cline{6-7} \\[-1.5ex]
	\colhead{KIC} & 
	\colhead{KOI}&  
	\colhead{\Per}&
	\colhead{\Rp}&   
	\colhead{} & 
	\colhead{\Per} &  
	\colhead{\Rp} & \\[-2.5ex]
}
\startdata
\input{terra50d/tab2.tex}
\enddata
\label{tab:planetsJOIN}
\tablecomments{All \cite{Batalha12} candidates with \Per~=~5--50~days
  belonging to stars in the Best12k sample are included. Candidates
  are considered equal if they belong to the same star and the periods
  in each catalog agree to better than 0.01 days. Eighty-two
  candidates appear in both catalogs, 33 appear in \cite{Batalha12}
  only, and 47 appear in this work only (although 10 were listed as
  false positives by the \Kepler team). Differences in \Rp between the
  two catalogs stem from different values of \Rstar. Most \TERRA
  planet candidates have \SpecMatch-derived stellar parameters which
  are more accurate than \cite{Batalha12} parameters, which were
  derived from KIC broadband photometry.}
\end{deluxetable}

\begin{figure}[htbp]
\begin{center}
\includegraphics[width=1\textwidth]{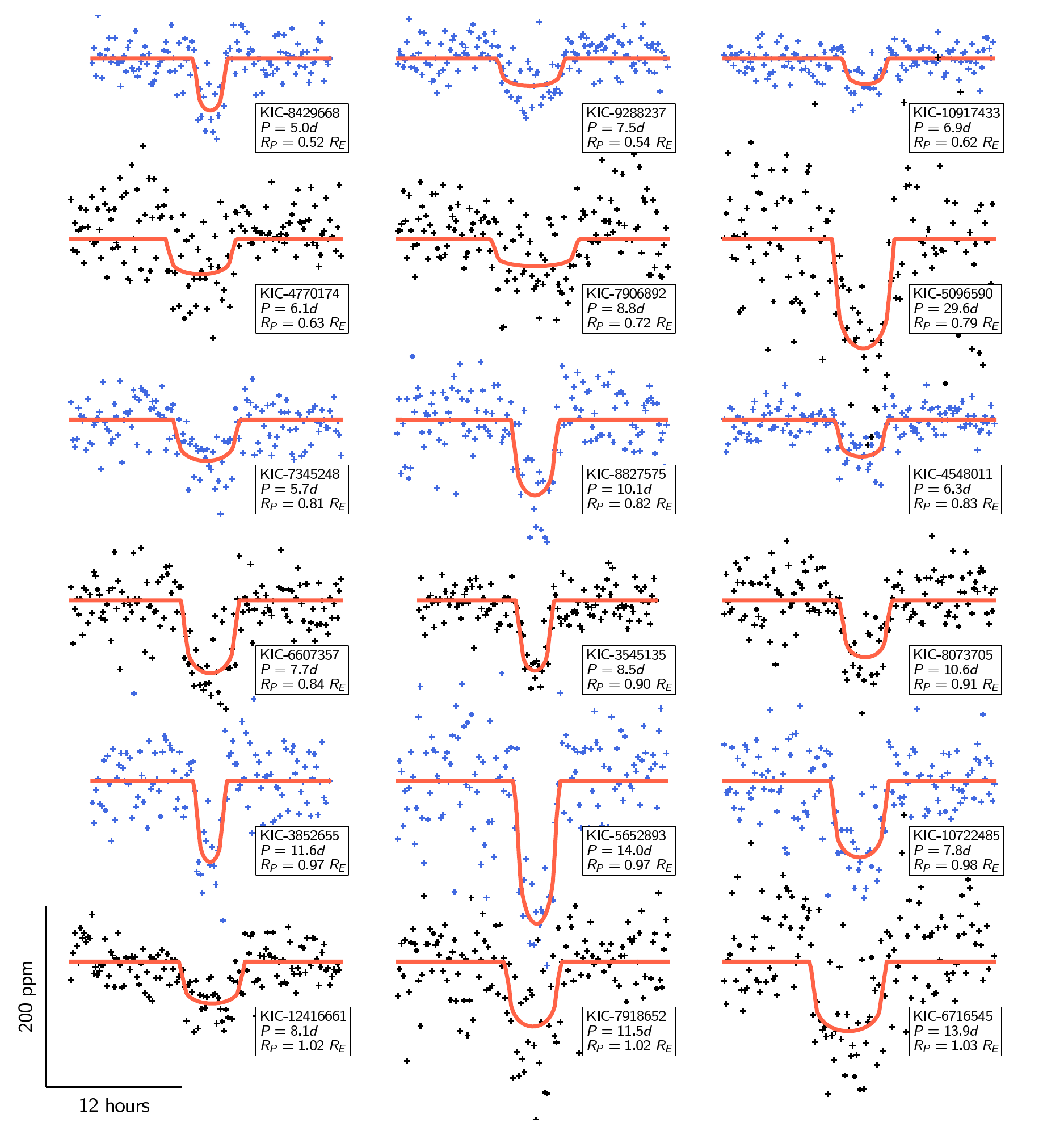}
\caption[Transits of \TERRA planet candidates, not in \cite{Batalha12}]{Phase-folded photometry for 18 of the 37 \TERRA planet
  candidates, not in \cite{Batalha12}, ordered according to size. For
  clarity, we show median photometric measurements in 10~min bins. The
  red lines are the best-fitting \cite{Mandel02} model.}
\label{fig:binned1}
\end{center}
\end{figure}

\begin{figure}[htbp]
\begin{center}
\includegraphics[width=1\textwidth]{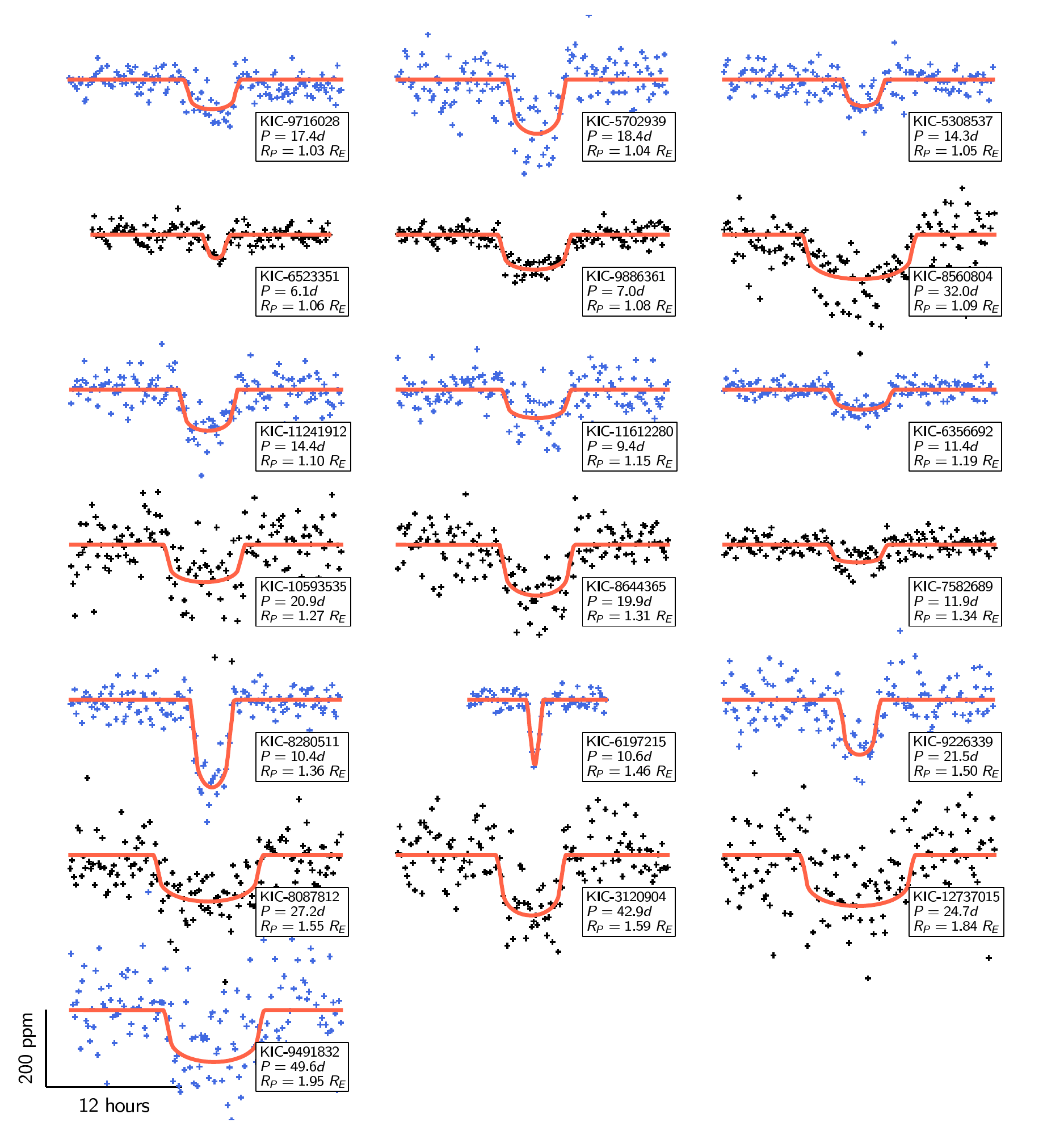}
\caption[Transits of \TERRA planet candidates, not in \cite{Batalha12}]{Same as Figure~\ref{fig:binned1}, but showing the remaining
  19 of the 37 \TERRA planet candidates, not in \cite{Batalha12}.}
\label{fig:binned2}
\end{center}
\end{figure}

%% file: terra50d/tab1.tex
  2142522 &  13.323 &     67.043 &  0.98 &  0.34 &  2.32 &  0.51 &  < 0.85 &       &        &  6046 &  4.40 &   1.04 &  1.11 &  0.39 &     P1 &     Y &   N \\
  2307415 &  13.122 &     66.396 &  1.26 &  0.13 &  1.90 &  0.12 &  < 0.52 &       &        &  6133 &  4.38 &   1.14 &  1.57 &  0.16 &      S &     N &   Y \\
  2441495 &  12.493 &     71.457 &  2.38 &  0.24 &  1.13 &  0.03 &  < 0.47 &       &        &  5192 &  4.56 &   0.76 &  1.99 &  0.20 &      S &     N &   Y \\
  2444412 &  14.911 &     74.337 &  3.38 &  0.34 &  4.24 &  0.31 &    0.93 &  0.01 &        &  5551 &  4.47 &   0.92 &  3.41 &  0.34 &      S &     N &   Y \\
  2571238 &   9.287 &     68.984 &  2.90 &  0.29 &  3.72 &  0.38 &    0.90 &  0.03 &        &  5544 &  4.50 &   0.89 &  2.82 &  0.28 &      S &     N &   Y \\
  2853446 &   7.373 &     70.613 &  1.39 &  0.14 &  0.74 &  0.06 &  < 0.70 &       &        &  5969 &  4.37 &   1.10 &  1.65 &  0.17 &      S &     N &   Y \\
  3098810 &  40.811 &     75.673 &  1.98 &  0.20 &  1.45 &  0.12 &  < 0.70 &       &        &  6071 &  4.31 &   1.27 &  2.75 &  0.28 &      S &     N &   Y \\
  3120904 &  42.915 &     72.866 &  1.15 &  0.12 &  2.96 &  0.36 &  < 0.66 &       &        &  6151 &  4.31 &   1.26 &  1.59 &  0.16 &      S &       &   N \\

%% file: terra50d/tab3.tex
  3120904 &  42.915 &     72.866 &  1.15 &  0.12 &  2.96 &  0.36 &  < 0.66 &       &        &  6151 &  4.31 &   1.26 &  1.59 &  0.16 &      S \\
  3545135 &   8.483 &     65.973 &  0.81 &  0.08 &  1.44 &  0.10 &  < 0.57 &       &        &  5794 &  4.40 &   1.02 &  0.90 &  0.09 &      S \\
  3852655 &  11.629 &     65.817 &  0.85 &  0.30 &  1.49 &  0.23 &  < 0.79 &       &        &  5822 &  4.36 &   1.05 &  0.97 &  0.34 &     P1 \\
  4548011 &   6.284 &     66.198 &  0.60 &  0.06 &  2.23 &  0.64 &  < 0.49 &       &        &  5991 &  4.30 &   1.26 &  0.83 &  0.09 &      S \\
  4770174 &   6.096 &     67.600 &  0.57 &  0.20 &  3.08 &  0.44 &  < 0.75 &       &        &  6013 &  4.44 &   1.01 &  0.63 &  0.22 &     P1 \\
  5096590 &  29.610 &     70.332 &  1.00 &  0.35 &  2.60 &  0.20 &  < 0.67 &       &        &  5623 &  4.63 &   0.73 &  0.79 &  0.28 &     P1 \\
  5308537 &  14.265 &     76.217 &  0.76 &  0.08 &  1.92 &  0.53 &  < 0.87 &       &        &  5831 &  4.29 &   1.26 &  1.05 &  0.11 &      S \\
  5652893 &  14.010 &     65.959 &  1.15 &  0.13 &  1.98 &  0.49 &  < 0.85 &       &        &  5150 &  4.55 &   0.78 &  0.97 &  0.11 &      S \\

%% file: terra50d/tab2.tex
  2142522 &          &        &         &        &  13.32 &   1.11 \\
  2307415 &  2053.01 &  13.12 &    1.65 &        &  13.12 &   1.57 \\
  2441495 &   166.01 &  12.49 &    2.70 &        &  12.49 &   1.99 \\
  2444412 &   103.01 &  14.91 &    2.97 &        &  14.91 &   3.41 \\
  2571238 &    84.01 &   9.29 &    2.53 &        &   9.29 &   2.82 \\
  2853446 &  1118.01 &   7.37 &    2.52 &        &   7.37 &   1.65 \\
  3098810 &  1878.01 &  40.81 &    3.24 &        &  40.81 &   2.75 \\
  3120904 &          &        &         &        &  42.91 &   1.59 \\

%% file: terra1yr/terra1yr.tex
\renewcommand{\nc}[2]{\newcommand{#1}{\ensuremath{#2}\xspace}}
\newcommand{\num}[2]{\newcommand{#1}{{#2}\xspace}}
\input{terra1yr/numbers.tex}

\input{terra1yr/nc.tex}

%%%%%%%%%%%%%%%%%%%%%%%%%%%%%%
%% Don't type in anything in the following section:
%%%%%%%%%%%%
%% For PNAS Only:
%\contributor{Submitted to Proceedings
%of the National Academy of Sciences of the United States of America}
%\url{www.pnas.org/cgi/doi/10.1073/pnas.0709640104}
%\copyrightyear{2008}
%\issuedate{Issue Date}
%\volume{Volume}
%\issuenumber{Issue Number}
%%%%%%%%%%%%

%\begin{document}

%%%%%%%%%%%%%%%%%%%%%%%%%%%%%%

%% For titles, only capitalize the first letter
\chapter{The Prevalence of Earth-Size Planets Orbiting Sun-Like Stars}
\label{c.terra1yr}

%% Enter authors via the \author command.  
%% Use \affil to define affiliations.
%% (Leave no spaces between author name and \affil command)

%% Note that the \thanks{} command has been disabled in favor of
%% a generic, reserved space for PNAS publication footnotes.

%\author{Erik A Petigura%
%	    \affil{1}{University of California, Berkeley}%
%	    \affil{2}{University of Hawaii, Manoa},
%        Andrew W Howard\affil{2}{},
%        \and
%        Geoffrey W Marcy\affil{1}{}
%        }
%
%\contributor{Submitted to Proceedings of the National Academy of Sciences
%of the United States of America}
%
%%% The \maketitle command is necessary to build the title page.
%\maketitle
%
%%%%%%%%%%%%%%%%%%%%%%%%%%%%%%%%%%%%%%%%%%%%%%%%%%%%%%%%%%%%%%%%

%% Optional for entering abbreviations, separate the abbreviation from
%% its definition with a comma, separate each pair with a semicolon:
%% for example:
%% \abbreviations{SAM, self-assembled monolayer; OTS,
%% octadecyltrichlorosilane}

% \abbreviations{}

%% The first letter of the article should be drop cap: \dropcap{}
%\dropcap{I}n this article we study the evolution of ''almost-sharp'' fronts

%% Enter the text of your article beginning here and ending before
%% \begin{acknowledgements}
%% Section head commands for your reference:
%% \section{}
%% \subsection{}
%% \subsubsection{}

\noindent A version of this chapter was previously published in the {\em Proceedings of the National Academy of Science}
(Erik~A.~Petigura, Andrew~W.~Howard, \& Geoffrey~W.~Marcy,  2013, PNAS 110, 19273).\\

NASA's \Kepler mission was launched in 2009 to search for
planets that transit (cross in front of) their host stars
\cite{Borucki10, Koch10, Borucki11,Batalha12}.  The resulting dimming
of the host stars is detectable by measuring their brightness, and
\Kepler monitored the brightness of 150,000 stars every 30 minutes for
four years.  To date, this exoplanet survey has detected more than
3000 planet candidates \cite{Batalha12}.

The most easily detectable planets in the \Kepler survey are those
that are relatively large and orbit close to their host stars,
especially those stars having lower intrinsic brightness fluctuations
(noise).  These large, close-in worlds dominate the list of known
exoplanets.  But the \Kepler brightness measurements can be analyzed
and debiased to reveal the diversity of planets, including smaller
ones, in our Milky Way Galaxy \cite{Howard12, Petigura13, Fressin13}.
These previous studies showed that small planets approaching
Earth-size are the most common, but only for planets orbiting
{\it close} to their host stars.  Here, we extend the planet survey to
{\it Kepler's} most important domain -- Earth-size planets orbiting far
enough from Sun-like stars to receive a similar intensity of light
energy as Earth.

\section{Planet Survey}
We performed an independent search of \Kepler photometry for
transiting planets with the goal of measuring the underlying
occurrence distribution of planets as a function of orbital period,
$\Per$, and planet radius, \Rp.  We restricted our survey to a set of
Sun-like stars (``GK-type'') that are the most amenable to the
detection of Earth-size planets.  We define GK-type stars as those
with surface temperatures \teff~=~4100--6100~K and gravities
\logg~=~4.0--4.9 (cgs) \cite{Pinsonneault12}.  Our search for planets
was further restricted to the brightest Sun-like stars observed by
\Kepler ($Kp$~=~10--15~mag).  These 42557 stars (``Best42k'') have the
lowest photometric noise making them amenable to the detection of
Earth-size planets.  When a planet crosses in front of its star, it
causes a fractional dimming that is proportional to the fraction of
the stellar disk blocked, $\df = (\Rp / \Rstar)^{2}$, where \Rstar is
the radius of the star.  As viewed by a distant observer, the Earth dims the Sun
by $\sim$100 parts per million (ppm) lasting 12 hours every 365 days.

% \section*{Planet Detection Pipeline}
We searched for transiting planets in \Kepler brightness measurements
using our custom-built \TERRA software package described in previous
work \cite{Petigura12,Petigura13} and in the Supporting Information
(SI).  In brief, \TERRA conditions \Kepler photometry in the
time-domain, removing outliers, long timescale variability (> 10
day), and systematic errors common to a large number of
stars.  \TERRA then searches for transit signals by evaluating the
signal-to-noise ratio (SNR) of prospective transits over a
finely-spaced three-dimensional grid of orbital period, \Per, time of
transit, \ep, and transit duration, \tdur.  This grid-based search
extends over the orbital period range 0.5--400~days.

\TERRA produced a list of ``Threshold Crossing Events'' (TCEs) that
meet the key criterion of a photometric dimming signal-to-noise ratio,
SNR > 12.  Unfortunately, an unwieldy 16227 TCEs met this criterion,
many of which are inconsistent with the periodic dimming profile from
a true transiting planet.  Further vetting was performed by
automatically assessing which light curves were consistent with
theoretical models of transiting planets \cite{Mandel02}.  We also
visually inspected each TCE light curve, retaining only those
exhibiting a consistent, periodic, box-shaped dimming, and rejecting
those caused by single epoch outliers, correlated noise, and other
data anomalies.  The vetting process was applied homogeneously
to all TCEs and is described in further detail in the SI.

To assess our vetting accuracy, we evaluated the 235 Kepler Objects of
Interest (KOIs) among Best42k stars having \Per~>~50~days which had been found by the
\Kepler Project and identified as planet ``candidates'' in the
official Exoplanet Archive.%
\footnote{URL:exoplanetarchive.ipac.caltech.edu (accessed 19 September
  2013)}
Among them, we found four whose light curves are not
consistent with being planets.  These four KOIs (364.01, 2224.02,
2311.01, and 2474.01) have long periods and small radii (see SI).
This exercise suggests that our vetting process is robust, and that
careful scrutiny of the light curves of small planets in long period
orbits is useful to identify false positives.

Vetting of our TCEs produced a list of \neKOI eKOIs, which are
analogous to KOIs produced by the \Kepler Project. Each light curve is
consistent with an astrophysical transit, but could be due to an
eclipsing binary (EB), either in the background or gravitationally
bound, instead of a transiting planet.  If an EB resides within the
software aperture of a \Kepler target star (within $\sim$10 arcsec),
the dimming of the EB can masquerade as a planet transit when diluted
by the bright target star.  We rejected as likely EBs any eKOIs with
these characteristics: radii larger than 20 \Re, observed secondary
eclipse, or astrometric motion of the target star in and out of
transit (SI).  This rejection of EBs left 603 eKOIs in our catalog.

\Kepler photometry can be used to measure \rrat with high precision,
but the extraction of planet radii is compromised by poorly known
radii of the host stars \cite{Brown11}.  To determine \Rstar and
\teff, we acquired high-resolution spectra of \neKOISM eKOIs using
HIRES spectrometer on the 10-m Keck I telescope.  Notably, we obtained spectra of
all 62 eKOIs that have \Per > 100~days.   For these stars, the
$\sim$35\% errors in \Rstar were reduced to $\sim$10\% by matching
spectra to standards.

To measure planet occurrence, one must not only detect planets but
also assess what fraction of planets were missed.  Missed planets are
of two types, those whose orbital planes are so tilted as to avoid
dimming the star and those whose transits were not detected in the
photometry by \TERRA.  Both effects can be quantified to establish a
statistical correction factor.  The first correction can be computed
as the geometrical probability that an orbital plane is viewed edge-on
enough (from Earth) so that the planet transits the star.  This
probability is, $\PT = \Rstar/a$, where $a$ is the semi-major axis of
the orbit.  

\begin{figure}
\centering
\includegraphics[width=6in]{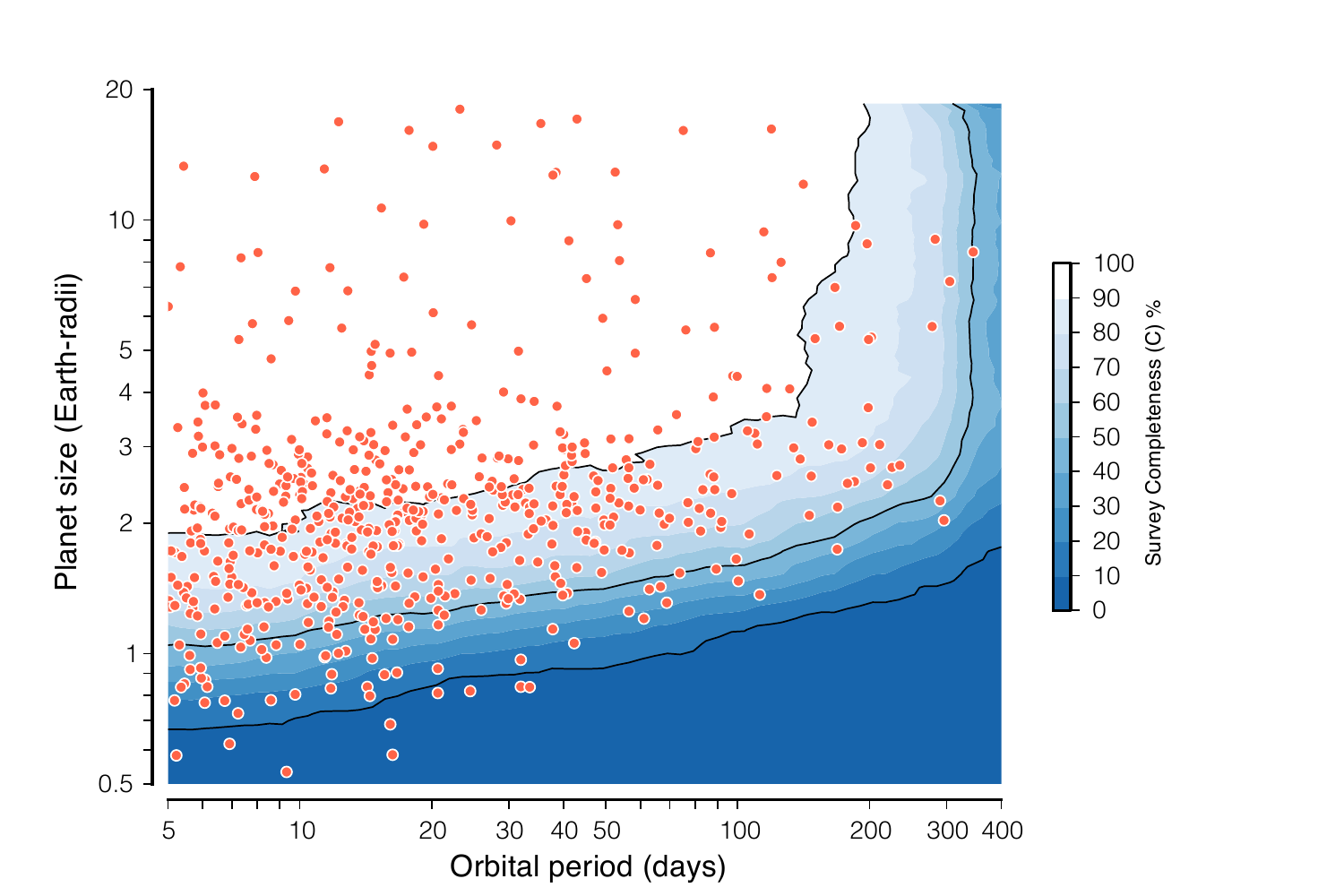}
\caption[Detected planets and survey completeness]{Two-dimensional domain of orbital period and planet size, on
  a logarithmic scale.  Red circles show the \nPlnt detected planets
  in our survey of \nSamp bright, Sun-like stars (\Kp~=~10--15~mag, GK
  spectral type). The color scale shows survey completeness measured
  by injection and recovery of synthetic planets into real
  photometry. Dark regions represent (\Per,\Rp) with low completeness,
  $C$, where significant corrections for missed planets must be made
  to compute occurrence. The most common planets {\em detected} have
  orbital $\Per < 20$~days and \Rp~$\approx$~1--3~\Re (at middle-left
  of graph).  But their detectability is favored by orbital tilt and
  detection completeness, $C$, that favors detection of such close-in,
  large planets.}
\label{fig:PlanetSample}
\end{figure}

The second correction is computed by the injection and recovery of
synthetic (mock) planet-caused dimmings into real \Kepler photometry.
We injected 40,000 transit-like synthetic dimmings having randomly
selected planetary and orbital properties into the actual photometry
of our Best42k star sample, with stars selected at random.  We
measured survey completeness, $C(\Per,\Rp)$, in small bins of
(\Per,\Rp), determining the fraction of injected synthetic planets
that were ``discovered'' by \TERRA (SI).  Fig.~\ref{fig:PlanetSample}
shows the 603 detected planets and the survey completeness, $C$, color-coded as a function of \Per and \Rp.

The survey completeness for small planets is a complicated function of
\Per and \Rp. It decreases with increasing \Per and decreasing \Rp as
expected due to fewer transits and less dimming, respectively.  It is
dangerous to replace this injection and recovery assessment with noise
models to determine $C$.  Such models are not sensitive to the
absolute normalization of $C$, only providing relative completeness.
Models also may not capture the complexities of a multistage
transit-finding pipeline that is challenged by correlated,
non-stationary, and non-Gaussian noise.  Measuring the occurrence of
small planets with long periods requires injection and recovery of
synthetic transits to determine the absolute detectability of the
small signals buried in noise.

\section{Planet Occurrence}
We define planet occurrence, $f$, to be the fraction of stars having a
planet within a specified range of orbital period, size, and perhaps other criteria.  We report planet occurrence as a function of planet size
and orbital period, $f(\Per,\Rp)$ and as a function of planet size and
the stellar light intensity (flux) incident on the planet, $f(\Fp,\Rp)$.

\subsection{Planet Occurrence and Orbital Period}
\begin{figure}
\centering
\includegraphics[width=6in]{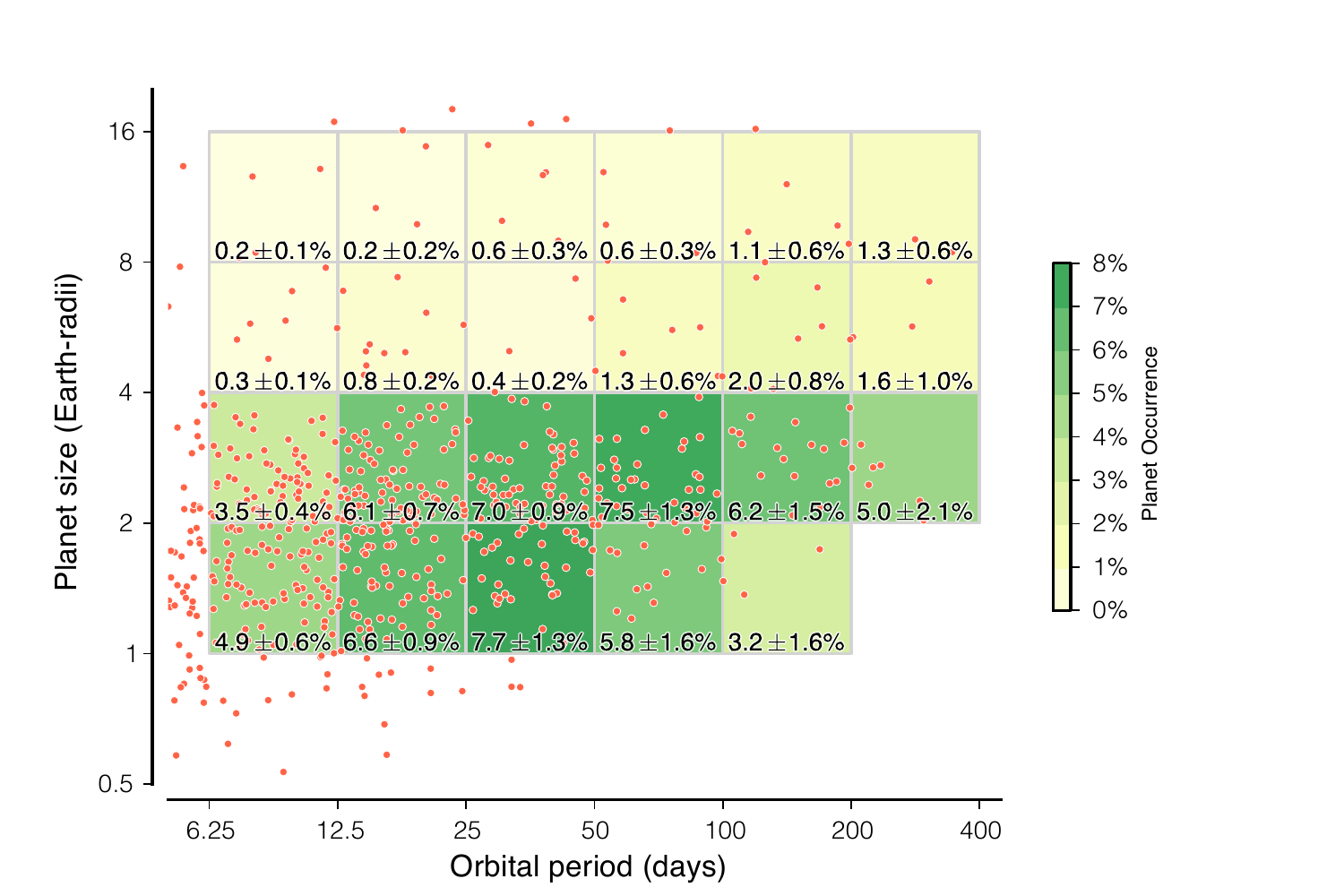}
\caption[Planet occurrence]{Planet occurrence, $f(\Per,\Rp)$, as a function of orbital
  period and planet radius for \Per~=~6.25--400~days and
  \Rp~=~0.5--16~\Re. As in Fig.~\ref{fig:PlanetSample}, detected
  planets are shown as red circles. Each cell spans a factor of two in
  orbital period and planet size. Planet occurrence in a cell is given
  by $f(\Per,\Rp) = 1/\nstar \sum_i a_i/(R_{\star,i}C_i)$ where the sum
  is over all detected planets within each cell.  Here, $a_i/R_i$ is
  the number of non-transiting planets (for each detected planet) due
  to large tilt of the orbital plane, $C_i = C(P_i,R_{P,i})$ is the
  detection completeness factor, and \nstar = \nSamp stars in the
  Best42k sample. Cells are colored according to planet occurrence
  within the cell. We quote planet occurrence within each cell.  We do
  not color cells where the completeness is less than 25\%. Among the
  small planets, 1--2 and 2--4 \Re, planet occurrence is constant
  (within a factor of two level) over the entire range of orbital
  period.  This uniformity supports mild extrapolation into the 
  \Per~=~200--400~day, \Rp~=~1--2~\Re domain.}
\label{fig:CheckerBoard}
\end{figure}

We computed $f(\Per,\Rp)$ in a $6\times4$ grid of \Per and \Rp shown
in Fig.~\ref{fig:CheckerBoard}. We start by first counting the number
of detected planets, \ncell, in each \Per-\Rp cell.  Then we computed
$f(\Per,\Rp)$ by making statistical corrections for planets missed
because of non-transiting orbital inclinations and because of the
completeness factor, $C$.  The first correction augments each detected
transiting planet by $1/\PT = a/\Rstar$, where \PT is the geometric
transit probability, to account for planets missed in inclined orbits.
Accounting for the completeness, $C$, the occurrence in a cell is
$f(\Per,\Rp)~=~1/\nstar \sum_i a_i/(R_{\star,i}C_i$), where
\nstar~=~\nSamp stars and the sum is over all detected planets within
that cell. Uncertainties in the statistical corrections for $a/R_{\star}$ and for completeness may cause errors in the final occurrence rates of $\sim$10\%. Such errors will be smaller than the Poisson uncertainties in the occurrence of Earth-size planets in long period orbits.

Fig.~\ref{fig:CheckerBoard} shows the occurrence of
planets, $f(\Per,\Rp)$, within the \Per-\Rp plane.  Each cell is
color-coded to indicate the final planet occurrence: the fraction of
stars having a planet with radius and orbital period corresponding to
that cell (after correction for both completeness factors).  For
example, $7.7\pm1.3\% $ of Sun-like stars have a planet with periods
between 25 and 50~days and sizes between 1 and 2 \Re.

\begin{figure}
\centerline{\includegraphics[width=1\textwidth]{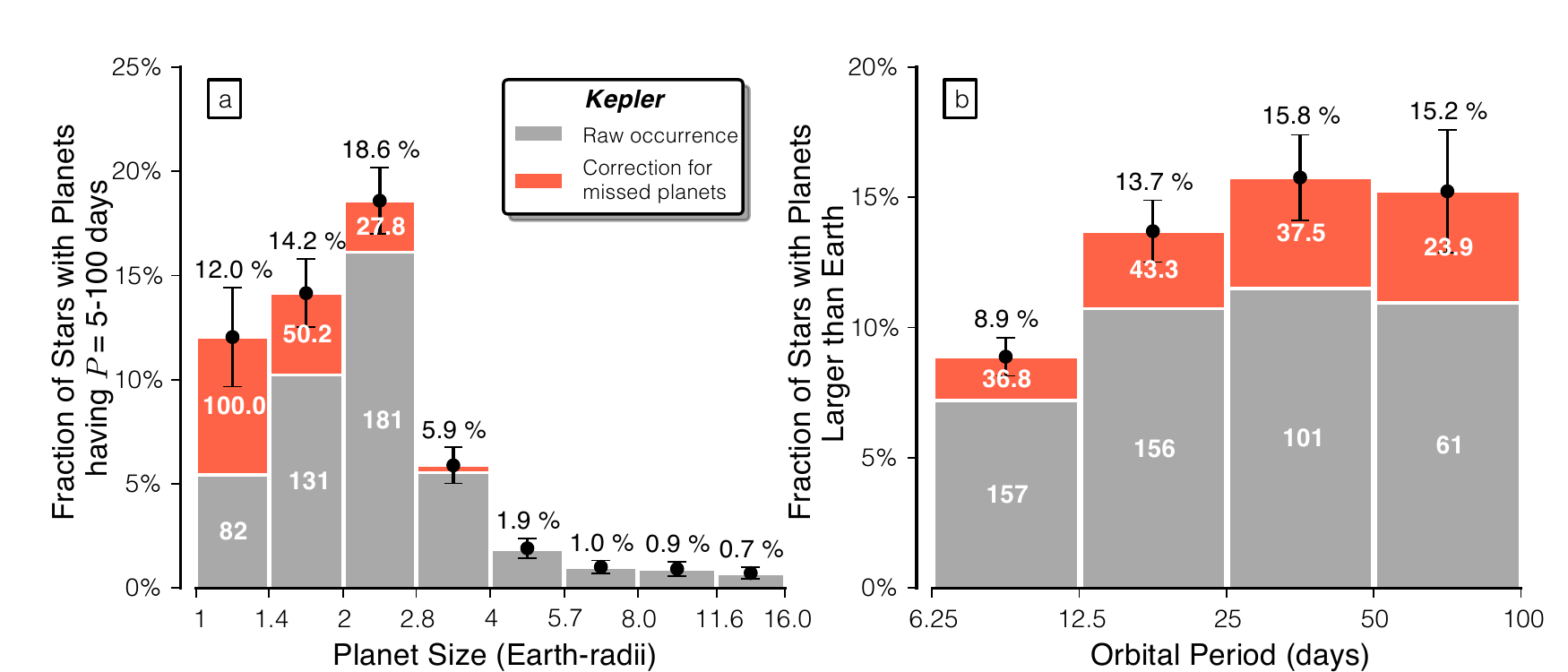}}
\caption[Distributions of planet sizes and orbital periods]{The measured distributions of planet sizes and orbital
  periods for \Rp~>~1~\Re and \Per~=~5--100 days.  Heights of the bars
  represent the fraction of Sun-like stars harboring a planet within a
  given \Per or \Rp domain. The gray portion of the bars show planet
  occurrence without correction for survey completeness, i.e. for
  $C=1$. The red region shows the correction to account for missed
  planets, 1/$C$. Bars are annotated to reflect the number of planets detected (gray bars) and missed (red barss)  The occurrence of planets of different sizes rises
  by a factor of 10 from Jupiter-size to Earth-sized planets.  The
  occurrence of planets with different orbital periods is constant,
  within 15\%, between 12.5 and 100~days. Due to the small number of
  detected planets with \Rp~=~1--2~\Re and \Per~>~100~days (four detected
  planets), we do not include \Per~>~100~days in these marginalized
  distributions.}
\label{fig:MargDistr}
\end{figure}

We compute the distribution of planet sizes, including all orbital
periods \Per~=~5--100~days, by summing $f(\Per,\Rp)$ over all periods.
The resulting planet size distribution is shown in
Fig~\ref{fig:MargDistr}a. Planets with orbital periods of 5--100 days
have a characteristic shape to their size distribution (Fig 3b).
Jupiter-sized planets (11~\Re) are rare, but the occurrence of planets
rises steadily with decreasing size down to about 2 \Re.  The
distribution is nearly flat (equal numbers of planets per $\log \Rp$
interval) for 1--2 \Re planets. {\em We find that $26\pm3\%$ of
  Sun-like stars harbor an Earth-size planet (1--2~\Re) with
  \Per~=~5--100~days,} compared to $1.6\pm0.4\%$ occurrence of
Jupiter-size planets (8--16~\Re).

We also computed the distribution of orbital periods, including all
planet sizes, by summing each period interval of $f(\Per,\Rp)$ over
all planet radii. As shown in Fig.~\ref{fig:MargDistr}b, the occurrence of planets larger than Earth
rises from $8.9\pm0.7\%$ in the \Per~=~6.25--12.5~day domain to
$13.7\pm1.2\%$ in the \Per~=~12.5--25~day interval and is consistent with constant for larger periods. This rise and plateau feature was observed
for $\gtrsim2~\Re$ planets in earlier work~\cite{Youdin11,Howard12}.

Two effects lead to minor corrections to our occurrence estimates.
First, some planets in multi-transiting systems are missed by \TERRA.
Second, a small number of eKOIs are false detections.  These two
effects are small, and they provide corrections to our occurrence
statistics with opposite signs.  To illustrate their impact, we
consider the small and long period (\Per > 50 days) planets that are the
focus of this study.

\TERRA detects the highest SNR transiting planet per system, so
additional transiting planets that cause lower SNR transits are not
included in our occurrence measurement.  Using the \Kepler Project
catalog (Exoplanet Archive), we counted the number of planets within
the same cells in \Per and \Rp as Fig.~\ref{fig:CheckerBoard}, noting
those that did not yield the highest SNR in the system.  Inclusion of
these second and third transiting planets boosts the total
number of planets per cell (and hence the occurrence) by 21--28\% over the 
\Per~=~50--400, \Rp~=~1--4~\Re domain (SI).

Even with our careful vetting of eKOIs, the light curves of some false
positives scenarios are indistinguishable from
planets. Fressin~et~al.~\cite{Fressin13} simulated the contamination
of a previous KOI~\cite{Batalha12} sample by false positives that were not removed by
the \Kepler Project vetting process.  They determined that the largest
source of false positives for Earth-size planets are physically bound
stars with a transiting Neptune-size planet with an overall false
positive rate of 8.8--12.3\%. As we have shown
(Fig.~\ref{fig:CheckerBoard}), the occurrence of Neptune-size planets
is nearly constant as a function of orbital period, in log \Per
intervals.  Thus, this false positive rate is also nearly constant in
period.  Therefore, we adopt a 10\% false positive rate for planets
having \Per~=~50--400~days and \Rp~=~1--2~\Re. Planet occurrence,
shown in Figs~\ref{fig:CheckerBoard} and \ref{fig:MargDistr}, has not
been adjusted to account for false positives or planet multiplicity.
The quoted errors reflect only binomial counting
uncertainties. Note that for Earth-size planets in the 50--100~day and
100--200~day period bins, planet occurrence is $5.8\pm1.8\%$ and
$3.2\pm1.6\%$, respectively. Corrections due to false positives or
planet multiplicity are smaller than fractional
uncertainties due to small number statistics.

\subsection{Planet Occurrence and Stellar Light Intensity}
The amount of light energy a planet receives from its host star
depends on the luminosity of the star (\Lstar) and the planet-star
separation ($a$). Stellar light flux, \Fp, is given by $\Fp~=~\Lstar /
4 \pi a^{2}$. The intensity of sunlight on Earth is \FE =
1.36~kW~m$^{-2}$. We compute \Lstar using $\Lstar~=~4 \pi \Rstar^{2}
\sigma \teff^{4}$ where $\sigma~=~5.670\times
10^{-8}$~W~m$^{-2}$~K$^{-4}$ is the Stefan-Boltzmann constant. The
dominant uncertainty in \Fp is due to \Rstar. Using spectroscopic
stellar parameters, we determine \Fp to 25\% accuracy, and to 80\%
accuracy using photometric parameters. We obtained spectra for all 62 
stars hosting planets with \Per~>~100~days, allowing more
accurate light intensity measurements.

\begin{figure}
\centering
\includegraphics[width=1\textwidth]{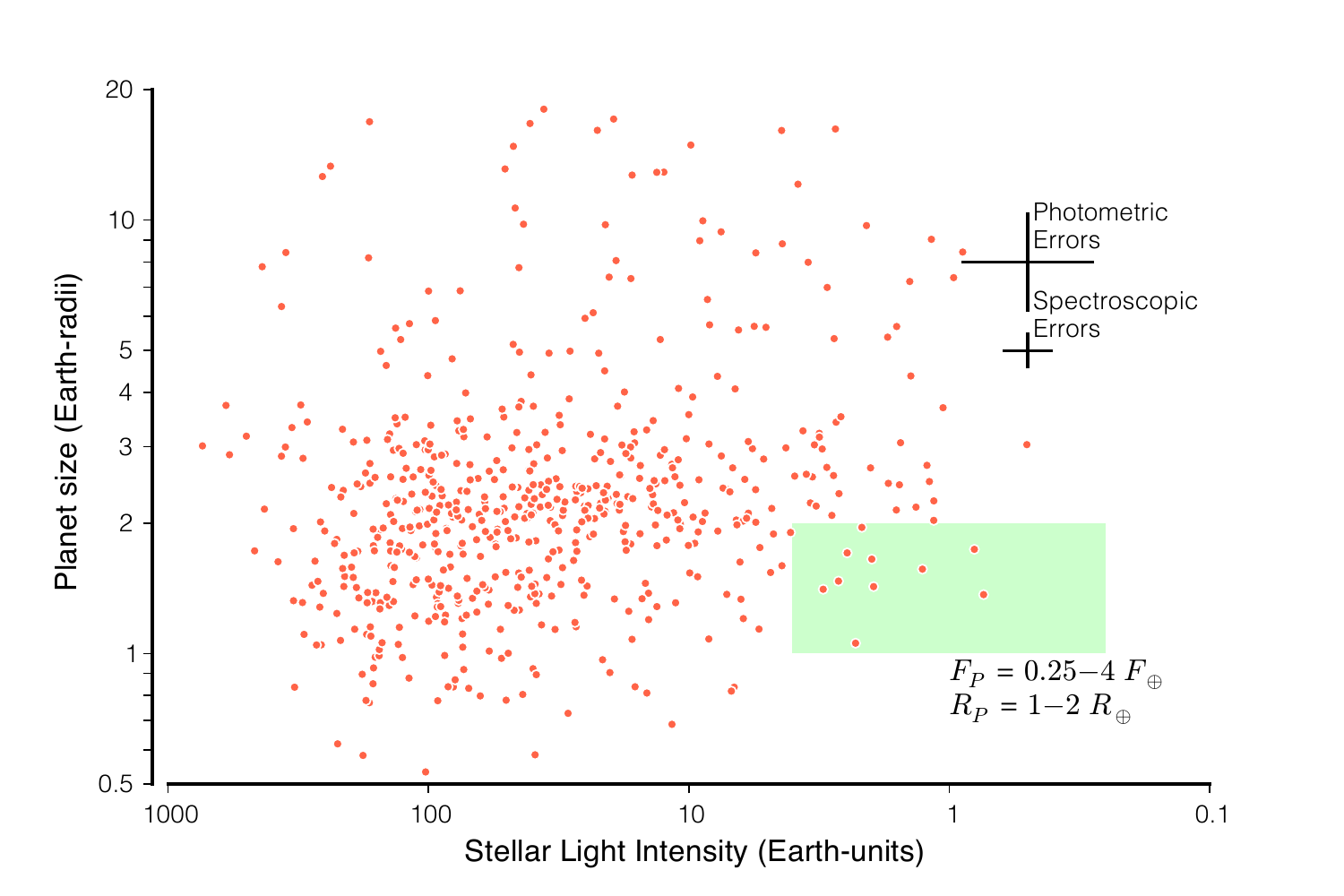}
\caption[Detected planets and stellar light intensity]{The detected planets (dots) in a two-dimensional domain
  similar to Figures \ref{fig:PlanetSample} and
  \ref{fig:CheckerBoard}.  Here, the two-dimensional domain has
  orbital period replaced by stellar light intensity, ``incident
  flux,'' hitting the planet. The highlighted region shows the 10
  Earth-size planets that receive a incident stellar flux comparable
  to the Earth: flux = 0.25--4.0$\times$ the flux received by Earth
  from the Sun.  Our uncertainties on stellar flux and planet radii
  are indicated at top right.}
\label{fig:FluxRp}
\end{figure}

Fig.~\ref{fig:FluxRp} shows the two-dimensional domain of stellar
light flux incident on our 603 detected planets, along with planet
size. The planets in our sample receive a wide range of flux from
their host stars, ranging from 0.5 to 700 \FE. We highlight the 10 small
(\Rp~=~1--2~\Re) planets that receive stellar flux comparable to
Earth, \Fp~=~0.25--4~\FE.

Since only two 1--2~\Re planets have \Fp < 1 \FE, we measure planet
occurrence in the domain, 1--2~\Re and 1--4~\FE. \textit{Correcting for survey
completeness we find that $11\pm4\%$ of Sun-like stars have a
\Rp~=~1--2~\Re planet that receives between 1 and 4 times the incident
flux as the Earth (SI)}.

\section{Interpretation}
\subsection{Earth-Size Planets with Year-Long Orbital Periods}
Detections of Earth-size planets having orbital periods of \Per~=~200--400~days are expected to be rare in this survey.  Low survey
completeness ($C \approx 10\%$) and low transit probability ($P_T$ =
0.5\%) imply that only a few such planets would be expected, even if
they are intrisically common.  Indeed, we did not detect any such
planets with \TERRA.%
\footnote{Although the radii of three planets (KIC-4478142, KIC-8644545, and KIC-10593626) have $1~\sigma$ confidence intervals that extend into the \Per~=~200--400~day \Rp~=~1--2~\Re domain.}
We can place an upper limit on their
occurrence: $f < 12\%$ with 95\% confidence using binomial statistics.
We would have detected one or two such planets if their occurrence
were higher than 12\%.

\begin{figure}
\centering
\centerline{\includegraphics[width=.7\textwidth]{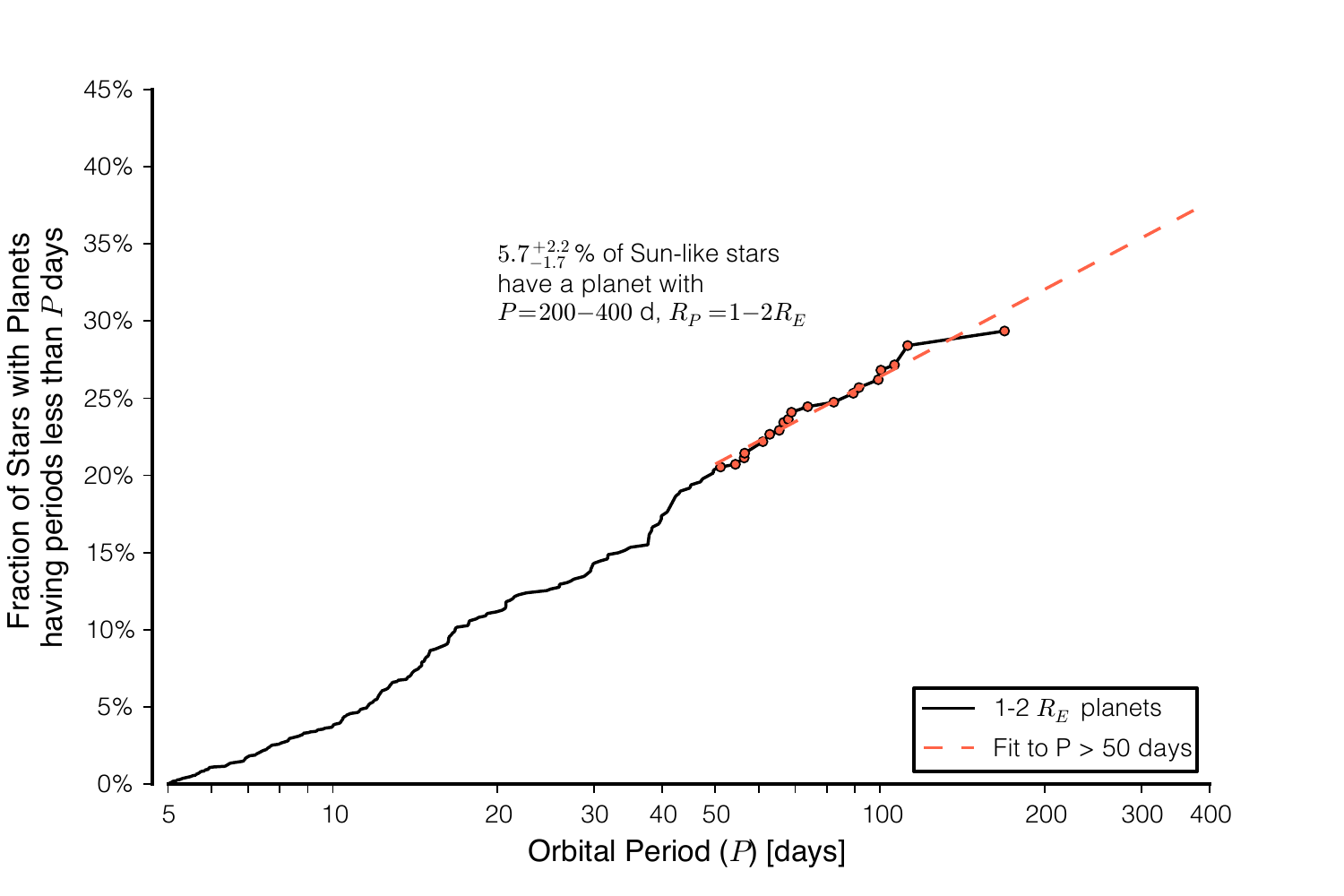}}
\caption[Cumulative distribution of small planets]{The fraction of stars having nearly Earth-size planets
  (1--2~\Re) with any orbital period up to a maximum period, \Per, on
  the horizontal axis.  Only planets of nearly Earth-size (1--2~\Re)
  are included.  This cumulative distribution reaches 20.2\% at
  \Per~=~50~days, meaning 20.4\% of Sun-like stars harbor a 1--2~\Re
  planet with an orbital period, \Per~$<$~50~days. Similarly, 26.2\% of
  Sun-like stars harbor a 1--2~\Re planet with a period of \Per $<$ 100
  days.  The linear increase in this cumulative quantity corresponds to planet
  occurrence that is constant in equal intervals of $\log \Per$.  One
  may perform a modest extrapolation into the \Per~=~200--400~day
  range, equivalent to assuming constant occurrence per $\log \Per$
  interval, using all planets with \Per~>~50~days. Such an extrapolation predicts that
  \EtaEarthErr of Sun-like stars have a planet with size, 1--2 \Re,
  with an orbital period between \Per~=~200--400~days.}
\label{fig:CDF}
\end{figure}

However, one may estimate the occurrence of 1--2 \Re planets with
periods of 200--400 days by a modest extrapolation of planet occurrence
with \Per.  Fig.~\ref{fig:CDF} shows the fraction of stars with
1--2~\Re planets, whose orbital period is less than a maximum period,
\Per, on the horizontal axis. This cumulative period distribution shows
that 20.4\% of Sun-like stars harbor a 1--2~\Re planet with an orbital
period, \Per~$<$~50~days.  Similarly, 26.2\% of Sun-like stars harbor a
1--2~\Re planet with a period less than 100 days.  The linear increase
in cumulative occurrence implies constant planet occurrence per
$\log \Per$ interval. Extrapolating the cumulative period distribution
predicts \EtaEarthErr occurrence of Earth-size (1--2~\Re) planets with
orbital periods of $\sim1$~year (\Per~=~200--400~days).  The details
of our extrapolation technique are explained in the SI.  {\em
  Extrapolation based on detected planets with \Per~<~200~days predicts
  that \EtaEarthErr of Sun-like stars have an Earth-size planet on an
  Earth-like orbit (\Per~=~200--400~days).}

Naturally, such an extrapolation caries less weight than a direct
measurement. However, the loss of \textit{Kepler's} second reaction wheel in
May 2013 ended observations shortly after the completion of the
nominal 3.5~year mission. We cannot count on any additional \Kepler
data to improve the low completeness to Eartha analog planets beyond
what is reported here. Indeed, low survey sensitivity to Earth analogs
was the primary reason behind a four-year extension to the \Kepler
mission. Modest extrapolation is required to understand the prevalence
of Earth-size planets with Earth-like orbits around Sun-like stars.

We offer empirical and theoretical justification for extrapolation out
to 400~days. As shown in Fig.~\ref{fig:CheckerBoard}, the prevalence
of small planets as a function of $\log \Per$ is remarkably uniform. To
test the reliability of our extrapolation into a region of low
completeness, we used the same technique to estimate occurrence in
more complete regions of phase space and compared the results with our
measured occurrence values.

Again, assuming uniform occurrence per $\log \Per$ interval,
extrapolating the occurrence of 2--4~\Re planets from 50--200~day orbits
out to 400~days predicts $6.4^{+0.5}_{-1.2}\%$ occurrence of planets in
the domain of \Rp~=~1--2~\Re and \Per~=~200--400~days. The extrapolation is
consistent with the measured value of $5.0\pm2.1\%$ to within 1~$\sigma$
uncertainty. Futhermore, extrapolation based on 1--2~\Re planets with
\Per~=~12.5--50~days predicts $6.5^{+0.9}_{-1.7}\%$ occurrence within the
\Rp~=~50--100~day bin. Again, the measured value of $5.8\pm1.6\%$ agrees to
better than 1~$\sigma$.

While planet size is governed by non-linear processes such as runaway
gas accretion \cite{Ida13}, which favors certain planet sizes over
others, no such non-linear processes occur as a function of orbital
period in the range, 200--400~days. Extrapolation out to orbital
periods of 200--400~days, while dangerous, seems unlikely to be
unrealistic by more than factors of two.

\subsection{Earth-Size Planets in the Habitable Zone}
While the details of planetary habitability are debated and depend on 
planet-specific properties as well as the stochastic nature 
of planet formation \cite{Seager13}, the
``habitable zone'' (HZ) is traditionally defined as the set of
planetary orbits that permit liquid water on the surface. The precise inner and outer edges of the HZ depend on details of the
model \cite{Kasting1993, Kopparapu2013, Zsom2013, Pierrehumbert2011}.
For solar analog stars, Zsom et al. (2013) estimated that the inner edge of the HZ could reside as close as 0.38 AU, for planets having either a reduced greenhouse effect due to low humidity or a high reflectivity
\cite{Zsom2013}. Pierrehumbert and Gaidos (2011) estimated that the outer
edge of the HZ may extend up to 10 AU for planets that are kept warm
by efficient greenhouse warming with an H$_2$ atmosphere
\cite{Pierrehumbert2011}.  

A planet's ability to retain surface liquid water depends, in large part,
on the energy received from its host star. We consider a planet to
reside in the HZ if it is bathed in a similar level of starlight as
Earth. One may adopt \Fp~=~0.25--4~\FE as a simple definition of the HZ, which corresponds orbital separations of 0.5 to 2.0~AU for solar analog stars. This definition is more conservative than the range of published HZ boundaries that extend from 0.38~AU to 10~AU~\cite{Zsom2013,Pierrehumbert2011}. This HZ includes Venus (0.7~AU) and Mars (1.5~AU) which
do not currently have surface liquid water. However, Venus may have had 
liquid water in its past, and there is strong geomorphological evidence
of liquid water earlier in Mars' history \cite{Seager13}.

Previously, we showed that $11\pm4\%$ of stars harbor a planet having
an \Rp~=~1--2~\Re and \Fp~=~1--4~\FE. Using the definition of \Fp and
Kepler's third law, \Fp is proportional to $P^{-4/3}$.  Therefore,
uniform occurrence in $\log \Per$ translates to uniform occurrence in
$\log \Fp$. We find that the occurrence of 1--2~\Re planets is
constant per $\log \Fp$ interval for \Fp~=~100--1~\FE. If one were to adopt \Fp~=~0.25--4~\FE as the HZ and extrapolate from the \Fp~=~1--4~\FE domain, then occurrence of HZ Earth-size planets is 22\% for Sun-like stars. 

One may adopt alternative definitions of both the properties of Earth-size planets and the domain of the HZ.  We showed previously that the occurrence of planets is approximately constant as a function of both \Rp (for \Rp~<~2.8 \Re) and \Per (in logarithmic intervals).  Thus, the occurrence of planets in this domain is proportional to a logarithmic area in the \Rp--\Fp parameter space being considered.  For example, the occurrence of planets of size 1.0--1.4 \Re in orbits that receive 0.25--1.0~\FE in stellar flux is 22\%/4~=~5.5\%. We offer a number of estimates for the prevalence of Earth-size planets in the HZ based different published definitions of the HZ in Table~\ref{tab:HZ}.

Cooler, M dwarf stars also have a high occurrence of Earth-size
planets.  Based on the \Kepler planet catalog, Dressing et al.\
\cite{Dressing2013} found that $15^{+15}_{-6}$\% of early M dwarfs
have an Earth-size planet (0.5--1.4 \Re) in the HZ using a
conservative definition of 0.5--1.1 \FE \cite{Kasting1993}, and three
times that value when the HZ is expanded to 0.25--1.5 \FE
\cite{Kopparapu2013b}.  This result is consistent with a Doppler
survey that found that $41^{+54}_{-13}$\% of nearby M dwarfs have
planets with masses 1--10 Earth masses ($M_{\oplus}$) in the HZ
\cite{Bonfils2013}.  Thus, Earth-size planets appear to be common in
the HZs of a range of stellar types.

\section{Conclusions}

Using \Kepler photometry of Sun-like stars (GK-type), we have measured the prevalence of planets having different orbital 
periods and sizes, down to the size of the Earth and out to orbital periods of one year. We gathered Keck spectra of 
all host stars of planets having periods greater than 100 days to accurately determine their radii. The detection of 
planets with periods longer than 100 days is challenging, and we have characterized our sensitivity to such planets
by using injection and recovery of synthetic planets in the photometry. After correcting for orbital tilt and detection completeness, we find 
that $26\pm3\%$ of Sun-like stars have an Earth-size (1--2~\Re) planet with \Per~=~5--100 days. We also find that $11\pm4\%$ of 
Sun-like stars harbor an Earth-size planet that receives nearly Earth-levels of stellar energy (\Fp~=~1--4~\FE).  

We have shown that small planets far outnumber large ones. Only
$1.6\pm0.4\%$ of Sun-like stars harbor a Jupiter-size (8--16~\Re) planet with \Per~=~5--100~days compared to $23\pm3\%$ occurrence of
Earth-size planets. This pattern supports the
core accretion scenario in which planets form by the accumulation of
solids first and gas later in the protoplanetary disk
\cite{Lissauer93,Pollack96,Ida13,Mordasini12}.  The details of
this family of models are hotly debated, including the movement of
material within the disk, the timescale for planet formation, and the
amount of gas accretion in small planets.  Our measurement of a
constant occurrence of 1--2.8~\Re planets per $\log$ \Per interval
establishes an important observational constraint for these models.

The occurrence of Earth-size planets is constant with decreasing stellar light intensity from 100~\FE down to 1~\FE. If one were to assume that this pattern continues down to 0.25~\FE, then the occurrence of planets having flux levels of 1--0.25~\FE is also $11\pm4\%$.

Earth-size planets are common in the \Kepler field. If the stars in the \Kepler field are representative of stars in the solar neighborhood, then Earth-size planets are common around nearby Sun-like stars. If one were to adopt a 22\% occurrence rate of Earth-size planets in habitable zones of Sun-like stars, then the nearest such planet is expected to orbit a star that is less than 12 light-years from Earth and can be seen by the unaided eye. Future instrumentation to image and take spectra of these Earths need only observe a few dozen nearby stars to detect a sample of Earth-size planets residing in the habitable zones of their host stars.

%% For Tables, put caption above table
%%
%% Table caption should start with a capital letter, continue with lower case
%% and not have a period at the end
%% Using @{\vrule height ?? depth ?? width0pt} in the tabular preamble will
%% keep that much space between every line in the table.

%% \begin{table}
%% \caption{Repeat length of longer allele by age of onset class}
%% \begin{tabular}{@{\vrule height 10.5pt depth4pt  width0pt}lrcccc}
%% table text
%% \end{tabular}
%% \end{table}

\renewcommand{\nd}{...}
%\begin{table*}
\begin{table}
\caption{Occurrence of Small Planets the Habitable Zone}
\label{tab:HZ}
\begin{tabular*}{\hsize}{@{\extracolsep{\fill}}llllll}
  HZ Definition & $a_{\text{inner}}$ & $a_{\text{outer}}$ 
                & $F_{P,\text{inner}}$ & $F_{P,\text{outer}}$ 
                & $f_{HZ}$ \cr
  \hline

  Simple                         & 0.5   & 2     & 4    & 0.25 & 22\% \cr

  Kasting (1993)                 & 0.95  & 1.37  & 1.11 & 0.53 & 5.8\% \cr

  Kopparapu et al. (2013)        & 0.99  & 1.70  & 1.02 & 0.35 & 8.6\% \cr

  Zsom et al. (2013)             & 0.38  & \nd   & 6.92 & \nd  & 26\% \cr
%  \tablenote{
%    If a model does not quote an inner or outer edge, 
%    we adopt boundaries from the ``Simple'' model} 
  
  Pierrehumbert \& Gaidos (2011) & \nd   & 10    & \nd  & 0.01 & $\sim$50\%$^{\ast}$\cr
%
%  \tablenote{
%    Extrapolation out to 10 AU is severely under-constrained. 
%    This estimate is highly uncertain and is included for completeness.} \\

  \hline
\end{tabular*}
\end{table}
%\end{table*}

%% For two column figures and tables, use the following:

%% \begin{figure*}
%% \caption{Almost Sharp Front}\label{afoto}
%% \end{figure*}

%% \begin{table*}
%% \caption{Repeat length of longer allele by age of onset class}
%% \begin{tabular}{ccc}
%% table text
%% \end{tabular}
%% \end{table*}

%% file: terra1yr/numbers.tex
\num{\nPinKp}{98,471}
\num{\nSamp}{42,557}
\num{\nOnSi}{188,329}
\num{\nPinKpTeff}{63,915}
\num{\nPin}{155,046}
\num{\neKOI}{836}
\num{\nPlntMyDV}{650}
\num{\nPlntMyDVCentInfo}{609}
\num{\nPlnt}{603}
\num{\nFPR}{115}
\num{\nFPSE}{44}
\num{\nFPV}{11}
\num{\nFPplnt}{603}
\num{\nFPTTV}{5}
\num{\nCentInfo}{654}
\num{\nFPC}{31}
\num{\nTCE}{2184}
\num{\nFPVD}{27}

%% file: terra1yr/nc.tex
%\nc{\tdur}{\Delta T}

%\newcommand{\hlf}{\tfrac{1}{2}}
\nc{\tdurCirc}{ \Delta T_{\text{circ}} }
%\nc{\ep}{t_0}
%\nc{\Per}{P}

%\newcommand{\df}{\delta F}
%\newcommand{\mdf}{\overline{\df}}
%\newcommand{\mtx}[1]{\ensuremath{\mathbf{#1}}}
%\nc{\teff}{T_{\rm eff} }
%\nc{\logg}{\log g}
%\nc{\sm}{\sim}
%\nc{\Pcad}{ P_{\text{cad}} }
\nc{\NstarEff}{N_{\star,\text{eff}}}

\nc{\neKOISM}{274} % Number of eKOIs for which we have SM parameters
%\nc{\Np}{ n_{\text{pl,cell}} }
%\nc{\NpAug}{ n_{\text{pl,aug,cell}} }
%\nc{\fcell}{f_{\text{cell}} }

%\nc{\fcellBa}{f_{\text{cell,Batalha}}}
%\nc{\flogA}{d^{2}\fcell / d\log{P} / d\log{R_P} }
%\nc{\NstarAmen}{ n_{\star,\text{amen}} }

%\nc{\Rsun}{ R_{\odot} }
%\nc{\Kepler}{ \textit{Kepler} }
\nc{\Kp}{ \textit{Kp} }
%\nc{\SpecMatch}{\tt SpecMatch}
%\nc{\TERRA}{\tt TERRA}

%\renewcommand{\Re}{\ensuremath{ R_{E} }\xspace} 
%\nc{\Rp}{ R_P }
%\nc{\Rstar}{R_\star} 

%\nc{\rrat}{\Rp / \Rstar}  
%\nc{\rratfrac}{ \frac{\Rp}{\Rstar} } 

\nc{\EtaEarthErr}{5.7^{+1.7}_{-2.2}\%}

\nc{\fBigYearExtrap}{6.4^{+0.5}_{-1.2}\%}
\nc{\fBigYearMeas}{5.0\pm2.1\%}
\nc{\fSmMercExtrap}{6.5^{+0.9}_{-1.7}\%}
\nc{\fSmMercMeas}{5.8\pm1.6\%}

%% file: terra1yr/terra1yr_si.tex
\chapter{Supporting Information for Prevalence of Earth-size Planets Orbiting Sun-like Stars}
\label{c.terra1yr_si}

\renewcommand{\nc}[2]{\newcommand{#1}{\ensuremath{#2}\xspace}}
\renewcommand{\num}[2]{\newcommand{#1}{{#2}\xspace}}
%\usepackage[font=small]{caption}

%\pagestyle{plain}

%\begin{document} 
%\pagenumbering{arabic}

%\makeatletter 
%\renewcommand{\thefigure}{S\@arabic\c@figure}
%\makeatother

%\renewcommand{\thepage}{S\arabic{page}}  % add 'S' prefix to SOM page numbers
%\renewcommand{\thesection}{S\arabic{section}}  % add 'S' prefix to SOM section numbers
%\renewcommand{\thefigure}{S\arabic{figure}}  % add 'S' prefix to SOM figure numbers
%\baselineskip24pt
%\maketitle 

%\noindent
%{\LARGE\textbf{Supporting Information (SI)}}
%\vspace{0.5in}
\noindent A version of this chapter was previously published in the {\em Proceedings of the National Academy of Science}
(Erik~A.~Petigura, Andrew~W.~Howard, \& Geoffrey~W.~Marcy,  2013, PNAS 110, 19273).\\

In this supplement to ``The Prevalence of Earth-size Planets 
Orbiting Sun-Like Stars'' by Petigura et al., we elaborate on the
technical details of our analysis. In Section~\ref{sec:Best42k}, we
define our sample of \nSamp Sun-like stars that are amenable to the
detection of small planets --- the ``Best42k'' stellar sample. In
Section~\ref{sec:TERRA}, we describe the algorithmic components of
\TERRA, our custom pipeline that we used to
find transiting planets within \Kepler
photometry.  Section~\ref{sec:DV} describes ``data validation,'' how
we prune the large number of ``Threshold Crossing Events'' into a list
of \neKOI eKOIs, analogous to KOIs from the \Kepler
Project. Section~\ref{sec:Bogus4} shows four KOIs in the current
online Exoplanet Archive ~\cite{Akeson13} that failed the data
validation step. Section~\ref{sec:AFP} describes the procedure by
which we remove astrophysical false positives from our list of
eKOIs. Section~\ref{sec:RpRefine} describes how we refine our initial
estimate of planet radii using using spectra of the eKOIs coupled
with MCMC-based light curve fitting.  Section~\ref{sec:Completeness}
contains a description of the fundamental component of this study:
measuring the completeness of our planet search by injecting synthetic
transit light curves, caused by planets of all sizes and orbital
periods, directly into the \Kepler photometry, and analyzing the
photometry with our pipeline to determine the fraction of planets
detected. In Section~\ref{sec:Occurrence} we provide details
describing our calculation of planet occurrence and discuss the
effects of multiplanet systems and false positives.

\section{The Best42k Stellar Sample}
\label{sec:Best42k}

We restrict our planet search to Sun-like stars with well-determined
photometric properties and low photometric noise. We select stars
having revised \Kepler Input Catalog (KIC) parameters. Effective temperatures are based on the Pinsonneault et al.~\cite{Pinsonneault12} revisions to the KIC effective temperatures. Surface gravities are based on fits to Yonsei-Yale stellar evolution models~\cite{Yi01} assuming [Fe/H] = $-0.2$. Further details regarding isochrone fitting can be found in Batalha~et~al.~\cite{Batalha12}; Burke~et~al., submitted; and Rowe~et~al., in prep. These revised stellar parameters are tabulated on the Exoplanet Archive with the {\tt prov\_prim} flag set to ``Pinsonneault.'' Out of the \nOnSi stars observed at some point during Q1--Q15, we selected stars that:

\begin{enumerate}
\item Have revised KIC stellar properties. (\nPin stars),
\item \Kp~=~10--15 mag (\nPinKp stars), 
\item \teff~=~4100--6100~K (\nPinKpTeff stars), and
\item \logg~=~4.0--4.9 (cgs) (\nSamp stars).
\end{enumerate}

\begin{figure}
\centering
\includegraphics[width=0.8\textwidth]{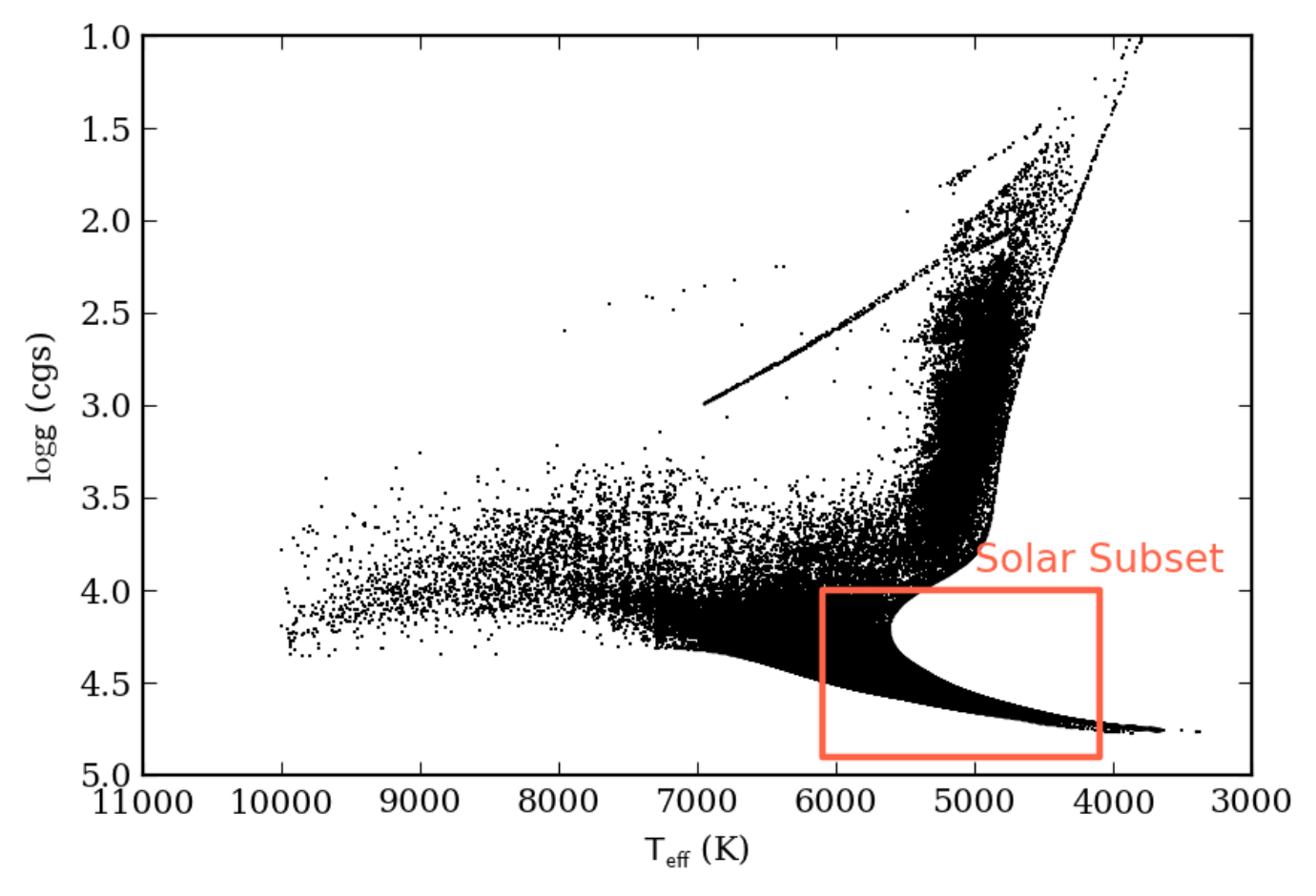}
\caption[Effective temperatures and surface gravities of stellar sample]{Distribution of \nPin stars with revised photometric 
  stellar parameters. The Best42k sample of
  42000 stars is
  made up of Solar-type stars with \teff~=~4100--6100~K,
  \logg~=~4.0--4.9 (cgs), and Kepmag = 10-15 (brighter half of targets).}
\label{fig:SampHR}
\end{figure}

Figure~\ref{fig:SampHR} shows the position of the 155,046 stars with
revised stellar properties along with
the ``solar subset'' corresponding to G and K dwarfs. Figure~\ref{fig:SampKpNoise} shows
the distribution of brightness and noise level of the Best42k stellar
sample.

\begin{figure}
\centering
\includegraphics[width=0.8\textwidth]{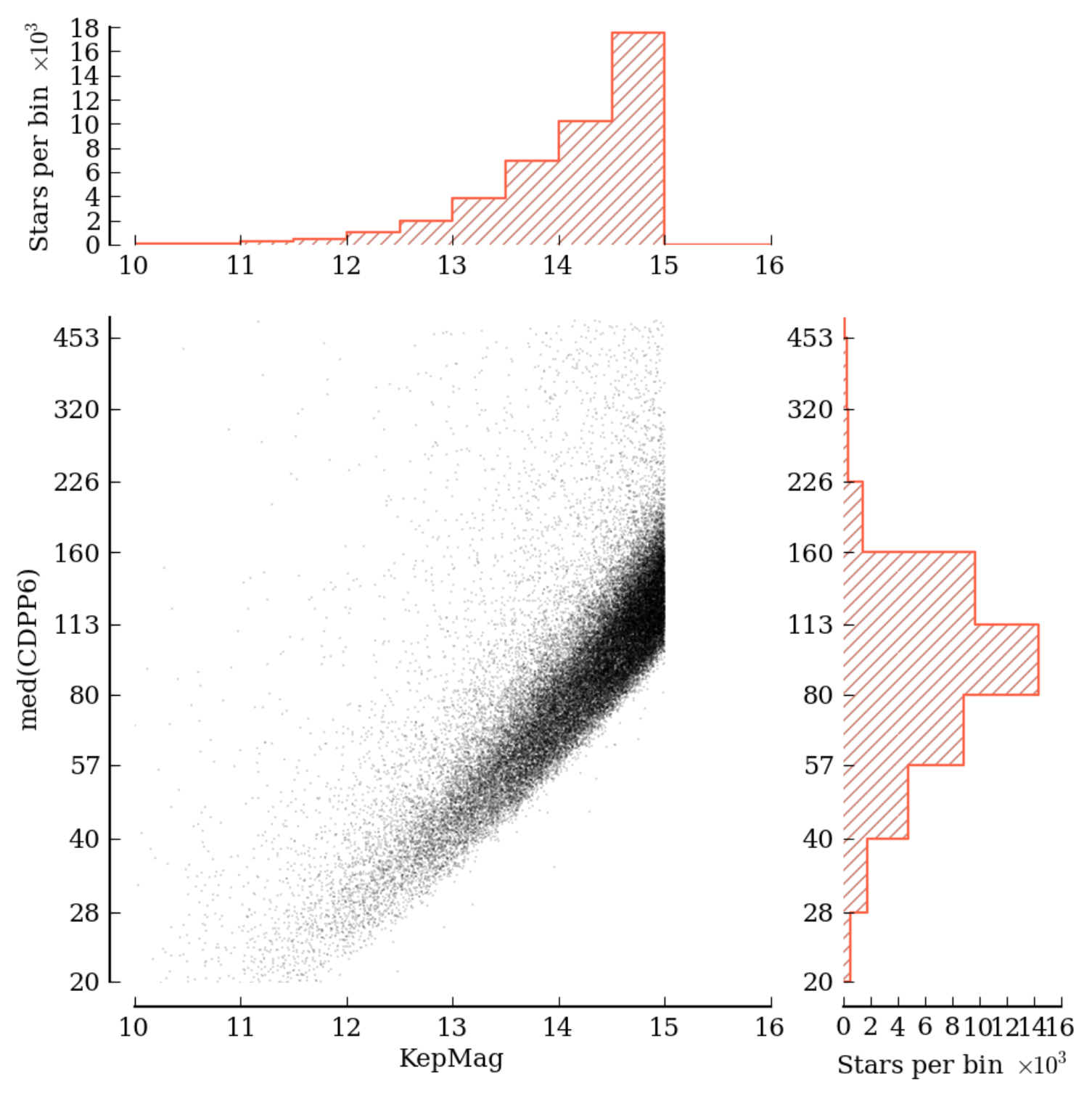}
\caption[Brightness and photometric variability of stellar sample]{Distribution of photometric noise (median quarterly 6 hour CDPP) and brightness \Kp for the \nSamp stars in the Best42k stellar sample. }
\label{fig:SampKpNoise}
\end{figure}
\clearpage

\section{Planet Search Photometric Pipeline}
\label{sec:TERRA}
We search for planet candidates in the Best42k stellar sample using
the \TERRA pipeline described in detail in Petigura \& Marcy (2012)
and in Petigura, Marcy, and Howard (2013; P13, hereafter)
\cite{Petigura12,Petigura13}. We review the major components of \TERRA
below, noting the changes since P13.

\subsection{Time-domain pre-processing of raw \Kepler Photometry}
\TERRA begins by conditioning the photometry in the
time-domain. \TERRA first searches for single cadence outliers, mostly
due to cosmic rays. \TERRA also searches for abrupt drops in the raw
photometry known as Sudden Pixel Sensitivity Drops (SPSDs) discussed
by Stumpe et al. \cite{Stumpe12}. SPSDs are particularly challenging
since they mimic transit ingress, and aggressive attempts to remove
them run the risk of removing real transits. \TERRA removes the
largest SPSDs, but they remain a source of non-astrophysical false
positives that we remove during manual triage
(Section~\ref{sec:ManualTriage}).

\TERRA also removes trends longer than $\sim$10 days. In P13, this
high-pass filtering was implemented by fitting a spline to the raw
photometry with the knots of the spline separated by 10~days. But in
this work we employ high-pass filtering using Gaussian Process
regression~\cite{Rasmussen06}, which gives finer control over the
timescales removed. We adopt a squared exponential kernel with a 5-day
correlation length. After this high-pass filter, \TERRA identifies
systematic noise modes via principle components analysis on large
number of stars.

\subsection{Grid-based transit search}
We search for periodic box-shaped dimmings by evaluating the
signal-to-noise ratio (SNR) of a putative transit over a finely-spaced
grid of period, \Per; epoch, \ep; and transit duration, \tdur. In P13,
we searched over a period range, \Per~=~5--50~days, and over transit
durations ranging from 1.5--8.8~hr. But in this work, we extend our
search in orbital period to \Per~=~0.5--400~days. Since we search over
nearly three decades in orbital period, and because transit duration
is proportional to $\Per^{1/3}$, we let the range of trial transit
durations vary with period. We break our period range into 10 equal
logarithmic intervals. Then, using photometrically determined
parameters for each star, namely \Mstar and \Rstar, we compute an approximate, expected
transit duration (\tdurCirc) for the simple case of circular orbits
with impact parameter, $b=1$. However, we actually search over \tdur =
0.5--1.5 \tdurCirc to account for a range of impact parameters and
orbital eccentricities and for mis-characterized \Mstar and \Rstar. As
an example, Table~\ref{tab:tdurGrid} shows our trial \tdur for a star
with solar mass and radius.

\begin{deluxetable}{rrr}
\tabletypesize{}
%\tablenum{S1}
\tablecaption{\TERRA Grid Search Parameters}
\label{tab:tdurGrid}
\tablewidth{0pt}
\tablehead{
	\colhead{$\Per_1$} &
	\colhead{$\Per_2$} &
	\colhead{Trial Transit Duration (\tdur)}\\
	\colhead{days} &
	\colhead{days} &
	\colhead{long cadence measurements}
 }
\startdata
      5.0 &     7.7 &      [3, 4, 5, 7, 9, 11] \\
      7.7 &    12.0 &         [4, 5, 7, 9, 13] \\
     12.0 &    18.6 &     [4, 5, 7, 9, 13, 15] \\
     18.6 &    28.9 &    [5, 7, 9, 12, 16, 17] \\
     28.9 &    44.7 &   [6, 8, 11, 14, 19, 20] \\
     44.7 &    69.3 &   [7, 9, 12, 16, 22, 23] \\
     69.3 &   107.4 &  [8, 11, 14, 19, 25, 26] \\
    107.4 &   166.5 &  [9, 12, 16, 21, 28, 31] \\
    166.5 &   258.1 & [10, 13, 18, 24, 31, 35] \\
    258.1 &   400.0 & [12, 16, 21, 28, 38, 41] \\
\enddata
\end{deluxetable}
\clearpage

\section{Data Validation}
\label{sec:DV}
If \TERRA detects a (\Per, \ep, \tdur) with SNR > 12, we flag the
light curve for additional scrutiny. While the grid-based component of
\TERRA is well-matched to exoplanet transits, there are other
phenomena that can produce SNR > 12 events and contaminate our planet
sample. We distinguish between two classes of contaminates:
``astrophysical false positives'' such as diluted eclipsing binaries
(EBs), and ``non-astrophysical false positives'' such as noise that
can mimic a transit. We establish a series of quality control measures
called ``Data Validation'' (DV), designed to remove formally strong
dimmings (i.e. SNR > 12) found by the blind photometric pipeline that are
not consistent with an astrophysical transit. DV consists of two
steps:

\begin{enumerate}
\item {\em Machine triage}: Select potential transits by automated cuts.
\item {\em Manual triage}: Manually remove light curves that are inconsistent with a Keplerian transit.
\end{enumerate}

Manual triage is accomplished by inspection of DV summary plots which
contain numerous useful diagnostics necessary to warrant planet
status.  The diagnostics permit a multi-facted evaluation of the
integrity (as a potential planet candidate) of a given dimming
identified by the photometric pipeline. Figure~\ref{fig:eKOIpass}
shows an sample DV report, this for KIC-5709725 that passed
examination.

\newcommand{\dvwid}{8.5in}

\begin{landscape}
\begin{figure}
\includegraphics[width=\dvwid]{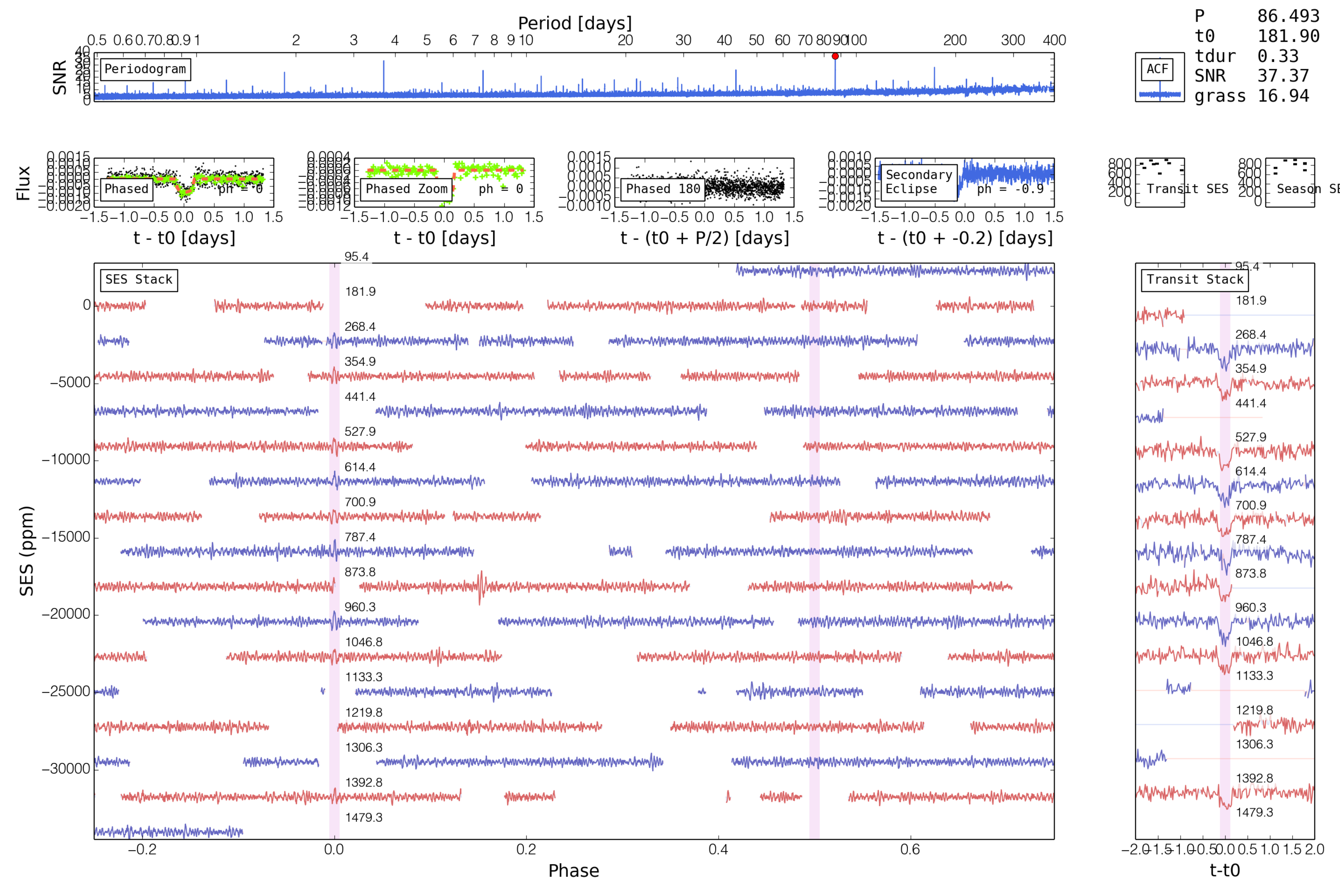}

  \caption[DV summary plots for KIC-5709725]{(Continued on the following page)}% First caption

\label{fig:eKOIpass}
\end{figure}

\begin{figure}[t]
  \contcaption{DV summary plots for KIC-5709725. Top row: SNR
  ``periodogram'' of box-car photometric search for transiting
  planets, ranging from 0.5 to 400 days. The red dot shows the \Per
  and SNR of the most significant peak in the periodogram at
  \Per~=~86.5~days (also found by the \Kepler Project as
  KOI-555.02). Also visible is a second set of peaks corresponding to
  KOI-555.01 at \Per~=~3.702~days. \TERRA does not search for more than
  one planet per system. Moreover, KOI-555.01 would be excluded from
  our planet sample since \Per~<~5~days. The autocorrelation
  function,``ACF'', plot at upper right shows the circular
  auto-correlation function from the phase folded photometry, used to
  identify secondary eclipses and correlated noise in the
  photometry. Second row: at left ``Phase'' shows the phase--folded
  photometry, where black points represent detrended photometry near
  the time of transit, green symbols show median flux over 30~min
  bins, and the dashed line shows the best-fitting Mandel-Agol transit
  model~\cite{Mandel02}. At second from left, ``Phased Zoom'' shows a
  zoomed y-axis to highlight the transit itself.  For this TCE, the
  transit model is a good match to the photometry. At third from left,
  ``Phased 180'' shows phase folded photometry 180\deg~in orbital
  phase from transit center. At fourth from left, ``Secondary
  eclipse'' shows how \TERRA notches out the putative transit and
  searches for secondary eclipses. We show the photometry folded on
  the second most significant dimming. For KIC-5709725, this phase is
  0.9\deg~relative to the primary transit, so close in phase that the
  primary transit is still visible. This transit does not show signs
  of a secondary eclipse. Transit SES --- transit single event
  statistic as a function of transit number. Conceptually, SES is the
  depth of the transit in ppm, as described by Petigura and Marcy
  \cite{Petigura12}. ``Season SES'' shows the SES statistic grouped
  according to season. Bottom row: At left, ``SES stack'' shows SES
  for the entire light curve, split on the best-fitting transit period
  and stacked so that transit number increases downward. Compelling
  transits appear as a sequence of SES peaks at phase =
  0\deg. ``Transit stack'' shows for TCEs with fewer than 20 transits
  a plot of the \TERRA-calibrated photometry of each transit (transit
  number increases downward).}% Continued caption
\end{figure}

\end{landscape}

The product of the DV quality control is a list of ``eKOIs,'' for
which most instrumental events identified preliminarily
and erroneously by the photometric pipeline have been rejected.  The
resulting planet candidates are analogous to the KOIs \Kepler Project.
Astrophysically plausible causes (i.e. transiting planets and
background eclipsing binaries) are retained among our eKOIs.  We
address astrophysical false positives in Section~\ref{sec:AFP}.

\subsection{Machine Triage}
\label{sec:MachineTriage}
\begin{landscape}
\begin{figure*}
\includegraphics[width=\dvwid]{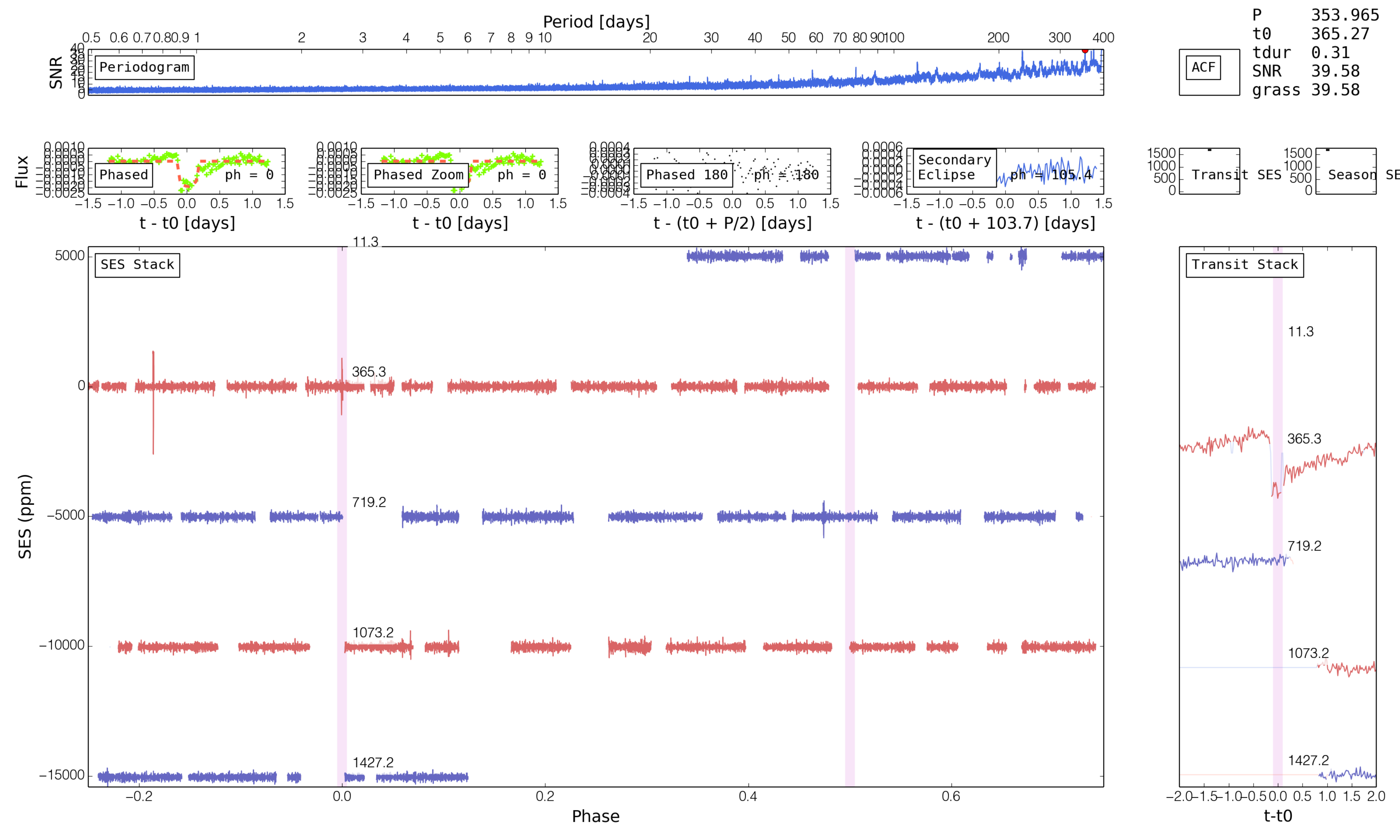}
\caption[DV summary plots for KIC-1570270]{DV summary plots (defined in Figure~\ref{fig:eKOIpass}) for
  KIC-1570270 showing a non-astrophysical false positive removed in
  the machine triage step. Here, an uncorrected SPSD at $t = 365.3$~days
  resulted in a SNR$\sim$40 event with \Per~=~353.965 days, seen in
  the SNR periodogram. SES stack plot shows this high SNR TCE is due
  to a single spike in SES due to the SPSD. \Per~=~353.965 days is favored
  over nearby periods because the anomaly aligns with gaps in the
  photometry. We flag cases like this with our {\tt s2n\_on\_grass}
  statistic.  We find several nearby peaks with nearly equal
  SNR. For KIC-1570270, {\tt s2n\_on\_grass} is less than our
  threshold of 1.2 and does not pass machine triage.}
\label{fig:MachineTriage}
\end{figure*}
\end{landscape}

Prior to the identification of final eKOIs, we carry out machine
triage to identify a set of ``Threshold Crossing Events'' (TCEs) that can be
classified by a human in a reasonable amount of time. TCE status
requires a SNR > 12; however, we find that 16227 light curves (out of
the 42000 target stars) meet this criterion. Outliers and correlated
noise are responsible for the majority of SNR > 12 events. We show set
of diagnostic plots for such an outlier in
Figure~\ref{fig:MachineTriage}. Here, an uncorrected sudden pixel
sensitivity dropoff at $t$~=~365.3~days, raises the noise floor to
SNR$\sim$15 for $\Per \lesssim 100$~days. Its contribution to SNR is
averaged down for shorter periods.

We flag such outliers by comparing the most significant period,
\PerMax, to nearby periods. We call the ratio of the maximum SNR to
the median of the next tallest five peaks between $[\PerMax/1.4 ,
\PerMax \times 1.4$] the {\tt s2n\_on\_grass} statistic. We require
{\tt s2n\_on\_grass} > 1.2 for TCE status. After that cut, 3438 TCEs
remain.  We also require \PerMax > 5 d, which leaves \nTCE TCEs.

\subsection{Manual Triage}
\label{sec:ManualTriage}
The sample of \nTCE TCEs has a significant degree of contamination
from non-astrophysical false positives. In P13, we relied on
aggressive automatic cuts that removed nearly all of the
non-astrophysical false positives (final sample was $\sim 90\%$
pure). However, by comparing our sample to that of Batalha et al. \cite{Batalha12},
we found that these automatic cuts were removing a handful of
compelling planet candidates.

\begin{landscape}
\begin{figure*}
\includegraphics[width=\dvwid]{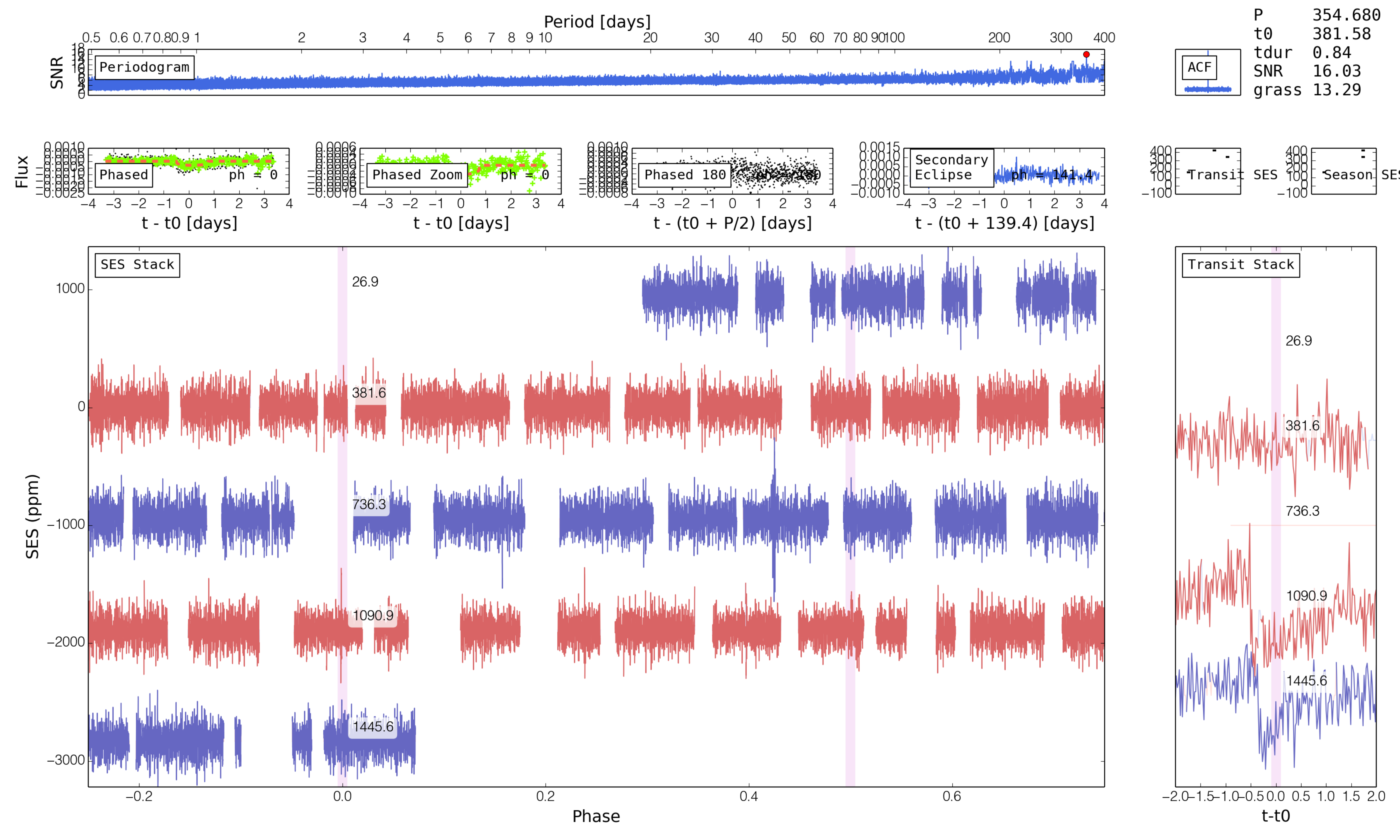}
\caption[DV summary plots for KIC-1724842]{DV summary plots (defined in Figure~\ref{fig:eKOIpass}) for
  the KIC-1724842 TCE at \Per~=~354.680~days that we removed during the
  manual triage. This photometry contains two pixel sensitivity drops
  spaced by 354.680~days. These two data anomalies combine to produce
  SNR~=~16.035 event, which is substantially higher than the
  background (``grass'' = 13.294) Since {\tt s2n\_on\_grass} = 1.206 >
  1.2, this event passed our \TERRA software-based triage. However,
  such data anomalies are easily identified by eye.}
\label{fig:ManualTriage}
\end{figure*}
\end{landscape}

In this work, we aim for higher completeness and rely more heavily on
visual inspection of light curves. We assess whether a TCE is due to a
string of three or more transits or instead caused by outlier(s) such
as SPSDs. Figure~\ref{fig:ManualTriage} shows an example of a light
curve that passed machine triage, but was removed manually. During
manual triage, we do not attempt to distinguish between planets and
astrophysical false positives. The end product is a list of \neKOI
eKOIs, which are analogous to KOIs produced by the \Kepler Project in
that they are highly likely to be astrophysical in origin but false
positives have not been ruled out.

\clearpage

\section{KOIs That Fail Data Validation}
\label{sec:Bogus4}
\begin{figure}
\includegraphics[width=1\textwidth]{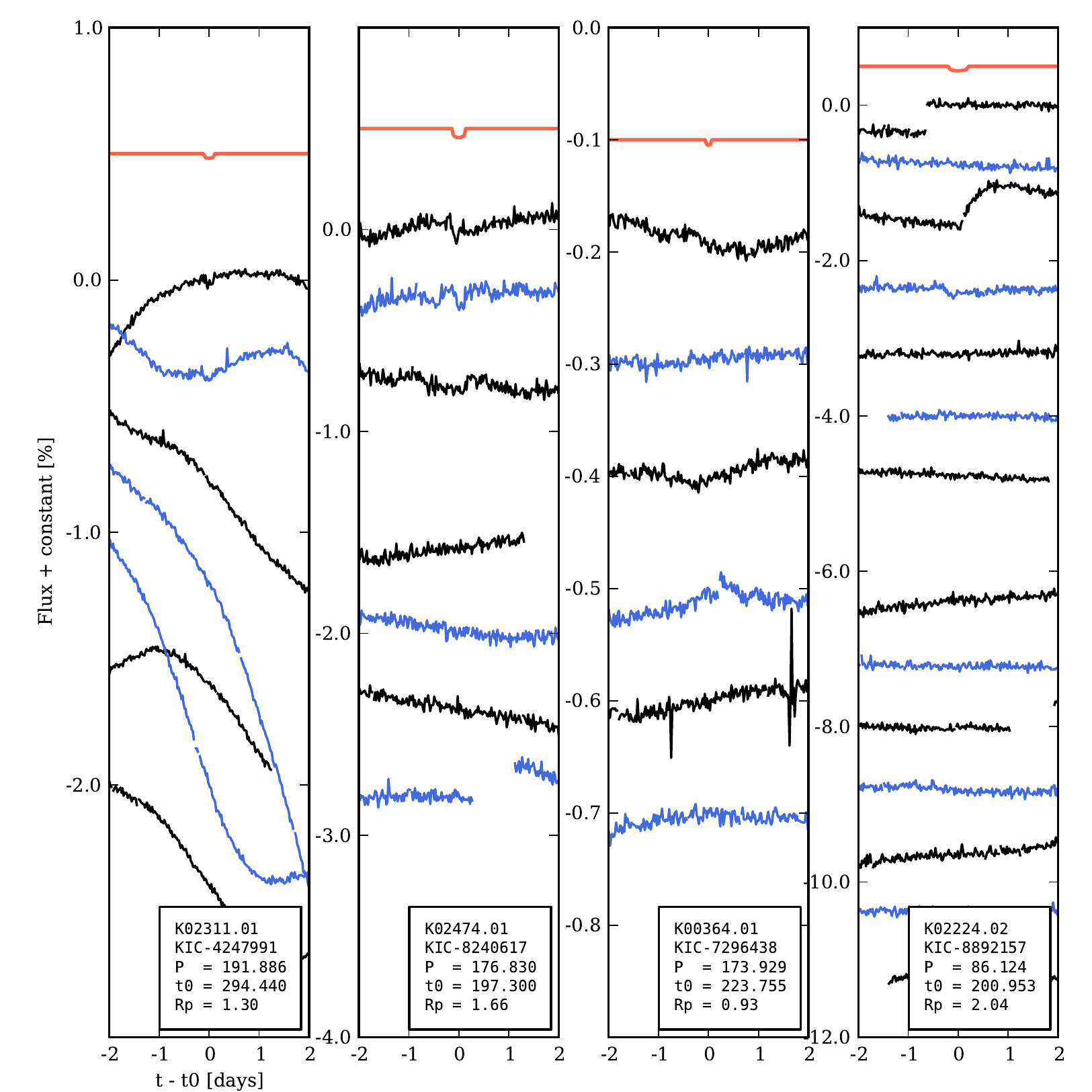}
\caption[Four KOIs that fail manual vetting]
{Four small KOIs with long orbital periods that fail our
  manual vetting. We show 4-day chunks of raw photometry (``{\tt
    SAP\_FLUX}'' in fits table) around the supposed transits. Transit
  number increases downward. We alternate the use of black and blue
  lines for clarity. The red lines show a Mandel Agol light curve
  model synthesized according to the published transit parameters.}
\label{fig:Bogus4}
\end{figure}

As a cross-check of our DV quality control methods, we performed the
same inspection on 235 KOIs that had been identified by the \Kepler
Project and which appear currently in the online Exoplanet Archive
~\cite{Akeson13}. These KOIs have periods longer than 50 days,
representative of long period transiting planets that enjoy a reduced
number transits (compared to short-period planets) during the 4-year
lifetime of the \Kepler mission.  We found four KOIs, 2311.01,
2474.01, 364.01, and 2224.01, that are not consistent with an
astrophysical transit. We show the raw light curves around the
published ephemerides in Figure~\ref{fig:Bogus4}. All four have
$\Rp~\leq~2.04~\Re$, and three have $\Per~\geq~173$~days. Due to the
small number of KOIs near the habitable zone, inclusion of these KOIs
would bias occurrence measurements upward by a large amount. Vetting
all 3000 KOIs in the Exoplanet Archive requires a substaintial effort,
and is beyond the scope of this paper. However, these four KOIs are a
reminder than detailed, expert, and visual vetting of DV diagnostics
existing KOIs is useful.

\clearpage

\section{Removal of Astrophysical False Positives}
\label{sec:AFP}
\begin{landscape}
\begin{figure*}
\includegraphics[width=\dvwid]{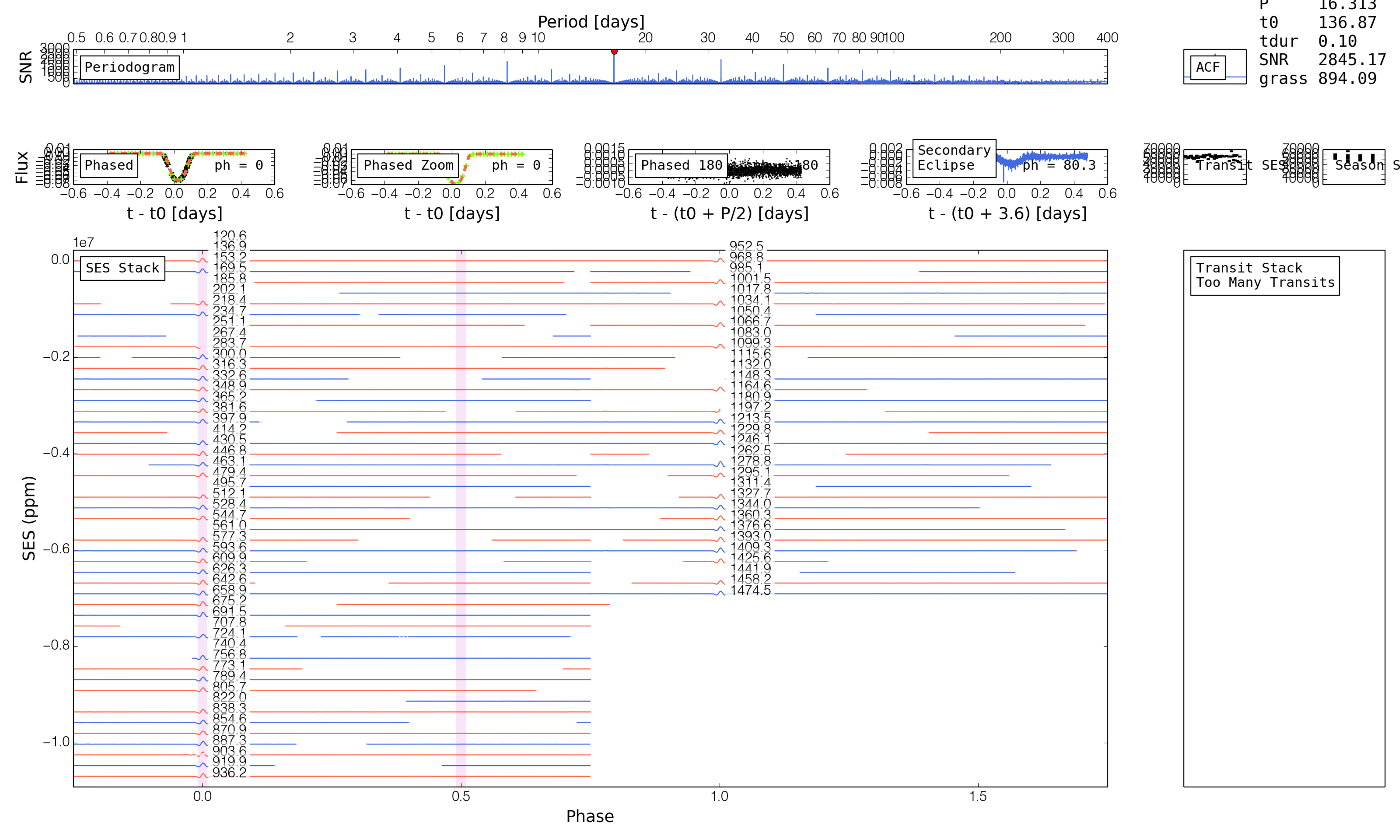}
\caption[DV summary plots for KIC-8879427]
  {DV summary plots (defined in Figure~\ref{fig:eKOIpass}) for
  eKOI KIC-8879427 (\Per = 16.313). The ``Secondary Eclipse'' plot
  shows the second most significant dimming at \Per~=~16.313, offset
  from the primary transit in phase by 80.3\deg. The ratio of the
  primary to secondary eclipses (\dfpri~=~0.07; \dfsec~=~0.002) along
  with the effective temperature of the primary (\teff~=~5995~K) imply
  the transiting object is 2343~K -- too high to be consistent with a
  planet with \Per~=~16.313~days orbital period.}
\label{fig:SecondaryEclipse}
\end{figure*}
\end{landscape}

\begin{landscape}
\begin{figure*}
\includegraphics[width=\dvwid]{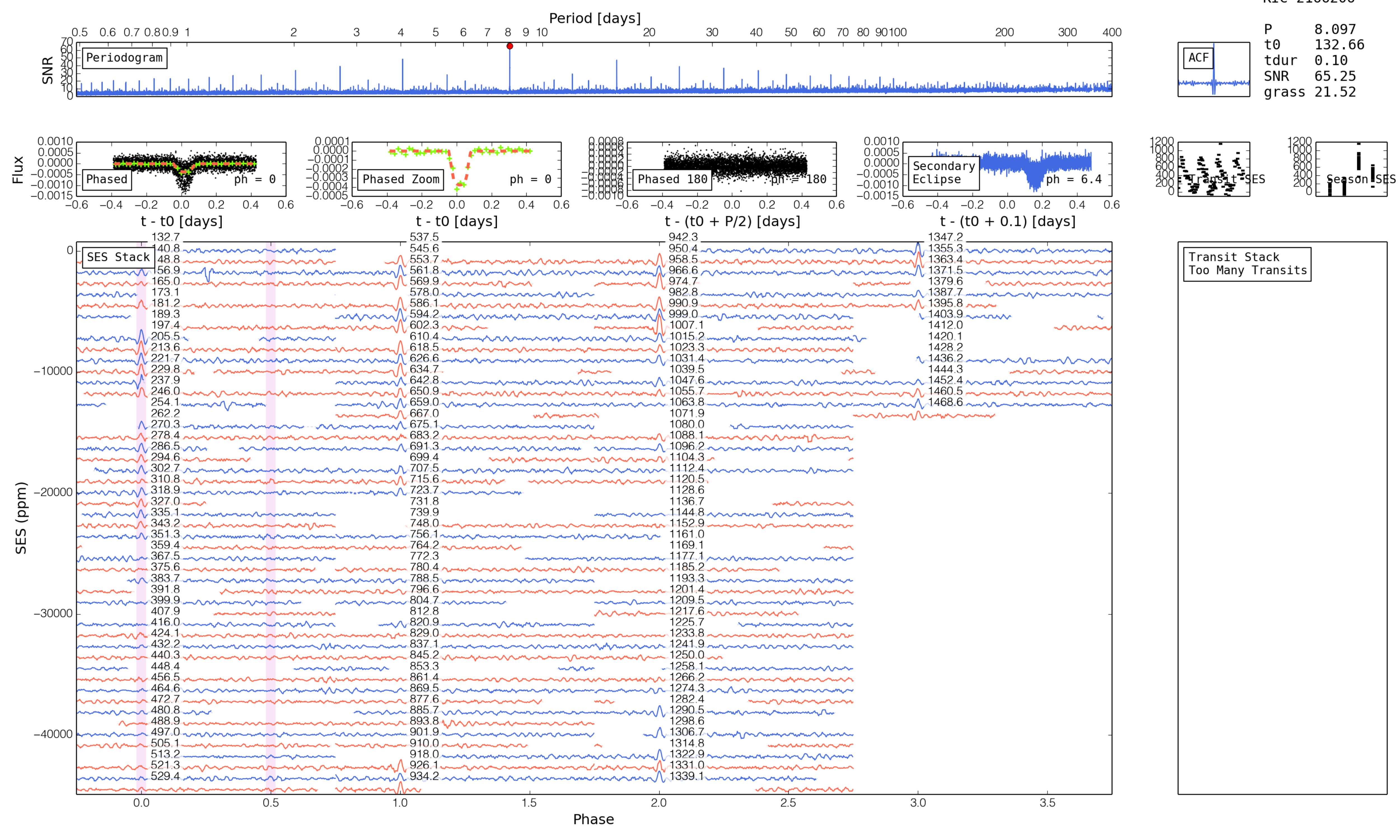}
\caption[DV summary plots eKOI KIC-2166206 ]
  {DV summary plots (defined in Figure~\ref{fig:eKOIpass}) for
  eKOI KIC-2166206 with season-dependent transit depths. Season SES
  plot shows that the transit depths vary significantly for different
  observing seasons. Since the apparent dimming is a strong function
  of the orientation of the spacecraft, the dimming is likely not
  associated with KIC-2166206, but rather an EB displaced from the
  target by several arc seconds.}
\label{fig:SeasonDependent}
\end{figure*}
\end{landscape}

We take great care to cleanse our sample of false positives
(FPs). Some transits are so deep ($\df \gtrsim10\%$) that they can
only be caused by an EB. However, if an EB is close enough to a
\Kepler target star, the dimming of the EB can be diluted to the point
where it resembles a planetary transit. For each eKOI, we assess four
indicators of EB status. Here, we list the indicators along with the
number of eKOIs removed from our planet sample due to each cut:
\begin{enumerate}
\item {\em Radius too large (\nFPR)}. We consider any transit where
  the best fit planet radius is larger than 20~\Re to be
  stellar. Planets are generally smaller than
  $1.5~R_{\text{J}}~=~16~\Re$ especially for \Per~>~5~days, where
  planets are less inflated. Our cut at 20 \Re allows some margin of
  safety to account for mis-characterized stellar radii.

\item {\em Secondary eclipse (\nFPSE)}. The expected equilibrium
  temperature for a planet with \Per~>~5~days is too small to produce a
  detectable secondary eclipse. Therefore, the presence of a secondary
  eclipse indicates the eclipsing body is stellar. We search for
  secondary eclipses by masking out the primary transit and searching
  for additional transits at the same period. If an eKOI, such as
  KIC-8879427 shown in Figure~\ref{fig:SecondaryEclipse}, has a
  secondary eclipse, we designate it an EB.

\item {\em Variable depth transits (\nFPVD)}. Since \Kepler
  photometric apertures are typically two or three pixels (8 or 12
  arcsec) on a side, light from neighboring stars can contribute to
  the overall photometry. A faint EB, when diluted with the target
  star's light, can produce a dimming that looks like a planetary
  transit. If the angular separation between the two stars is large
  enough, the EB will contribute a different amount of light at each
  \Kepler orientation. For eKOIs like KIC-2166206 shown in
  Figure~\ref{fig:SeasonDependent}, the contribution of a nearby EB
  results in a season-dependent transit depths. Since the target 
  apertures are defined to include nearly all ($\gtrsim90\%$) of 
  the light from the target star, variations between
  quarters produce a negligible effect  on transits associated with the target star, i.e. fractional changes of $\lesssim 1\%$.

\item {\em Centroid offset (\nFPC)}. \Kepler project DV reports exist
  for nearly all (\nPlntMyDVCentInfo/\nPlntMyDV) of the eKOIs that
  survive the previous cuts and are available on the Exoplanet Archive.  We inspect the
  transit astronomy diagnostics~\cite{Bryson13} for significant motion
  of the transit photocenter in and out of transit. eKOIs with
  significant motion are designated false positives.

\end{enumerate}

We remove a small number of eKOIs (11) with V-shaped transits. Since
planets are so much smaller than their host stars, ingress/egress
durations are short compared to the duration of the transit,
i.e. planetary transits are box-shaped. Stellar eclipses tend to be
V-shaped. Limb-darkening, the 30-minute integration time, and the
possibility of grazing incidence blur this distinction. We assessed
transit shape visually rather than using more detailed approaches
based on light curve fitting and models of Galactic
structure~\cite{Torres11,Morton12}. Only 1.3\% of eKOIs are removed in
this way and are a small effect compared to other uncertainties in our
occurrence measurements.

We also remove five eKOIs with large TTVs. Since \TERRA's light curve
fitting assumes constant period, fits are biased toward smaller planet
radii in the presence of transit timing variations $\gtrsim \tdur$. If
the resulting error is $\gtrsim 25\%$, we remove that eKOI. While
these eKOIs are likely planets, our constant period model results in a
significant bias in derived planet radii. Given the small number of
eKOIs with such large TTVs, our decision to remove them has small
effect on our statistical results based off of hundreds of planets.

\begin{figure}
\centering
\includegraphics[width=0.8\textwidth]{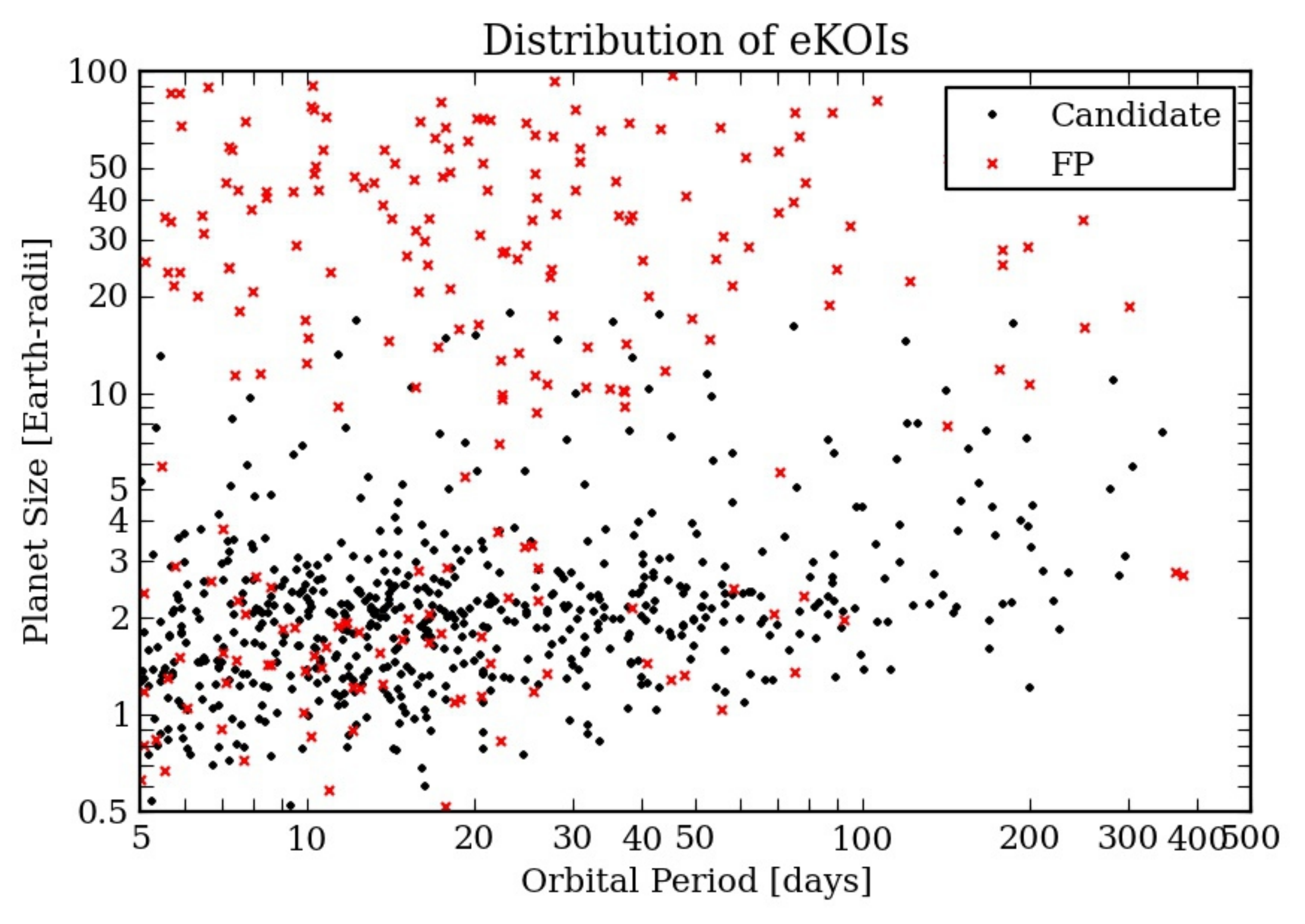}
\caption[Sizes and orbital periods of \TERRA planet candidates]
{Distribution of \TERRA planet candidates as a function of
  planet size and period. Candidates are labeled as black points; FPs
  are labeled as red Xs.}
\label{fig:Catalog}
\end{figure}

We compute planet occurrence from the \nPlnt eKOIs that survive the
above cuts. We show the distribution of \TERRA candidates and FPs on
the \Per--\Rp plane in Figure~\ref{fig:Catalog}. All \neKOI eKOIs are
listed in Table~\ref{tab:eKOI}. For each eKOI, Table~\ref{tab:eKOI}
lists KIC identifier, transit ephemeris, FP designation, Mandel-Agol
fit parameters, adopted host star parameters, and size. We also
crossed checked our eKOIs against the catalog \Kepler team KOIs
accessed from the NASA Exoplanet Archive~\cite{Akeson13} on 13
September 2013. If the \Kepler Project KOI number exists for an
eKOI, we include it in Table~\ref{tab:eKOI}.

\clearpage

\section{Planet Radius Refinement}
\label{sec:RpRefine}
We fit the phase folded transit photometry of each eKOI with a
Mandel-Agol model~\cite{Mandel02}. That model has three free
parameters: \rrat, the planet to star radius radio, $\tau$, the time
for the planet travel a distance \Rstar during during transit; and
$b$, the impact parameter. Following P13, we account for the
covariance among the three parameters using an MCMC exploration of the
parameter posteriors. The error on \rrat in Table~\ref{tab:eKOI}
incorporates the covariance with $\tau$ and $b$.

Because photometry alone only provides the \rrat, knowledge of the
planet population depends heavily on our characterization of their
stellar hosts. We obtained spectra of \neKOISM eKOIs with HIRES on the
Keck I telescope using the standard configuration of the California
Planet Survey (Marcy et al. 2008). These spectra have resolution of
$\sim$50,000 and SNR of $\sim$45/pixel at 5500 \AA. We obtained spectra for
all 62 eKOIs with \Per~>~100~days.

We determine stellar parameters using a routine called {\tt
  \SpecMatch} (Petigura et al., in prep). {\tt \SpecMatch} compares a
target stellar spectrum to a library of $\sim800$ spectra from stars
that span the HR diagram (\teff~=~3500--7500~K; \logg =
2.0--5.0). Parameters for the library stars are determined from LTE
spectral modeling. Once the target spectrum and library spectrum are
placed on the same wavelength scale, we compute $\chi^2$, the sum of
the squares of the pixel-by-pixel differences in normalized
intensity. The weighted mean of the ten spectra with the lowest
$\chi^2$ values is taken as the final value for the effective
temperature, stellar surface gravity, and metallicity. We estimate
{\tt \SpecMatch}-derived stellar radii are uncertain to 10\% RMS,
based on tests of stars having known radii from high resolution
spectroscopy and asteroseismology.

\section{Completeness}
\label{sec:Completeness}

\begin{figure*}
\centering
\includegraphics[width=0.8\textwidth]{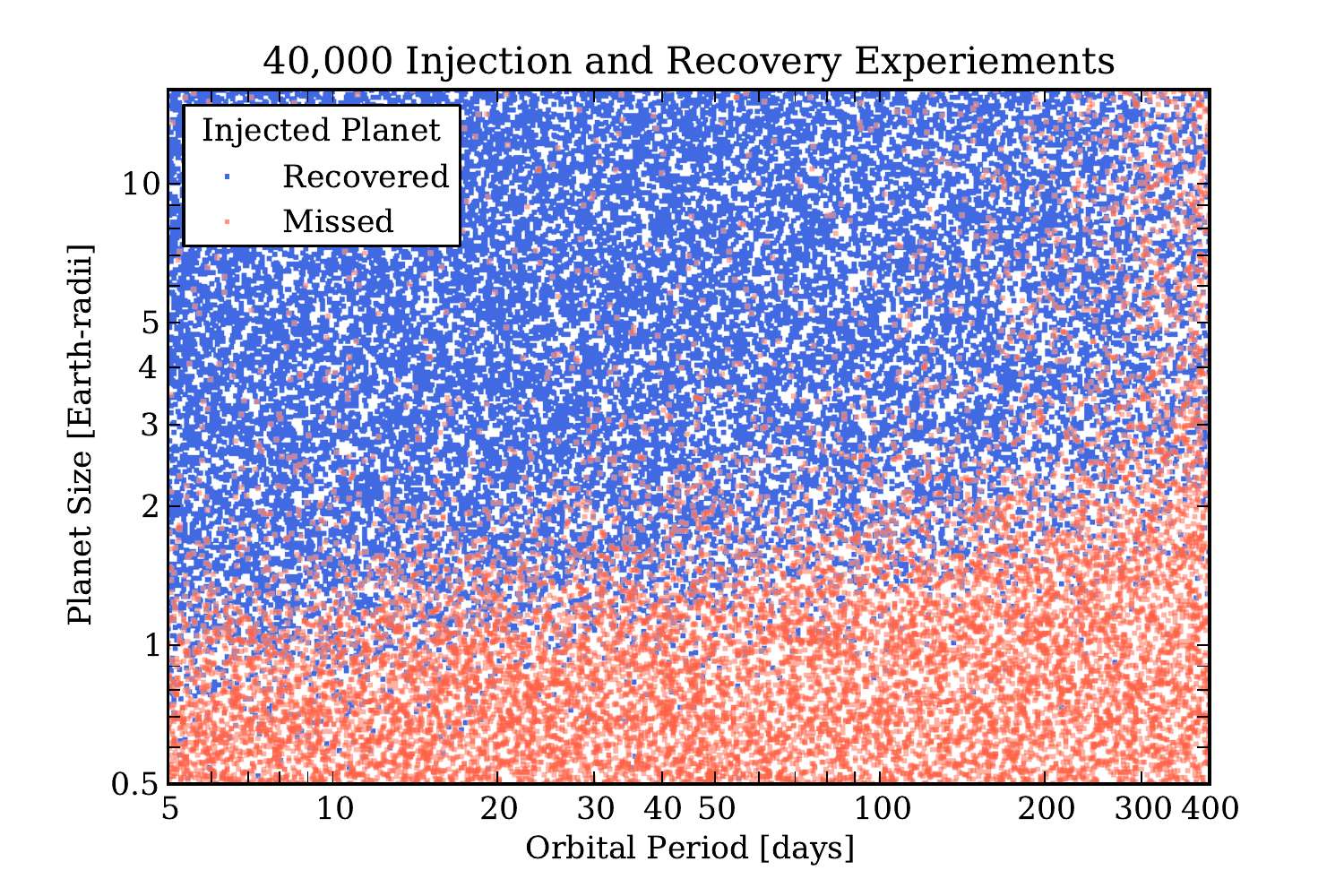}
\caption[Results of injection and recovery experiments]
  {\Per and \Rp of 40,000 injected planets color coded by
  whether they were recovered by \TERRA. Completeness over a small
  range in \Per--\Rp is computed by dividing the number of
  successfully recovered transits (blue points) by the total number of
  injected transits (blue and red points). For planets larger than
  2~\Re, completeness is > 50\% out to 400 d. Completeness rapidly
  falls over 1--2~\Re and is $\lesssim$ 10\% for planets smaller than
  1~\Re.}
\label{fig:Completeness}
\end{figure*}

When measuring planet occurrence, understanding the number of missed
planets is as important as the planet catalog itself. We measure
\TERRA's planet finding efficiency as a function of \Per and \Rp using
the injection/recovery framework developed for P13. We briefly review
the key aspects of our pipeline completeness study; for more detail,
please see P13. We generate 40,000 synthetic light curves according to
the following steps:
\begin{enumerate}
\item Select a star randomly from the Best42k sample,
\item draw (\Per,\Rp) randomly from log-uniform distributions over
  5--400 d and 0.5--16 \Re,
\item draw impact parameter and orbital phase randomly from uniform
  distributions over 0--1,
\item synthesize a Mandel-Agol model \cite{Mandel02}, and
\item inject the model into the ``simple aperture photometry'' of a
  random Best42k star.
\end{enumerate}
We process the synthetic photometry with the calibration, grid-based
search, and DV components of \TERRA. We consider a synthetic light
curve successfully recovered if the injected (\Per, \ep) agree with
the recovered (\Per,\ep) to 0.1 days. Figure~\ref{fig:Completeness}
shows the distribution of recovered simulations as a function of
injected planet size and orbital period.

Pipeline completeness is determined in small bins in (\Per,\Rp)-space
by dividing the number of successfully recovered transits by the total
number of injected transits on a bin-by-bin basis. This ratio is
\TERRA's recovery rate of putative planets within the Best42k
sample. Pipeline completeness is higher among a more rarefied sample
of low noise stars. However, a smaller sample of stars yields fewer
planets.

\begin{figure*}
\centering
\includegraphics[width=0.8\textwidth]{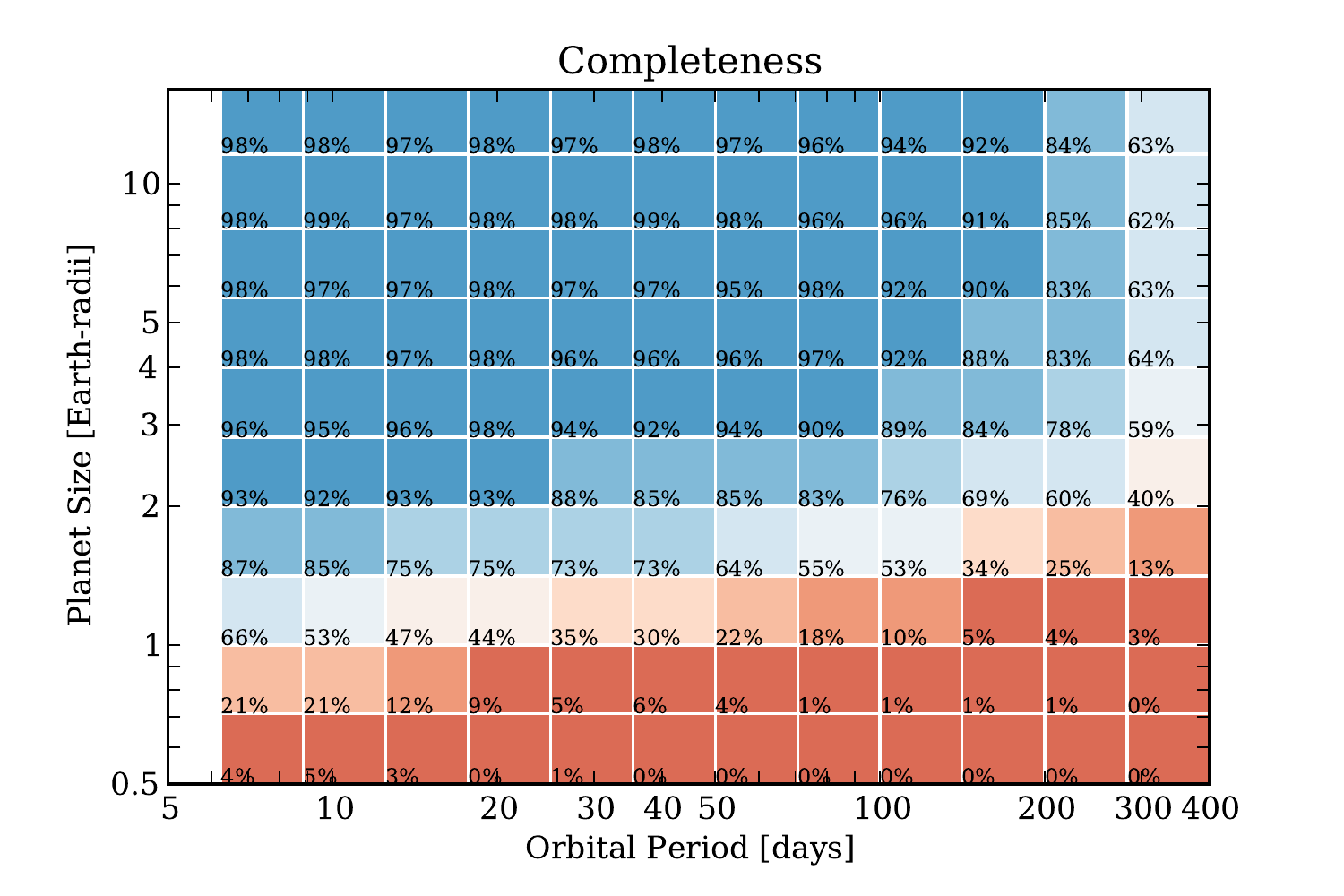}
\caption[Completeness computed over small bins in \Per and \Rp]{Completeness computed over small bins in \Per and \Rp.}
\label{fig:CompletenessBinned}
\end{figure*}

We show survey completeness for a dense grid of \Per and \Rp cells in
Figure~\ref{fig:CompletenessBinned}. Completeness falls toward smaller
\Rp and longer \Per. Above 2 \Re, completeness is greater than 50\%
even for the longest periods searched (except for the \Rp~=~2--2.8
\Re, \Per~=~283--400~days bin). Completeness falls precipitously toward
smaller planet sizes; very few simulated planets smaller than Earth
are recovered. Compared to a 1 \Re planet, a 2 \Re planet produces a
transit with 4 times the SNR and is much easier to detect. For planets
larger than 2 \Re, we note a gradual drop in completeness toward
longer periods, that steepens at $\sim$300 days. Above $\sim$300 days,
the probability that a two or more transits land in data gaps becomes
appreciable, and the completeness falls off more rapidly.

Measuring completeness by injection and recovery captures the vagaries
in planet search pipeline. Real and synthetic transits are treated the
same way, up until the manual triage section. Recall from
Section~\ref{sec:ManualTriage} that \neKOI of \nTCE TCEs pass machine
triage. We perform no such manual inspection of TCEs from the
injection and recovery simulations. A potential concern is that a
planet may pass machine triage, but is accidentally thrown out in
manual triage. Such a planet would be missing from our planet catalog,
but not properly accounted in the occurrence measurement by lower
completeness. However, because our SNR > 12 threshold for TCE status
is high, distinguishing non-astrophysical false positives and eKOIs is
easy. Therefore, we consider it unlikely that they are cut during the
manual triage stage, and do not expect the lack of manual vetting of
the injected TCEs to bias our completeness measurements.

\clearpage

\section{Planet Occurrence}
\label{sec:Occurrence}
Here, we expand on the key planet occurrence results presented in the
main text. We describe our method for extrapolation into the
\Rp~=~1--2~\Re, \Per~=~200--400 day domain. We give additional details
regarding our measurement of the prevalence of Earth-size planets in
the HZ. We also discuss two minor corrections to our occurrence
measurements due to planets in multiplanet systems and false positives
(FPs).

\subsection{Occurrence of Earth-Size Planets on Year-Long Orbits}
In the main text, we reported \EtaEarthErr occurrence of planets with
\Rp~=~1--2~\Re and \Per~=~200--400~days based on extrapolation from
shorter periods. The use of such extrapolation is supported by uniform
planet occurrence per $\log \Per$ interval. Cumulative Planet
Occurrence (CPO) is helpful to understand the detailed shape of the
planet period distribution.  If planet occurrence is constant per
$\log \Per$ interval, CPO is a linear function in $\log \Per$. The
slope of the CPO conveys planet occurrence: the higher the planet
occurrence, the steeper the slope of the CPO.

\begin{figure*}
\centering
\includegraphics[width=0.8\textwidth]{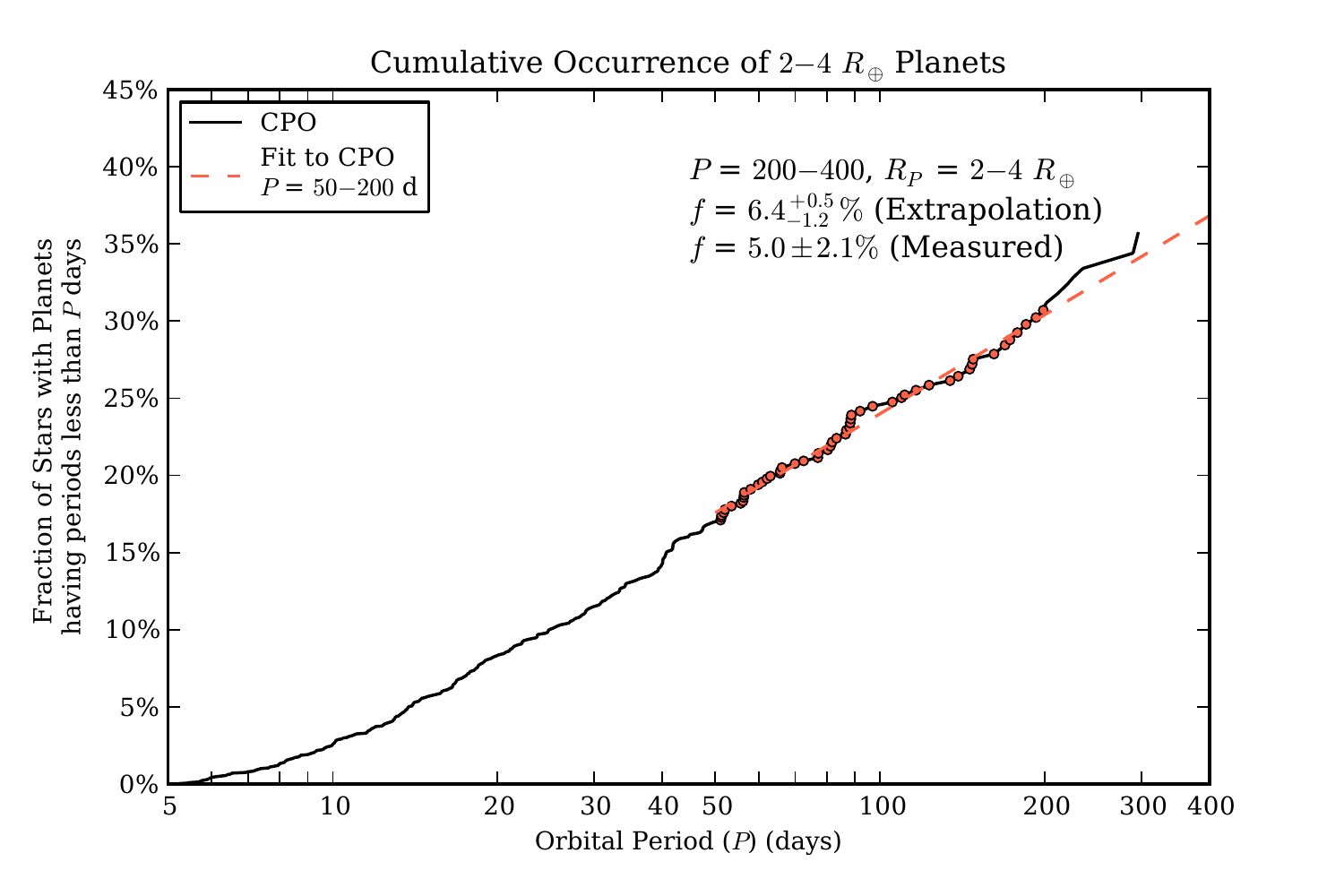}
\caption[Cumulative occurrence of 2--4~\Re planets with increasing orbital period]
  {The fraction of stars having 2--4~\Re planets with any
  orbital period up to a maximum period, \Per, on the horizontal
  axis. This is the Cumulative Planet Occurrence (CPO). A linear
  increase in CPO corresponds to planet occurrence that is constant in
  equal intervals of $\log \Per$.  The CPO steepens from 5 to
  $\sim10$~days, corresponding to increasing planet occurrence in the
  5--10~day range. For $\Per~\gtrsim~10$~days, the CPO has a constant
  slope, reflecting uniform planet occurrence per $\log \Per$
  interval. Planet occurrence in the \Rp~=~2--4~\Re,
  \Per~=~200--400~day domain is predicted to be \fBigYearExtrap by
  extrapolation from shorter periods, which is consistent with our
  measured value of \fBigYearMeas.}
\label{fig:CPO24}
\end{figure*}

Figure~\ref{fig:CPO24} shows CPO for \Rp~=~2--4~\Re planets. Planet
occurrence increases with period from 5 days up to $\sim10$~days, and
is consistent with uniform for larger periods. This change in the
planet period distribution was noted in previous
work~\cite{Youdin11,Howard12,Dong12}. We fit a line to the CPO from
50--200~days and extrapolate into the 200--400~day range. The
extrapolation predicts \fBigYearExtrap occurrence, which agrees with
our measured value of \fBigYearMeas to 1~$\sigma$. We estimate errors
on our extrapolation by fitting subsets of the CPO that span half the
original period range.  We fit 100 subsections ranging from
\Per~=~50--100~days up to \Per~=~100--200~days.

\begin{figure*}
\centering
\includegraphics[width=0.8\textwidth]{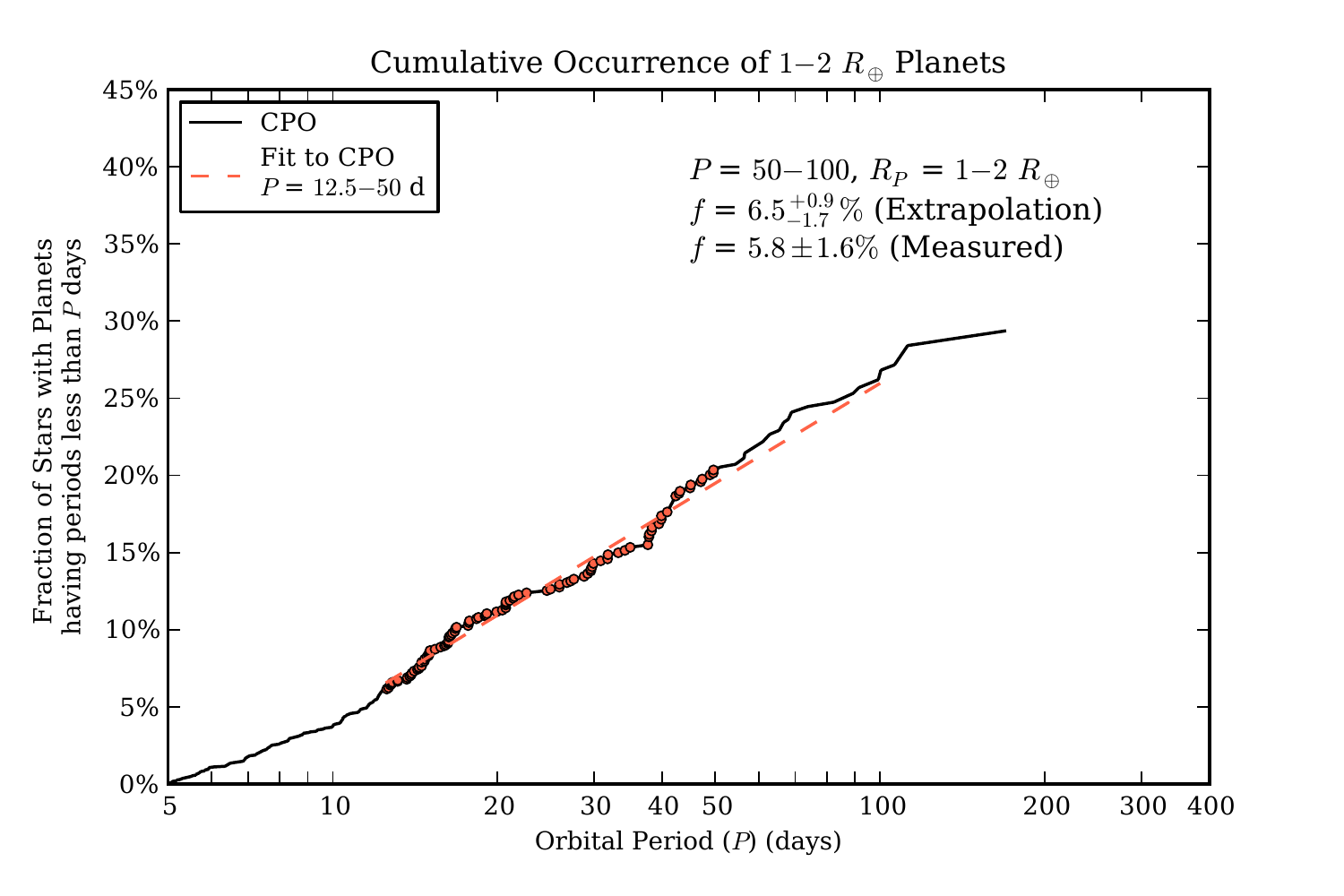}
\caption[Cumulative occurrence of 1--2~\Re planets with increasing orbital period]
  {Same as Figure~\ref{fig:CPO24}, but showing the CPO for 1--2
  \Re planets. Planet occurrence in the \Rp~=~1--2~\Re,
  \Per~=~50--100~day bin is predicted to be \fSmMercExtrap by
  extrapolation from shorter periods, which is consistent with our
  measured value of \fSmMercMeas.}
\label{fig:CPO12}
\end{figure*}

We also compare occurrence in the \Per~=~50--100~day, \Rp~=~1--2~\Re
domain based on extrapolation to our measured value. Figure~\ref{fig:CPO12}
shows the CPO for \Rp~=~1--2~\Re planets. We fit the CPO from
\Per~=~12.5--50~days. This fit predicts an occurrence of
\fSmMercExtrap in the 50--100~day range, in good agreement with our
measured value of \fSmMercMeas. The uniformity in the occurrence of
small planets as a function of period, lends support to the same kind
of modest extrapolation into the \Rp~=~1--2~\Re, \Per~=~200--400~day
domain.

\subsection{Planet Occurrence in the Habitable Zone}
We consider a planet to reside in the habitable zone if it receives a
similar amount of light flux, \Fp, from its host star as does the
Earth. As described in the main text, we consider the most recent
theoretical work on habitability of planets following the seminal work
by Kasting \cite{Kasting1993, Seager13, Kopparapu2013, Zsom2013,
 Pierrehumbert2011}.

We adopt an inner edge of the HZ
at 0.5 AU for a Sun-like star where a planet would receive four times
the light flux that Earth does. This inner edge is slightly more
conservative than that found by Zsom et al.~\cite{Zsom2013}.
The outer edge of the HZ less well
understood.  Kasting found the outer edge to be at 1.7~AU
\cite{Kasting1993}; Pierrehumbert and Gaidos \cite{Pierrehumbert2011}
found it could extend to 10~AU for planets with thick H$_2$
atmospheres. Here, we adopt an intermediate value of 2 AU for solar
analogs where the stellar flux is 1/4 that incident on the Earth. This
outer edge is consistent with the presence of liquid water on Mars in
its past. Mars might still have liquid water today, if it were more
massive.  Thus following the theory of planetary habitability, we
adopt a habitable zone for stars in general based on stellar flux
between 4x and 1/4 the solar flux falling on the Earth:
\Fp~=~0.25--4~\FE.

The stellar light flux hitting a planet, \Fp, depends linearly on
stellar luminosity, \Lstar, and inversely as the square of the
distance between the planet and the star.  Stellar luminosity, \Lstar,
is given by:
\[
\Lstar = 4\pi\Rstar^{2}\sigma\teff^{4},
\]
where $\sigma = 5.670 \times 10^{-8}$~W~m$^{-2}$~K$^{-4}$ is the
Stefan-Boltzmann constant.  In our study, the stellar radii and
temperatures, \teff, are computed two ways.  We obtained high SNR
spectra with high spectral resolution using the Keck Observatory HIRES
spectrometer for all of the 62 stars that host planets with periods
over 100 days, approaching the HZ.  For those 62 stars, we performed a
{\tt SpecMatch} analysis \cite{Petigura13} to determine \teff and the
surface gravity, \logg, and metalicity, [Fe/H].  These stellar values
were matched to stellar evolution models (Yonsei-Yale) to yield the
radii and masses of the stars.  The resulting values of stellar radii
are uncertain by 10\%, as determined by calibrations with nearby stars
having parallaxes and hence having more accurately determined stellar
radii.  The values of \teff are accurate to within 2\%.  Thus, summing
the fractional errors in quadrature, the resulting stellar
luminosities for the 62 stars (having \Per~>~100~days) are measured but
carry uncertainties of 25\%. For those stars without Keck spectra, 
we adopted photometric stellar radius and mass, for which the stellar radii
are in error by 35\% and the \teff values are 
uncertain by 4\%, giving errors in luminosity of 80\%.
We estimated the star-planet separation ($a$) using \Per, $\Mstar$,
and Kepler's third law.  The stellar light flux falling on a planet is
now easily calculated from \Fp $\propto$ \Lstar $/a^2$.  In what
follows, we quote the flux falling on a planet relative to that falling on the
Earth.

\begin{figure}
\centering
\includegraphics[width=0.8\textwidth]{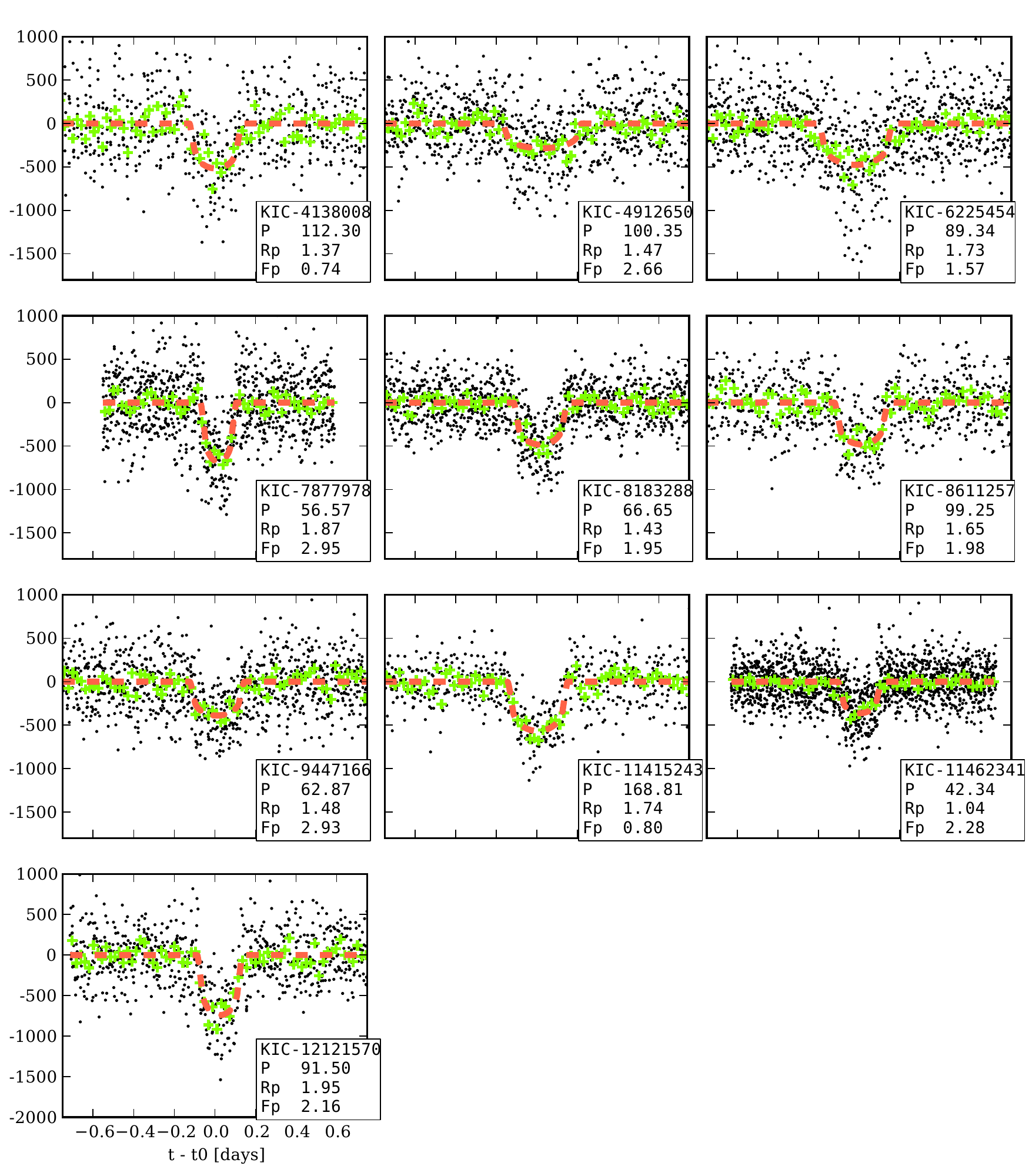}
\caption[Phase folded photometry for ten Earth-size habitable zone candidates]{Phase folded photometry for ten Earth-size HZ candidates. Black point
  shows show \TERRA-calibrated photometry folded on the best fit
  ephemeris listed in Table~\ref{tab:eKOI}. The green symbols show the
  median flux value in 30-minute bins. The red dashed lines shows the
  best-fit Mandel-Agol model. We have annotated each plot with the KIC
  identifier, period, planet size (Earth-radii), incident flux level
  (relative to Earth). All measurements of planet size and incident flux
  are based on spectra taken with the Keck 10~m telescope. Spectra for 
  KIC-6225454, KIC-7877978, KIC-9447166, and KIC-11462341 were obtained during
  peer-review and were added in proof (see Table~\ref{tab:eKOIinpress}).}
\label{fig:HZ10}
\end{figure}

\begin{figure*}
\centering
\includegraphics[width=0.8\textwidth]{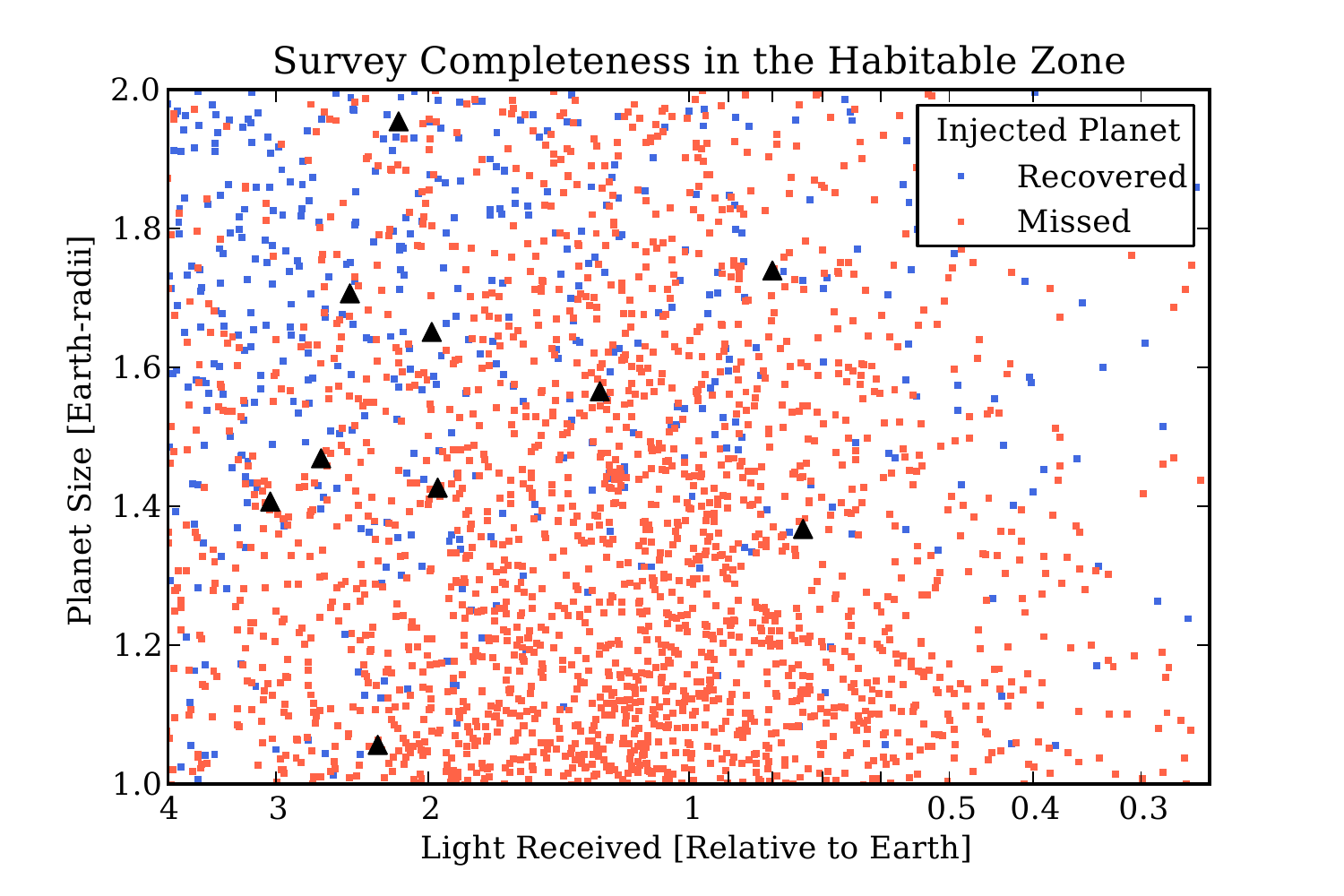}
\caption[Survey completeness for small planets in the habitable zone]
  {Ten small (\Rp~=~1--2~\Re) planets (black triangles) fall
  within our adopted habitable zone of \Fp~=1/4--4~\FE. We also plot
  the injected planets over this same domain. Survey completeness $C$
  is computed locally for each planet by dividing the number of
  injected planets that were recovered by the total number of injected
  planets in a small box centered on the real planet.}
\label{fig:HZComp}
\end{figure*}

We find 10 planets having radii 1--2 \Re that fall within the stellar
incident flux domain of the habitable zone, 0.25--4 \FE. As a
reference, we plot their phase folded light curves in
Figure~\ref{fig:HZ10} along with the KIC identifier, period, radius,
and stellar light flux. To compute the prevalence of such planets
within the HZ, we apply the usual geometric correction for orbital
tilts too large to cause transits, augmenting the counting of each
transiting planet by $a/\Rstar$ total planets. We compute \Fp for each
synthetic planet in our completeness measurement
study. Figure~\ref{fig:HZComp} shows stellar flux level and radii of
the 10 habitable zone planets, having size 1--2~\Re, along with the
synthetic HZ planets from our completeness study. Because the
number of synthetic trials is small for \Fp~<~1~\FE, we compute
occurrence using the 8 planets with \Fp~=~1--4 \FE. We find $11\pm4\%$
of Sun-like stars have a \Rp~=~1--2~\Re planet that receives
\Fp~=~1--4~\FE light energy from their host star.

\begin{figure*}
\centering
\includegraphics[width=0.8\textwidth]{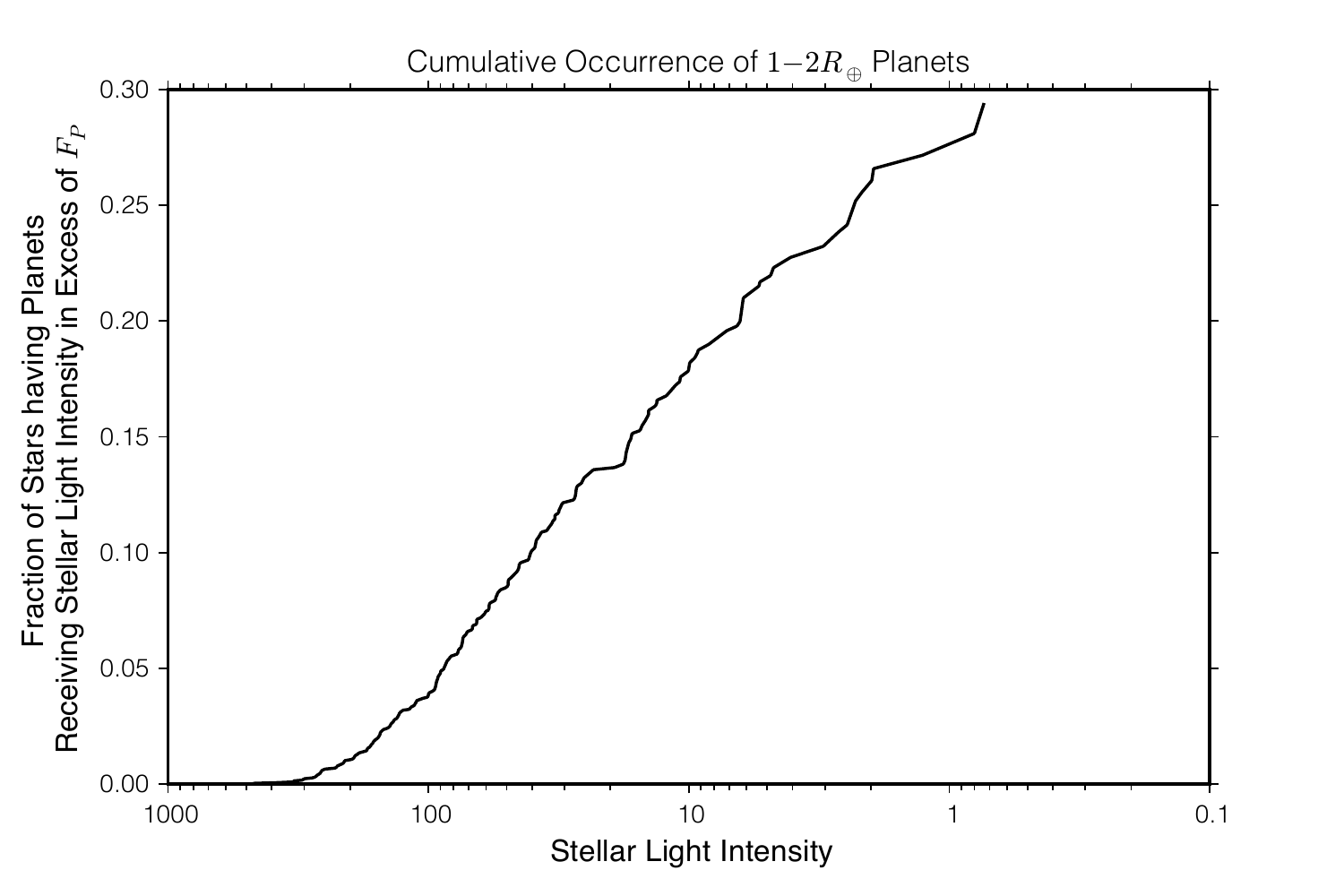}
\caption[Cumulative occurrence of 1--2
  \Re planets with decreasing stellar light intensity]
  {Same as Figure~\ref{fig:CPO24}, but showing the CPO for 1--2
  \Re planets as a function of decreasing flux. Planet occurrence is
  constant over a wide range of incident flux values,
  \Fp~=~100--4~\FE, which supports extrapolation to the outer edge of
  our adopted HZ, \Fp~=~0.25~\FE.}
\label{fig:CPOFp}
\end{figure*}

We account for the entire HZ (extending out to 0.25~\FE) by
extrapolating occurrence in \Fp, assuming constant planet occurrence
per $\log \Per$ interval. Figure~\ref{fig:CPOFp} shows the CPO as a
function of \Fp. Planet occurrence is constant from $\sim100~\FE$ down
to $\sim4~\FE$, beyond which, small number fluctuations are
significant. Assuming the occurrence of planets is constant in $\log
\Fp$ implies that the same number of 1--2 \Re planets have incident
fluxes of 1--0.25 \FE as have fluxes of 1--4 \FE where we computed
directly the occurrence of planets to be 11\%.  Thus, $22\pm8\%$ of
Sun-like stars have a \Rp~=~1--2~\Re planet within our adopted
habitable zone with fluxes of 0.25-4.0 \FE.

\clearpage

\subsection{Occurrence Including Planets in Multi-planet Systems}
For systems harboring more than one planet, \TERRA only detects the
planet with highest SNR, i.e. the most significant planet. The actual
rate of planet occurrence is higher than we report when the missed
planets in these multi-transiting systems is included.  (Note that
this correction only applies to multi-transiting system and not all
multi-planet systems.) We estimate the size of this effect using the
Q12 sample of KOIs from the \Kepler project, which includes stars with
multiple planets. We selected the 1190 ``candidates'' with
well-determined periods ($\sigma(\Per) < 0.1$~days) that orbit stars in
the Best42k. In order to make a fair comparison between our planet
sample and the Q12 sample, we computed the SNR of each of the 1190
candidates using \TERRA. We excluded 82 KOIs with SNR < 12,
i.e. candidates that would have been deemed sub-significant by \TERRA.

\begin{figure}
\centering
\includegraphics[width=0.8\textwidth]{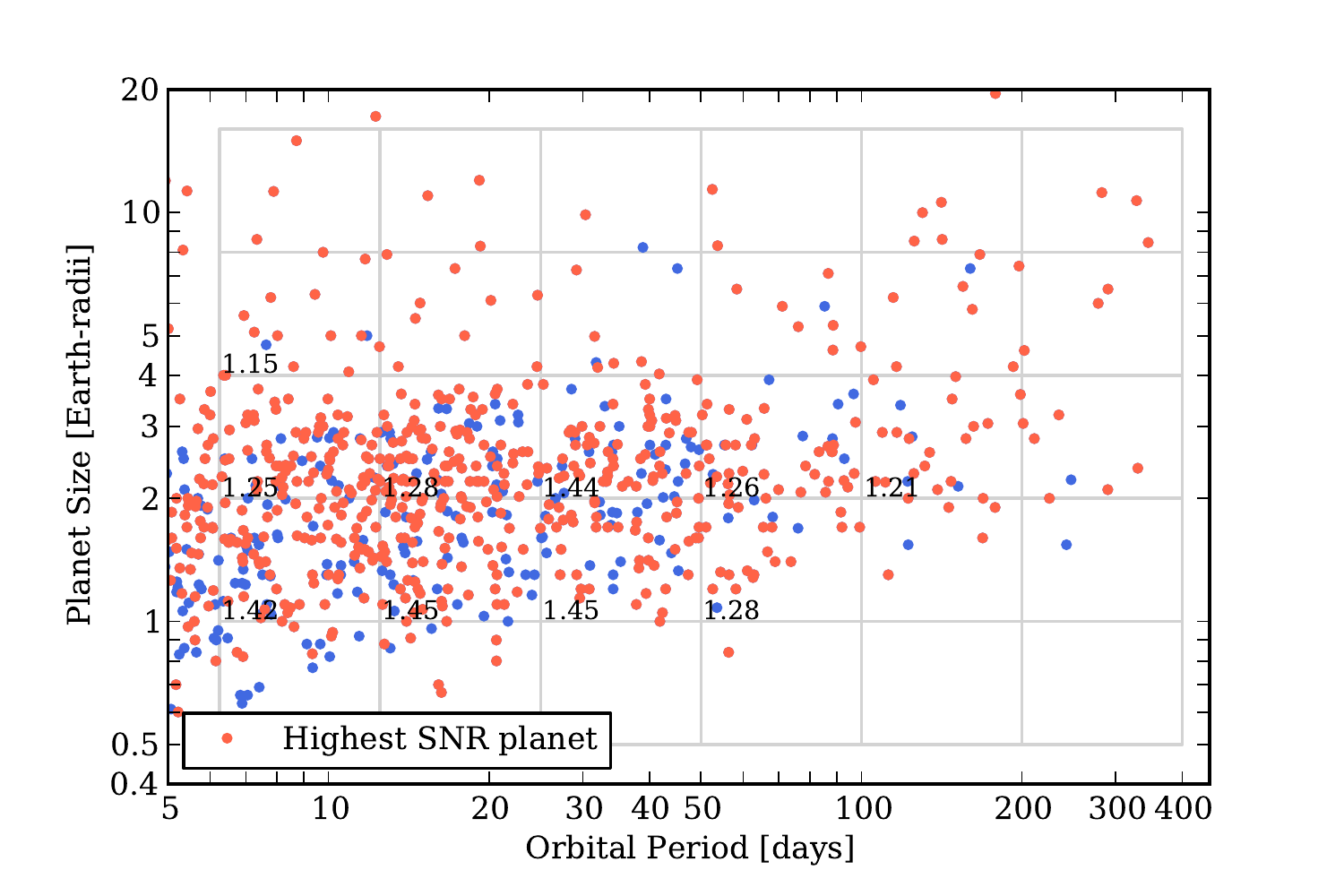}
\caption[Planet occurrence including multi-planet systems]
  {Distribution of \Per and \Rp for \Kepler team candidates
  from the Q12 catalog. We include planets with \TERRA SNR > 12 and
  well-determined orbital periods ($\sigma(\Per) < 0.1$~days) that around
  ``Best42k'' stars. Planets that are either single or are the most
  significant planet in a multi-planet system are shown in red. Blue
  points correspond to additional planets in multi planet systems. For
  each cell with 10 or more planets, we compute the boost in planet
  counts due to multiplanet systems, the total number of planets
  divide by the number of most significant planets. For planets
  smaller than 4~\Re and \Per~50~days, the boost ranges from
  21--28\%. Thus, including multis, we expect $\sim25\%$ higher
  occurrence.}
\label{fig:MultiBoost}
\end{figure}

For each planet in a multi-transiting system, we rank order each
candidate by its ``Relative SNR'' defined as:
\[
\text{Relative SNR} = \df \sqrt{ \frac{\tdur}{\Per} }.
\]
Figure~\ref{fig:MultiBoost} shows the distribution of Q12 candidates in
the Best42k as points on the \Per--\Rp plane. We highlight points
corresponding to the most significant planet. We assess the boost in
planet counts due to multi-transiting systems for different domains in
\Per and \Rp. For \Per~>~50~days and \Rp~<~4~\Re, this multi-boost factor
ranges from 21 to 28\%, neglecting bins with fewer than 8 detected
planets that suffer from small number fluctuations. Had we included
additional planets, our occurrence measurements would rise by
$\sim25\%$, which is comparable to or slightly smaller than the
fractional occurrence error for small planets in long-period orbits.

\subsection{Correction due to False Positives}
As discussed earlier, the sample of eKOIs is polluted by astrophysical
false positives. Like the \Kepler team, we do our best to identify and
remove transits that are clearly due to eclipsing binaries, but cannot
remove all eclipsing binary configurations. Thus, our sample, as well
as those produced by the Kepler team, still contain a false positive
component.

Fressin et al. (2013) addressed the contamination of the February 2012
\Kepler Project sample of KOIs~\cite{Batalha12} by FPs that were not
removed by the \Kepler Project vetting process. FPs include background
eclipsing binaries, physically associated eclipsing binaries
(hierarchical triples), and physically associated stars, which
themselves have a transiting planet. We consider the last scenario to
be a FP because even though the transiting object is a planet, the
radius is at least 1.4 times larger (1.4 corresponds stars of equal
brightness). Fressin et al. (2013) added FPs to the Batalha et al
(2012) sample of KOIs according to models of galactic structure,
stellar binarity, and assumptions about the distributions of
planets. Simulated FPs that would exhibit a detectable secondary
eclipse or a significant centroid offset were removed, assuming the
Kepler Project vetting process catches these FPs.

Fressin et al. (2013) found an overall FP rate of $8.8\pm1.9\%$
for 1.25--2.0~\Re planets and $12.3\pm3.0\%$ for 0.8--1.25~\Re
planets. Again, note that this fractional occurrence correction is
small compared to our reported errors for small planets in long-period
orbits. Stars with bound companions with transiting planets are the
dominant fraction of FPs for small planets (76\% for 1.25--2.0~\Re planets and
66\% for 0.8--1.25~\Re planets). FPs of this type are very difficult
to identify. A Sun-like star with $V$~=~14.7~mag (typical for our
sample) is 1~kpc away. The binary star separation distribution peaks at
50 AU~\cite{Raghavan10} or 0.05 arcsec assuming a face-on
orbit. Detecting companions separated by 0.05 arcsec is near the
limits of current ground-based AO. Even if a companion was detected,
we still wouldn't know which star harbored the transiting planet.

We adopt a 10\% FP rate for planets having \Per~=~50--400~days and
\Rp~=~1--2~\Re. Adopting a false positive rate that is constant with
period is justified because the occurrence of Neptune-sized planets is
approximately constant with period, as shown in the main text. In the
context of the occurrence of Earth-size planets with
\Per~=~200--400~days (\EtaEarthErr) and Earth-size planets in the HZ
($22\pm8\%$), FPs contribute 10\% fractional uncertainty and are
secondary compared to statistical uncertainty.

%\Long-Tables
%\begin{landscape}
\begin{deluxetable}{rrrrlrrrrrrlrrrr}
%\tablenum{S2}
\tabletypesize{\footnotesize}
\setlength{\tabcolsep}{0.035in}
\tablecaption{Properties of \neKOI eKOIs}
\tablewidth{0pt}
\tablehead{
  \colhead{KIC} &
  \colhead{KOI} &
  \colhead{\Per} &
  \colhead{\ep} &
  \colhead{Disp.} &
  \colhead{$p$} &
  \colhead{$\sigma(p)$} &
  \colhead{\teff} &
  \colhead{\logg} &
  \colhead{\Rstar} &
  \colhead{$\sigma(\Rstar)$} &
  \colhead{Prov.} &
  \colhead{\Rp} &
  \colhead{$\sigma(\Rp)$} &
  \colhead{\Fp}\\
%%%%
  \colhead{} &
  \colhead{} &
  \colhead{days} &
  \colhead{days} &
  \colhead{} &
  \colhead{\%} &
  \colhead{} &
  \colhead{K} &
  \colhead{cgs} &
  \colhead{\Rsun} &
  \colhead{} &
  \colhead{} &
  \colhead{\Re} &
  \colhead{} &
  \colhead{\FE} 
}
\startdata
\input{terra1yr/table_ekoi836.tex}
\enddata
\label{tab:eKOI}
\tablecomments{For each of the \neKOI eKOIs, we list the target star
  identifier, ephemeris, false positive status, transit fit
  parameters, host star properties, and planet radius. KIC --- \Kepler
  Input Catalog~\cite{Brown11} identifier. KOI --- \Kepler team
  identifier, if eKOI appears in 13 September 2013 cumulative list of
  candidates from the NASA Exoplanet
  Archive~\cite{Akeson13}. Disp. --- disposition according to the
  false positive vetting described in~\ref{sec:AFP}. eKOIs may be
  designated as a false positives for any of the following reasons:
  `SE' --- secondary eclipse, `VD' --- variable depth transits, `TTV'
  --- large transit timing variations, `V' --- V-shaped transit, `C'
  --- centroid offset. If an eKOI passes all the vetting steps, it is
  considered a planet, `P.' $p$ --- planet to star radius ratio,
  \rrat. Prov. --- provenience of stellar parameters: SpecMatch `SM'
  or photometric `P'.}
\end{deluxetable}
%\end{landscape}

%\LongTables
\begin{deluxetable}{rrrrlrrrrrrlrrrr}
\tabletypesize{\footnotesize}
%\tablenum{S3}
\setlength{\tabcolsep}{0.035in}
\tablecaption{Spectroscopic properties of 13 eKOIs (added in proof)}
\tablewidth{0pt}
\tablehead{
  \colhead{KIC} &
  \colhead{KOI} &
  \colhead{\Per} &
  \colhead{\ep} &
  \colhead{Disp.} &
  \colhead{$p$} &
  \colhead{$\sigma(p)$} &
  \colhead{\teff} &
  \colhead{\logg} &
  \colhead{\Rstar} &
  \colhead{$\sigma(\Rstar)$} &
  \colhead{Prov.} &
  \colhead{\Rp} &
  \colhead{$\sigma(\Rp)$} &
  \colhead{\Fp}\\
%%%%
  \colhead{} &
  \colhead{} &
  \colhead{days} &
  \colhead{days} &
  \colhead{} &
  \colhead{\%} &
  \colhead{} &
  \colhead{K} &
  \colhead{cgs} &
  \colhead{\Rsun} &
  \colhead{} &
  \colhead{} &
  \colhead{\Re} &
  \colhead{} &
  \colhead{\FE} 
}
\startdata
\input{terra1yr/table_ekoi_inpress13.tex}
\enddata
\label{tab:eKOIinpress}
\tablecomments{We obtained 13 additional spectra of long period eKOIs during peer-review. In this work, we used photometric properties for these 13 eKOIs, but we include them here for completeness. Column descriptions are the same as Table~\ref{tab:eKOI}.}
\end{deluxetable}

%% file: terra1yr/table_ekoi836.tex
  1026032 &          &   8.460 & 133.78 &    R & 40.6 &  0.09 & 6044 &   4.5 &  0.95 &   0.29 &    P &  42.12 & 12.64 & 159.7 \\
  1026957 &   958.01 &  21.762 & 144.77 &    P &  3.1 &  0.42 & 4861 &   4.6 &  0.72 &   0.07 &   SM &   2.46 &  0.41 &  13.4 \\
  1571511 &   362.01 &  14.023 & 135.52 &   SE & 12.8 &  0.00 & 6055 &   4.4 &  1.04 &   0.31 &    P &  14.46 &  4.34 &  98.7 \\
  1718189 &   993.01 &  21.854 & 144.32 &    P &  1.7 &  0.10 & 5827 &   4.5 &  0.88 &   0.26 &    P &   1.65 &  0.50 &  34.6 \\
  1849702 &  2538.01 &  39.833 & 155.62 &    P &  1.7 &  0.37 & 5023 &   4.6 &  0.73 &   0.07 &   SM &   1.37 &  0.33 &   7.1 \\
  1872821 &  2351.01 &  10.274 & 134.64 &    P &  1.8 &  0.11 & 5706 &   4.6 &  0.85 &   0.25 &    P &   1.69 &  0.52 &  82.7 \\
  2141783 &  2201.01 & 116.521 & 144.14 &    P &  2.5 &  0.37 & 6063 &   4.2 &  1.53 &   0.15 &   SM &   4.09 &  0.74 &  10.9 \\
  2164169 &  1029.01 &  32.312 & 133.88 &    P &  2.3 &  0.07 & 5975 &   4.5 &  0.96 &   0.29 &    P &   2.39 &  0.72 &  26.5 \\

%% file: terra1yr/table_ekoi_inpress13.tex
  4478142 &          & 219.909 & 329.43 &    P &  2.5 & 0.34 & 5648 &   4.3 &  1.06 &   0.11 &   SM &  2.92 &  0.49 & 2.3 \\
  4820550 &  3823.01 & 202.121 & 292.04 &    P &  5.7 & 0.65 & 5594 &   4.5 &  0.94 &   0.09 &   SM &  5.79 &  0.88 & 1.8 \\
  6225454 &          &  89.338 & 218.54 &    P &  2.4 & 0.58 & 4652 &   4.6 &  0.68 &   0.07 &   SM &  1.73 &  0.46 & 1.6 \\
  6307083 &  2050.01 &  75.378 & 167.11 &   SE &  1.7 & 0.09 & 5109 &   4.6 &  0.77 &   0.08 &   SM &  1.43 &  0.16 & 3.4 \\
  7101828 &   455.01 &  47.878 & 145.35 &   VD &  2.2 & 0.10 & 4328 &   4.7 &  0.55 &   0.05 &   SM &  1.33 &  0.15 & 2.0 \\
  7866914 &  3971.01 & 366.020 & 182.66 &   SE &  4.0 & 1.15 & 5713 &   4.3 &  1.06 &   0.11 &   SM &  4.67 &  1.40 & 1.2 \\
  7877978 &  2760.01 &  56.573 & 146.59 &    P &  2.5 & 0.35 & 4675 &   4.6 &  0.68 &   0.07 &   SM &  1.87 &  0.32 & 3.0 \\
  8044608 &  3523.01 & 106.176 & 234.62 &    R & 55.2 & 2.76 & 6056 &   4.4 &  1.12 &   0.11 &   SM & 67.58 &  7.56 & 7.7 \\

%% file: co/co.tex
% This sample file is dedicated to the public domain.
\chapter{Carbon and Oxygen in Nearby Stars: Keys to Protoplanetary Disk Chemistry}
\label{c.co}

\noindent A version of this chapter was previously published in the {\em Astrophysical Journal}\\
\noindent (Erik~A.~Petigura \& Geoffrey~W.~Marcy, 2011, ApJ 735, 41).\\

\input {co/texcmd.tex}

  We present carbon and oxygen abundances for \nStarsTot~FGK
  stars---the largest such catalog to date.  We find that
  planet-bearing systems are enriched in these elements.  We
  self-consistently measure $N_{C}/ N_{O}$, which is thought to play a key role in planet formation.  We identify~\ncoGtThresh~stars with $N_{C}/ N_{O}$ $\geq$ \coThresh~as potential hosts of carbon-dominated exoplanets. We measure a downward trend in [O/Fe] versus [Fe/H] and find
  distinct trends in the thin and thick disk, supporting the work
  of~\cite{Bensby04}.
  Finally, we measure sub-solar $N_{C}/ N_{O}$ = \waspCO~for WASP-12, a
  surprising result as this star is host to a transiting hot Jupiter
  whose dayside atmosphere was recently reported to have $N_{C}/ N_{O}$ $\geq$ 1
  by \cite{Madhu11}.
  Our measurements are based on \nSpecTot~high signal-to-noise spectra
  taken with the Keck 1 telescope as part of the California Planet
  Search.  We derive abundances from the [OI] and CI absorption lines
  at $\lambda = $ 6300 and 6587~\AA~using the {\tt SME} spectral
  synthesizer.

\section{Introduction}
After primordial hydrogen and helium, carbon and oxygen are the most
abundant elements in the cosmos.  Life on earth is built upon the
versatility of carbon's four valence electrons and is powered by
metabolizing nutrients with oxygen.

The prevalence of carbon and oxygen gives them a prominent role in stellar interiors, opacities, and energy generation.  As a result, studying their abundances helps to reveal the nucleosynthetic chemical evolution of galaxies.  

The interstellar medium is thought to be enriched with oxygen by Type
II supernovae.  Taken with iron, which is produced in both Type Ia and
Type II supernovae, oxygen provides a record of galactic chemical
enrichment and star formation rate~\citep{Bensby04}.  It is well known
that stars synthesize helium into carbon through the triple alpha
reaction.  However, it is still unclear which stars dominate carbon
production in the galaxy.  For a discussion of the possible sites of
carbon synthesis see ~\cite{Gustafsson99}.

The ratio of carbon to oxygen ($N_{C}/ N_{O}$) is thought to play a critical role
in the bulk properties of terrestrial extrasolar
planets. ~\cite{Kuchner05} and~\cite{Bond10} predict that above a
threshold ratio of $N_{C}/ N_{O}$ near unity, planets transition from silicate-
to carbide-dominated compositions.

We present the oxygen and carbon abundances derived from the [OI] line
at 6300~\AA~and the CI line at 6587~\AA~for \nStarsTot~stars in the
California Planet Search (CPS) catalog.  We compute the abundances
with the {\tt Spectroscopy Made Easy (SME)} spectral
synthesizer~\citep{Valenti96}.  Using {\tt SME}, we self-consistently
account for the NiI contamination in [OI] and report detailed Monte
Carlo-based errors.  Others have measured stellar carbon and oxygen
before. \cite{Edvardsson93} measured oxygen in 189 F and G dwarfs,
and~\cite{Gustafsson99} measured carbon in 80 of these stars.  More
recent studies include,~\cite{Bensby05},~\cite{Luck06},
and~\cite{Ramirez07}.  However, the shear number (\nSpecTot) of CPS
spectra give us a unique opportunity to measure the distributions of
these important elements in a large sample.

\section{Observations}
\subsection{Stellar Sample}
The stellar sample is drawn from the Spectroscopic Properties of Cool
Stars (SPOCS) catalog \citep[hereafter VF05]{Valenti05} and from the N2K
(``Next 2000'') sample~\citep{N2K}.  We include 533 N2K stars and 537
VF05 stars for a total of~\nSampStars~stars.

We adopt stellar atmospheric parameters for each star from VF05 and
from the identical analysis for the N2K targets (D. Fischer 2008,
private communication).  These parameters are: effective temperature,
$T_{eff}$; gravity, $\log g$; metallicity, [M/H]; rotational
broadening, $v \sin i$; macroturbulent broadening, $v_{mac}$;
microturbulent broadening, $v_{mic}$; and abundances of Na, Si, Ti,
Fe, and Ni.  Metallically includes all elements heavier than helium.
A star's abundance distribution is the solar abundance pattern
from~\cite{Grevesse98} scaled by the star's metallicity.  Na, Si, Ti,
Fe, and Ni abundances are computed independently from [M/H] and are
allowed vary from scaled solar [M/H].

\subsection{Spectra}
Our spectra were taken with HIRES, the High Resolution Echelle
Spectrograph~\citep{Vogt94} between August, 2004 and April, 2010 on the
Keck 1 Telescope. The spectra were originally obtained by the CPS to
detect exoplanets.  For a more complete description of the CPS and its
goals, see~\cite{Marcy08}.  The CPS uses the same detector setup each
observing run and employs the HIRES exposure meter (Kibrick et
al. 2006) to set exposure times, ensuring consistent and high quality
spectra across years of data collection. The spectra have resolution
R~=~50,000 and S/N~$\sim$~200 at 6300 and 6587~\AA.  This analysis
deals with three classes of observations:

\begin{enumerate}
\item {\em Iodine cell in}.  For the majority of its observations, the
  CPS passes starlight through an iodine cell~\citep{Marcy92}, which
  imprints lines between 5000 and 6400~\AA~that serve as a wavelength
  fiducial.  We discuss how we remove these lines and their effect on
  oxygen measurements in \S \ref{sec:iodine}.
\item {\em Iodine cell out}.  Calibration spectra taken without the
  iodine cell.
\item {\em Iodine reference}.  At the beginning and end of each
  observing night, the CPS takes reference spectra of the iodine cell
  using an incandescent lamp.
\end{enumerate}

\section{Spectroscopic Analysis}
\subsection{Line Synthesis}
We use the {\tt SME} suite of routines to fine-tune line lists based
on the solar spectrum, determine global stellar parameters, and
measure carbon and oxygen.  To generate a synthetic spectrum, {\tt
  SME} first constructs a model atmosphere by interpolating between
the~\cite{Kurucz92} grid of model atmospheres.  Then, {\tt SME} solves
the equations of radiative transfer assuming Local Thermodynamic
Equilibrium (LTE).  Finally, {\tt SME} applies line-broadening to
account for photospheric turbulence, stellar rotation, and instrument
profile.  For a more complete description of {\tt SME}, please
consult~\cite{Valenti96} and VF05.  We emphasize that {\tt SME} solves
molecular and ionization equilibrium for a core group (around 400) of
species that includes CO (N. Piskunov 2011, private communication).

\subsection{Atomic Parameters}
Measuring stellar oxygen is notoriously difficult because of the
limited number of indicators in visible wavelengths.  The general
consensus is that the weak, forbidden [OI] transition at 6300~\AA~is
the best indicator because it is less sensitive to departures from
local thermodynamic equilibrium than other indicators.  In dwarf stars, this line suffers from a significant NiI blend, which is 
an isotopic splitting of $^{58}$Ni and $^{60}$Ni \citep{Johansson03}.  
The NiI feature was first noted by~\cite{Lambert78}, but only recently
included in abundance studies~\citep{Prieto01}.  Carbon is more generous to visual spectroscopists.  We select the CI line at
6587~\AA~because it sits relatively far from neighboring lines and is
in a wavelength region with weak iodine lines (see \textsection~\ref{sec:iodine}).

Line lists are initially drawn from the Vienna Astrophysics Line
Database~\citep{VALD}.  We tune line parameters by fitting the 
disk-integrated National Solar Observatory (NSO) solar spectrum of~\cite{Kurucz84} with the {\tt SME} model of the solar atmosphere.  
Table~\ref{tab:solpar} lists the atmospheric parameters adopted when 
modeling the sun. We fit a broad spectral range from 6295 to 6305~\AA~surrounding the
[OI] line and 6584 to 6591~\AA~surrounding the CI line.  We adopt the solar abundances of~\cite{Grevesse98} except for O and Ni where we adopt $%
\logepso$\footnote{$\log \epsilon_X = \log_{10} (N_X /N_H ) + 12$}
= 8.70 and $\logepsni = 6.17$ ~\citep{Scott09} and $\logepsc = 8.50$
~\citep{Caffau10}.  We adjust line centers, van der Waals broadening
parameters ($\Gamma_6$), and oscillator strengths ($\log gf$) so our
synthetic spectra best match the NSO atlas.  Table~\ref{tab:atomic}
shows the best fit atomic parameters after fitting the NSO solar
atlas.

Given the high quality of the solar spectrum, solar abundances
and line parameters are often measured using sophisticated 
three-dimensional, hydrodynamical, non-LTE codes.  
For this work, however, we are more interested in 
self-consistently determining line parameters using {\tt SME}
than from a more sophisicated solar model.
As a result, the line parameters in Table~\ref{tab:atomic} are not in tight
agreement with the best laboratory measurements.  For example, 
\cite{Johansson03} measured $\log gf = -2.11$ for the NiI blend
in contrast to $\log gf = -1.98$ in this work.  The purpose of fitting the
atomic parameters in the sun is to determine the best parameters given our
atmospheric code and our adopted solar abundance distribution.

\begin{figure}
\begin{center}
\includegraphics{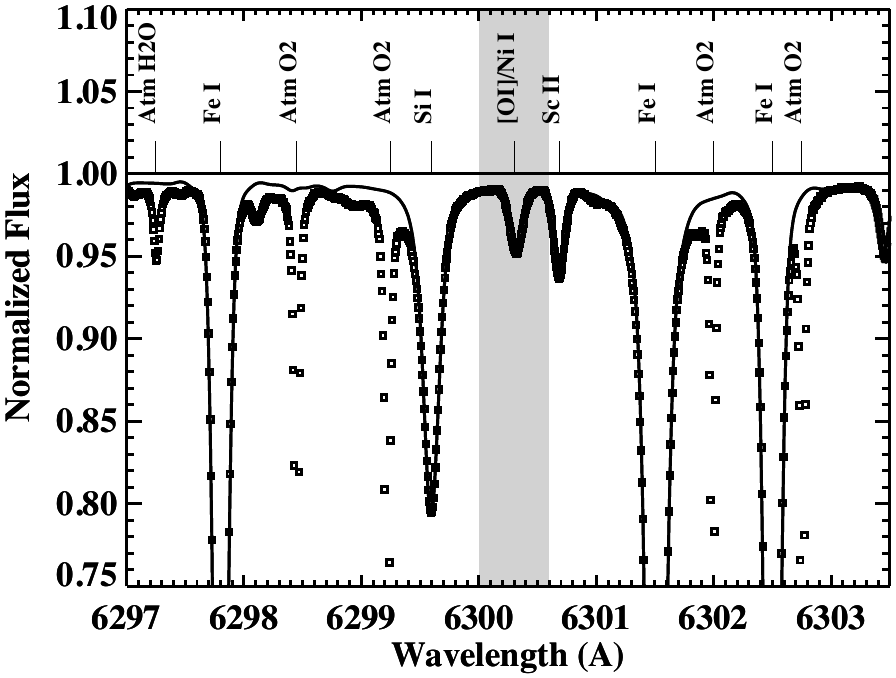}
\end{center}
\caption[Solar spectrum in the vicinity of the $\mathrm{[OI]}$ line at 6300.312~\AA]
{Solar spectrum in the vicinity of the [OI] line at 6300.312~\AA.  The points are from the NSO solar atlas, and the solid line is the {\tt SME} fit.  The shaded region marks the region that is included in the $\chi^2$ fit to the [OI]/NiI blend.}
\label{fig:solar_6300}
\end{figure}

\begin{figure}
\begin{center}
\includegraphics{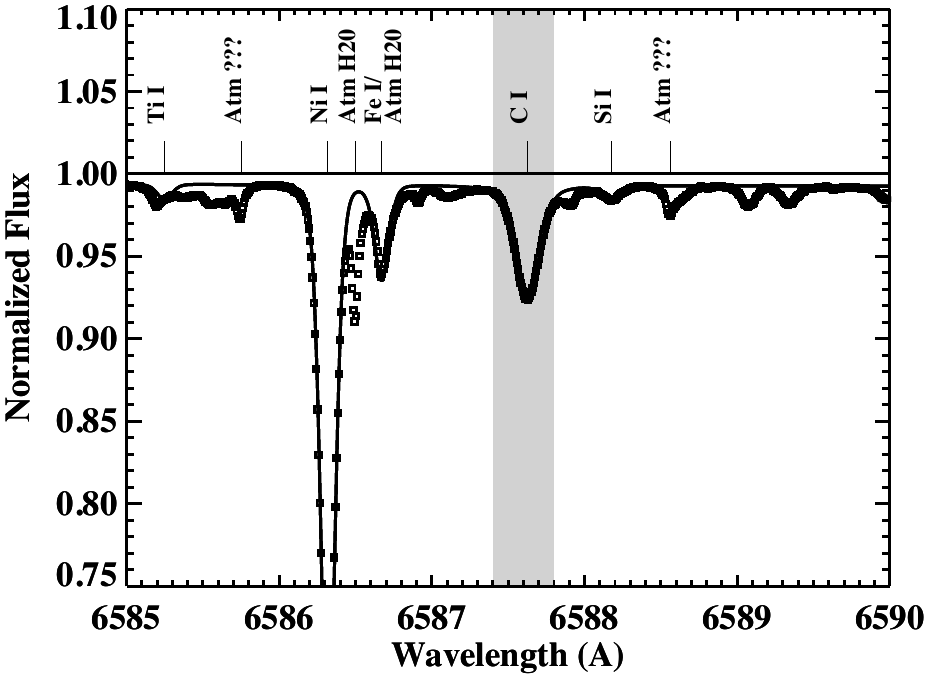}
\end{center}
\caption[Solar spectrum in the vicinity of the CI line at 6587.625~\AA]{Solar spectrum in the vicinity of the CI line at 6587.625~\AA.  The points are from the NSO solar atlas, and the solid line is the {\tt SME} fit.  The shaded region marks the region that is included in the $\chi^2$ fit to the CI line.}
\label{fig:solar_6587}
\end{figure}

We show the fitted NSO spectrum for both wavelength regions in
Figures~\ref{fig:solar_6300} and~\ref{fig:solar_6587}.  The shaded
regions (6300.0-6300.6 and 6587.4-6587.8~\AA) represent the fitting
region.  Only points in the fitting region are used in the $\chi^2$
minimization routines (see \S~\ref{sec:fitting}).

\begin{figure}
\begin{center}
\includegraphics{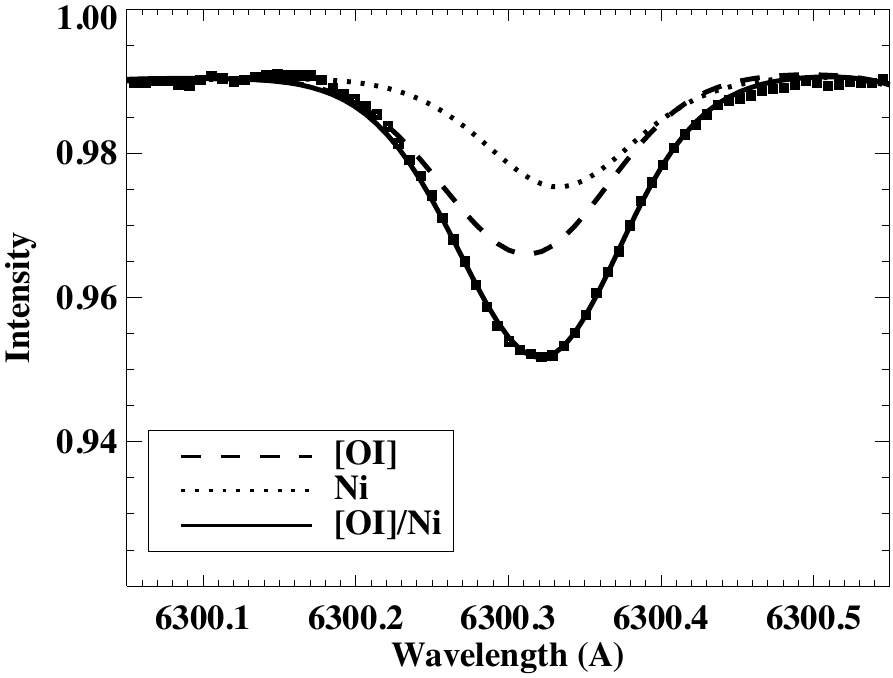}
\end{center}
\caption[The $\mathrm{[OI]/NiI}$ blend in the NSO spectrum.]
{The [OI]/NiI blend in the NSO spectrum.  The points are the from NSO solar atlas, and the solid line is the {\tt SME} fit.  The relative contribution of [OI] and Ni are shown by the dashed and dotted lines respectively.}
\label{fig:solarzoom_6300}
\end{figure}

Figure~\ref{fig:solarzoom_6300} shows a close up view of the [OI]/NiI
blend in the sun.  To help the reader visualize the relative contributions
of each line in the sun, we synthesize the oxygen and nickel lines individually.
To compute the relative strength of [OI], we remove all Ni in our solar
model and re-synthesize the spectrum in {\tt SME}.  
To calculate the NiI contribution we remove all oxygen.  
Since the both lines are weak ($<$ 5 \% of continuum), the line profile
for the [OI]/Ni blend is nearly the product of the individual [OI] and Ni lines.
This would not be true in the case of deeper lines.
In the sun, the [OI] and NiI contributions to the blend are comparable.  
In some stars, the blend is decidedly nickel-dominated, while in others,
oxygen dominates.

\begin{figure}
\begin{center}
\includegraphics{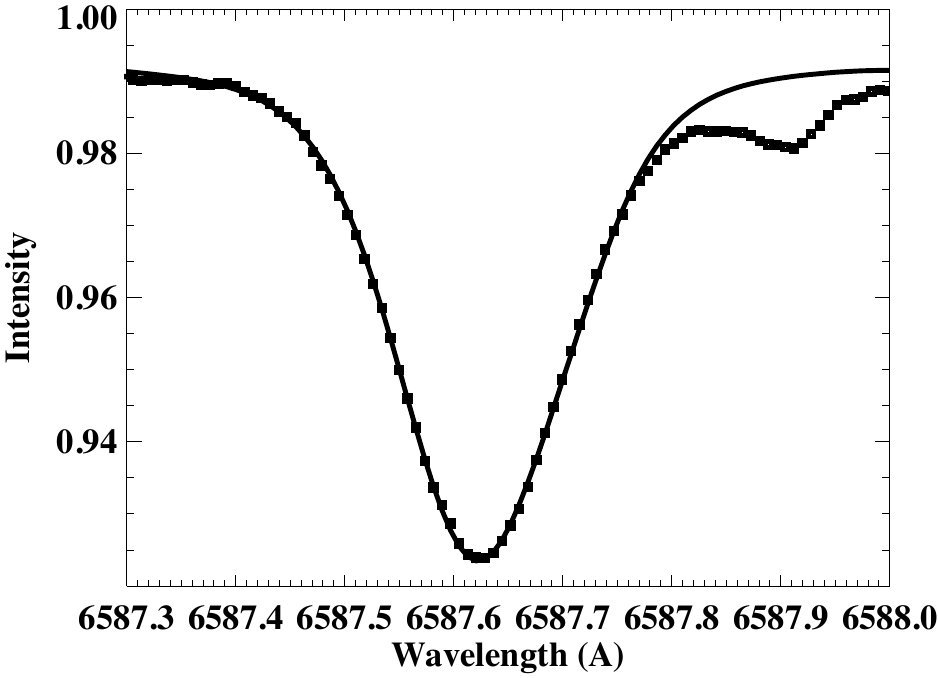}
\end{center}
\caption[The CI line in the NSO spectrum]
{The CI line in the NSO spectrum.  The points are the from NSO solar atlas, and the solid line is the {\tt SME} fit.  The fitting region for this line is 6587.4-6587.8~\AA~and excludes the unidentified feature at 6587.9.}
\label{fig:solarzoom_6587}
\end{figure}

Figure~\ref{fig:solarzoom_6587} shows the carbon indicator plotted on the same intensity scale as the oxygen detail shown in Figure~\ref{fig:solarzoom_6300}.  There is an unknown line on the red wing of the carbon indicator.  We exclude the mystery line from the fitting region.  

\begin{figure}
\begin{center}
\includegraphics{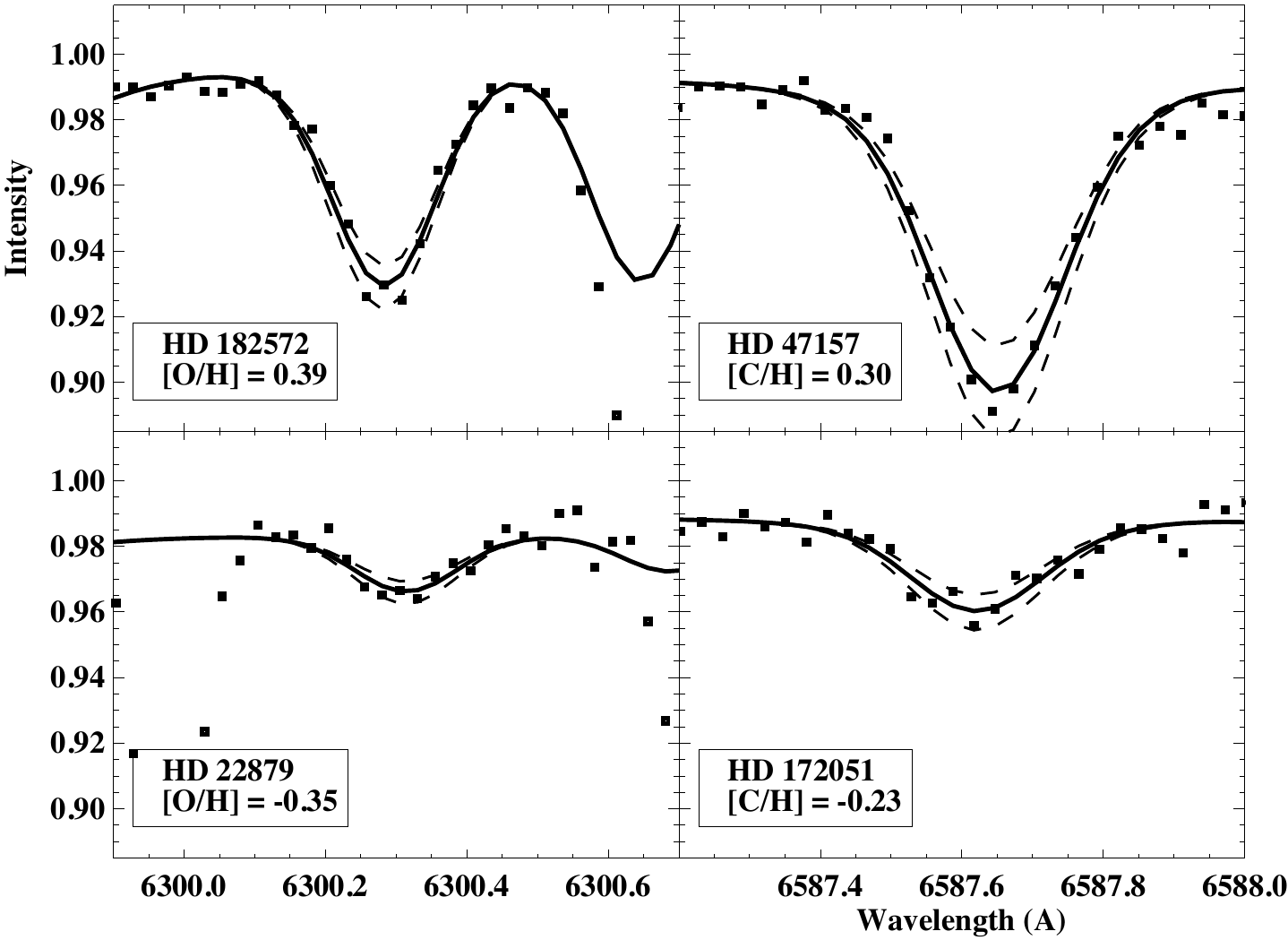}
\end{center}
\caption[Sample spectra of stars with low and high carbon and oxygen abundances]
{Sample spectra of stars with low and high carbon and oxygen abundances.  The solid line shows the best fit {\tt SME} spectrum.  The dashed lines are the {\tt SME} spectra with [X/H] increased and decreased by 0.1 dex $\sim 25$\% from the best fit value.  The [OI] line in the HD~22879 spectrum sits between two telluric lines.
 }
\label{fig:starsamp}
\end{figure}

As a point of reference for the reader, we include stellar counterparts 
to Figures~\ref{fig:solarzoom_6300} and ~\ref{fig:solarzoom_6587} in 
Figure~\ref{fig:starsamp}.  We show stars with low and high carbon and oxygen 
abundance along with the best fit {\tt SME} spectrum.

\begin{deluxetable}{lc}
\tablewidth{0pt}
\setlength{\tabcolsep}{0.3in}
\tablecaption{Adopted solar atmospheric parameters}
\tablehead{
\colhead{Parameter}& 
\colhead{Value}    \\
}
\startdata
\teff     & 5770 K      \\
$\log g$  & 4.44 (cgs) \\
$v_{mic}$ & 1.00 km/s   \\
$v_{mac}$ & 3.60 km/s   \\
\vsini    & 1.60 km/s   \\
$v_{rad}$ & 0.02 km/s   \\
\enddata
\tablecomments{Adopted atmospheric parameters in the {\tt SME} solar model}
\label{tab:solpar}
\end{deluxetable}

\begin{deluxetable}{l c c c}
\tablewidth{0pc}
\tablecaption{Atomic parameters from fitting the NSO atlas}
\tablehead{
\multirow{2}{*}{Element} &
\colhead{$\lambda$}	&
\multirow{2}{*}{$\log gf$} &
\multirow{2}{*}{$\Gamma_6$}  		   \\
&
\colhead{(\AA) }	&
&
\\
}
\startdata
\sidehead{[OI] region}
\hline
Fe 1 & 6297.801 & -2.766 & -7.89 \\
Si 1 & 6297.889 & -2.899 & -6.88 \\
O  1 & 6300.312 & -9.716 & -8.89 \\
Ni 1 & 6300.335 & -1.983 & -7.12 \\
Sc 2 & 6300.685 & -2.041 & -8.01 \\
Fe 1 & 6301.508 & -0.793 & -7.53 \\
Fe 1 & 6302.501 & -0.972 & -7.99 \\
\sidehead{CI region}
\hline
Ti 1 & 6585.249 & -0.399 & -7.56 \\
Ni 1 & 6586.319 & -2.775 & -7.68 \\
Fe 2 & 6586.672 & -2.247 & -7.76 \\
C  1 & 6587.625 & -1.086 & -7.19 \\
Si 1 & 6588.179 & -3.082 & -7.12 \\
\enddata
\tablecomments{Best fit line center ($\lambda$), oscillator strengths
  ($\log gf$), and van der Waals broadening parameter ($\Gamma_6$) for
  our [OI] and CI indicators and nearby lines.  They are derived by
  fitting the NSO atlas.}
\label{tab:atomic}
\end{deluxetable}

\subsection{Telluric Rejection}
There are several telluric lines from O$_2$ and H$_2$O in the vicinity
of our indicators including the 6300.3~\AA~airglow (see
Figures~\ref{fig:solar_6300} and~\ref{fig:solar_6587}).  These lines
are produced in the rest frame of the Earth and contaminate different
parts of a star's spectrum depending on the relative line of sight
velocity between the Earth and the star.  We compute this velocity
directly from the spectra, by cross-correlating the stellar spectra
with the NSO solar atlas.  Based on this velocity, we account for any
shift in the location of the telluric line in the stellar rest frame.

\begin{figure}
\begin{center}
\includegraphics{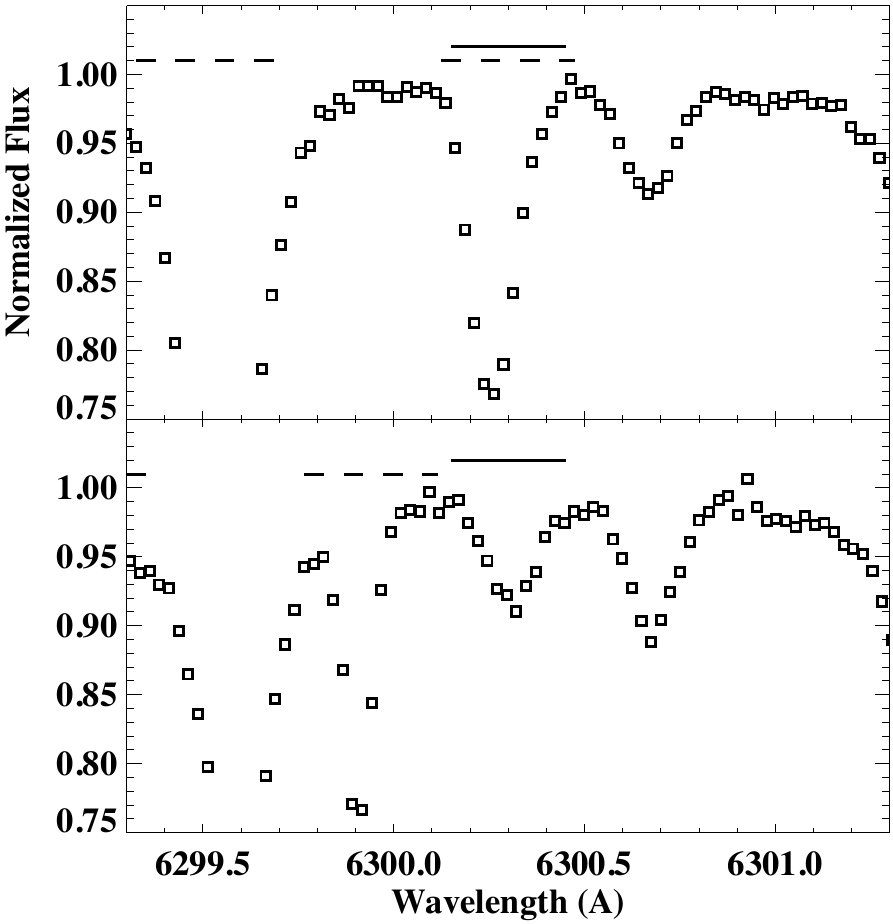}
\end{center}
\caption[Two spectra of HIP~92922 in the star's rest frame]
{Two spectra of HIP~92922 in the star's rest frame.  The [OI]/NiI region (marked with the solid line) in the upper panel is contaminated with an atmospheric O2 (marked with a dashed line).  The same telluric absorption line is shifted 0.3~\AA~blueward in the lower panel and does not contaminate the [OI]/NiI blend.}
\label{fig:tell}
\end{figure}

If a telluric line enters the fitting region, we discard that
observation.  Figure~\ref{fig:tell} shows the [OI]/NiI blend from two
different observations of HIP~92922: one where the blend is
contaminated by a telluric absorption line and one where the blend is
free from telluric contamination.  We reject~\fTellFailO\% of our [OI]
spectra and~\fTellFailC\% of our CI spectra because of telluric
contamination.  Telluric lines affect the [OI] region more strongly
due to the airglow at 6300~\AA.

\subsection{Iodine Removal}
\label{sec:iodine}
\begin{figure}
\centering
\includegraphics[width=0.6\textwidth]{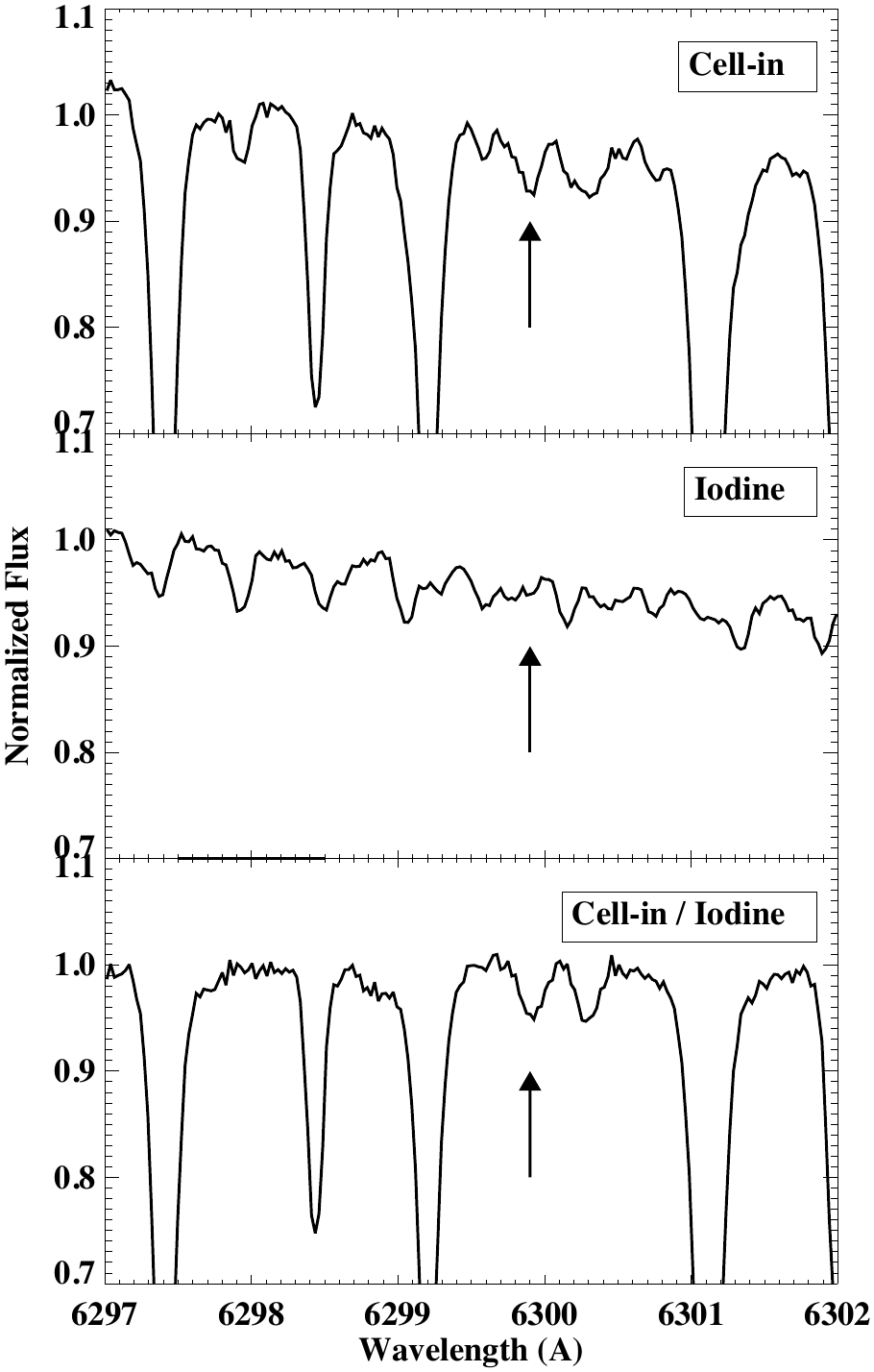}
\caption[HD~148284 spectrum with iodine contamination]{HD~148284 spectrum with iodine contamination (top), reference
  iodine observation (middle), and stellar spectrum divided by iodine
  spectrum (bottom).  The arrow marks the [OI]/NiI blend.  The iodine
  is removed to a level of $\sim$1\% of the continuum intensity.}
\label{fig:idiv}
\end{figure}

The majority of the spectra in the CPS catalog were taken through the
iodine cell.  Iodine lines are $\sim 0.5$\% deep in the CI
region---comparable to the photon noise.  In the [OI] region, they are
a $\sim 5$\% effect and must be removed.  For a given iodine cell in
observation, we locate the most recent iodine observation (usually at
the beginning of the night).  We account for any shift of the CCD
between the two observations by cross-correlating spectral orders 8,
9, and 10 ($\lambda =$ 5608 - 5895\AA, where the iodine lines are
strongest).  After removing any shift, we divide the iodine cell in
observations by the iodine reference
observations. Figure~\ref{fig:idiv} shows a stellar spectrum, an
iodine spectrum, and the ratio of the two.

\begin{figure}
\centering
\includegraphics[width=0.6\textwidth]{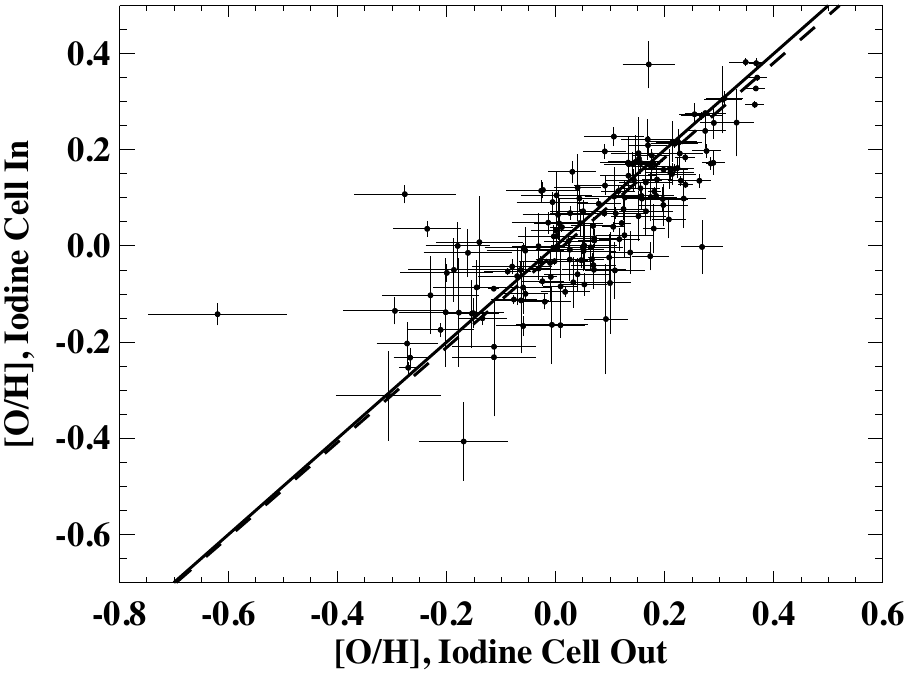}
\caption[Oxygen abundances from spectra taken
  with and without the iodine cell]{Comparison of oxygen abundances derived from spectra taken
  with the iodine cell in (vertical axis) to those taken with the
  iodine cell out (horizontal axis).  The solid line corresponds to an
  equality of the two, while the dashed line shows the best fit to the
  points.  Apparently the oxygen abundances derived from both types of
  spectra show no systematic difference, indicating that the removal
  of the iodine spectrum works without introducing a systematic
  error.}
\label{fig:calib}
\end{figure}

Dividing the iodine spectrum from an iodine cell in spectrum cannot be
done to within photon statistics.  On nights of good seeing, a star's
image may be narrower than the HIRES slit.  The reference iodine
spectra are produced with a lamp that fills the slit uniformly, so the
iodine lines from the iodine cell in observations can be narrower than
the reference iodine lines.  The result is artifacts from the division
at the $\sim1$\% level.  Iodine cell in spectra for a single star
generally yield a larger spread in derived oxygen abundance compared
to iodine cell out observations.  However, when we plot oxygen
abundances derived from iodine cell in observations against abundances
from iodine cell out observations in Figure~\ref{fig:calib}, we see no
systematic trend.

%%% FIG - IDIV %%%

%%% FIG - CALIB %%%

\subsection{Fitting Abundances}
\label{sec:fitting}
We converge on abundance by iterating three $\chi^2$ minimization
routines that fit the continuum, line center, and abundance.  Only
points within the fitting range are used to calculate the $\chi^2$
statistic.  Any point deviating from the fit by more than five times
the photon noise is not included in calculating $\chi^2$.  A short
description of each routine is given below:

\begin{enumerate}
\item {\em Continuum}.  Given the shallowness of our indicators, a
  small error in the continuum level will have a significant effect on
  the derived abundance.  We refine the continuum value by registering
  the level of the spectrum so that $\chi^2$ is minimized.

\item {\em Line center}.  The wavelength zero point and dispersion is
  initially determined from a thorium lamp calibration taken each
  night and refined by cross-correlating the observed spectrum with
  the solar spectrum.  We adjust the radial velocity of the model
  spectrum to minimize $\chi^2$.

\item {\em Abundance}.  We begin with the solar oxygen abundance
  scaled by the star's metallicity.  We refine this value by searching
  over 2 dex of abundance space and minimizing $\chi^2$.
\end{enumerate}

We terminate the iteration when the fits arrive at a stable solution
or when we exceed 10 iterations.

\section{Results}
\subsection{Carbon and Oxygen Abundances}
\label{sec:abundances}
We report [O/H]%
\footnote{[X/H] =  $\log \epsilon_X - \log \epsilon_{X,\odot}$}
and [C/H] for \nStarsO~and \nStarsC~stars respectively.  These are
subsamples of our initial \nSampStars~star sample and arise after we
apply the following global cuts:

\begin{enumerate}
\item {\em \vsini}. In rapidly rotating stars, our indicators can be
  polluted by the wings of neighboring lines due to rotational
  broadening.  When this happens, the abundances of our elements of
  interest become degenerate with that of the polluting line.  This
  effect sets in earlier for the [OI] line, which sits shoulder to
  shoulder between SiI and ScII features. We do not report oxygen or
  carbon abundances for stars with \vsini~greater than \vsiniCutO~and
  \vsiniCutC~km/s respectively.
\item {\em \teff}.  The high excitation energy of the CI line
  (\chiexC~eV) means the line is very weak in cool stars.  For
  example, at 5000K, the line depth in a solar analog is 1\%.  We do
  not report carbon abundances for stars cooler than 5300 K.
\item {\em Statistical scatter}.  We choose to report abundances for
  stars where the scatter in derived abundance is less than
  \scatterCut~dex or, in other words, stars where our measurements are
  precise to a factor of 2.  Our estimates of measurement precision
  are based on empirical scatter and a Monte Carlo analysis, which we
  describe in sections \S~\ref{sec:staterr} \&~\S~\ref{sec:nierr}.
  While our measurement precision is based on a variety of factors
  including line depth and signal to noise, stars that fail this cut
  generally have sub-solar carbon and oxygen abundances.
\end{enumerate}

\begin{figure}
\centering
\includegraphics[width=0.8\textwidth]{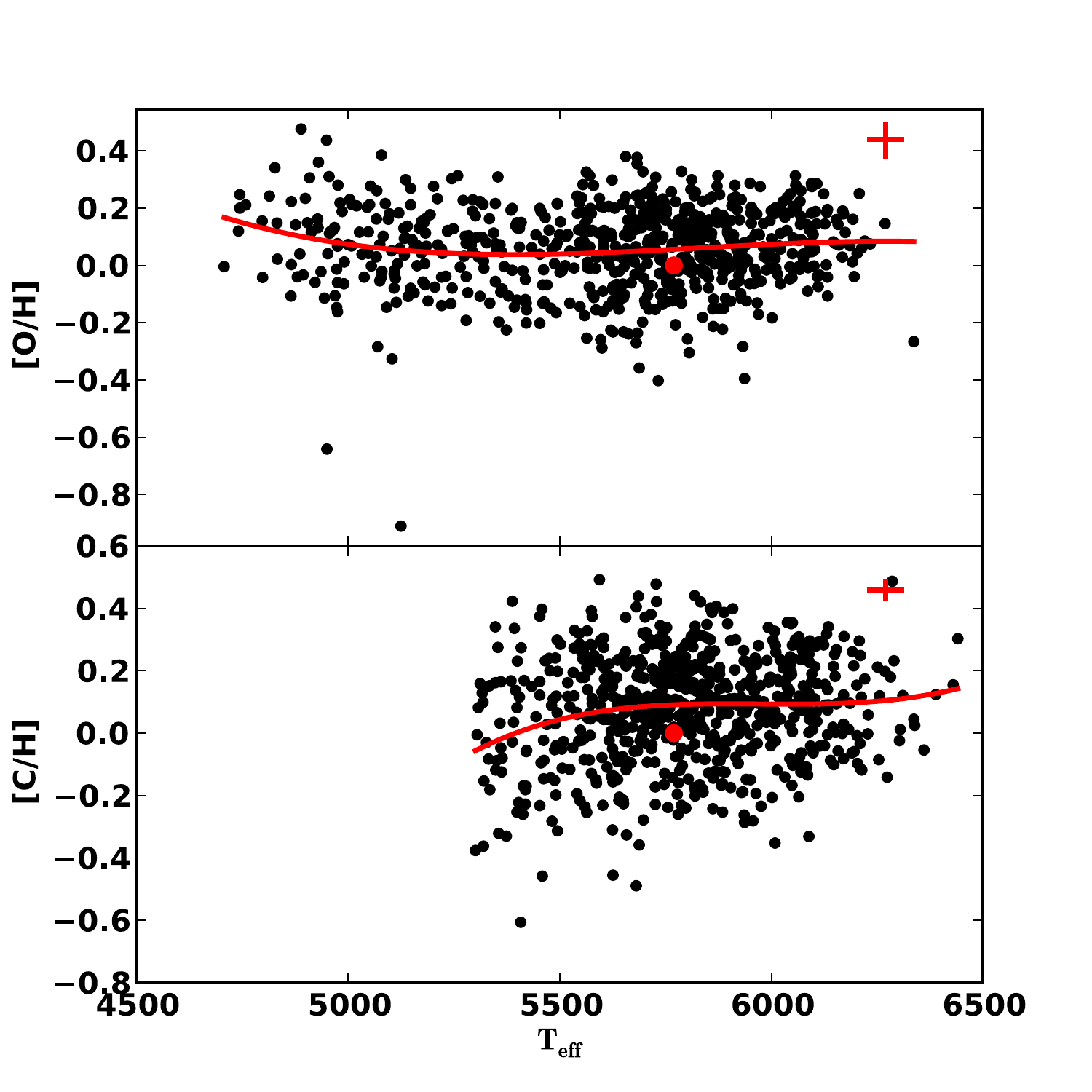}
\caption[Systematic trends of $\mathrm{[O/H]}$ and $\mathrm{[C/H]}$ with
  temperature]{Plots showing systematic trends of [O/H] and [C/H] with
  temperature.  The red line is the best fit cubic.  Our correction for
  the \teff trend is this cubic with the constant term chosen so that
  the correction is zero at $T_{eff} =$ \teffSol~K (large red dot).  The
  crosses show the median errors.}
\label{fig:teff}
\end{figure}

With our large stellar sample, it is possible to detect and correct
for systematic trends that would be invisible in smaller samples.
Figure~\ref{fig:teff} shows carbon and oxygen abundances plotted
against temperature.  We believe that the~\cite{Kurucz92} model atmospheres
are most accurate for solar analogs and that errors in the atmosphere
profile grows as we move away from $T_{eff} = T_{\odot}$ = ~\teffSol~K.

We model the systematic behavior of implied abundance
with \teff~by fitting a cubic to the data.  Simply subtracting
out the cubic would artificially force the mean [X/H] to zero, but there
is no reason why the mean disk abundance should be solar. 
Therefore, we let the solar abundance fix the constant term in the cubic by
requiring the systematic correction be zero at~\teffSol~K.  
This correction reaches \maxTO~dex for oxygen and
\maxTC~dex for carbon.  We have removed the temperature trend for all
abundances quoted henceforth.

By removing abundance trends with \teff~for the sake of correcting errors in
atmosphere models, we may
have erased a real astrophysical trend of [X/H] with~\teff.  For example,
hotter stars are more massive and have shorter main sequence lifetimes than
cool stars.  Therefore, the hotter stars in our sample are on average younger
and formed at a later time in the galactic chemical enrichment history.
However, we chose to remove the~\teff~trends because we believe 
uncertainties in atmospheric models are the dominant effect. 

\begin{figure}
\centering
\includegraphics[width=0.8\textwidth]{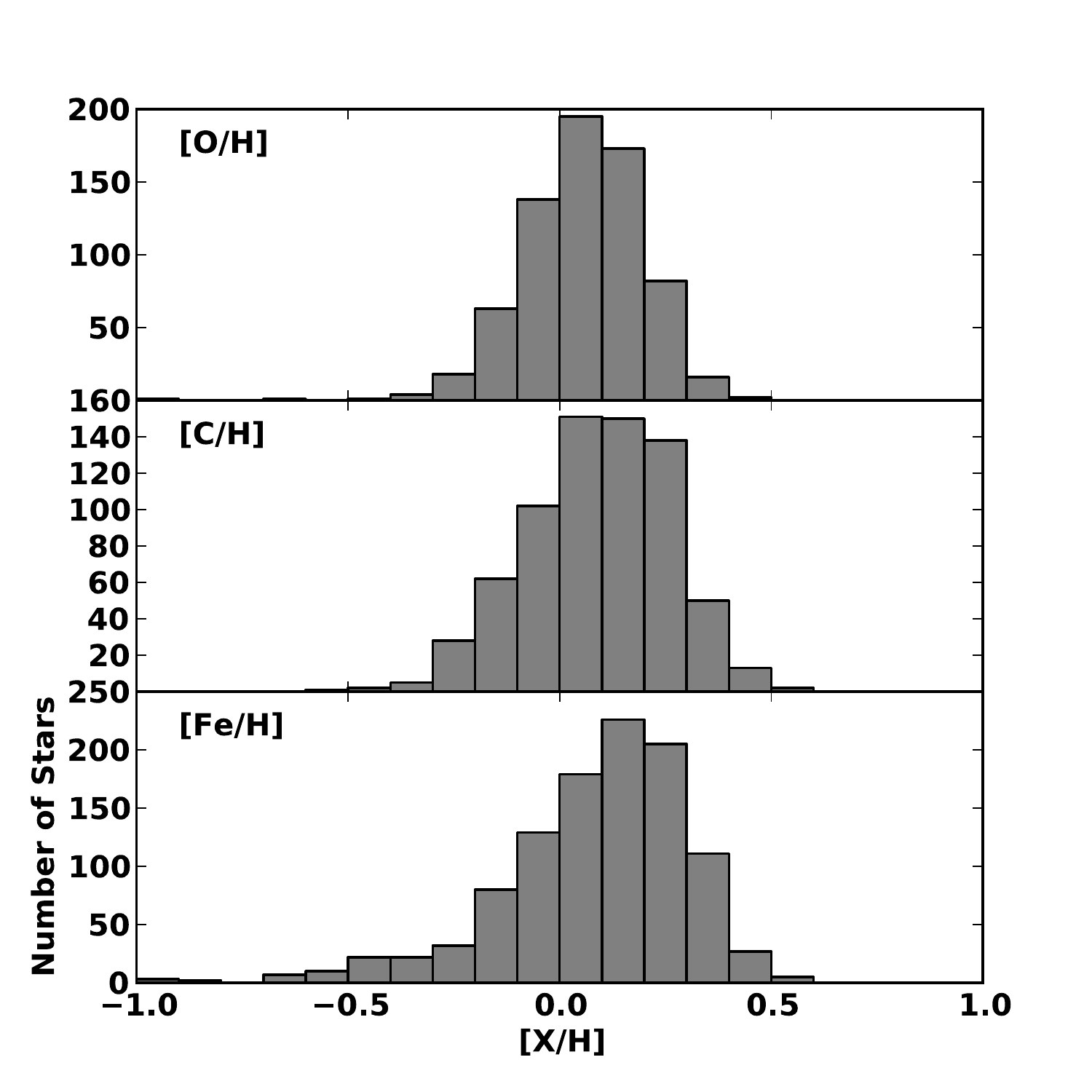}
\caption[Distributions of $\mathrm{[O/H]}$, $\mathrm{[C/H]}$, and $\mathrm{[Fe/H]}$]
{Distributions of [O/H], [C/H], and [Fe/H] for comparison.}
\label{fig:abundhist}
\end{figure}

We report our temperature-corrected values for [O/H] and [C/H] with
85\% and 15\% confidence limits along with other stellar data in the
Appendix.  We summarize the statistical properties of derived
abundances in Table~\ref{tab:abundhist} and show their distributions
in Figure~\ref{fig:abundhist}.

%%% TAB - ABUNDHIST %%%
\begin{deluxetable}{l c c c c c}
\tablewidth{0pc}
%\tablenum{2}
\tablecaption{Summary of Derived Abundances.}
\tablehead{
\colhead{}      &
 \multirow{2}{*}{N} &
\colhead{m}	&
\colhead{S}	&
\colhead{Min}	&
\colhead{Max} \\
\colhead{}      &
\colhead{}	&
\colhead{(dex)}	&
\colhead{(dex)}	&
\colhead{(dex)}	&
\colhead{(dex)} \\
}
\startdata
\input{co/abundhist.tex}
\enddata
\tablecomments{Here, N is the number of stars with determined
  abundances, m is the mean abundance, and S is the standard deviation
  of abundance distribution.}
\label{tab:abundhist}
\end{deluxetable}

%%% FIG - TEFF %%%

%%% FIG - ABUNDHIST %%%

\subsection{Random Errors}
\label{sec:staterr}
We use Monte Carlo bootstrapping to estimate random errors.  We
generate Monte Carlo spectra by scrambling the residuals from our fits
and adding them back to the synthetic spectra.  For each star we
generate and refit 1000 Monte Carlo realizations of the spectrum.  The
resulting abundance distribution provides a good estimate of the true
error distribution.

\begin{figure}
\begin{center}
\includegraphics[width=0.8\textwidth]{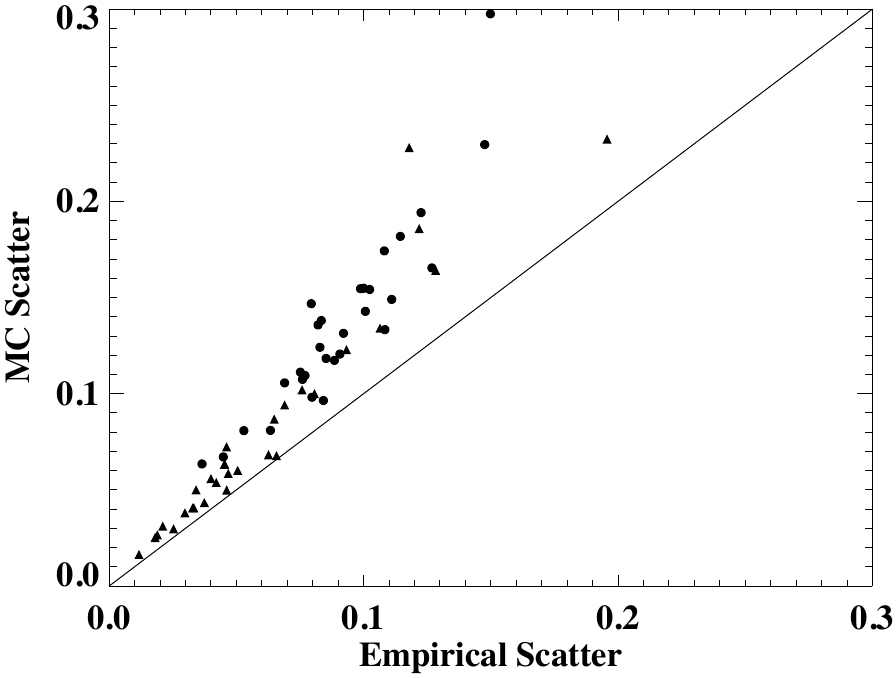}
\end{center}
\caption[Statistical uncertainties in derived abundances]{Scatter in MC simulations as a function of empirical scatter
  for stars with more than 50 observations.  Oxygen and carbon
  measurements are represented by circles and triangles respectively.
  The solid line represents a 1:1 correlation.}
\label{fig:mcstaterr}
\end{figure}

For some stars we have many independent spectra, allowing us to
compute confidence limits of the oxygen abundances from them as an
{\em empirical} measure of our internal errors.
Figure~\ref{fig:mcstaterr} shows the length of the error bars computed
empirically and from Monte Carlo for stars with more than 50 empirical
fits.  The error estimate from Monte Carlo tracks the empirical
scatter well, slightly overestimating it.  This is due to systematic
errors in our fits that appear as random errors when we scramble the
residuals.

For stars with fewer than 20 observations, we adopt the Monte Carlo confidence intervals as our statistical error; for stars with 20 or more observations, we adopt the empirical confidence intervals.  We diminish these errors by $\sqrt{N_{obs}}$.  Futhuremore, we impose an error floor of \errFloor~dex.

%%% FIG - ERRORFUNC %%%

\subsection{Nickel Systematics}
\label{sec:nierr}
Since we are deriving oxygen from a line that is blended with nickel,
the errors in nickel abundance are covariant with errors in oxygen
abundance.  FV05 quote a uniform error of 0.03 dex for their nickel
measurements.  The amount that [OI] and NiI contribute to the blend is
different for every star.  Therefore, we evaluate the effect of the
0.03 dex error in nickel abundance on oxygen abundance on a
star-by-star basis.  We begin with a synthetic spectrum at our quoted
oxygen abundance.  We then refit the oxygen line to a spectrum with
0.03 dex more and 0.03 dex less nickel.  These errors are added in
quadrature to the statistical errors.

There are many other sources of systematic error in our abundance
measurements such as inaccurate solar reference abundances, additional
blends, and our assumption of LTE.  However, these effects should be
largely consistent between stars, so we expect them to contribute
little to errors in our differential abundances.

\subsection{Comparison with Literature}
\begin{figure}
\begin{center}
\includegraphics{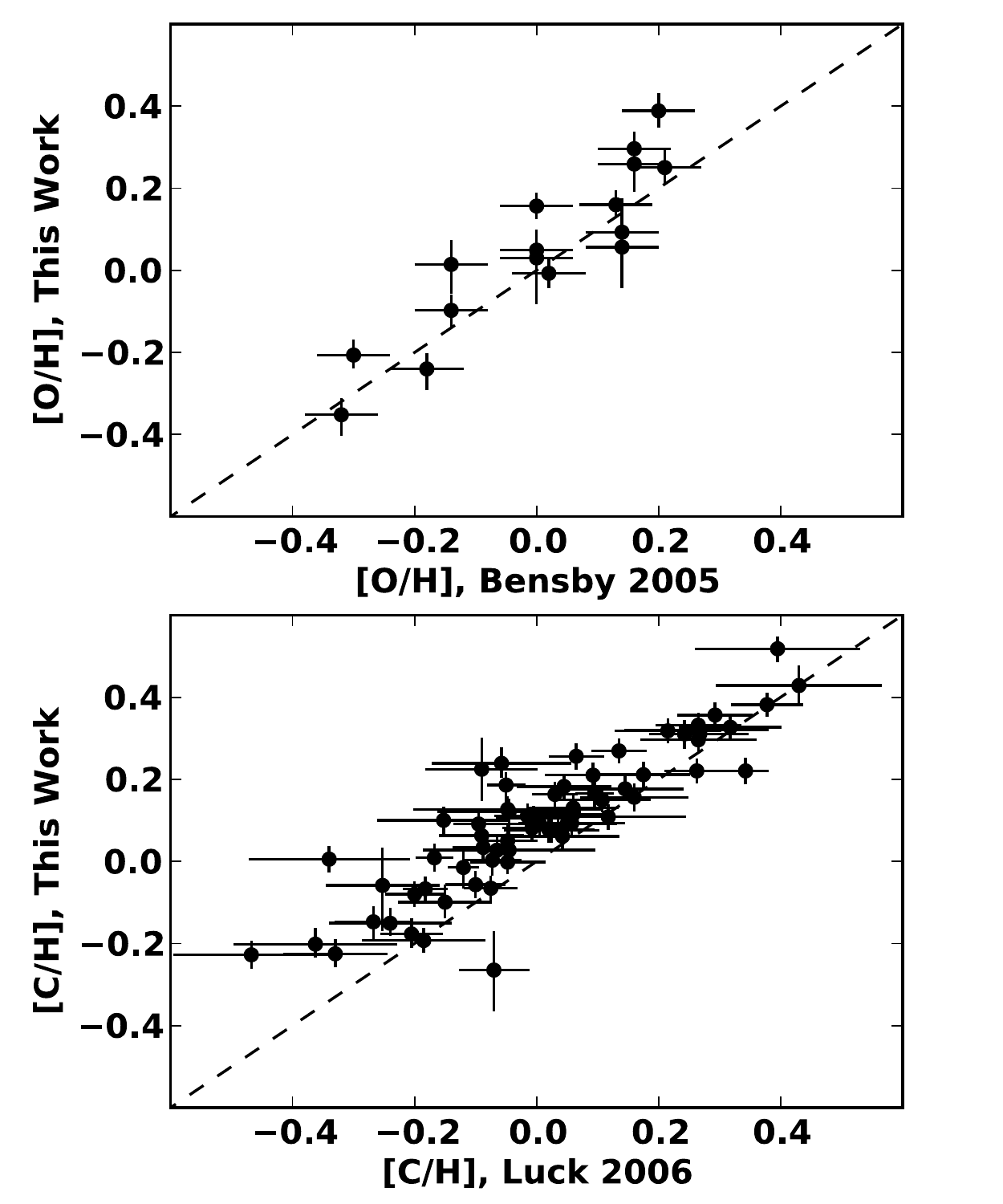}
\end{center}
\caption[Comparison to \cite{Bensby05} and \cite{Luck06}]
{Comparison plots of abundances from this work against those
  of~\cite{Bensby05} and~\cite{Luck06}.  The line shows a 1:1
  correlation.  We note a systematic offset in the carbon comparison.
  This may stem from the fact that the two works use different
  indicators.}
\label{fig:comp}
\end{figure}

We compare our results with~\cite{Bensby05} and~\cite{Luck06}.  We
report oxygen abundances for~\nCompO~stars analyzed by~\cite{Bensby05}
and \nCompC~stars analyzed by~\cite{Luck06}. We plot the comparison in
Figure~\ref{fig:comp}.  Our results track these comparison studies
well.  We recognize that the agreement is poorest for low values of
[C/H].  This likely the result of less robust fits to stars with
weaker carbon features.

The standard deviation of the differences in derived abundances is
\StdCompO~dex for oxygen and \StdCompC~dex for carbon.  Since
\cite{Bensby05} and \cite{Luck06} use different instruments,
spectral synthesizers, and fitting algorithms, it is unlikely there
are common systematic errors.  Therefore, the scatter in the
differences can be interpreted as a measure of the typical combined
statistical and systematic error.  We cannot say how much of the
observed scatter is due to our errors and those of the comparison
studies.

%%% FIG - COMP %%%

\subsection{Abundance Trends in the Thin and Thick Disks}
The Milky Way is thought to be made up of three distinct star
populations: The thin disk, thick disk, and the halo.  Most of the
stars in the local neighborhood belong to the thin disk, which has a
scale height of 300 pc.  The thick disk has a scale height of 1450 pc
and is comprised of older, metal-poor stars.  

\cite{PeekThesis} combined proper motion measurements from the {\em Hipparcos}
catalog~\citep{ESA97} with radial velocity measurements from the 
\cite{Nidever02}, SPOCS, and N2K catalogs into three-dimensional space motions
for \nThreeD~of our \nSampStars~program stars.  \cite{PeekThesis} computed
the probability of membership to each of the three populations in the manner
of \cite{Bensby03}, \cite{Mishenina04}, and \cite{Reddy06} for \nPop~of our 
\nStarsTot~stars with measured carbon and oxygen.
Our sample contains \nThin~thin disk stars, \nThick~thick disk stars, 
\nHalo~halo stars, and \nBdr~borderline stars (all three membership
probabilities less than 0.7).

\begin{figure}
\centering
\includegraphics[width=0.8\textwidth]{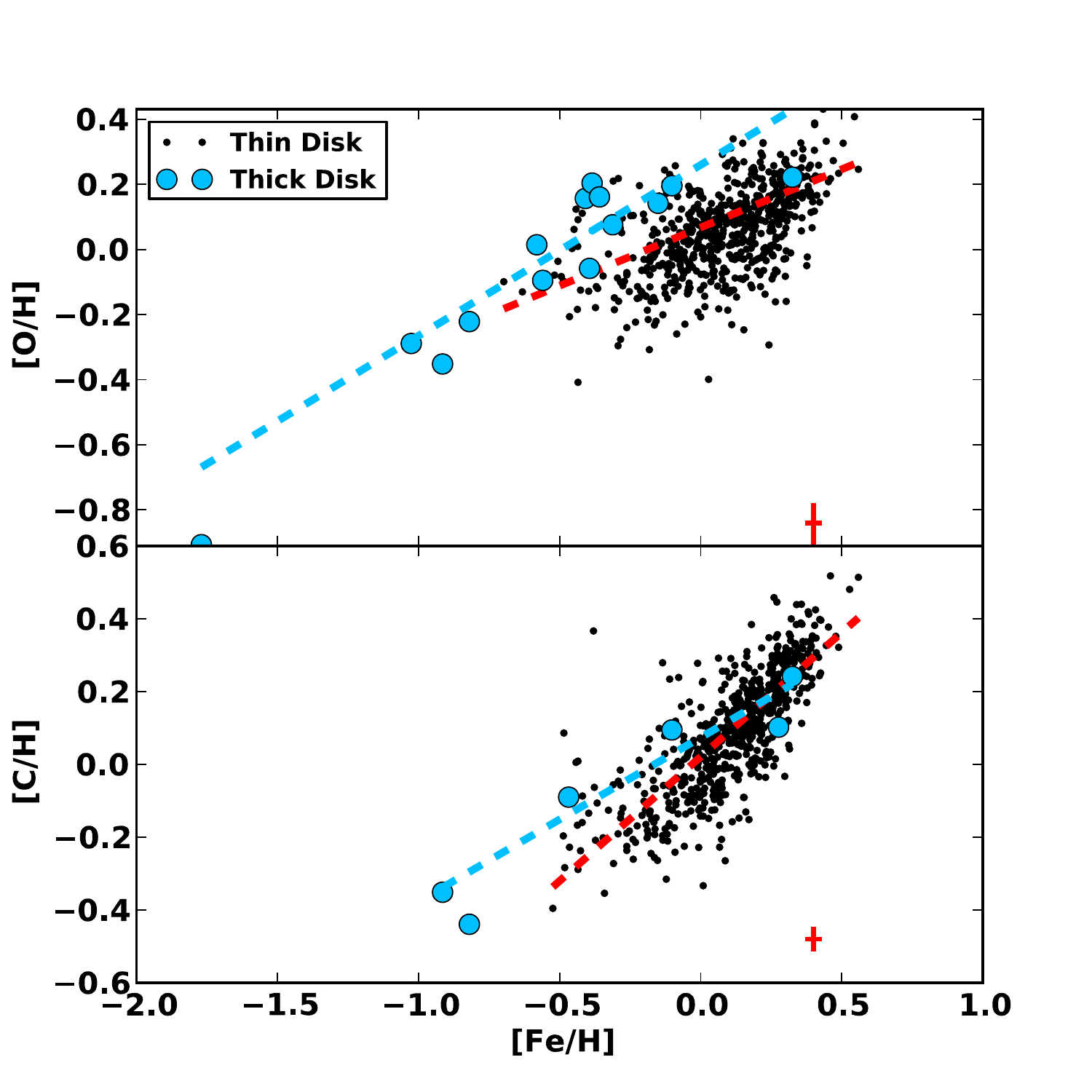}
\caption[Carbon and oxygen abundance versus iron abundance]{Carbon and oxygen abundance plotted against iron abundance.
  The \colorone\ points are the thin disk stars; the \colortwo\ points
  are the thin disk stars.  The line shows the abundance ratios in 0.1
  dex bins. The crosses show median uncertainties.}
\label{fig:xonh}
\end{figure}

We plot [O/H] and [C/H] against [Fe/H] in Figure~\ref{fig:xonh}.  We
fit the trends with a line and list the best fit parameters in
Table~\ref{tab:xonh}. If the scatter was purely statistical, we would
expect our fits to have a reduced-$\sqrt{\chi^2}\sim1$.  Our fits have
reduced-$\sqrt{\chi^2}\sim2$, which suggests that some of the observed
scatter is astrophysical.  These main sequence stars have not begun to
process heavy elements, so the ranges of C, O, and Fe ratios reflect
the heterogeneous interstellar medium from which they formed.

%%% TAB - XONH %%%
\begin{deluxetable}{l c c c c c}
\tablewidth{0pc}
%\tablenum{2}
\tablecaption{Best fit parameters to abundance trends.}
\tablehead{
\colhead{}    &
\colhead{Pop.}&
\colhead{$m$} &
\colhead{$b$} &
\colhead{$\sqrt{\chi^2}$} \\
}
\startdata
\input{co/xonh.tex}
\enddata
\tablecomments{We fit thin and thick disk abundance trends with the
  following function [X/H] = $m$ [X/Fe] + $b$.  The best fit
  parameters are listed above along with the reduced-$\sqrt{\chi^2}$.}
\label{tab:xonh}
\end{deluxetable}

\begin{figure}
\centering
\includegraphics[width=0.8\textwidth]{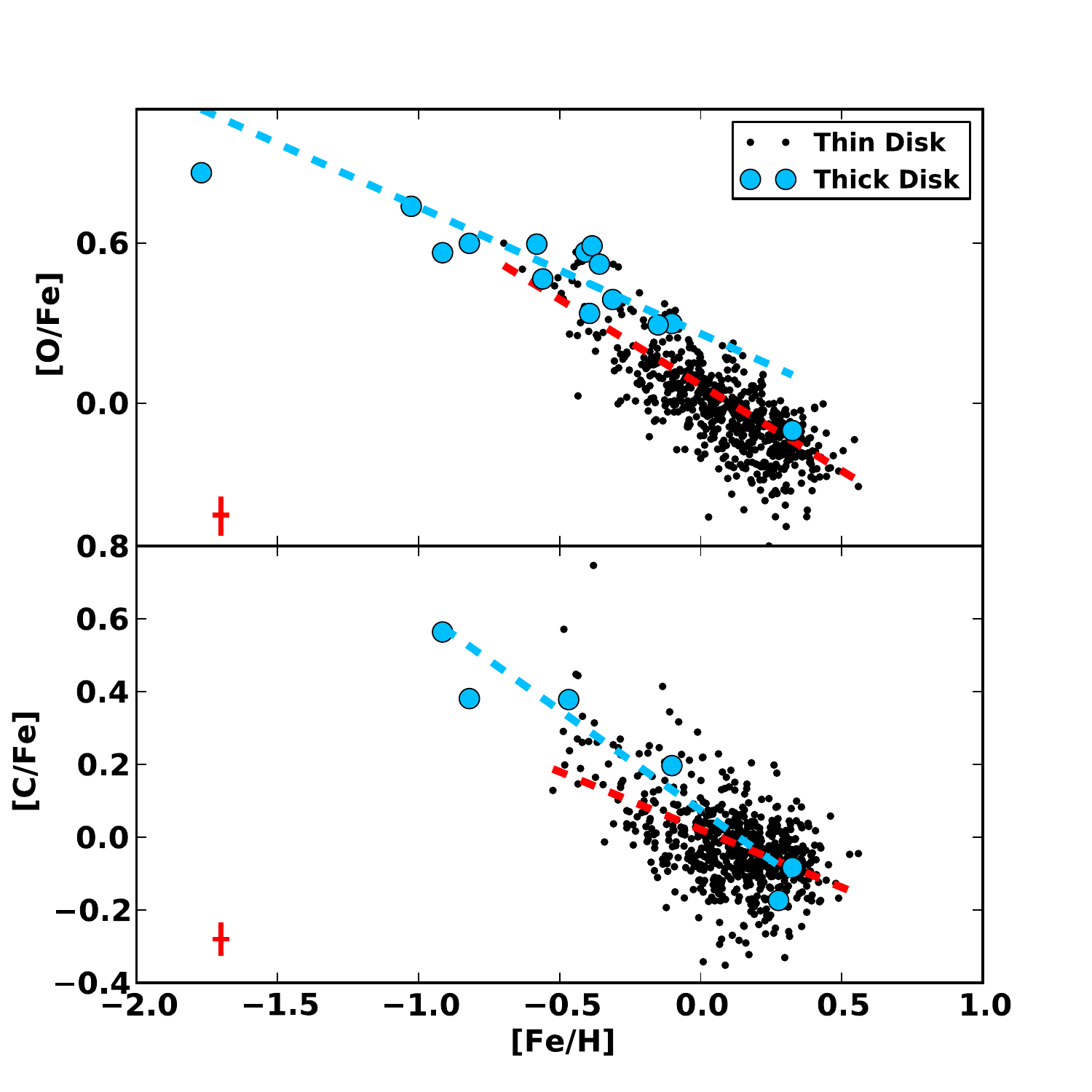}
\caption[$\mathrm{[C/Fe]}$ and $\mathrm{[O/Fe]}$ versus iron
  abundance]
  {The ratios of carbon and oxygen to iron plotted against iron
  abundance.  The \colorone\ points are the thin disk stars; the
  \colortwo\ points are the thin disk stars.  The line shows the
  average ratios in 0.1 dex bins.  The crosses show median
  uncertainties.}
\label{fig:xonfe}
\end{figure}

We also plot [O/Fe] and [C/Fe] against [Fe/H] in
Figure~\ref{fig:xonfe}. The trends suggest that carbon and oxygen
lagged behind iron production for much of the period of galactic
chemical enrichment.  These trends flatten out for high [Fe/H].  Due
to the paucity of thick disk stars in our sample, we are cautious in
interpreting its abundance trends.  However, in the thick disk, oxygen
seems to be enhanced relative to iron, a result also reported by
\cite{Bensby04}.  This enhancement in oxygen suggests that type II
supernova played a more active role in enriching the thick disk.

%%% FIG - XONH-V-FEH %%%

%%% FIG - XONFE-V-FEH %%%

\subsection{Exoplanets}
\hostFen~stars in our initial~\nSampStars~sample are known to host
planets.~\cite{Gonzalez97} measured relatively high stellar
metallically in the first four exoplanet host stars,
and \cite{Santos04} and \cite{Fischer05} showed that the fraction of stars bearing planets
increases rapidly above solar metallicity.  In light of the
correlation between C, O, and Fe, it is not surprising that hosts to
exoplanets are enriched in carbon and oxygen relative to the
comparison sample.

As shown in Table~\ref{tab:exo}, the mean [O/H] of the planet host and
comparison sample is~\hostOm~dex and~\compOm~dex respectively.  If we
take the error on the mean abundance to be the standard deviation of
derived abundances divided by the square root of the
number of stars in each sample i.e.
$\sigma_{mean} = \frac{\mathrm{Std. \ Dev.}}{\sqrt{N}}$, $\sigma_{mean}$
for [O/H] is 0.01 dex.  Carbon is also enriched in planet hosts 
where the mean [C/H]
is \hostCm~dex ($\sigma_{mean} = 0.02$ dex)
compared to \compCm~dex in the comparison sample with.
For both carbon and oxygen, the mean abundance
of the planet host sample is enriched by $\sim 5 \sigma $ compared
to the non-host sample.

\begin{figure}
\begin{center}
\includegraphics[width=0.8\textwidth]{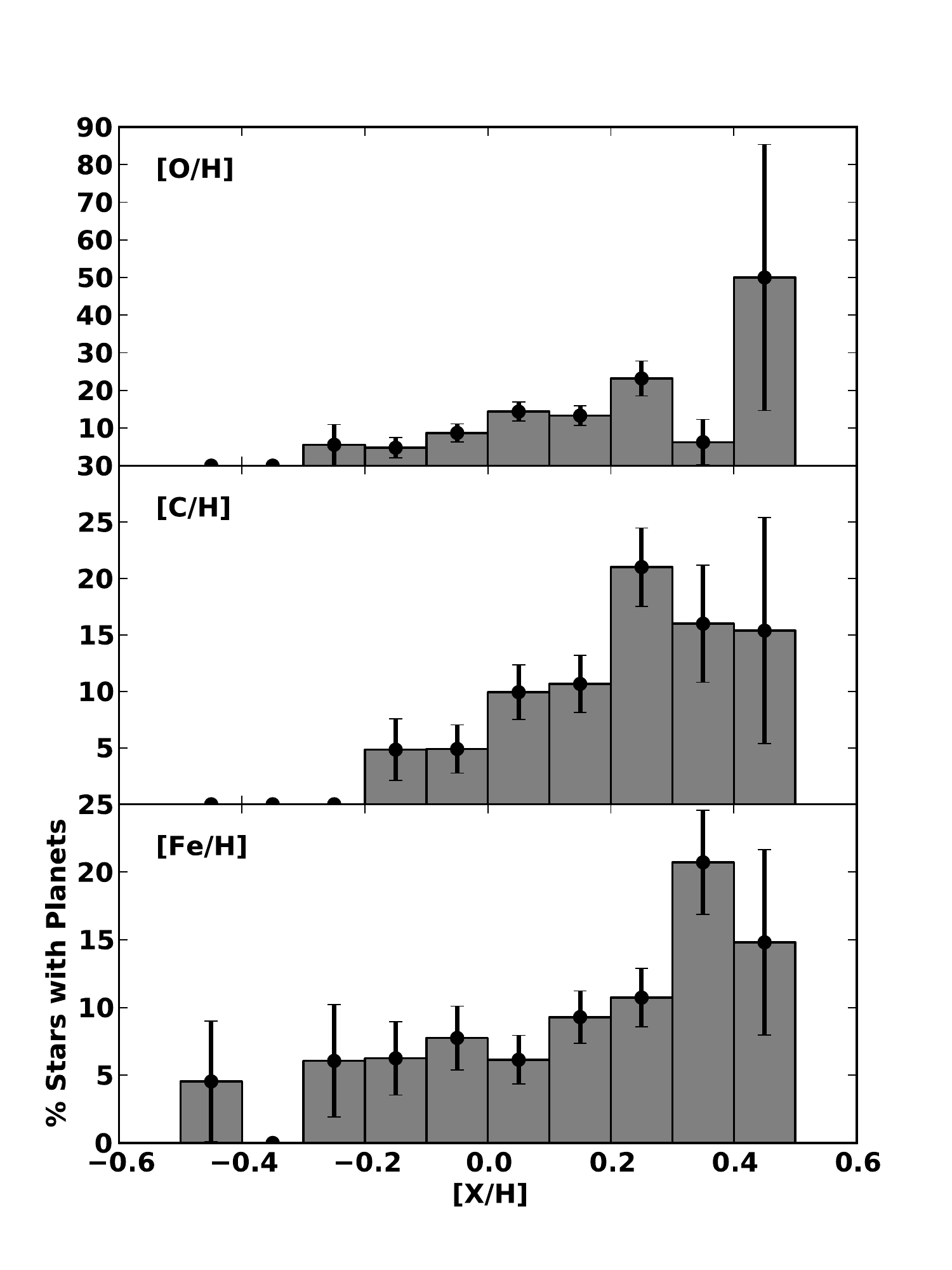}
\end{center}
\caption[Planet occurrence as a function of oxygen, carbon, and iron abundance]
{The percentage of stars with known planets for 0.1 dex bins in oxygen, carbon, and iron.  The histograms are constructed from the \nStarsO, \nStarsC, and \nSampStars~stars with reliable measurements of [O/H], [C/H], and [Fe/H] respectively.}
\label{fig:exo}
\end{figure}

In Figure~\ref{fig:exo}, we divide the stars into 0.1 dex bins in
[X/H].  For each bin, we divide the number of planet-bearing stars by
the total number of stars in the bin.  As with iron, we observe an
increase in planet occurrence rate as carbon and oxygen abundance
increases.  While there is a hint of a possible plateau or turnover at
the highest abundance bins, these bins are dominated by small number
statistics. The data are not inconsistent with a monotonic rise,
within the errors.  The possibility that very enriched systems inhibit
planet formation is intriguing, and this parameter space warrants
further exploration.

%%% TAB - Exo %%%
\begin{deluxetable}{l  c c c c  c c c c c c}
\tablewidth{0pc}
\tablecaption{Statistical abundance properties of stars with planets. }
\tablehead{
\multicolumn{1}{c}{} &
\multicolumn{4}{c}{Hosts} &
\multicolumn{1}{c}{} &
\multicolumn{4}{c}{Non-Hosts} &
\multicolumn{1}{c}{}
 \\
 \cline{2-5} \cline{7-10}
&
\colhead{N} &
\colhead{mean} &
\colhead{Std. Dev.} &
\colhead{$\sigma_{mean}$} &
&
\colhead{N} &
\colhead{mean} &
\colhead{Std. Dev.} &
\colhead{$\sigma_{mean}$}
}
\startdata
\input{co/exo.tex}
\enddata
\label{tab:exo}
\tablecomments{We list the number of stars, mean abundance (dex),
  standard deviation (dex), and error on the mean abundance (dex) for
  the host and non-host populations.  The error on the mean abundance
  is computed by $\sigma_{mean} =
  \frac{\mathrm{Std. \ Dev.}}{\sqrt{N}}$.  }
\end{deluxetable}

%%% FIG - EXO - Planet Occurrence Rate %%%

\subsection{$N_{C}/ N_{O}$}
We present the ratio of carbon to oxygen atoms, $N_{C}/ N_{O}$%
\footnote{$N_{C}/ N_{O} = 10^{\logepsc -\logepso}$}
for~\ncoStarsTot~stars with reliable
carbon and oxygen measurements as listed in the last column of
Table~\ref{tab:stellardata} of the Appendix.
Since we do not report carbon for stars cooler than~\teffCutClo~K,
our $N_{C}/ N_{O}$ measurements apply only to F and G spectral types.
While there is a weak
correlation between $N_{C}/ N_{O}$ and Fe at high [Fe/H], we note the large
degree of scatter in $N_{C}/ N_{O}$, which spans a wide range
from~\coMin~to~\coMax.

We emphasize that our measurements of [C/H] and [O/H] are differential
relative to solar and should be insensitive to revisions in the solar
abundance distribution.  $N_{C}/ N_{O}$ depends on our adopted solar
abundances of of oxygen \citep{Scott09} and carbon \citep{Caffau10}.  
We believe the abundances of carbon and oxygen are known at the $\sim 0.1$ dex
level.  Therefore, we expect revisions to the solar abundance distribution to 
systematically shift our $N_{C}/ N_{O}$ measurements by roughly $\sim 10^{0.1}$ or $\sim 25\%$ 

We measure~\ncoGtThresh~stars with $N_{C}/ N_{O}$ greater than~\coThresh.
Given the size of our random errors as determined by the Monte Carlo analysis of 
\S~\ref{sec:staterr}, very few of these stars are $1 \sigma$ detections of 
$N_{C}/ N_{O} > 1.$  However, since these errors are random, we believe our
measurements accurately reflect the {\em distribution} of $N_{C}/N_{O}$ in 
nearby disk stars.  Neglecting the zero-point offsets discussed earlier, we 
measure $N_{C}/ N_{O} > 1$ for roughly 10\% of nearby FG stars.

As noted by our anonymous reviewer, the CO molecule controls the equilibrium between carbon and oxygen in M dwarfs.  It is believed that $N_{C}/N_{O} > 1$ in M dwarfs results in an atmosphere rich in C$_{2}$, while $N_{C}/N_{O} < 1$ 
gives rise to TiO.  We are unaware of M dwarfs with strong C$_{2}$ bands indicating $N_{C}/N_{O} > 1$. This suggests such a
population is rare, assuming we understand the behavior of carbon-rich
M dwarf atmospheres.
We also note the additional complexities involved in modeling M star atmospheres.  Abundance estimates in cool stars rely on opacity tables of H$_{2}$O and 
other molecules that are not well understood at the temperatures
probed by M star atmospheres.
The fact that M stars are fully convective and have strong magnetic fields 
introduce additional complexities into model atmospheres.

Despite the uncertainties in accurately measuring $N_{C}/ N_{O}$, we have characterized the distribution of $N_{C}/ N_{O}$ for an unprecedented number of FG stars.  Furthermore, we have identified~\ncoGtThresh\ stars have high $N_{C}/ N_{O}$.
Given the predictions regarding exotic planets that form in a carbon rich
environment, these stars constitute important hosts for future work on
their exoplanets and exozodiacal dust.  Observations of dust with ALMA
and JWST may be particularly valuable.

%%% FIG - COFE - The C/O Ratio %%%
\begin{figure}
\begin{center}
\includegraphics{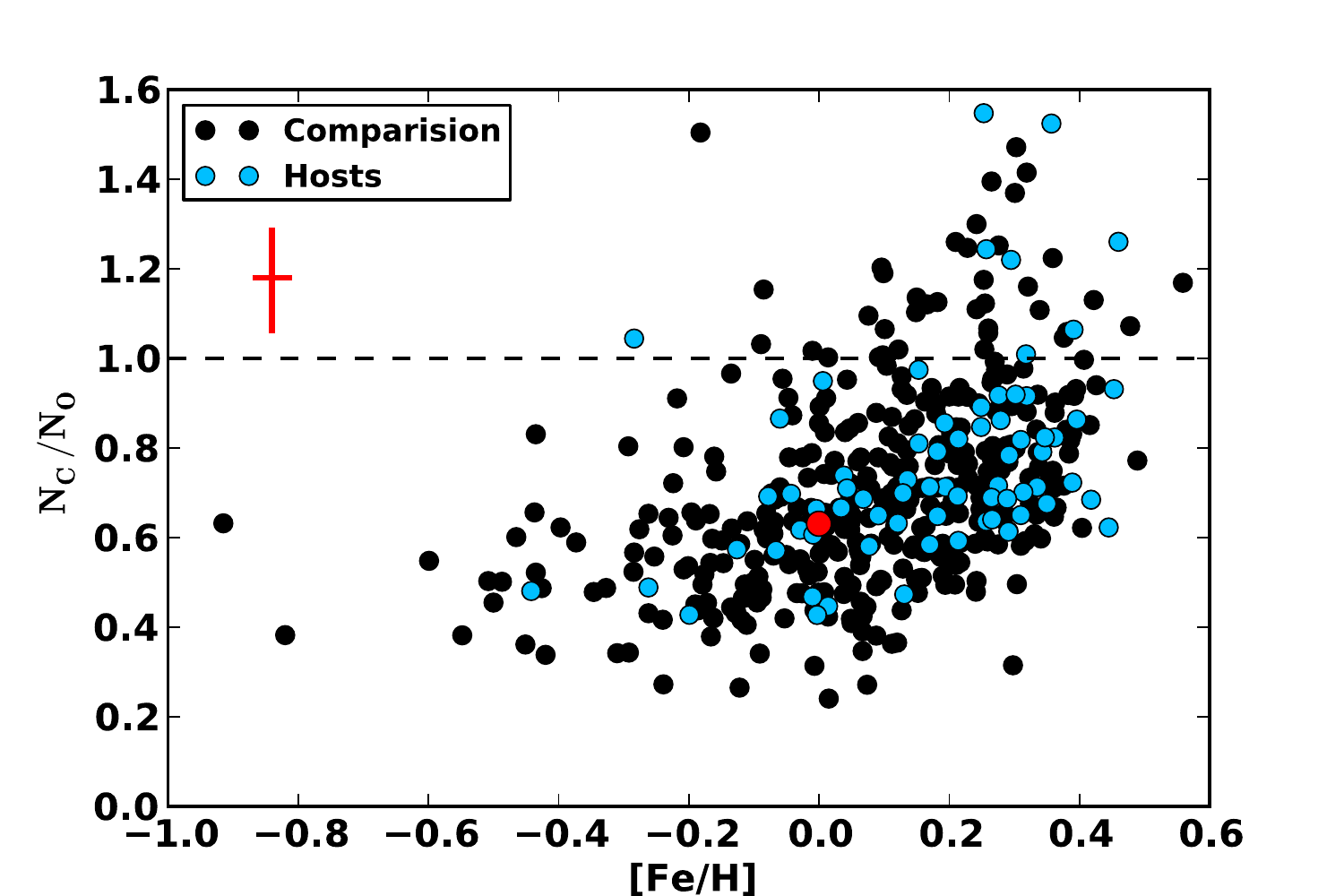}
\end{center}
\caption[$N_{C}/ N_{O}$ as a function of iron
  abundance]
  {$N_{C}/ N_{O}$ as a function of iron
  abundance.  The large \colorthree\ dot shows the solar values. The
  horizontal line shows equal carbon and oxygen.}
\label{fig:cofe}
\end{figure}

\subsection{WASP-12}
WASP-12b, discovered by~\cite{Hebb09}, is a transiting gas giant and a
favorable target for atmosphere studies.  \cite{Campo11} and
\cite{Croll11} measured secondary eclipses of WASP-12b at wavelengths
ranging from 1-8 \um, which can be used to characterize the planet's
dayside emission spectrum.  In a recent study, \cite{Madhu11} found
that these measurements are best-described by atmosphere models with
$N_{C}/ N_{O}$ $\geq$ 1 at 3 sigma significance.

We analyze WASP-12 identically to the~\nSampStars~star sample.  With a
V-mag of~\waspVmag~\citep{Hebb09}, WASP-12 is dimmest star in this
work and our spectra have S/N $\sim$ 50.  However, we measured oxygen
and carbon based on~\waspNobsO~and~\waspNobsC~spectra respectively.
We find [O/H] = \waspO, [C/H] = \waspC, and $N_{C}/ N_{O}$ = \waspCO\ i.e.,
sub-solar.

If the composition of the host star truly reflects the material from
which WASP-12b formed, our measurements suggest that WASP-12b does not
have a carbon-dominated bulk composition.  It is possible that the
planet acquired extra carbon at some point during its formation, or
that the planet's nightside and/or interior are acting as a sink for
oxygen, creating a carbon-rich dayside atmosphere while maintaing bulk
$N_{C}/ N_{O}$ less than unity.  In any case, this planet and its host star
warrant further study.

\section{Conclusion}
We have presented oxygen and carbon abundances for \nStarsTot~stars
based on HIRES spectra gathered by the Keck telescope.  We measure
oxygen by fitting the reliable 6300~\AA~forbidden line with {\tt SME}
and self-consistently account for the significant nickel blend.  Our
carbon abundances are derived from the 6587~\AA~ CI line.  Our errors
are based on a rigorous Monte Carlo treatment, and our measurements
agree with values in the literature.  Our sample is large enough to
characterize and remove systematic trends due to \teff.  We see that
carbon and oxygen are both enriched in stars with known planets.  We
see a significant number of stars with $N_{C}/ N_{O}$ exceeding unity, which
supports the possibility that some stars host exotic carbon-rich
planets.  However, our measurement of sub-solar $N_{C}/ N_{O}$ for WASP-12,
complicates the recent claim by \cite{Madhu11} that WASP-12b is a
carbon world.

\begin{landscape}
\begin{deluxetable}{rrrrrrrrrrrrrrrrr}
\tabletypesize{\footnotesize}
\setlength{\tabcolsep}{0.1in}
\renewcommand{\arraystretch}{1.2}
\tablecaption{Stellar Data.}
\tablewidth{0pt}
\tablehead{
  \colhead{Name\tablenotemark{a}}
  & \colhead{$V$\tablenotemark{b}}
  & \colhead{$d$\tablenotemark{b}}
  & \colhead{T\tablenotemark{c}}
  & \colhead{$\log{g}$\tablenotemark{c}}
  & \multirow{2}{*}{Pop\tablenotemark{d}}	
  & \multirow{2}{*}{Host}
  & \multirow{2}{*}{[M/H]\tablenotemark{c,e}}
  & \multirow{2}{*}{[Ni/H]\tablenotemark{c}}	
  & \multicolumn{1}{c}{\multirow{2}{*}{$N_{O}$}}
  & \multicolumn{1}{c}{\multirow{2}{*}{[O/H]\tablenotemark{f}}}
  & \multicolumn{1}{c}{\multirow{2}{*}{$N_{C}$}}
  & \multicolumn{1}{c}{\multirow{2}{*}{[C/H]\tablenotemark{f}}}
  & \multicolumn{1}{c}{\multirow{2}{*}{C/O\tablenotemark{f}}}\\
  \colhead{(HD)}
  & \colhead{(mag)}
  & \colhead{(pc)}
  & \colhead{(K)}
  & \colhead{(cgs)}
  &
  &
  & 
  & 
  &
  &
  &
  &
}
\startdata
\input{co/bigtable.tex}
\enddata
\tablenotetext{a}{Names are HD numbers unless otherwise indicated.}
\tablenotetext{b}{Johnson $V$-magnitude and distance from the Hip
   catalog; where stars were not in the Hip
   catalog, $d$ and $V$ are from the Geneva-Copenhagen Catalog 
   \citep{Nordstrom04}; $d$ and $V$ are taken from SIMBAD 
   as a last resort.}
\tablenotetext{c}{Stellar parameters from VF05 and D.~Fischer.}
\tablenotetext{d}{Membership identification taken from \cite{PeekThesis}}
\tablenotetext{e}{[M/H] is computed from all elements heavier than He.}
\tablenotetext{f}{We have corrected these abundances a for a systematic trend in temperature (see \S~\ref{sec:abundances}).}

\label{tab:stellardata}
\end{deluxetable}
\end{landscape}

%% file: co/texcmd.tex
\newcommand{\nThin}{847} % # of stars in Thin pop.
\newcommand{\nThick}{16} % # of stars in Thick pop.
\newcommand{\nHalo}{12} % # of stars in Halo pop.
\newcommand{\nBdr}{25} % # of stars in Bdr pop.
\newcommand{\nPop}{900} % # of stars with pop prob
\newcommand{\vsiniCutO}{7} % vsinicut for O
\newcommand{\teffCutOlo}{4700} % teff for O
\newcommand{\teffCutOhi}{6500} % teff for O
\newcommand{\maxTO}{0.11} % max temp correction O
\newcommand{\nStarsO}{694} % Number of stars with O analysis
\newcommand{\vsiniCutC}{15} % vsinicut for C
\newcommand{\teffCutClo}{5300} % teff for C
\newcommand{\teffCutChi}{6500} % teff for C
\newcommand{\maxTC}{0.15} % max temp correction C
\newcommand{\nStarsC}{704} % Number of stars with C analysis
\newcommand{\coThresh}{1.00} % Theshhold for high co
\newcommand{\scatterCut}{0.30} % Cut on the scatter
\newcommand{\teffSol}{5770} % Solar Effective Temp
\newcommand{\StdCompC}{0.09} 
\newcommand{\StdCompO}{0.08} 
\newcommand{\nCompC}{67}   
\newcommand{\nCompO}{16}   
\newcommand{\compCsigm}{0.00} % comp - C - sigm
\newcommand{\compCs}{0.17} % comp - C - s
\newcommand{\compCm}{0.08} % comp - C - m
\newcommand{\compCn}{625} % comp - C - n
\newcommand{\compFesigm}{0.00} % comp - Fe - sigm
\newcommand{\compFes}{0.27} % comp - Fe - s
\newcommand{\compFem}{0.06} % comp - Fe - m
\newcommand{\compFen}{970} % comp - Fe - n
\newcommand{\compOsigm}{0.00} % comp - O - sigm
\newcommand{\compOs}{0.14} % comp - O - s
\newcommand{\compOm}{0.05} % comp - O - m
\newcommand{\compOn}{606} % comp - O - n
\newcommand{\hostCsigm}{0.02} % host - C - sigm
\newcommand{\hostCs}{0.14} % host - C - s
\newcommand{\hostCm}{0.17} % host - C - m
\newcommand{\hostCn}{79} % host - C - n
\newcommand{\hostFesigm}{0.02} % host - Fe - sigm
\newcommand{\hostFes}{0.18} % host - Fe - s
\newcommand{\hostFem}{0.17} % host - Fe - m
\newcommand{\hostFen}{100} % host - Fe - n
\newcommand{\hostOsigm}{0.01} % host - O - sigm
\newcommand{\hostOs}{0.12} % host - O - s
\newcommand{\hostOm}{0.10} % host - O - m
\newcommand{\hostOn}{88} % host - O - n
\newcommand{\nStarsTot}{941} % 
\newcommand{\ncoGtThresh}{46} % 
\newcommand{\coMax}{1.55} % 
\newcommand{\ncoStarsTot}{457} % 
\newcommand{\coMin}{0.24} % 
\newcommand{\ncoLtThresh}{411} % 
\newcommand{\nPlanetTot}{97} % 
\newcommand{\colorone}{black}
\newcommand{\colortwo}{blue}
\newcommand{\colorthree}{red}
\newcommand{\waspO}{$0.29_{-0.10}^{+0.06}$} 
\newcommand{\waspC}{$0.10_{-0.06}^{+0.04}$}
\newcommand{\waspCO}{$0.40_{-0.07}^{+0.11}$}
\newcommand{\waspVmag}{11.69}
\newcommand{\waspNobsO}{9}
\newcommand{\waspNobsC}{7}

%% file: co/abundhist.tex
$ {[}O/H] $& 694 & 0.06 & 0.14 & -0.91 & 0.43\\$ {[}C/H] $& 704 & 0.09 & 0.17 & -0.52 & 0.52\\$ {[}Fe/H] $& 1070 & 0.07 & 0.27 & -1.95 & 0.56\\

%% file: co/xonh.tex
[C/H] & thick & 0.450 $\pm$ 0.074 & 0.074 $\pm$ 0.035 & 2.19 \\

[C/H] & thin & 0.682 $\pm$ 0.019 & 0.021 $\pm$ 0.004 & 2.52 \\

[O/H] & thick & 0.525 $\pm$ 0.081 & 0.260 $\pm$ 0.047 & 2.41 \\

[O/H] & thin & 0.358 $\pm$ 0.017 & 0.067 $\pm$ 0.004 & 1.66 \\

%% file: co/exo.tex
{[}O/H] & 88 & 0.10 & 0.12 & 0.01 & 
         & 606 & 0.05 & 0.14 & 0.01 & \\

{[}C/H] & 79 & 0.17 & 0.14 & 0.02 & 
         & 625 & 0.08 & 0.17 & 0.01 & \\

{[}Fe/H] & 100 & 0.17 & 0.18 & 0.02 & 
         & 970 & 0.06 & 0.27 & 0.01 & \\

%% file: co/bigtable.tex
 105 & 7.51 & 40 & 6126 & 4.65 & dn & no &-0.02 & 0.02 & 0 & \nd & 4 & $-0.04_{-0.04}^{+0.04}$ & \nd \\ 
 166 & 6.07 & 13 & 5577 & 4.58 & dn & no &0.12 & 0.19 & 11 & $-0.09_{-0.08}^{+0.07}$ & 7 & $0.12_{-0.04}^{+0.03}$ & $1.02_{-0.19}^{+0.21}$ \\ 
 377 & 7.59 & 39 & 5873 & 4.28 & dn & no &0.11 & 0.12 & 0 & \nd & 10 & $-0.09_{-0.03}^{+0.03}$ & \nd \\ 
 449 & 8.58 & 95 & 5860 & 4.40 & dn & no &0.12 & 0.27 & 1 & $0.23_{-0.07}^{+0.07}$ & 3 & $0.38_{-0.03}^{+0.03}$ & $0.90_{-0.15}^{+0.16}$ \\ 
 457 & 7.72 & 54 & 5897 & 4.07 & dn & no &0.22 & 0.40 & 1 & $0.10_{-0.10}^{+0.08}$ & 1 & $0.35_{-0.03}^{+0.03}$ & $1.11_{-0.22}^{+0.26}$ \\ 
 531B & 8.65 & 50 & 5707 & 4.62 & \nd & no &0.03 & 0.12 & 0 & \nd & 2 & $-0.09_{-0.06}^{+0.05}$ & \nd \\ 
 691 & 7.95 & 34 & 5633 & 4.66 & dn & no &0.20 & 0.31 & 0 & \nd & 2 & $0.04_{-0.05}^{+0.06}$ & \nd \\ 
 804 & 8.18 & 51 & 5753 & 4.26 & dn & no &-0.02 & 0.05 & 4 & $-0.05_{-0.05}^{+0.04}$ & 0 & \nd & \nd \\

%% file: sm/smcat.tex
% This sample file is dedicated to the public domain.
\chapter{SpecMatch: Accurate Stellar Characterization with Optical Spectra}

\label{c.sm}

\newcommand{\etal}{\mbox{\rm et al.~}}
\newcommand{\ms}{\mbox{m s$^{-1}~$}}
\newcommand{\mse}{\mbox{m s$^{-1}$}}
\newcommand{\ks}{\mbox{km s$^{-1}~$}}
\newcommand{\kse}{\mbox{km s$^{-1}$}}
\newcommand{\msy}{\mbox{m s$^{-1}$ yr$^{-1}~$}}
\newcommand{\msye}{\mbox{m s$^{-1}$ yr$^{-1}$}}
\newcommand{\msun}{M$_{\odot}~$}
\newcommand{\msune}{M$_{\odot}$}
\newcommand{\rsun}{R$_{\odot}~$}
\newcommand{\rsune}{R$_{\odot}$}
\newcommand{\msat}{M$_{\rm SAT}$}
\newcommand{\msate}{M$_{\rm SAT}~$}
\newcommand{\mnep}{M$_{\rm NEP}$}
\newcommand{\mnepe}{M$_{\rm NEP}~$}
\newcommand{\mearth}{M$_{\oplus}~$}
\newcommand{\mearthe}{M$_{\oplus}$}
\newcommand{\rjup}{R$_{\rm JUP}~$}
\newcommand{\msini}{$M \sin i~$}
\newcommand{\msinie}{$M \sin i$}
\newcommand{\mbsini}{$M_b \sin i~$}
\newcommand{\mcsini}{$M_c \sin i~$}
\newcommand{\mdsini}{$M_d \sin i~$}
\newcommand{\chisq}{$\chi_{\nu}^2$}
\newcommand{\arel}{$a_{\rm rel}$}
\renewcommand{\feh}{\ensuremath{[\mbox{Fe}/\mbox{H}]}}
\newcommand{\rphk}{\ensuremath{R'_{\mbox{\scriptsize HK}}}}
\newcommand{\lrphk}{\ensuremath{\log{\rphk}}}
\newcommand{\caii}{\ion{Ca}{2} H \& K}
\newcommand{\mv}{\ensuremath{M_{\mbox{\scriptsize V}}}}
\newcommand{\cs}{$\sqrt{\chi^2_{\nu}}$}

\renewcommand{\nc}[2]{\newcommand{#1}{\ensuremath{#2}\xspace}}
%\newcommand{\num}[2]{\newcommand{#1}{{#2}\xspace}}
%\nc{\Kp}{ \textit{Kp} }
\nc{\fe}{ \text{[Fe/H]}}
\nc{\fchi}{\chi_f^2}
\renewcommand{\Re}{\ensuremath{ R_{\oplus} }\xspace} 

\nc{\rhostar}{\rho_{\star}} 
\renewcommand{\SpecMatch}{\text{SpecMatch}\xspace} 
\renewcommand{\vsini}{\ensuremath{v \sin i}\xspace}
\newcommand{\K}{\text{K}\xspace}

\nc{\sigip}{\sigma_{\rm{IP}}}
\nc{\Kelvin}{\mathrm{K}}
\nc{\numkoi}{1039} % Total number of KOIs with Spectroscopic parameters.
\renewcommand{\kms}{\ensuremath{\mathrm{km~s}^{-1}}\xspace} 
\renewcommand{\ms}{\ensuremath{\mathrm{m~s}^{-1}}\xspace} 

% Huber stats
\nc{\nhubercomp}{75}
\nc{\rmsteffhuber}{64}
\nc{\rmslogghuber}{0.087}
\nc{\rmslogghuberout}{0.075}

\nc{\rmsfehuber}{0.09}
\nc{\rmslogguncal}{0.106}
\nc{\rmsloggcal}{0.067}

\newcommand{\huberc}[1]{%
    \IfEqCase{#1}{%
        {0}{$c_0 = 0.040$\xspace}%
        {1}{$c_1 = 0.025$\xspace}%
        {2}{$c_2 = -0.084$~dex\xspace}
        }
}

\newcommand{\meandlogggiant}{0.014}

% Torres Stats
\nc{\ntorrescomp}{43}
\nc{\rmstefftorres}{89}
\nc{\rmsloggtorres}{0.077}
\nc{\rmsfetorres}{0.11}

% SPOCS stats
\nc{\nspocscomp}{352}
\nc{\rmsteffspocs}{66}
\nc{\rmsloggspocs}{0.161}
\nc{\rmsfespocs}{0.04}

\newcommand{\spocsfec}[1]{%
    \IfEqCase{#1}{%
        {0}{$c_0 = 0.0065$\xspace}%
        {1}{$c_1 = -0.015$~dex\xspace}%
        }
}

\newcommand{\spocsteffc}[1]{%
    \IfEqCase{#1}{%
        {0}{$c_0 = -4.23$\xspace}%
        {1}{$c_1 = -7.6$~K\xspace}%
        }
}

\newcommand{\siglogg}{\ensuremath{\sigma(\logg)\xspace}}

% Albrecht
\newcommand{\rmsalbrechtvsini}{1.1}

% CKS
\nc{\ncks}{1039}
\nc{\nbinary}{52}
\nc{\nvsini}{59}

%Adopted uncertanties
\nc{\uloggas}{0.08}
\nc{\uloggms}{0.10}
\nc{\uteff}{60}
\nc{\ufe}{0.04}
\nc{\uvsini}{1.0}

\nc{\nfe}{n_{\mathrm{Fe}}}
\nc{\nfesun}{n_{\mathrm{Fe},\odot}}

\newcommand{\ttablehead}[1]{
  \tablehead{
    \colhead{Name} &	 
    \multicolumn{3}{c}{\teff} & 
    \multicolumn{3}{c}{\logg} & 
    \multicolumn{3}{c}{\fe} & 
    \colhead{\vsini} \\[0.5ex]
    \cline{2-4}\cline{5-7}\cline{8-10} \\[-1.5ex]
    \colhead{} & 
    \colhead{SM} &  
    \colhead{#1} &
    \colhead{$\Delta$} &
    \colhead{SM} &  
    \colhead{#1} &
    \colhead{$\Delta$} &
    \colhead{SM} &  
    \colhead{#1} &
    \colhead{$\Delta$} &
    \colhead{SM}
  }
}

\citestyle{aa}

%%%%%%%%%%%%%%%%%%%%%%%%%%
% \section{Introduction} %
%%%%%%%%%%%%%%%%%%%%%%%%%%
\section{Introduction}
The ability to extract fundamental stellar parameters from spectra is essential in understanding a wide range of astrophysical phenomena, from exoplanet host star characterization to the study of galactic stellar kinematics and chemical enrichment history. Recently, research in extrasolar planets has spawned a renewed interest in the fundamental properties (e.g. masses, radii, effective temperates, ages) of stars, particularly for M-dwarfs \citep{Boyajian12,Mann13}. New observational techniques such as asteroseismology using space-based photometers like \Kepler \citep{Huber13,Chaplin14} and interferometric measurements of stellar radii \citep{Boyajian12,Boyajian13,von-Braun14} are providing new empirical touchstones that serve to improve spectroscopic methods.

The observation, analysis, and interpretation of exoplanets is closely linked to the fundamental properties of their host stars. In some cases, the presence of an extrasolar planet can improve our knowledge of the host star beyond that from spectroscopy or photometry alone. For example, the transit profile can constrain mean stellar density, \rhostar. \cite{Torres12} used transit-constrained \rhostar to refine host star properties beyond existing spectroscopic analyses. In most cases, however, knowledge of planetary properties like planet mass, size, and equilibrium temperature is limited by our knowledge of stellar mass, radius, and effective temperature.

In this paper, we present a new technique called \SpecMatch that extracts the following properties from high-resolution optical spectra:  effective temperatures, \teff; surface gravities, \logg; metallicities, \fe; and projected rotational velocities, \vsini. Throughout this paper, \teff is measured in Kelvin; \logg is $\log_{10} (g) = G \Mstar / \Rstar^2$, measured in cgs units; metallicity is $\fe~=~\log_{10} (n_{\mathrm{Fe}} / n_{\mathrm{H}}) - \log_{10} (n_{\mathrm{Fe},\odot} / n_{\mathrm{H},\odot})$, where \nfe and $n_{\mathrm{H}}$ are the number densities of iron and hydrogen, respectively; and \vsini is measured in \kms.

Here, we briefly review the processes that connect an observed stellar spectrum to the physical properties of the star (for a more thorough review, see \citealt{Gray05}). Stellar effective temperature, surface gravity, metallicity, and rotation all affect the depths and detailed shapes of spectral lines. The dependence is perhaps the most straightforward; stars with higher \fe, have deeper iron lines. The energy level populations of absorber species is set by the Boltzmann equation for neutral species, and by the Saha equation for ionized species. Both the Boltzmann and Saha equations have an exponential sensitivity to temperature. Thus the relative strength of different lines having different excitation potentials is a good diagnostic of temperature.

Surface gravity is much harder to measure. The Saha equation has a linear dependence on the local electron pressure which is a probe of the surface gravity. One diagnostic of \logg is the relative strength of lines corresponding to different ionization states of the same element. Fe I and Fe II are often used. Another approach involves modeling the wings of so-called ``pressure-sensitive'' lines like Na~I~D doublet, Ca~II~H\&K, Mg~I~b triplet, and Ca~I at 4227 \AA. However, modeling in detail how, as an example, the Mg~I~b line profile changes with differing surface gravities is challenging, especially considering the covariant effects of temperature and Mg abundance. Temperature, surface gravity and metallicity all influence a star's spectrum in different, but non-orthogonal ways. Measurements of \teff, \logg, \fe are often complicated by covariances between all three parameters.

Lastly, rotation broadens spectral lines due to the Doppler effect. Lines formed at the receding limb of the star are red-shifted and blue shifted if they are formed on the approaching limb. Rotation is relatively easy to measure in the case of moderate stellar rotation (\vsini~$\gtrsim 5$~\kms) by fitting the profiles of unsaturated lines. For slowly rotating stars, rotational broadening becomes sub-dominant compared to other broadening terms like, macroturbulence, microturbulence, thermal broadening, pressure broadening, and the instrumental profile of the spectrometer.

Determining \teff, \logg, \fe, and \vsini for planet-hosting stars has a wide variety of applications. Combining spectroscopic \teff, \logg, and \fe with isochrone-fitting offers dramatic improvements over photometrically determined stellar masses, radii, and luminosities, which result in more precise planet masses, radii, and equilibrium temperatures. For example, photometric radii from the \Kepler Input Catalog are uncertain at at the $\sm35\%$ level \citep{Brown11,Batalha13,Burke14} while spectroscopically-derived radii have uncertainties ranging from 1--10\%, depending on the type of star \citep{Valenti05}. Correlations between planet occurrence and stellar metallicity probe the connection between planet formation efficiency and the composition of the protoplanetary disk. Projected rotational velocity can constrain the inclination of the stellar spin axis (and star-planet spin orbit alignment) if the star's equatorial velocity is known (by, for example, rotation modulation).

Previous spectroscopic studies of nearby stars often enjoy high SNR spectra. For example, \cite{Valenti05} measured \teff, \logg, \fe, and \vsini for 1040 stars using spectra where SNR~>~100/pixel. While \SpecMatch is a general purpose tool, it was designed to measure spectroscopic properties of faint \Kepler stars with the HIRES spectrometer \citep{Vogt94} on the Keck I telescope. At 1~kpc, a typical distance to a \Kepler target star, a solar analog has $V~=~14.7$. Obtaining a spectrum with SNR~=~100/pixel would take 2.5~hours and is not feasible for large samples of stars. We designed \SpecMatch in order to yield reliable stellar parameters when SNR~$\lesssim$~40/pixel. 

This chapter is organized as follows: In Section~\ref{sec:sm-algorithm} we describe how we condition observed spectra and extract spectroscopic parameters. In Section~\ref{sec:sm-reliability} we assess the reliability of \SpecMatch parameters. We present a detailed Monte Carlo study of the precision \SpecMatch as a function of spectral SNR and assess systematic uncertainties by analyzing spectra of touchstone stars from the literature. We summarize our results in Section~\ref{sec:sm-conclusion}, and expand on several compelling applications of \SpecMatch.

% VF05 wavelengths 5165-5190, 6000-6030, 6030-6050, 6050-6070 6101-6117, 6125-6139, 6144-6158, 6160-6180 = 160 angstroms.
% SpecMatch walengths
% 5200-5280, 5360-5440, 5530-5610, 6100-6190, 6210-6260 = 380 Ang

% Median error of stars with Pinsonneault parameters 38%
% Spectroscopic parameters
% For the batalha table median koi_ror is 5%

\section{\SpecMatch}
\label{sec:sm-algorithm}

In brief, \SpecMatch works by comparing an observed high-resolution optical spectrum to a library of synthetic model spectra from \cite{Coelho05} (C05, hereafter) that span a range of \teff, \logg, and \fe. In its current implementation, \SpecMatch accepts spectra taken with HIRES on the Keck I telescope. However, \SpecMatch can be easily adapted to work with other high resolution optical spectrometers. In this section we describe how we condition observed spectra and extract stellar parameters.

\subsection{Reduction and Calibration of Target Spectra}
We remove the blaze function by dividing the target spectrum by the
spectrum of a rapidly rotating B star. The California Planet Search \citep{Marcy08} routinely takes spectra of rapidly rotating B stars which have nearly featureless spectra that, when observed with HIRES, provide a good description of the shape of the blaze
function. We divide the target spectrum by the average of 20 B stars
having weak stellar absorption lines. The resulting spectra still show
variations at the few percent level due to differences in slit
illumination and changes optics of HIRES. A low-order polynomial is
fit to the 95 percentile level of the spectrum to remove residual
curvature.

After normalizing out the continuum, we place the star's spectrum onto
its rest wavelength scale. We cross-correlate chunks of the target
spectrum with a model template spectrum, taken from the
C05 library of synthetic spectra. We use
one of six model spectra with parameters listed in \ref{tab:coelho-rw}
as our wavelength standard. We cross-correlate a segment%
\footnote{The middle chip on HIRES has 16 orders. We label the orders
  starting from zero, i.e. 0, 1, ... 15, from bluest to reddest. Order
  number 2 is used to select the model spectrum acting as wavelength
  standard.}
of the target star spectrum with each of the 6 models and select the
model spectrum with the highest cross-correlation peak as the
wavelength standard.

For each of the 16 orders, we cross-correlate seven segments of 1000 pixels wide.%
\footnote{A shift of one HIRES pixel is equivalent to a velocity
  displacement of$\sim$1.3~km s$^{-1}$. The seven segments are
  evenly-spaced, starting at pixel number 0, 500, 1000, 1500, 2000,
  2500, and 3000. }
\nc{\dv}{\delta v}
Each of these seven segments gives an apparent velocity shift. Taking all the orders together, we construct $\mtx{\dv}$, a matrix of velocity shifts:
\[
\mtx{\dv} =  
\begin{pmatrix}
  \dv_{1,1} & \hdots & \dv_{1,7} \\
  \vdots  & \ddots & \vdots  \\
  \dv_{16,1} & \hdots & \dv_{16,7} \\
\end{pmatrix},
\]
where $\dv_{i,j}$ corresponds to the velocity shift of segment $j$ in
order $i$. We compute the average shift (with sigma clipping) over
each of the columns of $\mtx{\dv}$ and fit a linear relationship to
derive \dv as a function of pixel number. This average velocity shift
is applied to all orders to place the target spectrum on its rest
wavelength scale.

\subsection{Comparison to Library Spectra}
We compare the target spectrum to a suite of model spectra from the
C05 library. C05 modeled spectra over a grid in \teff, \logg, \fe, and
[$\alpha$/Fe]. \SpecMatch uses a subset of C05 model spectra having
solar [$\alpha$/Fe], and \teff, \logg, and \fe listed in
Table~\ref{tab:coelho-grid}. The \SpecMatch library consists of a $15 \times 9 \times 7 $ regular grid of C05 model spectra. C05 used model atmospheres from \cite{Castelli03}. Line lists came
from \cite{Barbuy03} and \cite{Melendez99}. Oscillator strengths were
taken from the NIST database \citep{Reader02}, other works (see
references in C05), as well as empirically by fitting line profiles to
the solar spectrum. Damping constants ($\gamma$) for strong neutral
lines were taken from \cite{Anstee95,Barklem97}; and
\cite{Barklem98,Barklem00}. Damping constants for other lines were fit
manually to the solar spectrum or assumed to have an interaction
constant of $C_{6} = 0.3 \times 10^{-31}$. C05 synthesized model
spectra over $\lambda$~=~3000--18,000~\AA\ with 0.02~\AA\ sampling
using the {\tt PFANT} radiative transfer code
\citep{Spite67, Cayrel91,Barbuy03}. C05 used the following
prescription for microturbulence:

\begin{displaymath}
   v_t = \left\{
     \begin{array}{ll}
       1.0~\kms & : \logg \geq 3.0 \\
       1.8~\kms & : 1.5 \leq \logg \leq 3.0 \\
       2.5~\kms & : \logg \leq 1.0
     \end{array}
   \right.
\end{displaymath}

Each target-model comparison involves subtracting the model spectrum from the observed spectrum. We filter the residuals and remove trends longer than 400 pixels ($\sim8.6$~\AA) before computing
$\chi^{2}$. This high-pass filtering ensures that well-matched models
are not penalized on account of low-frequency noise (due to imperfect
continuum fitting). 

The comparisons are performed independently on different segments of stellar spectra. We found by trial and error that the five spectral regions listed in Table~\ref{tab:segdf} produced parameters that were in close agreement with touchstone stars from \cite{Huber13} (with asteroseismic constraints) and from \cite{Torres12} (with transit-constrained \rhostar). We also identified, by inspection, certain lines that poorly matched observed spectra and constructed a mask to exclude these regions from the computation of \fchi. Figures~\ref{fig:CK00001_5200}--\ref{fig:CK00001_6210} show the five spectral regions for a KOI-1, a star with nearly Solar parameters%
\footnote{
\teff~=~5850~K, \logg~=~4.46, and \fe~=~-0.15~dex \citep{Huber13}
}
The masked regions are grayed out. The \SpecMatch spectral segments are characterized by having relatively few saturated and overlapping lines. One aspect that makes \SpecMatch suitable for low SNR spectra is that it uses a wide region (380~\AA) of optical spectrum. As a point of comparison the VF05 analysis used $\approx170$~\AA. 

One surprise was that the HIRES order containing the Mg~I~b triplet produced \logg values in poor agreement with asteroseismology. The Mg~I~b line is a so-called ``pressure sensitive'' line and is often used as a \logg diagnostic. While the wings of the line are, in principle, good probes of stellar surface gravity, we excluded them because of the observed tension with asteroseismology. There may be issues with the C05 treatment of these lines or that strong covariances with \teff or magnesium abundance (which may not scale with \fe) are complicating the fit to the Mg~I~b lines.

The C05 model spectra incorporate natural, thermal, collisional, and
microturbulent broadening. Since the model spectra do not account for
broadening due to rotation, macroturbulence, or the instrumental
profile, lines in the C05 models are narrower than observed
spectra. During each target-model comparison we convolve the model
spectrum with a standard rotational broadening kernel (see p.~374 of 
\citealt{Gray92}) to account for
additional broadening. \vsini is allowed to float as a free
parameter. Note that \vsini is acting as a stand-in for other
broadening terms besides rotation like macroturbulence and the
instrumental profile. In a later polishing step (Section~\ref{sec:polish}),
macroturbulence and the instrumental profile are treated individually.
The instrumental profile is determined empirically for each spectrum using
telluric lines, following a procedure described in Section \ref{ssec:telluric}.

\begin{deluxetable}{llp{2in}}
  \tabletypesize{\footnotesize}
  \tablecaption{Parameters for Model Spectra in C05 Library}
  \tablewidth{0pt}
  \tablehead{Parameter & Library Values}
  \startdata
  \teff & 3500--7000~K, steps of 250~K\\
  \logg & 1.0--5.0 (cgs), steps of 0.5 dex.\\
  \fe & $\{-2.0, -1.5, -1.0, -0.5, +0.0, +0.2, +0.5\}$ dex \\
  $[\alpha/$Fe] & 0.0 dex
  \enddata
  \label{tab:coelho-grid}
\end{deluxetable}

\begin{deluxetable}{lll}
  \tabletypesize{\footnotesize}
  \tablecaption{\SpecMatch Wavelength Segments}
  \tablewidth{0pt}
  \tablehead{
    \colhead{$\lambda_{\rm{min}}$ (\AA)} & 
    \colhead{$\lambda_{\rm{max}}$ (\AA)} & 
    \colhead{HIRES Order}
  }
  \startdata
  5200  & 5280  & 3  \\
  5360  & 5440  & 5  \\
  5530  & 5610  & 7  \\
  6100  & 6190  & 13 \\ 
  6210  & 6260  & 14
  \enddata
  \label{tab:segdf}
\end{deluxetable}

\subsection{Stellar Parameters from Linear Combinations of Spectra}
\label{ssec:lincomb}
We zero in on a best-fit \teff, \logg, and \fe by taking linear
combinations of $m$ model spectra that have lowest $\fchi$, when compared to the target spectrum. This linear combination can be represented as,
$\mathbf{M A}$, where, $\mtx{M}$ is an $n \times m$ matrix of
best-match model spectra and $\mtx{A}$ is a $m \times 1$ matrix of
coefficients. We solve for the set of positive coefficients that
minimize $\chi^{2}$ using a non-negative least-squares solver,%
\footnote{As implemented in scipy.opimize.nnls algorithm from Lawson
  C., Hanson R.J., (1987) Solving Least Squares Problems, SIAM}
i.e.
\[
\text{argmin}_\mtx{A} ||\mtx{MA} - \mtx{S}||_2 \text{ for } \mtx{A} \geq 0.
\]
By trial and error, we found that using $m = 8$ model spectra to construct \mtx{M} gave the fitter adequate flexibility while keeping the number of free parameters manageable. We arrive at the target star's \teff, \logg, and \fe by taking a average of the \teff, \logg, and \fe associated with each of the best $m$ models, weighted by \mtx{A}. This averaging is done for each of the stellar parameters associated with each wavelength segment. The \teff, \logg, and \fe determined from each segment are averaged again to determine a single
set of \teff, \logg, and \fe for the target star spectrum.

\subsection{Measuring the Instrumental Profile with Telluric Lines}
\label{ssec:telluric}
\begin{figure*}
\centering
\includegraphics[width=1\columnwidth]{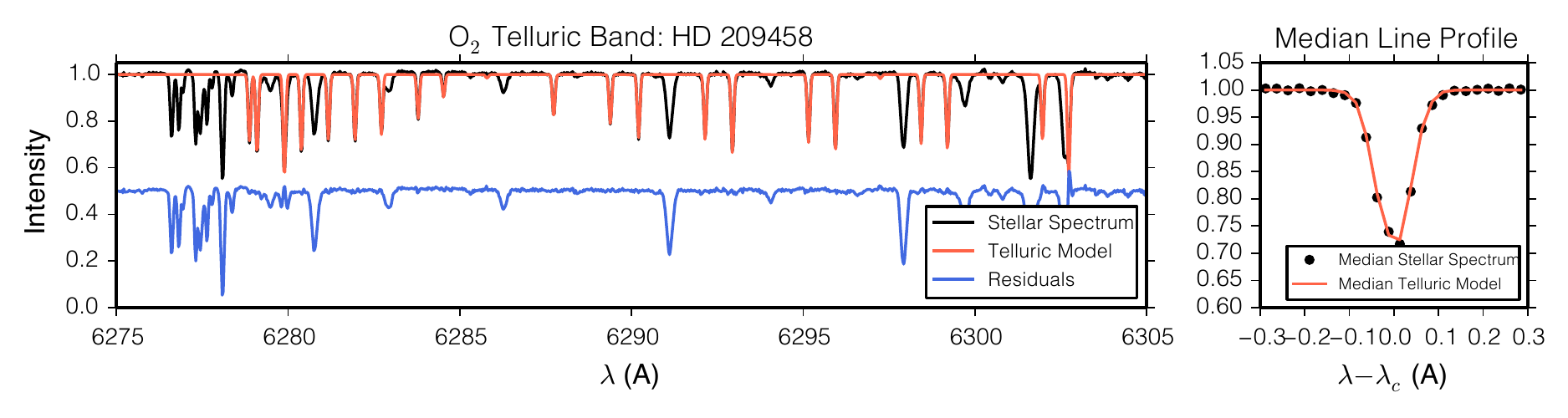}
\caption[Spectrum of HD~209458 around the O$_{2}$
  telluric band]
  {Left panel---Spectrum of HD~209458 around the O$_{2}$
  telluric band (black line). We fit the telluric lines with a comb of
  Gaussians (red line). Residuals are shown in blue. Right panel---median
  intensity (computed for 0.025~\AA\ bins) of the HD 209458 spectrum
  (black) and telluric model (red) within 0.3~\AA\ of the telluric
  line centers. The observed and model telluric line profiles have the
  same width. Since the telluric lines are not broadened by the
  thermal and convective velocities in the star's photosphere, they
  are a good diagnostic for the instrumental profile of HIRES. For HD
  209458, the best fit \sigip~=~1.41 HIRES pixels. In subsequent
  modeling of the HD 209458 spectrum, we describe the instrumental
  profile as a Gaussian with \sigip~=~1.41~pixels.}
\label{fig:209458_tell}
\end{figure*}

\begin{figure*}
\centering
\includegraphics[width=1\columnwidth]{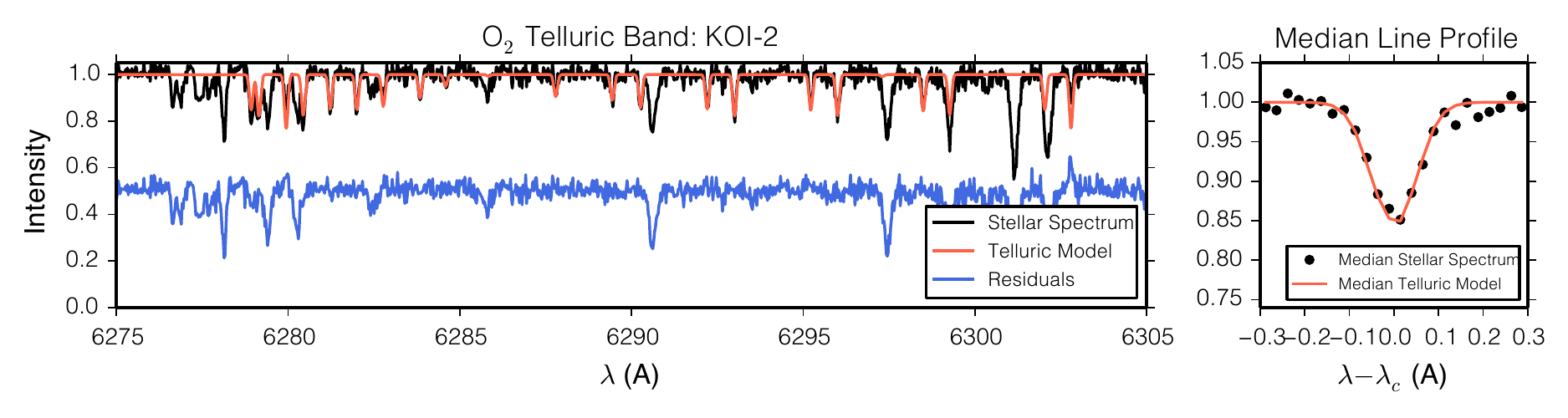}
\caption[Spectrum of KOI-2 around the O$_{2}$
  telluric band]{Same as Figure~\ref{fig:209458_tell}, except for KOI-2. This
  spectrum has SNR~=~45/pixel. This
  spectrum was taken using the wider C2 decker thus the
  instrumental profile is broader, \sigip~=~1.87~pixels.}
\label{fig:CK00002_tell}
\end{figure*}

Properly treating the width of the instrumental profile is especially important for
\vsini work. If our model of the instrumental profile is too narrow, \SpecMatch will
increase \vsini in order to correctly match the width of absorption lines. While telluric lines are broadened by turbulence in the Earth's atmosphere
to $\sim100$~\ms, telluric lines remain narrow compared to the
instrumental profile of HIRES (several \kms) and are good
diagnostics its width. For each target spectrum, fit the $\text{O}_2\text{~B-X}$ band of telluric lines with a comb of Gaussians:
\begin{equation}
\label{eqn:sigip}
I(\lambda) = 1 - \sum_{i} a_i e^{-\frac{1}{2}\left( \frac{\lambda - \lambda_{c,i}}{\sigip}\right)^2},
\end{equation}
where $a_i$ and $\lambda_i$ specifies line depths and centers
respectively and \sigip sets the width of the Gaussians. In the fit, \sigip is allowed to float as a free parameter. Figure~\ref{fig:209458_tell}
shows the fit to the telluric lines HD 209458 spectrum, a high SNR
spectrum (SNR \sm 160/pixel). We show the fits to the KOI-2 spectrum
in Figure~\ref{fig:CK00002_tell} with SNR~\sm~45/pixel.

\begin{figure}
\centering
\includegraphics[width=0.6\columnwidth]{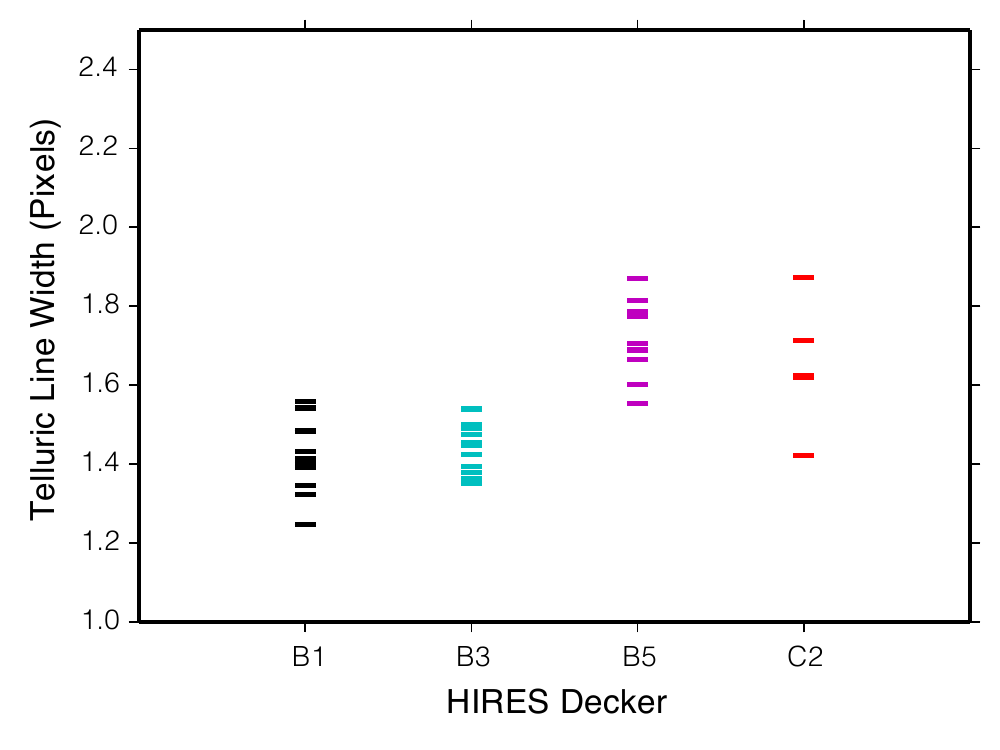}
\caption[Width of telluric lines as a function of HIRES decker]{Width of telluric lines in spectra of stars from the \cite{Albrecht12} sample for
  different HIRES deckers. On average, the wider B5 and C2 deckers
  have broader instrumental profiles (hence broader telluric
  lines) than the narrower B1 and B3 deckers. For any single decker,
  however, \sigip, the measured width of the telluric lines (see Equation~\ref{eqn:sigip}), varies by \sm~0.4~pixels = 0.5~\kms. Thus
  adopting a single instrumental profile width would introduce errors
  in \vsini as large as 0.5~\kms.}
\label{fig:psf}
\end{figure}

The width of the instrumental profile, \sigip, depends on seeing, 
the HIRES slit width, and the performance and focus of the spectrometer optics.
We illustrate this variability in PSF-width
in Figure~\ref{fig:psf}, where we show the measured line widths, for
43 spectra of stars from the \cite{Albrecht12} Rossiter-McLaughlin sample (see
Section~\ref{sec:albrecht}) organized by HIRES decker. The wider
deckers, B5 and C2, have a sky projected width of 0.861'' and larger
\sigip, than the B1 and B3 deckers (0.574'' projected width). For
different observations with the same decker, the instrumental profile
width varies by \sm~0.4~pixels~=~0.5~\kms. Thus, adopting a single
instrumental profile width would produce errors in \vsini as large as
\sm~0.5~\kms, hampering the measurement of small \vsini.

\begin{deluxetable}{lll}
\tabletypesize{\footnotesize}
\tablecaption{Parameters of Template Spectra Used for Wavelength Calibration}
\tablewidth{0pt}
\tablehead{
	\colhead{\teff} &	\colhead{\logg} & \colhead{\fe} \\
	\colhead{(K)}   &	\colhead{(cgs)} & \colhead{(dex)}
	}
\startdata
4000 & 5.0 & 0.0\\
4500 & 5.0 & 0.0\\
5000 & 4.5 & 0.0\\
5500 & 4.5 & 0.0\\
6000 & 4.5 & 0.0\\
6500 & 4.0 & 0.0
\enddata
\label{tab:coelho-rw}
\end{deluxetable}

\subsection{Polishing the Parameters by Forward Modeling}
\label{sec:polish}
The linear combination approach gives good initial guesses for the
parameters. We refine these parameters using a forward modeling
approach. Here, we synthesize model spectra with an arbitrary \teff,
\logg, \fe, \vsini, and instrumental profile according to the
following three steps:

\begin{enumerate}
\item  We select the eight model spectra with parameters that enclose the desired set of \teff, \logg, and \fe in a box in the 3D space of \teff, \logg, and \fe. We synthesize an intermediate spectrum within this box using trilinear interpolation.

\item We then account for rotation and macroturbulence (parametrized
  by $\xi$) by constructing a combined rotational-macroturbulent
  profile (equation 17 of \citealt{Hirano11}) and convolving it with
  the spectrum from step 1. For stars with moderate \vsini and $\xi$,
  it is possible to solve for \vsini and $\xi$ independently, with
  very high SNR and very high spectral resolution. For our CKS spectra
  there is a large degeneracy between \vsini and $\xi$, so we adopt
  the following relationship from \cite{Valenti05}:
  \[
  \xi = 3.98 + \left(\frac{\teff-5700~\K}{650~\K}\right)~\kms
  \]
  
\item We convolve the spectrum from step 2 with a Gaussian to model
  the effects of the HIRES instrumental profile, which is determined
  from telluric lines, as described in Section~\ref{ssec:telluric}. \end{enumerate}

After synthesizing the model spectrum, we re-compute \fchi according to the procedure outlined in Section~\ref{ssec:lincomb}. We 
refine the parameters determined during the linear combination step
(Section~\ref{ssec:lincomb}) by varying \teff, \logg, \fe, and
\vsini in search of the best match. We converge on the best match
using a Levenberg-Marquardt algorithm. As an example, we show the spectrum of KOI-1 along with the best-fitting model spectrum in
Figures~\ref{fig:CK00001_5200}--\ref{fig:CK00001_6210}.

\subsection{Calibrating Effective Temperature and Metallicity to Valenti \& Fischer (2005)}
\label{ssec:spocscal}
While spectroscopy routinely achieves effective temperature precision
of < 100~K, zero-point and temperature-dependent offsets persist at
the $\sim$100~K level. These offsets are observed when comparing
effective temperatures based on photometry and spectroscopy and when
comparing effective temperatures based on different spectroscopic
techniques \citep{Casagrande10}. Further work is needed to anchor the
spectroscopic effective temperatures to the absolute effective
temperatures from \Lstar and \Rstar.

\begin{figure}
\centering
\includegraphics[width=0.7\columnwidth]{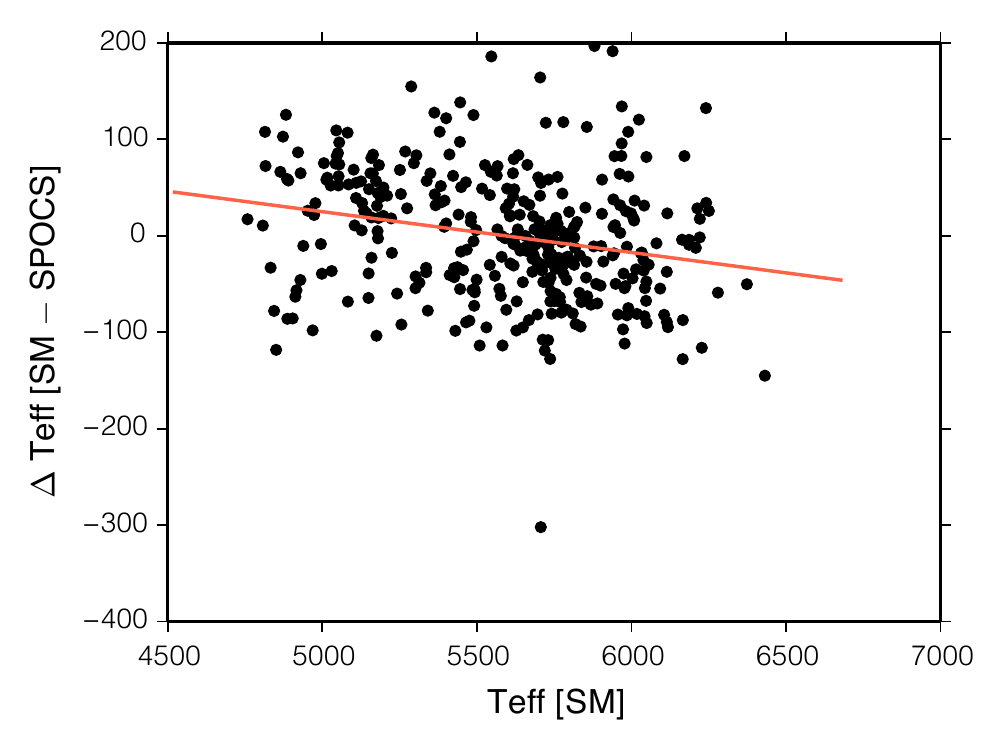}
\caption[Comparison to \cite{Valenti05} effective temperatures]
{Difference in effective temperatures from \SpecMatch and
  \cite{Valenti05}, $\Delta \teff = \teff \mathrm{[SM]} - \teff
  \mathrm{[VF05]}$. We have fit a straight line to these
  points. Subsequent effective temperatures presented in this work
  have had this trend removed.}
\label{fig:teff-spocs-cal.pdf}
\end{figure}

\cite{Valenti05}, VF05 hereafter, used the ``Spectroscopy Made Easy''
(SME) spectrum synthesis code \citep{Valenti96} to measure stellar
properties for 1040 nearby FGK stars. We choose to anchor \SpecMatch
effective temperatures to the VF05 scale because the VF05 catalog is
an important touchstone in the literature. We analyzed spectra of
\nspocscomp stars from the VF05 catalog using \SpecMatch. We use these
shared stars to link the \SpecMatch effective temperatures to the SME
scale. In Figure~\ref{fig:teff-spocs-cal.pdf}, we show the difference
between \SpecMatch and VF05 effective tempeartures, $\Delta \teff$. We fit these differences using a linear relationship,

\[
\Delta \teff = c_0 \left(\frac{\teff - 5770~K}{100~K}\right) + c_1,
\]
where \spocsteffc{0} and \spocsteffc{1}. Subsequent effective
temperatures presented in this work have been calibrated against this
relationship.

\begin{figure}
\centering
\includegraphics[width=0.7\columnwidth]{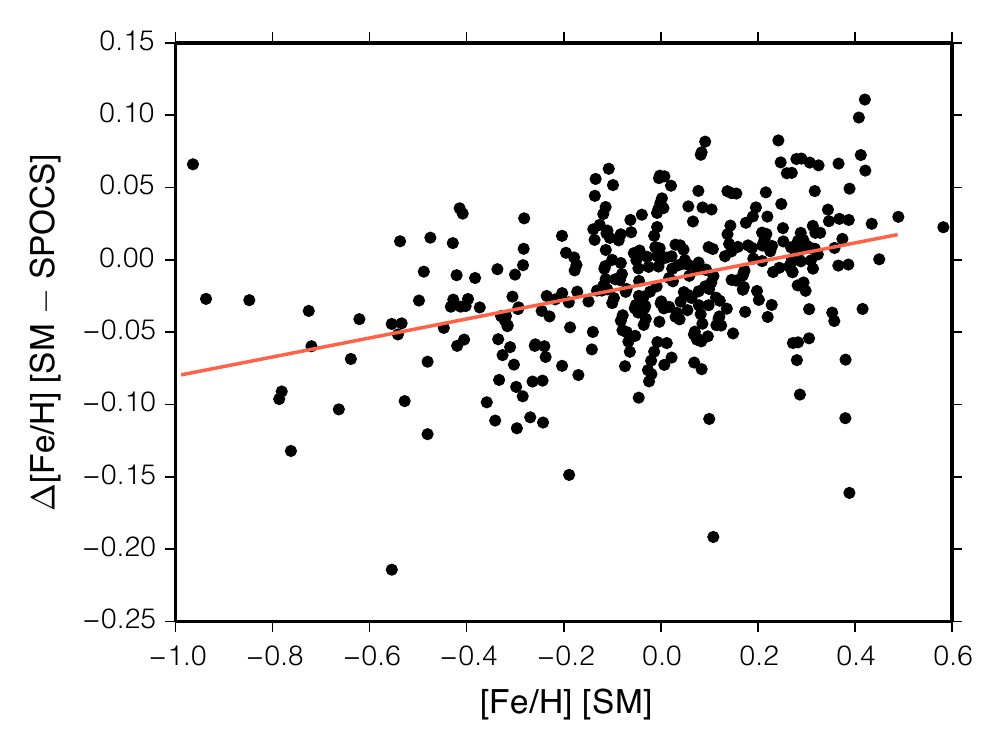}
\caption[Comparison to \cite{Valenti05} metallicities]
{Difference in metallicities from \SpecMatch and
  \cite{Valenti05}, $\Delta \fe = \fe \mathrm{[SM]} - \teff
  \mathrm{[VF05]}$. We have fit a straight line to these
  points. Subsequent metallicities presented in this work have had
  this trend removed.}
\label{fig:fe-spocs-cal.pdf}
\end{figure}

The VF05 catalog is also an important with respect to
planet-metallicity work. Using the VF05 catalog, \cite{Fischer05}
firmly established the correlation between Jovian planet occurrence
and host star metallicity. We elect to anchor \SpecMatch \fe to the
VF05 scale in order to facilitate comparisons between the two
studies. Figure~\ref{fig:fe-spocs-cal.pdf} shows the difference
between \SpecMatch and VF05 metallicities, $\Delta \fe$. We fit these differences using a linear relationship,
\[
\Delta \fe = c_0 \left(\frac{\fe}{0.1~\mathrm{dex}}\right) + c_1,
\]
where \spocsfec{0} and \spocsfec{1} are the best fit
coefficients. Subsequent metallicities presented in this work have
been calibrated according to this relationship.

\subsection{Calibrating Surface Gravities to Huber et al. (2013)}
\label{ssec:hubercal}

\cite{Huber13} (H13, hereafter) published stellar parameters of 77
\Kepler planet hosts where asteroseismic modes were detected. Analysis
of power spectra of \Kepler short cadence photometry revealed
solar-like oscillations in these 77 stars. H13 measured the large
frequency separation, $\delta \nu$, which depends on \Mstar and
\Rstar, and $\nu_{max}$, which depends on \Mstar, \Rstar, and
\teff---2 equations and three unknowns. Starting with an initial
temperature from spectroscopy, H13 solved for \Mstar and \Rstar and
hence \logg. \logg was then fixed in the spectroscopic analysis and
\teff and \fe were re-derived from modeling spectra. \teff was fed
back into the asteroseismic equations iteratively until convergence.

\begin{figure*}[h!]
\centering
\includegraphics[width=1\columnwidth]{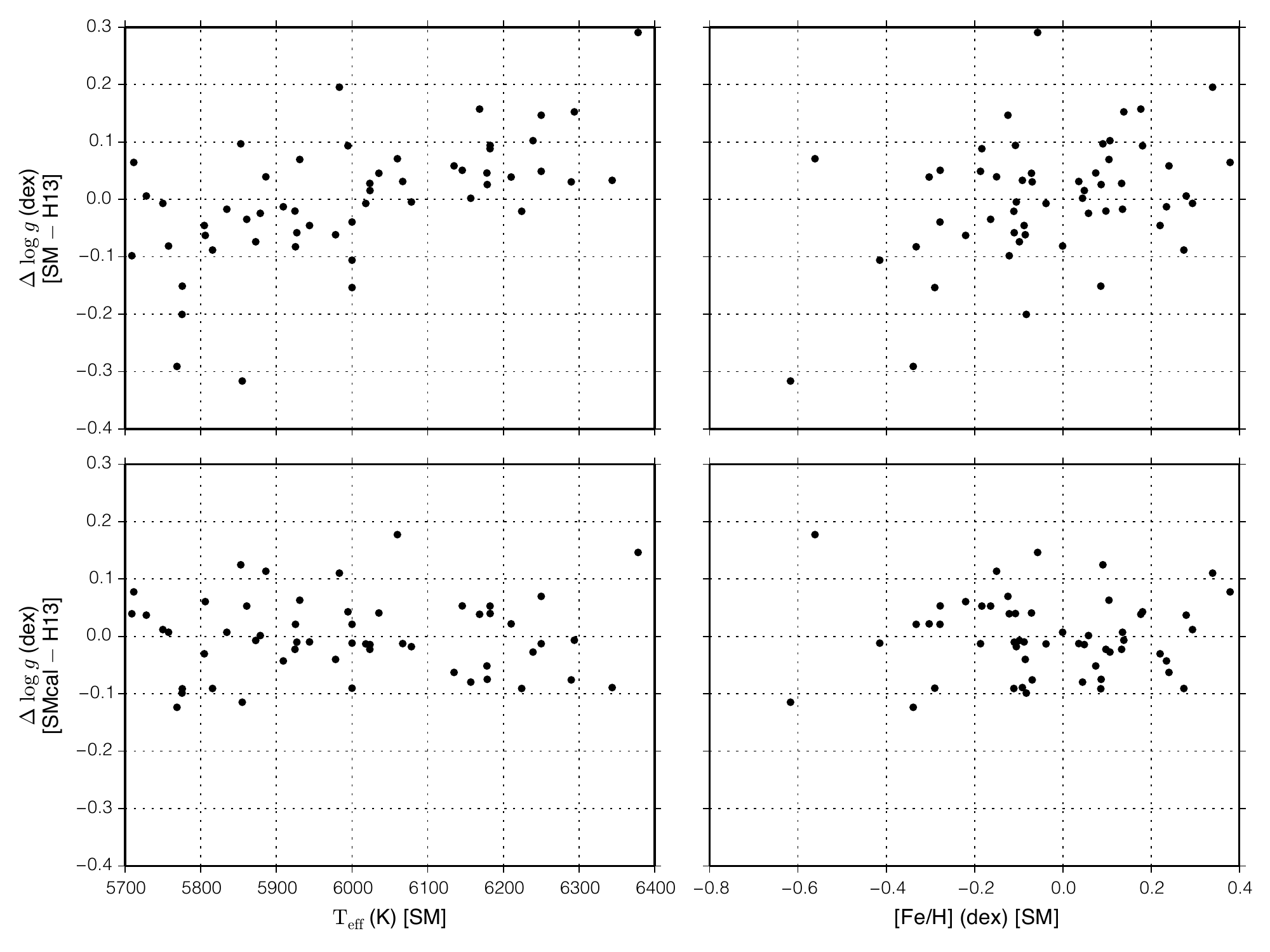}
\caption[Comparison to \cite{Huber13} surface gravities]{Top row: differences between \SpecMatch and H13 surface gravities,
  $\Delta \logg$, as a function of \SpecMatch \teff and \fe. The stars
  shown have H13 parameters within the following range:
  \logg~=~3.5--5.0~(cgs) and \teff~=~5700--6500~K. \SpecMatch tends to
  yield higher surface gravities than H13 for high temperature and
  high metallicity stars. We calibrate the \SpecMatch surface
  gravities to the H13 scale by fitting a plane to $\Delta \logg
  (\teff,\fe)$ and subtracting this relationship from the \SpecMatch
  \logg values. Bottom row: difference between calibrated \SpecMatch and H13
  surface gravities. Trends in $\Delta \logg$ have been removed, and
  scatter in $\Delta \logg$ is smaller using calibrated \SpecMatch
  \logg---\rmsloggcal~dex versus \rmslogguncal~dex.}
\label{fig:logg-huber-cal}
\end{figure*}

Because asteroseismology offers exquisite \logg precision for bright stars%
\footnote{The median \logg uncertainty from H13 is 0.025~dex.}
we use \nhubercomp stars%
\footnote{Two stars were omitted from our comparison: KOI-1054 and
  KOI-2481. KOI-1054 is a very evolved star and had a challenging
  seismic detection (D. Huber, private communication). The
  spectroscopic input parameters for KOI-2481 we based on a single low
  SNR spectrum and resulted in a poor fit during the SPC analysis.}
from the H13 sample to calibrate \SpecMatch surface gravities in the
following two regions of the HR diagram:

\begin{enumerate}
\item 
For stars with \logg~=~3.5--5.0~(cgs) and \teff~=~5700--6500~K, we
note a weak dependence of $\Delta$ \logg = \logg~(SM) $-$ \logg~(H13)
on \teff and \fe, which is shown in the top row of
Figure~\ref{fig:logg-huber-cal}. We calibrate the dependence of \logg
on \teff and \fe, by fitting a plane:

\[
\Delta \logg 
   = c_0 \left(\frac{\teff - 5770~\K}{100~\K}\right) +  
	 c_1 \left(\frac{\fe}{0.1~\mathrm{dex}}\right) +
	 c_2,
\]

where \huberc{0}, \huberc{1}, \huberc{2}. The bottom
panels of Figure~\ref{fig:logg-huber-cal} shows the difference between
the calibrated \SpecMatch and H13 surface gravities as a function of
\teff and \fe. The calibration decreased the \logg dispersion from
\rmslogguncal dex to \rmsloggcal dex.

\item 
For stars with \logg~=~1.0--4.0~(cgs) and \teff~=~4000--5500~K, we
subtract \meandlogggiant~dex from \SpecMatch surface gravities to place them on
the H13 scale.
\end{enumerate}
All subsequent surface gravities presented in this work have been
calibrated against H13, according to the above method.

%%%%%%%%%%%%%%%%%%%%%%%%%%%%%%%%%%%%%%%%%%%%%%%%%%
% \section{Reliability of \SpecMatch Parameters} %
%%%%%%%%%%%%%%%%%%%%%%%%%%%%%%%%%%%%%%%%%%%%%%%%%%

\section{Reliability of \SpecMatch Parameters}
\label{sec:sm-reliability}
\subsection{Photon-Limited Errors}
\label{ssec:photonlimited}
\begin{figure*}
  \centering
  \includegraphics[width=1\columnwidth]{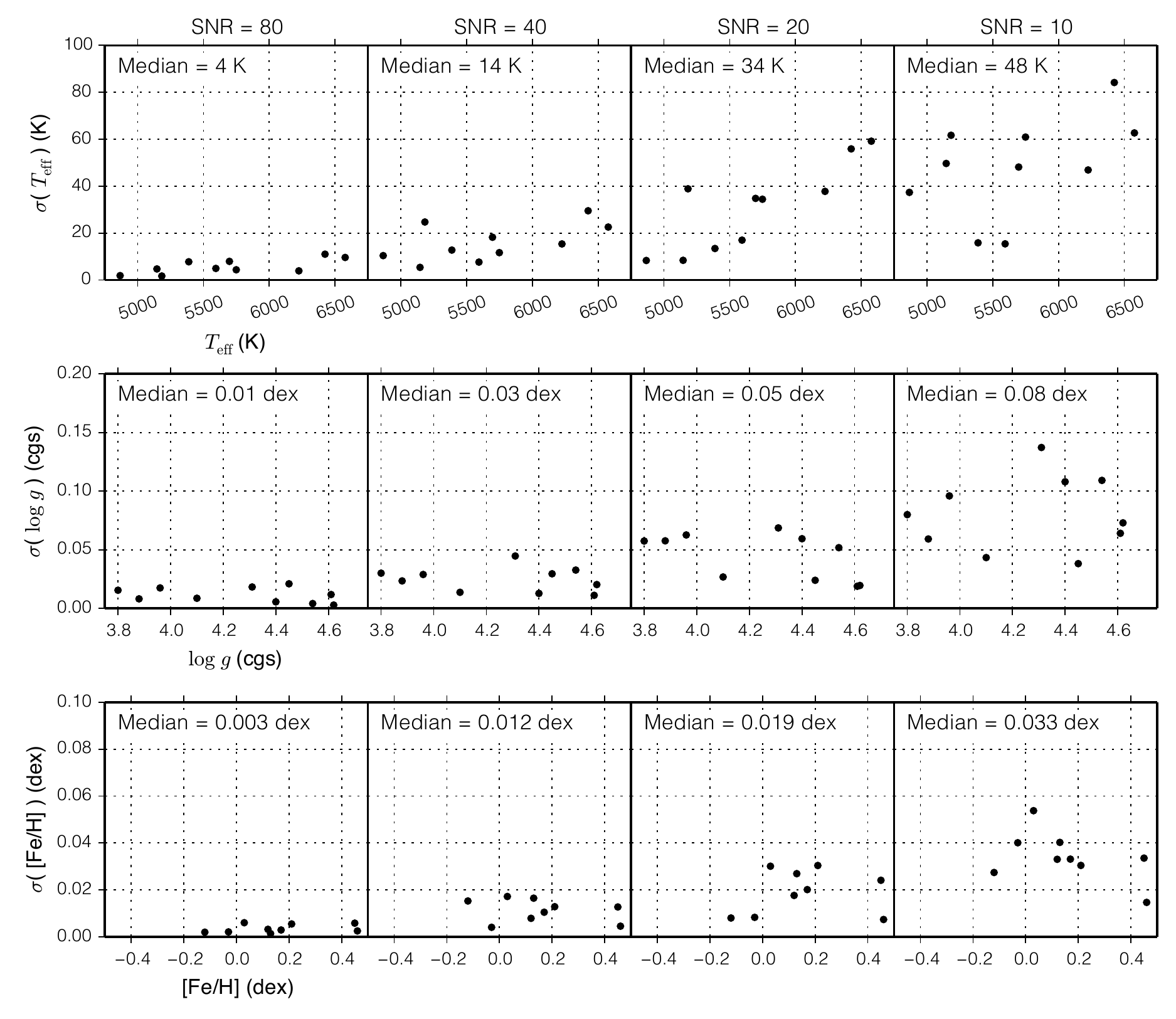}
  \caption[Statistical uncertainties for stellar parameters]
    {The scatter in best fit \teff, \logg, and \fe among the 20
    Monte Carlo realizations for each of the 10 diagnostic stars at 4
    different SNR levels: SNR/pixel~=~80, 40, 20, and 10. For example,
    the top left panel shows the standard deviation of the best fit
    temperatures computed from 20 SNR/pixel~=~80 realizations for 10
    diagnostic stars. The median standard deviation in temperature for
    all 10 stars 4~K. The SNR levels in columns 1, 2, 3, and 4 are 80,
    40, 20, and 10, respectively. As SNR declines, the scatter in the
    \SpecMatch-derived parameters increases for the Monte Carlo
    simulated spectra. This scatter is representative of
    photon-limited errors at different SNR levels. At SNR/pixel~=~40,
    slightly below the typical SNR/pixel~=~45 for CKS, photon-limited
    errors are 14~K in \teff, 0.03~dex in \logg, and 0.012~dex in \fe,
    and is second order compared to expected systematic
    uncertainties. Note that the statistical error in \teff is larger
    hotter \teff stars. At higher \teff, there are fewer lines, so
    pixel-by-pixel errors have more of an impact on the derived best
    fit parameters.}
  \label{fig:SNR}
\end{figure*}

We explored how photon noise affects our derived parameters using a
Monte Carlo approach. Starting with 10 high SNR spectra from the VF05
catalog (SNR/pixel$\sim$150), we injected noise to simulate lower
SNR spectra. Spectra were degraded to SNR/pixel of 80, 40,
20, and 10. To understand the degree to which pixel-to-pixel
fluctuations affect \teff, \logg, and \fe, we generated 20
realizations of each stellar spectrum at each SNR level
($20\times10\times4$ = 800 total realizations). We adopted the
standard deviation of the best fit values of \teff, \logg, and \fe for
the 20 realizations as the error associated with Poisson fluctuations
for each spectrum at each SNR level. Figure~\ref{fig:SNR} shows the
standard deviation of best fit \teff, \logg, and \fe for each star at
each SNR level. Scatter in the derived parameters grows with
decreasing SNR. Poisson errors tend be larger for hotter stars since
they have fewer lines in the \SpecMatch spectral regions. We list the
median photon-limited error for \teff, \logg, and \fe for different
SNR levels in Table~\ref{tab:SNR}. At SNR/pixel~=~40, photon-limited errors are 14~K in
\teff, 0.03~dex in \logg, and 0.012~dex in \fe, and as we will show in
Section~\ref{sec:touchstone}, photon-limited errors are second order
compared to systematic uncertainties. 

Photon-limited errors have the biggest impact on the PSF-fitting
component of \SpecMatch, described in Section~\ref{sec:polish}. The
HIRES PSF is determined empirically by fitting telluric lines in a
$\sim20$~\AA~spectral segment with a comb of Gaussians. Due to the
small number of telluric lines, pixel-to-pixel fluctuations have a
large effect on the measured PSF width. One could imagine implementing
a prior on the PSF width for the case of low SNR spectra. \SpecMatch
could likely maintain low statistical errors, even for low SNR spectra
at the expense of more uncertain \vsini values.

\begin{deluxetable}{lrrr}
  \tablecaption{Median photon-limited errors as a function of SNR}
  \tablewidth{0pt}
  \tabletypesize{\footnotesize}
  \tablehead{
    \colhead{} &
    \multicolumn{3}{c}{Median Scatter} \\ 
    \cline{2-4} \\[-2.0ex]
    \colhead{SNR} &
    \colhead{\teff (K)} & 
    \colhead{\logg (dex)} & 
    \colhead{\fe (dex)}
  }
  \startdata
  10  &       48 &    0.076 &  0.033 \\
  20  &       34 &    0.055 &  0.019 \\
  40  &       14 &    0.026 &  0.012 \\
  80  &        4 &    0.010 &  0.003 
  \enddata
  \label{tab:SNR}
\end{deluxetable}

\subsection{Systematic Errors}
\label{sec:touchstone}
We assessed these systematic errors by analyzing spectra of touchstone stars from \cite{Huber13}, \cite{Torres12}, \cite{Albrecht12}, and \cite{Valenti05}.

\subsubsection{Comparison with Huber et al. (2013)}
\begin{figure*}[h!]
\centering
\includegraphics[width=1\columnwidth]{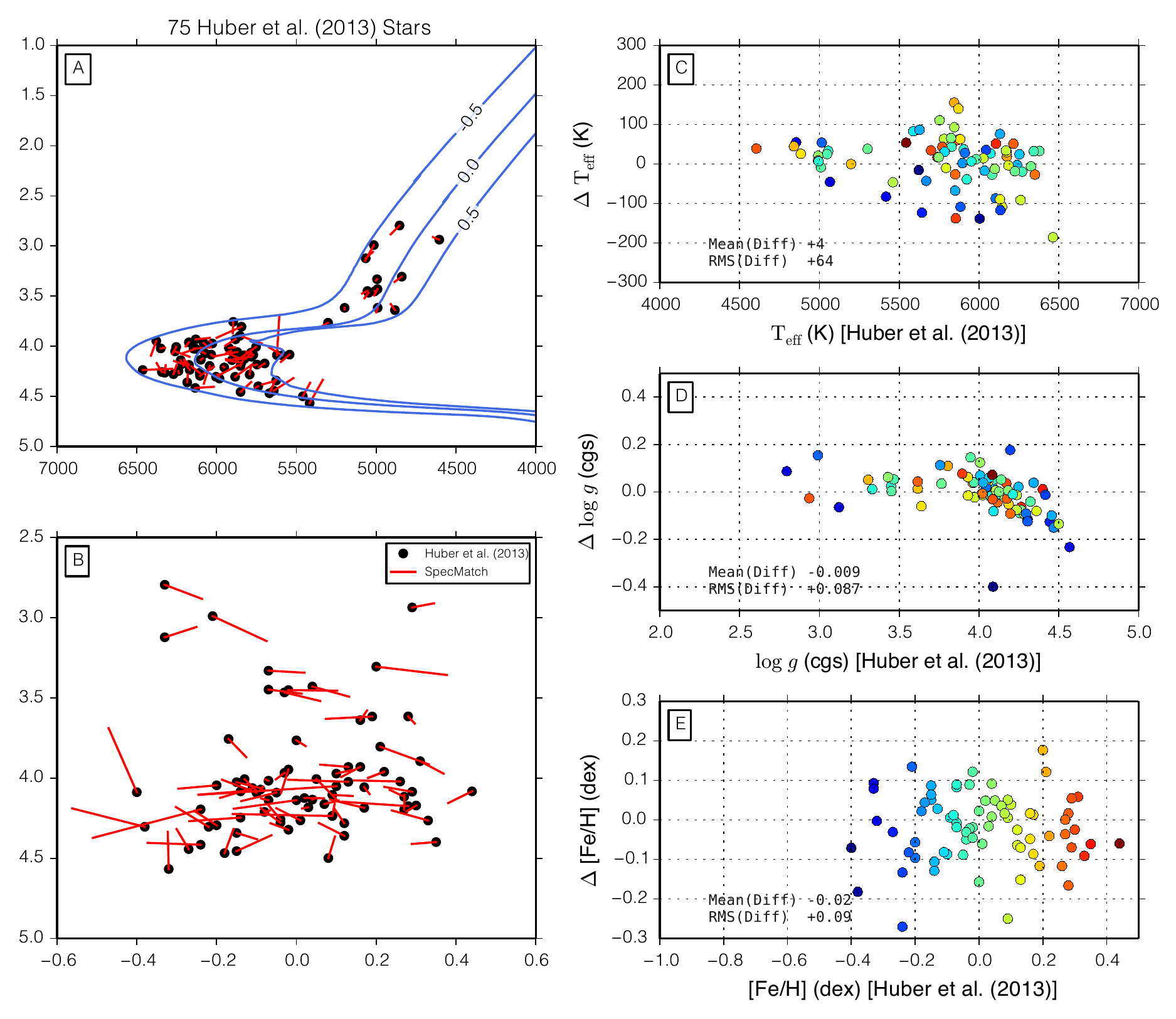}
\caption[Comparison to \cite{Huber13}]
  {Comparison of stellar parameters of \nhubercomp stars from
  the \cite{Huber13} asteroseismic analysis and \SpecMatch. Panel
  A---black points show \teff and \logg from \cite{Huber13} and red
  lines point to the \SpecMatch values. Shorter lines correspond to
  tighter agreement. We show several 5~Gyr Dartmouth isochrones
  \citep{Dotter08} having \fe of -0.5, 0.0, and 0.5~dex to indicate
  the range of \teff and \logg values allowed by stellar evolution
  models. Panel B---same as panel A, except showing \logg and
  \fe. Panel C---differences in \teff between \SpecMatch and
  \cite{Huber13}, i.e. $\Delta \teff =
  \teff(\rm{SM})-\teff(\rm{H13})$, as a function of \teff
  (H13). Points are colored according to \cite{Huber13} metallicity
  (see panel E for the mapping between color and metallicity). Panels
  D and E---same as panel C, except showing \logg and \fe,
  respectively. Dispersion (RMS) in $\Delta \teff$, $\Delta \logg$,
  $\Delta \fe$ is \rmsteffhuber~K, \rmslogghuber~(cgs),
  \rmsfehuber~(dex), respectively. If the largest outlier in $\Delta \logg$ is excluded the dispersion decreases to \rmslogghuberout~dex}
\label{fig:huber-fivepane}
\end{figure*}

As discussed in Section~\ref{ssec:hubercal}, asteroseismology offers
high precision surface gravity measurements that are largely independent of stellar spectra. We compared \SpecMatch
and H13 parameters for \nhubercomp stars and list these parameters in
Table~\ref{tab:huber}. We show the agreement between the two methods
graphically in Figure~\ref{fig:huber-fivepane}. Dispersion (RMS) about
the 1-to-1 line is \rmsteffhuber~K in \teff, \rmslogghuber~dex in
\logg, and \rmsfehuber~dex in \fe. If the star with the highest discrepancy in $\Delta \logg$ is excluded, the dispersion in \logg decreases to \rmslogghuberout~dex.

Assuming the H13 parameters represent the ground truth, this
dispersion is an upper limit to \SpecMatch errors. Solar-type
oscillations are detectable with \Kepler short cadence photometry for
stars solar-type and earlier, along with evolved stars. For regions of the HR diagram sampled by \cite{Huber13},%
\footnote{
\teff~=~5700--6500~\K; \logg~=~3.5--5.0 and \teff~=~4000--5700~\K; \logg~=~1.0--4.0}
we adopt $\siglogg~=~\uloggas~\mathrm{dex}$ as our uncertainty in surface gravity. Due to the nature of asteroseismic observations, we cannot use asteroseismology to assess the integrity of \SpecMatch surface gravities for main sequence stars earlier than $\sm$G2. However, because the agreement to \cite{Huber13} was 0.10~dex before any calibration, we adopt $\sigma(\logg)=0.10~\mathrm{dex}$ errors in surface gravity for stars with \teff~<~5700~\K.

\subsubsection{Comparison to Torres et al. (2012)}
An additional sample of comparison stars comes from the work of
\cite{Torres12} (T12, hereafter). T12 measured \teff, \logg, and \fe
for 56 stars with transiting planets. T12 measured these parameters
spectroscopically using three different techniques: SME \citep{Valenti96,Valenti05}; SPC \citep{Buchhave12nat}, and MOOG
\citep{Sneden73}. When making comparisons, we use the SME
parameters. Because these stars host transiting planets, mean stellar
density, \rhostar, may be extracted from the transit light
curve. Surface gravities with the additional constraint of \rhostar
from transit fitting may be more accurate than surface gravities based
on spectroscopy alone. However, as shown in Figure 8 in H13 and Figure
3 from \cite{Sliski14}, \rhostar from transit-fitting can disagree at
the >~50\%-level compared to \rhostar from asteroseismology. When
fitting a transit profile, strong degeneracies exist between impact
parameter and \rhostar, and H13 found that disagreements in \rhostar
were largest for high impact parameters.

\begin{figure*}[h!]
\centering
\includegraphics[width=1\columnwidth]{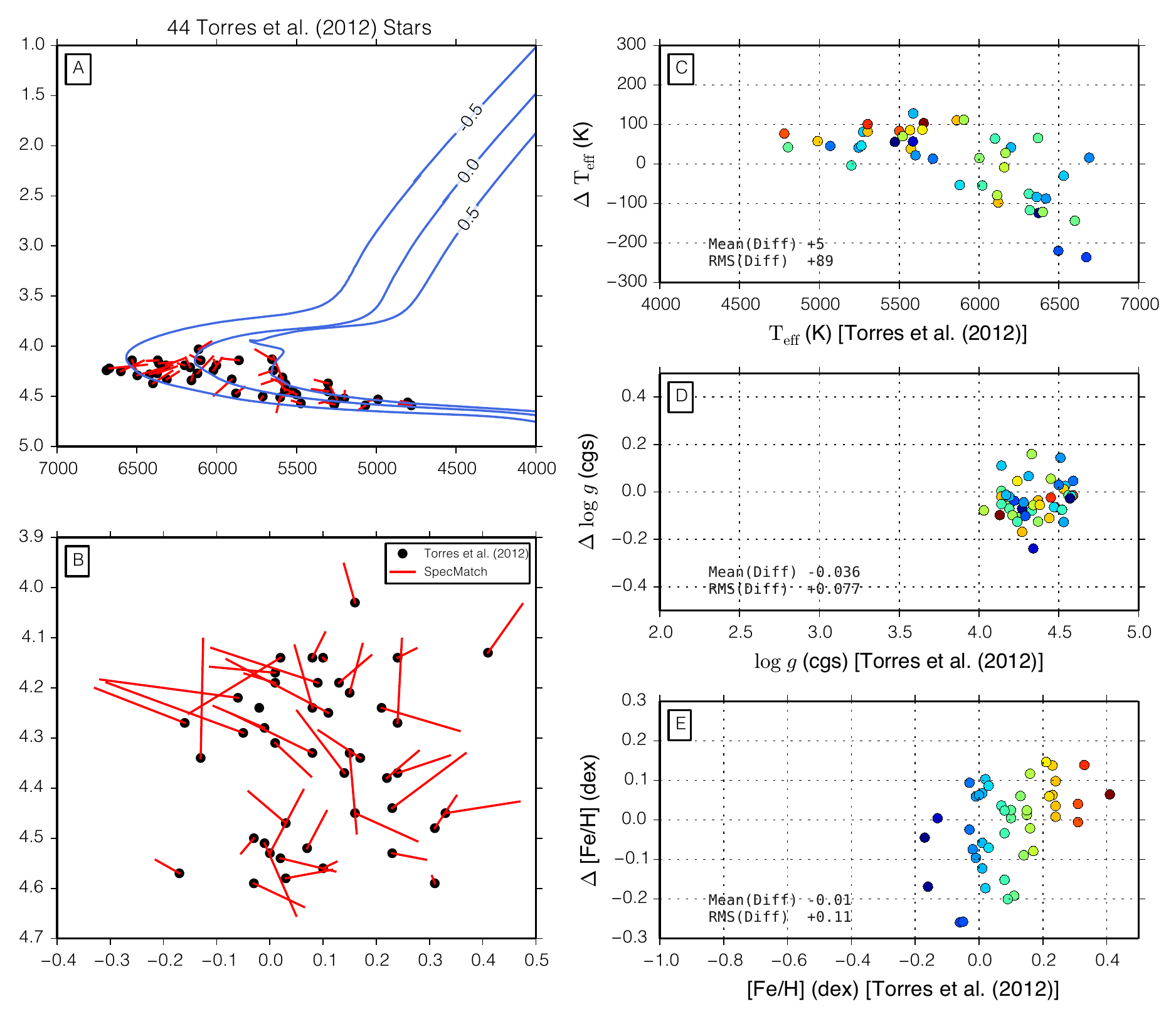}
\caption[Comparison to \cite{Torres12}]
 {Same as Figure~\ref{fig:huber-fivepane}, except comparing
  \SpecMatch parameters to the \cite{Torres12} sample. Dispersion in
  $\Delta \teff$, $\Delta \logg$, $\Delta \fe$ is \rmstefftorres~K,
  \rmsloggtorres~(cgs), \rmsfetorres~(dex), respectively.}
\label{fig:torres-fivepane}
\end{figure*}

Using \SpecMatch, we measured \teff, \logg, and \fe for \ntorrescomp
main-sequence stars from the T12 analysis having
\teff~=~4700--6700~K. We list the \SpecMatch and T12 parameters in
Table~\ref{tab:torres}. We show a graphical comparison in
Figure~\ref{fig:torres-fivepane}.  Dispersion about the 1-to-1 line is
\rmstefftorres~K in \teff, \rmsloggtorres~dex in \logg, and
\rmsfetorres~dex in \fe. The fact that surface gravities agree to \rmsloggtorres~dex, even when main sequence stars with \teff~<~5700~\K are included, supports our adoption of \siglogg~=~\uloggms for stars cooler than 5700~K.

We note a systematic trend in \fe, where above 0.2~dex, \SpecMatch
produces iron values that are higher compared to T12. Also, we note a
systematic trend in the \teff, below $\sim$6000 K, \SpecMatch runs hot
compared to T12, while above $\sim$6000~K, \SpecMatch runs cool. However, we are note able to tell whether these systematics are due to errors in the C05 models, the \SpecMatch fitting procedure, or the joint SME/\rhostar analysis of \cite{Torres12}.

\subsubsection{Comparison to Albrecht et al. (2012)}
\label{sec:albrecht}
Stellar rotation broadens stellar lines by an amount equal to twice
the projected rotational velocity of the star, or \vsini. Although
often treated as a nuisance parameter, \vsini contains information
about a star's spin axis (when combined with $v_{eq}$) and a star's
suitability to precise RV followup. Measuring accurate \vsini when
\vsini is small (< 5 km s$^{-1}$) is notoriously difficult because
rotational broadening competes with other broadening mechanisms like
microturbulence, macroturbulence, and the instrumental profile of the
spectrometer.

We adopt stars from \cite{Albrecht12} (A12 hereafter) as touchstones
to assess the accuracy of \SpecMatch-based \vsini. Stars with RM
measurements offer a robust calibration sample for
\vsini, given that the amplitude of the anomalous Doppler shift during
transit depends on \vsini. A12 compiled RM measurements from 53 stars,
from which we select 43 stars with \vsini uncertainties of less than
2~\kms as a calibration sample for our \vsini values. We list the
decker, measured instrumental profile width, and \vsini measurements
from \SpecMatch A12 \vsini for each of these stars in
Table~\ref{tab:albrecht}.

\begin{figure}
\centering
\includegraphics[width=0.7\columnwidth]{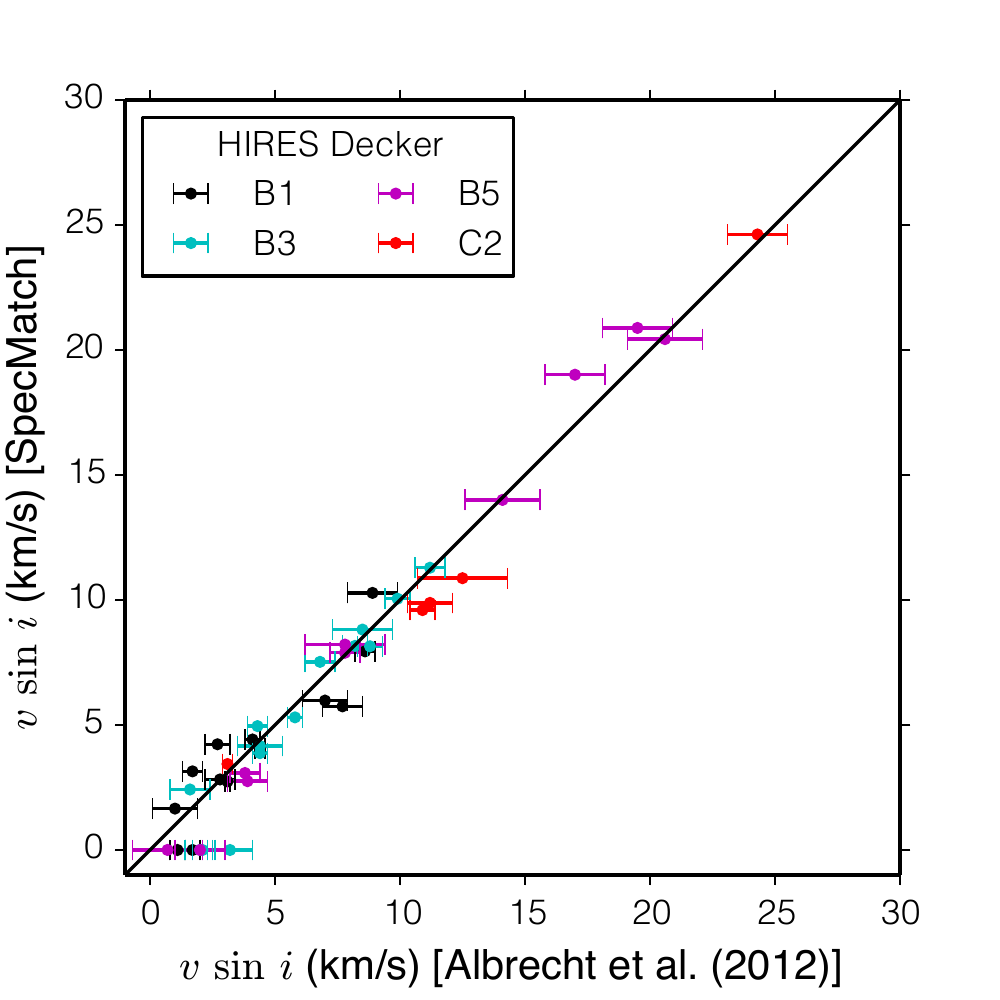}
\caption[Comparison to \cite{Albrecht12}]
{\SpecMatch \vsini versus \cite{Albrecht12} \vsini for stars
  where the uncertainty on the \cite{Albrecht12} \vsini (shown as
  horizontal bars) was less than 2~\kms.}
\label{fig:albrecht-vsini}
\end{figure}

Figure~\ref{fig:albrecht-vsini} compares \SpecMatch \vsini with the
RM-based \vsini from A12. There is good \vsini agreement down to
\vsini of \sm1~\kms. Below 2~\kms, \vsini is so small compared to other
broadening terms that we do not consider \SpecMatch \vsini values
reliable. \SpecMatch yeilds $\sigma(\vsini)~=~1~\kms$ down to \vsini~=~2~\kms.
For lower rotational velocity, we adopt an upper
limit on stellar \vsini at 2~\kms.

\subsubsection{Comparison to Valenti and Fischer (2005)}
\begin{figure*}[h!]
\centering
\includegraphics[width=1\columnwidth]{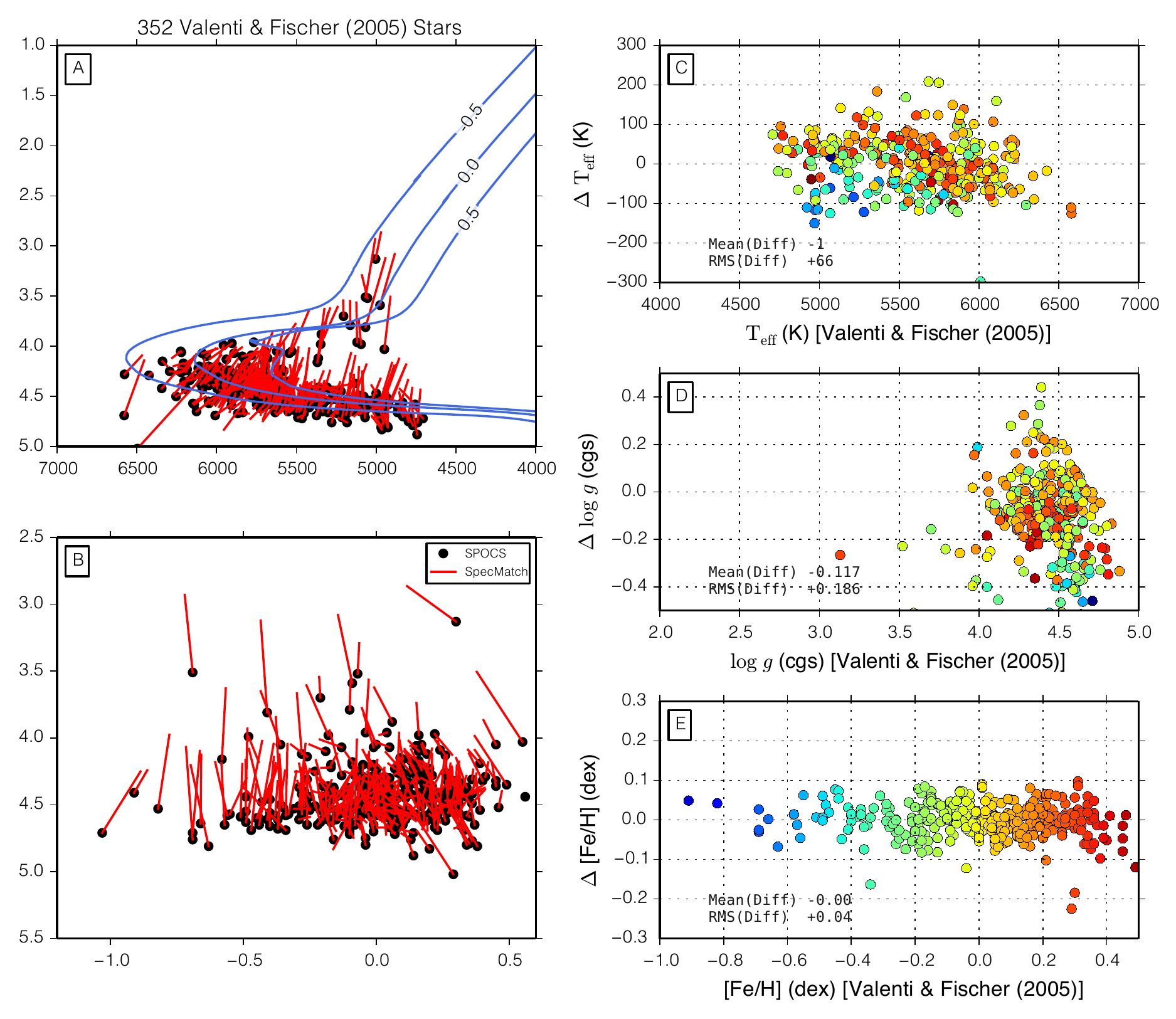}
\caption[Comparison to \cite{Valenti05}]
  {Same as Figure~\ref{fig:huber-fivepane}, but showing
  comparison between \SpecMatch and SPOCS
  \citep{Valenti05}. Dispersion in $\Delta \teff$, $\Delta \logg$,
  $\Delta \fe$ is \rmsteffspocs~K, \rmsloggspocs~(cgs),
  \rmsfespocs~(dex), respectively.}
\label{fig:spocs-fivepane}
\end{figure*}

\begin{figure*}[h!]
\centering
\includegraphics[width=1.0\columnwidth]{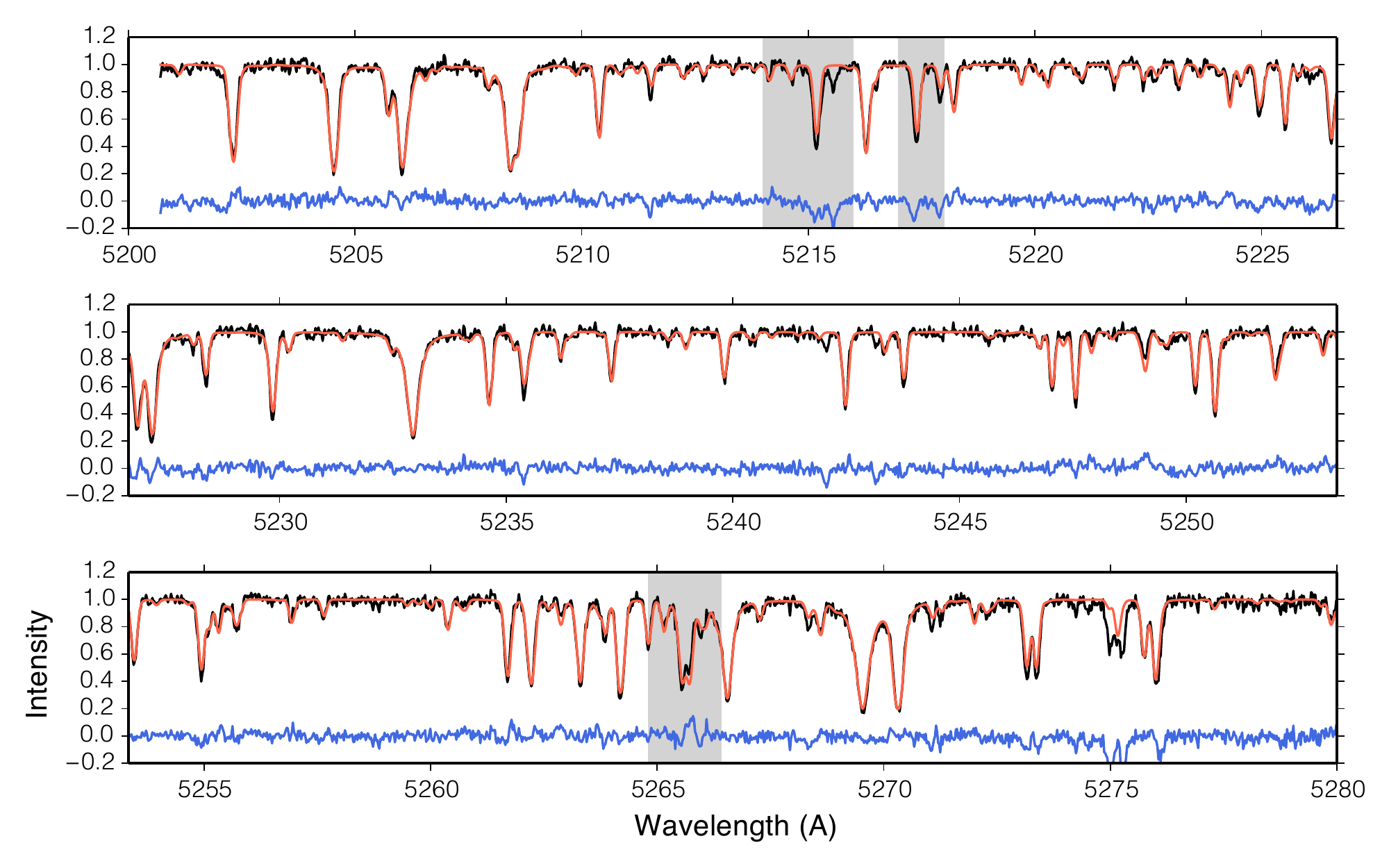}
\caption[Spectrum of KOI-1]
  {Spectrum of KOI-1 (black) with best fit model spectrum (red)
  and residuals (blue). The wavelength region 5200--5280~\AA\ is
  broken into three segments. The gray regions are excluded from
  $\chi^{2}$. Line depths and line widths generally well-matched, with
  a few exceptions. For example at 5275~\AA, several lines seem to be
  missing from the model. \cite{Huber13} determined the properties of
  KOI-1 and found that it was a close solar analog: \teff~=~5850~K,
  \logg~=~4.46 (cgs), and \fe~=~$-0.15$~dex.}
\label{fig:CK00001_5200}
\end{figure*}

\begin{figure*}[h!]
\centering
\includegraphics[width=1.0\columnwidth]{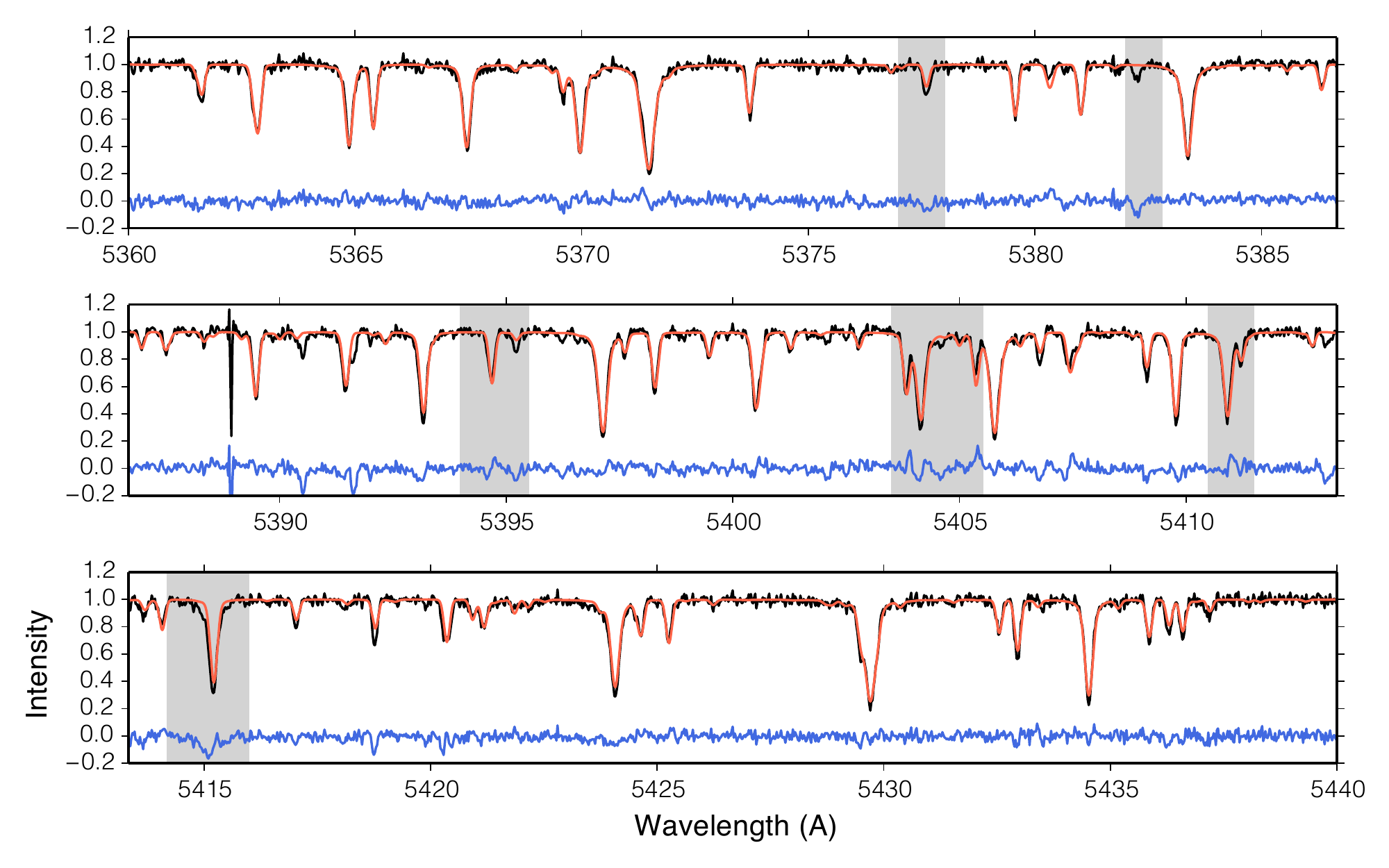}
\caption[Spectrum of KOI-1]{Same as Figure~\ref{fig:CK00001_5200} except for wavelength
  region beginning at 5360~\AA.}
\label{fig:CK00001_5360}
\end{figure*}

\begin{figure*}[h!]
\centering
\includegraphics[width=1.0\columnwidth]{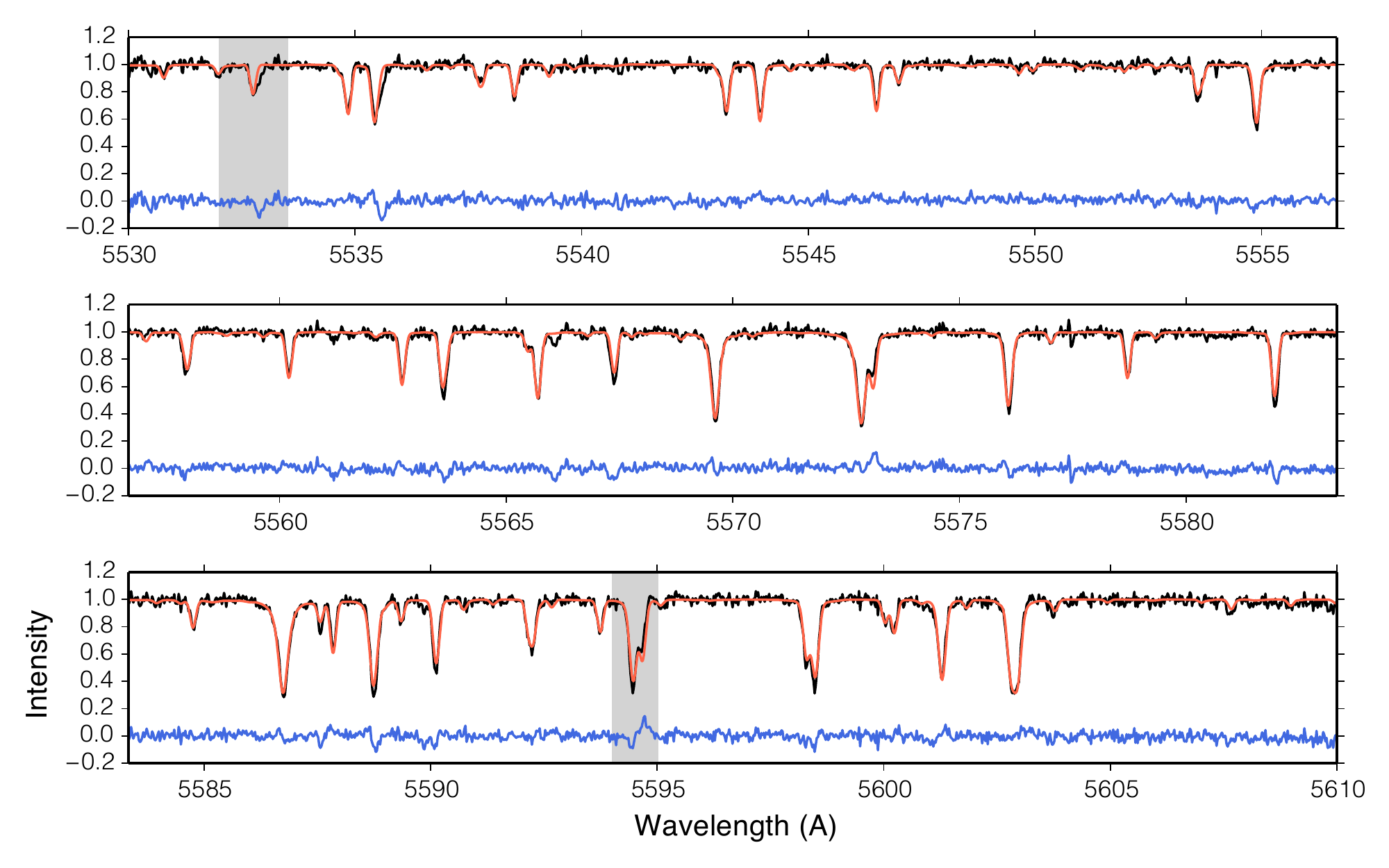}
\caption[Spectrum of KOI-1]{Same as Figure~\ref{fig:CK00001_5200} except for wavelength
  region beginning at 5530~\AA.}
\label{fig:CK00001_5530}
\end{figure*}

\begin{figure*}[h!]
\centering
\includegraphics[width=1.0\columnwidth]{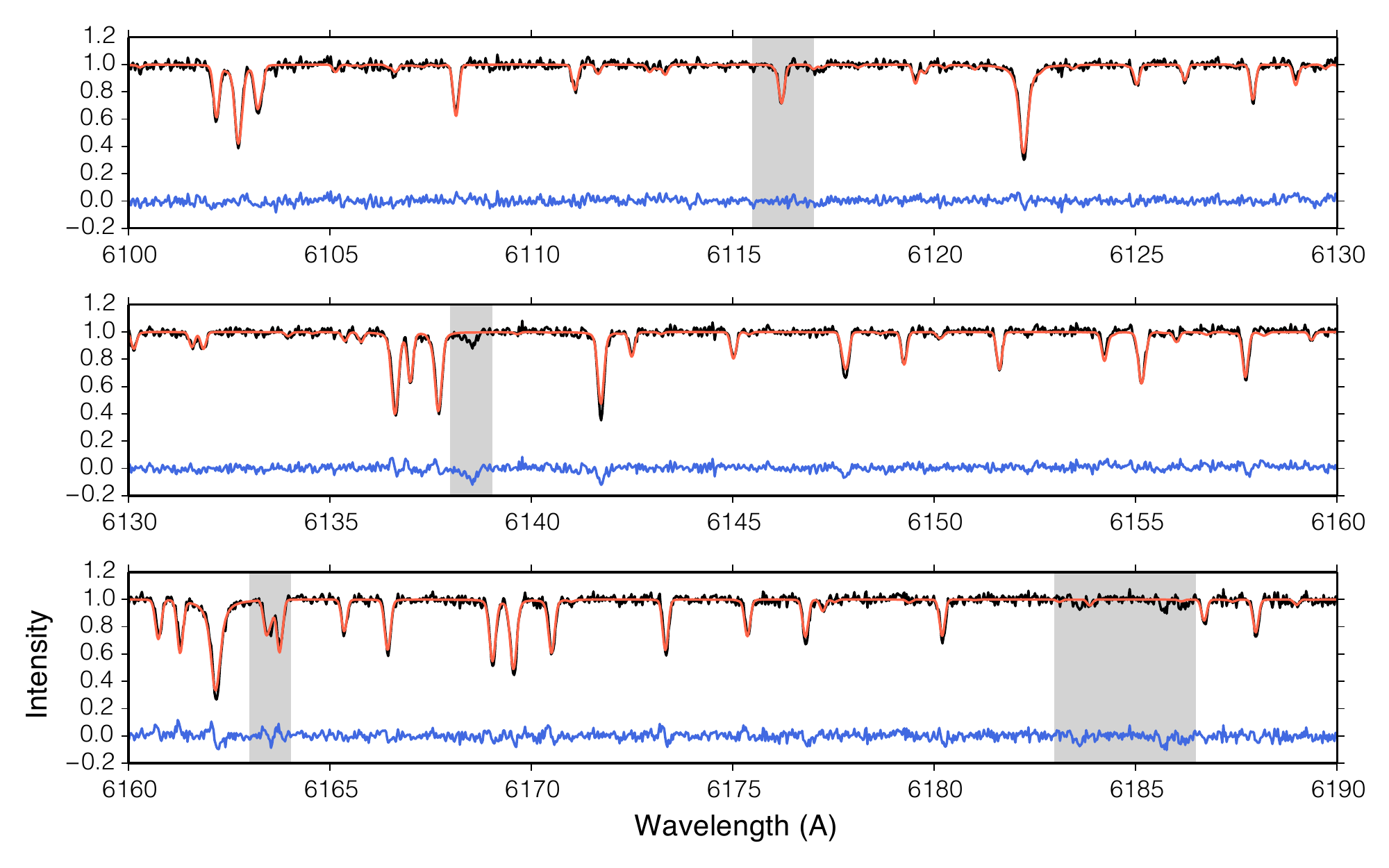}
\caption[Spectrum of KOI-1]{Same as Figure~\ref{fig:CK00001_5200} except for wavelength
  region beginning at 6100~\AA.}
\label{fig:CK00001_6100}
\end{figure*}

\begin{figure*}[h!]
\centering
\includegraphics[width=1.0\columnwidth]{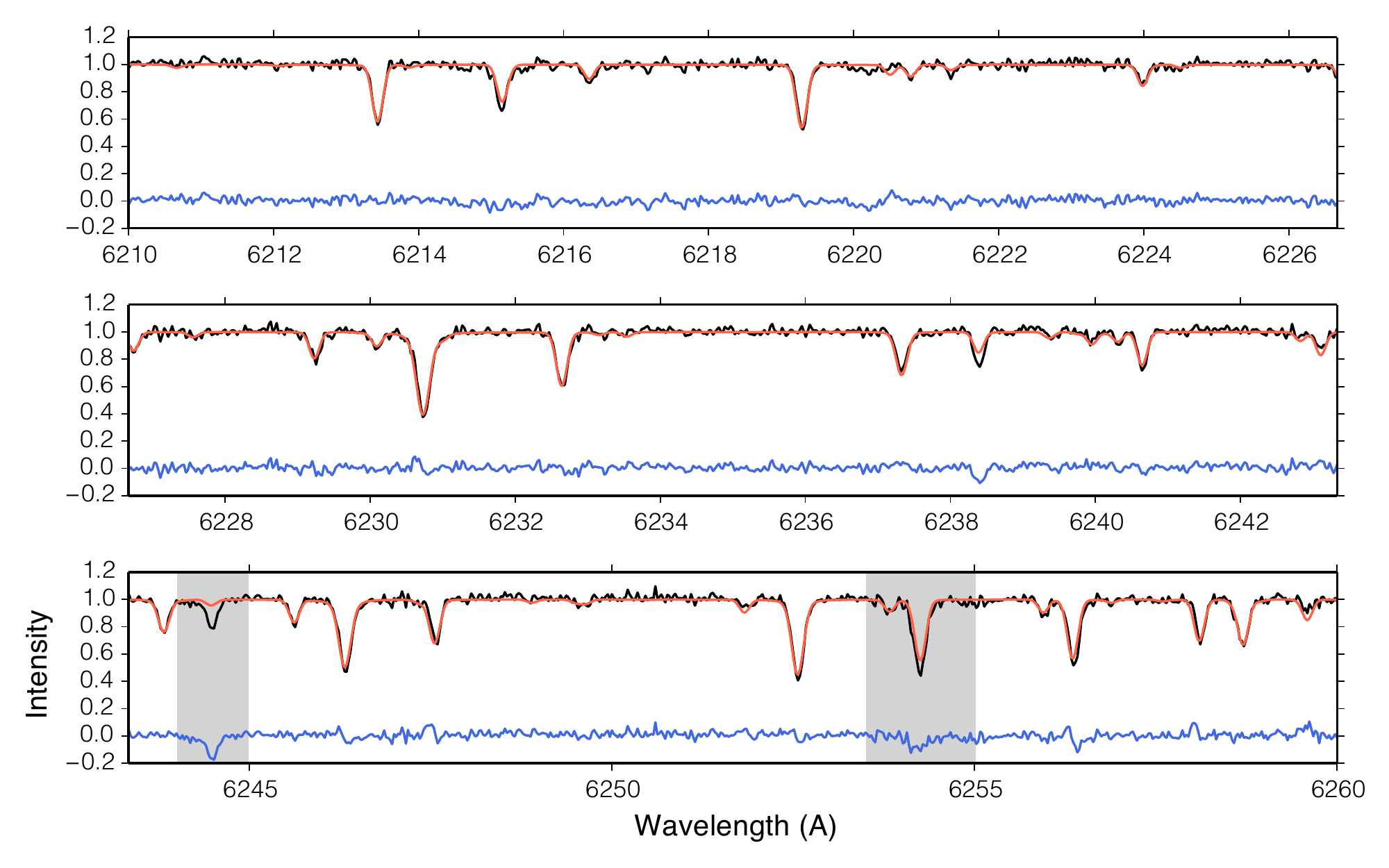}
\caption[Spectrum of KOI-1]{Same as Figure~\ref{fig:CK00001_5200} except for wavelength
  region beginning at 6210~\AA.}
\label{fig:CK00001_6210}
\end{figure*}

As discussed in Section~\ref{ssec:spocscal}, we ran \SpecMatch on
\nspocscomp stars from the VF05 catalog. We list the VF05 and
\SpecMatch parameters in Table~\ref{tab:spocs}, and compare the two
parameters graphically in Figure~\ref{fig:spocs-fivepane}.  VF05 list
two \logg values: a ``spectroscopic \logg,'' based solely on the SME
analysis and a ``isochrone \logg,'' which incorporates constraints on
\logg based on \Rstar (determined from \teff and absolute magnitudes)
and Yonsei-Yale isochrones. We choose to compare our \logg values to
the ``spectroscopic \logg'' because \SpecMatch does not impose any
constraints based on isochrones. Dispersion about the 1-to-1 line is
\rmsteffspocs~K in \teff, \rmsloggspocs~dex in \logg, and
\rmsfespocs~dex in \fe (recall that \teff and \fe have been calibrated
to the VF05 scale).

We use the scatter in the differences between \SpecMatch and VF05 effective temperatures and metallicities to assess our systematic uncertainties in these parameters. If systematics effects are independent during the \SpecMatch and VF05 fitting, the scatter in $\Delta \teff$ and $\Delta \fe$ represent systematic errors in \teff and \logg for both analyses, added in quadrature. We adopt $\sigma(\teff)~=~\uteff~\K$ and $\sigma(\fe)~=~\ufe$ which makes the conservative assumption that the VF parameters contribute negligibly to the errors. Zero-point offsets remain a concern given that both techniques relied on optical spectra and model atmospheres with similar provenience. SME used \cite{Kurucz92} atmospheres and \SpecMatch uses \cite{Castelli03} atmospheres. However, \SpecMatch and SME used different line lists and spectral regions.

The differences in \logg between \SpecMatch and VF05 were the highest
of any of the comparison samples. That we see lower dispersion and no
systematic offset comparing \SpecMatch to H13 and T12, which
incorporate additional constraints on \logg from asteroseismolgy and
transit light curves suggests lower \logg precision in VF05. In the
SME analysis of VF05, the measured \logg was largely driven by the
model fit to the wings of the Mg~I~b lines. Modeling these lines requires
careful treatment of the continuum and the effects of pressure
broadening, which could be hampering the \logg precision of SME. Recall that \SpecMatch \teff and \logg have been placed on the SPOCS scale (see Section~\ref{ssec:spocscal}) but that for most stars the correction is smaller than \uteff~K and \ufe~dex.

\subsection{Summary}
We have assessed systematic errors present in \SpecMatch parameters through comparisons to well-characterized stars from the literature. We summarize the these errors in Table~\ref{tab:sm-sys-err} for different domains of \teff, \logg, and \fe.

\begin{deluxetable}{@{\extracolsep{4pt}}l*{8}{r}}
  \tablecaption{\SpecMatch Systematic Uncertainties}
  \tablewidth{0pt}
  \tabletypesize{\footnotesize}
  \tablehead{

  \multicolumn{3}{c}{Region} & 
  \colhead{} & 

  \multicolumn{4}{c}{Systematic Uncertainty} \\
  \cline{1-3}\cline{5-8}\\[-1.5ex]

  \colhead{\teff} & 
  \colhead{\logg} & 
  \colhead{\fe} &
  \colhead{} & 
  \colhead{\teff} & 
  \colhead{\logg} & 
  \colhead{\fe} & 
  \colhead{\vsini} \\[0.5ex]
  \colhead{(K)} & 
  \colhead{(dex)} & 
  \colhead{(dex)} &
  \colhead{} & 
  \colhead{(K)} & 
  \colhead{(dex)} & 
  \colhead{(dex)} & 
  \colhead{(\kms)} \\[-2.5ex]

  }
  \startdata
  5700--6500 & 3.5--5.0 & $-1.0$--0.5 & & \uteff & \uloggas & \ufe & \uvsini \\
  4700--5700 & 1.0--4.0 & $-1.0$--0.5 & & \uteff & \uloggas & \ufe & \uvsini \\
  4700--5700 & 4.0--5.0 & $-1.0$--0.5 & & \uteff & \uloggms & \ufe & \uvsini
  \enddata
  \tablecomments{We have not assessed the reliability of \SpecMatch outside of the tabulated ranges of \teff, \logg, and \fe due to a lack of literature stars. Parameters outside of these ranges are uncertain by unknown amounts. When \SpecMatch returns \vsini~<~2~\kms, we treat the measurement as  an upper limit of 2~\kms.}\label{tab:sm-sys-err}
\end{deluxetable}

\section{Conclusions}
\label{sec:sm-conclusion}
We developed new tool, \SpecMatch, that can reliably extract stellar parameters from high-resolution optical spectra. We have assessed the systematic errors associated with \SpecMatch parameters by analyzing high SNR spectra of well-characterized touchstone stars from \cite{Huber13}, \cite{Torres12}, \cite{Albrecht12}, and \cite{Valenti05}. These errors are summarized in Table~\ref{tab:sm-sys-err}. The fact that \SpecMatch uses a robust model template fitting procedure involving a large spectral region (380~\AA) results in reliable parameters even for low SNR spectra. As we showed in Section~\ref{ssec:photonlimited}, photon-limited uncertainties become comparable to systematic uncertainties at SNR/pixel~=~10. Such a spectrum requires $100\times$ less integration time than the SNR/pixel~=~100 spectra often used for stellar parametrization. This corresponds to a reduction from 2.5~hours to 1.5~minutes of integration time for a $V$~=~14.7 star, typical of \Kepler planet hosts, observed with HIRES on Keck I. 

\SpecMatch is a powerful tool for the characterization of large samples of planet-bearing stars from \Kepler. Here, we elaborate on some of the science applications for \SpecMatch as applied to \Kepler planet-hosting stars. \SpecMatch can be used to improve the radii of large samples of \Kepler planets.  With \Kepler, the planet to star radius ratio, \Rp/\Rstar, which depends on the transit depth to first order, is typically measured with high precision.%
\footnote{The median $\sigma(\Rp/\Rstar)$ in the \cite{Batalha13} KOI catalog list is 5.4\%.}
Stellar radii based on photometry in the Kepler Input Catalog \citep{Brown11} are uncertain at the $\sim$35\% level \citep{Batalha13,Burke14}. Thus, our knowledge of the sizes of \Kepler planets is limited by stellar radius uncertainties. One of the key results of the \Kepler mission is the distribution of planet sizes \citep{Howard12,Fressin13,Petigura13} which provides important constraints for planet formation models. If the underlying radius distribution contains sharp features that would indicate an important size scale for planet formation are blurred by photometric radius errors. Combining \SpecMatch-derived parameters with Dartmouth isochrones \citep{Dotter08} yields stellar radii good to $\approx5\%$, for solar analog stars. By improving stellar radii, \SpecMatch will reveal smaller details in the planet radius distribution.

\SpecMatch can help probe the connection planet formation and the composition of protoplanetary disks. Among nearby stars, there is a well-established correlation between giant planet occurrence and host star metallicity \citep{Gonzalez97,Santos04,Fischer05}. Higher metallicity stars are thought to have more massive protoplanetary disks which appear to form giant planets more efficiently. However, planets the size of Neptune and smaller are found around stars with a wide range of metallicities, both in the solar neighborhood \citep{Mayor11} and in the \Kepler field \citep{Buchhave12nat}.

Finally, for transiting planets detected by \Kepler, \vsini can be combined with the stars equatorial velocity (derived from photometric rotational modulation) to derive the inclination of the star's spin axis with respect to the Earth. Transiting planets have orbital inclinations very close to 90\deg. The 1~\kms precision of \SpecMatch \vsini measurements, can probe star-planet spin-orbit misalignment for planets that are too small and stars that are too faint for current Rossiter-McLaughlin techniques.

\section*{Acknowledgements}
We thank Andrew Mann, Kelsey Clubb, Tabetha Boyajian, John
Brewer, Daniel Huber, and Marc Pinsonneault for enlightening
conversations that improved this manuscript. We thank the many
observers who took and reduced the spectra used in this work
including Geoffrey Marcy, Andrew Howard, John Johnson, Howard
Isaacson, Lauren Weiss, Lea Hirsch, Benjamin Fulton, and Evan
Sinukoff. We wish to acknowledge the staff of the Keck
Observatory for their outstanding support of the Keck Telescope
and HIRES spectrometer. We extend special thanks to those of
Hawai'ian ancestry on whose sacred mountain of Mauna Kea we are
privileged to be guests.  Without their generous hospitality, the
Keck observations presented herein would not have been
possible. This research used resources of the National Energy
Research Scientific Computing Center, which is supported by the
Office of Science of the U.S. Department of Energy under Contract
No. DE-AC02-05CH11231. This work made use of NASA's Astrophysics
Data System Bibliographic Services as well as the {\tt NumPy}
\citep{Oliphant07}, {\tt SciPy} \citep{scipy}, {\tt h5py}
\citep{Collette08}, {\tt IPython} \citep{ipython}, and {\tt
Matplotlib} \citep{matplotlib} Python modules.

\clearpage

%\acknowledgements 

%\bibliographystyle{apj}
%\bibliography{smcat}

\section{Appendix}

\clearpage
\begin{deluxetable}{@{\extracolsep{2pt}}lrrrrrrrrrr}
  \tabletypesize{\footnotesize}
  \tablecaption{\SpecMatch / Huber et al. (2013) Comparison}
  \tablewidth{0pt}
  \ttablehead{H13}
  \startdata
  \input{sm/table_huber.tex}
\enddata
  \tablecomments{}
  \label{tab:huber}
\end{deluxetable}

\clearpage
\begin{deluxetable}{@{\extracolsep{2pt}}lrrrrrrrrrr}
  \tabletypesize{\footnotesize}
  \tablecaption{\SpecMatch / Torres et al. (2012) Comparison}
  \tablewidth{0pt}
  \ttablehead{T12}
  \startdata
  \input{sm/table_torres.tex}
\enddata
  \tablecomments{For WASP-8, \vsini was misprinted as $20\pm0.6$~km
    s$^{-1}$, instead of $2.0\pm0.6$~km s$^{-1}$ in the original
    reference (Queloz 2010).}
 \label{tab:torres}
\end{deluxetable}

\clearpage
\begin{deluxetable}{@{\extracolsep{2pt}}lrrrrrrrrrr}
  \tabletypesize{\footnotesize}
  \tablecaption{\SpecMatch / \cite{Valenti05} Comparison}
  \tablewidth{0pt}
  \ttablehead{VF05}
  \startdata
  \input{sm/table_spocs.tex}
\enddata
  \tablecomments{}
 \label{tab:spocs}
\end{deluxetable}

\clearpage
\begin{deluxetable}{@{\extracolsep{2pt}}llllll}
  \tabletypesize{\footnotesize}
  \tablecaption{\SpecMatch / \cite{Albrecht12} Comparison}
  \tablewidth{0pt}
  \tablehead{
    \colhead{Name} &	
    \colhead{Decker} &
    \colhead{$\sigma_{IP}$} &
    \multicolumn{2}{c}{\vsini} \\[0.5ex]
    \cline{4-5} \\[-1.5ex]
    \multicolumn{3}{c}{} &
    \colhead{SM}&
    \colhead{A12}
  }
  \startdata
  \input{sm/table_albrecht.tex}
\enddata
  \tablecomments{For WASP-8, \vsini was misprinted as $20\pm0.6$~\kms, instead of $2.0\pm0.6$~\kms in the original reference (Queloz 2010).}
  \label{tab:albrecht}
\end{deluxetable}

%% file: sm/table_huber.tex
 CK00001 &    5783 &     5850 &       -66 &    4.36 &     4.46 &     -0.10 & -0.06 &  -0.15 &    0.09 &      0.0 \\
 CK00002 &    6323 &     6350 &       -26 &    4.01 &     4.02 &     -0.01 &  0.14 &   0.26 &   -0.12 &      4.4 \\
 CK00005 &    5864 &     5753 &       111 &    4.13 &     4.01 &      0.12 &  0.10 &   0.05 &    0.05 &      8.5 \\
 CK00007 &    5844 &     5781 &        63 &    4.11 &     4.10 &      0.01 &  0.14 &   0.09 &    0.05 &      1.9 \\
 CK00041 &    5890 &     5825 &        65 &    4.13 &     4.12 &      0.00 &  0.07 &   0.02 &    0.05 &      3.4 \\
 CK00042 &    6319 &     6325 &        -5 &    4.19 &     4.26 &     -0.08 & -0.05 &   0.01 &   -0.06 &     12.9 \\
 CK00064 &    5340 &     5302 &        38 &    3.80 &     3.77 &      0.04 &  0.02 &  -0.00 &    0.02 &      0.9 \\
 CK00069 &    5626 &     5669 &       -42 &    4.32 &     4.47 &     -0.15 & -0.16 &  -0.18 &    0.02 &      0.0 \\

%% file: sm/table_torres.tex
   CoRoT-2 &    5624 &     5602 &        22 &    4.65 &     4.51 &      0.14 &  0.05 &  -0.01 &    0.06 &      9.6 \\
   CoRoT-7 &    5355 &     5274 &        81 &    4.57 &     4.54 &      0.03 &  0.12 &   0.02 &    0.10 &      0.0 \\
  HAT-P-11 &    4857 &     4780 &        77 &    4.58 &     4.59 &     -0.01 &  0.30 &   0.31 &   -0.01 &      1.4 \\
  HAT-P-13 &    5756 &     5653 &       103 &    4.03 &     4.13 &     -0.10 &  0.47 &   0.41 &    0.06 &      3.2 \\
  HAT-P-14 &    6456 &     6600 &      -143 &    4.14 &     4.25 &     -0.11 & -0.08 &   0.11 &   -0.19 &      8.3 \\
  HAT-P-15 &    5654 &     5568 &        86 &    4.33 &     4.38 &     -0.05 &  0.28 &   0.22 &    0.06 &      0.0 \\
  HAT-P-16 &    6149 &     6158 &        -8 &    4.28 &     4.34 &     -0.06 &  0.09 &   0.17 &   -0.08 &      3.1 \\
  HAT-P-17 &    5287 &     5246 &        41 &    4.40 &     4.53 &     -0.13 &  0.06 &   0.00 &    0.06 &      0.0 \\

%% file: sm/table_spocs.tex
  100180 &    6024 &     5989 &        35 &    4.38 &     4.38 &      0.00 &  0.01 &  -0.01 &    0.02 &   2.5 \\
  100623 &    5131 &     5189 &       -57 &    4.37 &     4.68 &     -0.31 & -0.36 &  -0.37 &    0.01 &   0.3 \\
  101259 &    5006 &     5067 &       -60 &    2.93 &     3.51 &     -0.58 & -0.72 &  -0.69 &   -0.03 &   0.8 \\
   10145 &    5673 &     5638 &        35 &    4.34 &     4.44 &     -0.10 &  0.01 &  -0.01 &    0.02 &   1.2 \\
  101472 &    6211 &     6191 &        20 &    4.45 &     4.43 &      0.02 & -0.06 &  -0.03 &   -0.03 &   5.0 \\
  101959 &    6049 &     6049 &         0 &    4.39 &     4.39 &     -0.00 & -0.11 &  -0.11 &    0.00 &   3.0 \\
  102071 &    5392 &     5279 &       113 &    4.47 &     4.53 &     -0.06 &  0.09 &   0.01 &    0.08 &   0.0 \\
  102365 &    5633 &     5630 &         3 &    4.25 &     4.57 &     -0.32 & -0.30 &  -0.33 &    0.03 &   0.0 \\

%% file: sm/table_albrecht.tex
   HAT-P-1 &       B5 &  1.69 &       3.8 &      3.2 \\
  HAT-P-11 &       B1 &  1.40 &       1.0 &      1.8 \\
  HAT-P-13 &       B1 &  1.40 &       1.7 &      3.4 \\
  HAT-P-14 &       B3 &  1.36 &       8.2 &      8.3 \\
  HAT-P-16 &       B5 &  1.87 &       3.9 &      3.0 \\
   HAT-P-2 &       B5 &  1.71 &      19.5 &     21.1 \\
  HAT-P-23 &       B5 &  1.67 &       7.8 &      8.3 \\
  HAT-P-24 &       C2 &  1.42 &      11.2 &     10.0 \\

%% file: thesis.bbl
\begin{thebibliography}{170}
\expandafter\ifx\csname natexlab\endcsname\relax\def\natexlab#1{#1}\fi

\bibitem[{{Adams} {et~al.}(2008){Adams}, {Seager}, \&
  {Elkins-Tanton}}]{Adams08}
{Adams}, E.~R., {Seager}, S., \& {Elkins-Tanton}, L. 2008,
  \href{http://dx.doi.org/10.1086/524925}{\apj, 673, 1160}

\bibitem[{{Akeson} {et~al.}(2013){Akeson}, {Chen}, {Ciardi}, {Crane}, {Good},
  {Harbut}, {Jackson}, {Kane}, {Laity}, {Leifer}, {Lynn}, {McElroy}, {Papin},
  {Plavchan}, {Ram{\'{\i}}rez}, {Rey}, {von Braun}, {Wittman}, {Abajian},
  {Ali}, {Beichman}, {Beekley}, {Berriman}, {Berukoff}, {Bryden}, {Chan},
  {Groom}, {Lau}, {Payne}, {Regelson}, {Saucedo}, {Schmitz}, {Stauffer},
  {Wyatt}, \& {Zhang}}]{Akeson13}
{Akeson}, R.~L., {Chen}, X., {Ciardi}, D., {et~al.} 2013,
  \href{http://dx.doi.org/10.1086/672273}{\pasp, 125, 989}

\bibitem[{{Albrecht} {et~al.}(2012){Albrecht}, {Winn}, {Johnson}, {Howard},
  {Marcy}, {Butler}, {Arriagada}, {Crane}, {Shectman}, {Thompson}, {Hirano},
  {Bakos}, \& {Hartman}}]{Albrecht12}
{Albrecht}, S., {Winn}, J.~N., {Johnson}, J.~A., {et~al.} 2012,
  \href{http://dx.doi.org/10.1088/0004-637X/757/1/18}{\apj, 757, 18}

\bibitem[{{Allende Prieto} {et~al.}(2001){Allende Prieto}, {Lambert}, \&
  {Asplund}}]{Prieto01}
{Allende Prieto}, C., {Lambert}, D.~L., \& {Asplund}, M. 2001,
  \href{http://dx.doi.org/10.1086/322874}{\apjl, 556, L63}

\bibitem[{{Alonso} {et~al.}(2004){Alonso}, {Brown}, {Torres}, {Latham},
  {Sozzetti}, {Mandushev}, {Belmonte}, {Charbonneau}, {Deeg}, {Dunham},
  {O'Donovan}, \& {Stefanik}}]{Alonso04}
{Alonso}, R., {Brown}, T.~M., {Torres}, G., {et~al.} 2004,
  \href{http://dx.doi.org/10.1086/425256}{\apjl, 613, L153}

\bibitem[{{Anstee} \& {O'Mara}(1995)}]{Anstee95}
{Anstee}, S.~D., \& {O'Mara}, B.~J. 1995, \mnras, 276, 859

\bibitem[{{Bakos} {et~al.}(2007){Bakos}, {Noyes}, {Kov{\'a}cs}, {Latham},
  {Sasselov}, {Torres}, {Fischer}, {Stefanik}, {Sato}, {Johnson}, {P{\'a}l},
  {Marcy}, {Butler}, {Esquerdo}, {Stanek}, {L{\'a}z{\'a}r}, {Papp}, {S{\'a}ri},
  \& {Sip{\H o}cz}}]{Bakos07}
{Bakos}, G.~{\'A}., {Noyes}, R.~W., {Kov{\'a}cs}, G., {et~al.} 2007,
  \href{http://dx.doi.org/10.1086/509874}{\apj, 656, 552}

\bibitem[{{Barbuy} {et~al.}(2003){Barbuy}, {Perrin}, {Katz}, {Coelho},
  {Cayrel}, {Spite}, \& {Van't Veer-Menneret}}]{Barbuy03}
{Barbuy}, B., {Perrin}, M.-N., {Katz}, D., {et~al.} 2003,
  \href{http://dx.doi.org/10.1051/0004-6361:20030496}{\aap, 404, 661}

\bibitem[{{Barklem} \& {O'Mara}(1997)}]{Barklem97}
{Barklem}, P.~S., \& {O'Mara}, B.~J. 1997, \mnras, 290, 102

\bibitem[{{Barklem} {et~al.}(1998){Barklem}, {O'Mara}, \& {Ross}}]{Barklem98}
{Barklem}, P.~S., {O'Mara}, B.~J., \& {Ross}, J.~E. 1998,
  \href{http://dx.doi.org/10.1046/j.1365-8711.1998.01484.x}{\mnras, 296, 1057}

\bibitem[{{Barklem} {et~al.}(2000){Barklem}, {Piskunov}, \&
  {O'Mara}}]{Barklem00}
{Barklem}, P.~S., {Piskunov}, N., \& {O'Mara}, B.~J. 2000,
  \href{http://dx.doi.org/10.1051/aas:2000167}{\aaps, 142, 467}

\bibitem[{Batalha {et~al.}(2011)Batalha, Borucki, Bryson, Buchhave, Caldwell,
  Christensen-Dalsgaard, Ciardi, Dunham, Fressin, Gautier, Gilliland, Haas,
  Howell, Jenkins, Kjeldsen, Koch, Latham, Lissauer, Marcy, Rowe, Sasselov,
  Seager, Steffen, Torres, Basri, Brown, Charbonneau, Christiansen, Clarke,
  Cochran, Dupree, Fabrycky, Fischer, Ford, Fortney, Girouard, Holman, Johnson,
  Isaacson, Klaus, Machalek, Moorehead, Morehead, Ragozzine, Tenenbaum,
  Twicken, Quinn, VanCleve, Walkowicz, Welsh, Devore, \& Gould}]{Batalha:2011}
Batalha, N.~M., Borucki, W.~J., Bryson, S.~T., {et~al.} 2011,
  \href{http://dx.doi.org/10.1088/0004-637X/729/1/27}{ApJ, 729, 27}

\bibitem[{Batalha {et~al.}(2012)Batalha, Rowe, Bryson, Barclay, Burke,
  Caldwell, Christiansen, Mullally, Thompson, Brown, Dupree, Fabrycky, Ford,
  Fortney, Gilliland, Isaacson, Latham, Marcy, Quinn, Ragozzine, Shporer,
  Borucki, Ciardi, Gautier, Haas, Jenkins, Koch, Lissauer, Rapin, Basri, Boss,
  Buchhave, Charbonneau, Christensen-Dalsgaard, Clarke, Cochran, Demory,
  Devore, Esquerdo, Everett, Fressin, Geary, Girouard, Gould, Hall, Holman,
  Howard, Howell, Ibrahim, Kinemuchi, Kjeldsen, Klaus, Li, Lucas, Morris, Prsa,
  Quintana, Sanderfer, Sasselov, Seader, Smith, Steffen, Still, Stumpe, Tarter,
  Tenenbaum, Torres, Twicken, Uddin, Van~Cleve, Walkowicz, \&
  Welsh}]{Batalha12}
Batalha, N.~M., Rowe, J.~F., Bryson, S.~T., {et~al.} 2012,
  \href{http://adsabs.harvard.edu/cgi-bin/nph-data_query?bibcode=2012arXiv1202.5852B&link_type=ABSTRACT}{arXiv,
  1202, 5852}

\bibitem[{{Batalha} {et~al.}(2013){Batalha}, {Rowe}, {Bryson}, {Barclay},
  {Burke}, {Caldwell}, {Christiansen}, {Mullally}, {Thompson}, {Brown},
  {Dupree}, {Fabrycky}, {Ford}, {Fortney}, {Gilliland}, {Isaacson}, {Latham},
  {Marcy}, {Quinn}, {Ragozzine}, {Shporer}, {Borucki}, {Ciardi}, {Gautier},
  {Haas}, {Jenkins}, {Koch}, {Lissauer}, {Rapin}, {Basri}, {Boss}, {Buchhave},
  {Carter}, {Charbonneau}, {Christensen-Dalsgaard}, {Clarke}, {Cochran},
  {Demory}, {Desert}, {Devore}, {Doyle}, {Esquerdo}, {Everett}, {Fressin},
  {Geary}, {Girouard}, {Gould}, {Hall}, {Holman}, {Howard}, {Howell},
  {Ibrahim}, {Kinemuchi}, {Kjeldsen}, {Klaus}, {Li}, {Lucas}, {Meibom},
  {Morris}, {Pr{\v s}a}, {Quintana}, {Sanderfer}, {Sasselov}, {Seader},
  {Smith}, {Steffen}, {Still}, {Stumpe}, {Tarter}, {Tenenbaum}, {Torres},
  {Twicken}, {Uddin}, {Van Cleve}, {Walkowicz}, \& {Welsh}}]{Batalha13}
{Batalha}, N.~M., {Rowe}, J.~F., {Bryson}, S.~T., {et~al.} 2013,
  \href{http://dx.doi.org/10.1088/0067-0049/204/2/24}{\apjs, 204, 24}

\bibitem[{{Beaug{\'e}} \& {Nesvorn{\'y}}(2013)}]{Beauge13}
{Beaug{\'e}}, C., \& {Nesvorn{\'y}}, D. 2013,
  \href{http://dx.doi.org/10.1088/0004-637X/763/1/12}{\apj, 763, 12}

\bibitem[{{Bensby} {et~al.}(2003){Bensby}, {Feltzing}, \&
  {Lundstr{\"o}m}}]{Bensby03}
{Bensby}, T., {Feltzing}, S., \& {Lundstr{\"o}m}, I. 2003,
  \href{http://dx.doi.org/10.1051/0004-6361:20031213}{\aap, 410, 527}

\bibitem[{{Bensby} {et~al.}(2004){Bensby}, {Feltzing}, \&
  {Lundstr{\"o}m}}]{Bensby04}
---. 2004, \href{http://dx.doi.org/10.1051/0004-6361:20031655}{\aap, 415, 155}

\bibitem[{{Bensby} {et~al.}(2005){Bensby}, {Feltzing}, {Lundstr{\"o}m}, \&
  {Ilyin}}]{Bensby05}
{Bensby}, T., {Feltzing}, S., {Lundstr{\"o}m}, I., \& {Ilyin}, I. 2005,
  \href{http://dx.doi.org/10.1051/0004-6361:20040332}{\aap, 433, 185}

\bibitem[{{Billings}(2013)}]{Billings13}
{Billings}, L. 2013, Five Billion Years of Solitude: The Search for Life Among
  the Stars (Current)

\bibitem[{{Bond} {et~al.}(2010){Bond}, {Lauretta}, \& {O'Brien}}]{Bond10}
{Bond}, J.~C., {Lauretta}, D.~S., \& {O'Brien}, D.~P. 2010,
  \href{http://dx.doi.org/10.1017/S1743921310001079}{in IAU Symposium, Vol.
  265, IAU Symposium, ed. {K.~Cunha, M.~Spite, \& B.~Barbuy}}, 399

\bibitem[{{Bonfils} {et~al.}(2013){Bonfils}, {Delfosse}, {Udry}, {Forveille},
  {Mayor}, {Perrier}, {Bouchy}, {Gillon}, {Lovis}, {Pepe}, {Queloz}, {Santos},
  {S{\'e}gransan}, \& {Bertaux}}]{Bonfils2013}
{Bonfils}, X., {Delfosse}, X., {Udry}, S., {et~al.} 2013,
  \href{http://dx.doi.org/10.1051/0004-6361/201014704}{\aap, 549, A109}

\bibitem[{{Borucki} \& {Summers}(1984)}]{Borucki84}
{Borucki}, W.~J., \& {Summers}, A.~L. 1984,
  \href{http://dx.doi.org/10.1016/0019-1035(84)90102-7}{\icarus, 58, 121}

\bibitem[{{Borucki} {et~al.}(2003){Borucki}, {Koch}, {Lissauer}, {Basri},
  {Caldwell}, {Cochran}, {Dunham}, {Geary}, {Latham}, {Gilliland}, {Caldwell},
  {Jenkins}, \& {Kondo}}]{Borucki03}
{Borucki}, W.~J., {Koch}, D.~G., {Lissauer}, J.~J., {et~al.} 2003,
  \href{http://dx.doi.org/10.1117/12.460266}{in Society of Photo-Optical
  Instrumentation Engineers (SPIE) Conference Series, Vol. 4854, Future EUV/UV
  and Visible Space Astrophysics Missions and Instrumentation., ed. J.~C.
  {Blades} \& O.~H.~W. {Siegmund}}, 129

\bibitem[{{Borucki} {et~al.}(2010){Borucki}, {Koch}, {Basri}, {Batalha},
  {Brown}, {Caldwell}, {Caldwell}, {Christensen-Dalsgaard}, {Cochran},
  {DeVore}, {Dunham}, {Dupree}, {Gautier}, {Geary}, {Gilliland}, {Gould},
  {Howell}, {Jenkins}, {Kondo}, {Latham}, {Marcy}, {Meibom}, {Kjeldsen},
  {Lissauer}, {Monet}, {Morrison}, {Sasselov}, {Tarter}, {Boss}, {Brownlee},
  {Owen}, {Buzasi}, {Charbonneau}, {Doyle}, {Fortney}, {Ford}, {Holman},
  {Seager}, {Steffen}, {Welsh}, {Rowe}, {Anderson}, {Buchhave}, {Ciardi},
  {Walkowicz}, {Sherry}, {Horch}, {Isaacson}, {Everett}, {Fischer}, {Torres},
  {Johnson}, {Endl}, {MacQueen}, {Bryson}, {Dotson}, {Haas}, {Kolodziejczak},
  {Van Cleve}, {Chandrasekaran}, {Twicken}, {Quintana}, {Clarke}, {Allen},
  {Li}, {Wu}, {Tenenbaum}, {Verner}, {Bruhweiler}, {Barnes}, \&
  {Prsa}}]{Borucki10}
{Borucki}, W.~J., {Koch}, D., {Basri}, G., {et~al.} 2010,
  \href{http://dx.doi.org/10.1126/science.1185402}{Science, 327, 977}

\bibitem[{{Borucki} {et~al.}(2011){Borucki}, {Koch}, {Basri}, {Batalha},
  {Boss}, {Brown}, {Caldwell}, {Christensen-Dalsgaard}, {Cochran}, {DeVore},
  {Dunham}, {Dupree}, {Gautier}, {Geary}, {Gilliland}, {Gould}, {Howell},
  {Jenkins}, {Kjeldsen}, {Latham}, {Lissauer}, {Marcy}, {Monet}, {Sasselov},
  {Tarter}, {Charbonneau}, {Doyle}, {Ford}, {Fortney}, {Holman}, {Seager},
  {Steffen}, {Welsh}, {Allen}, {Bryson}, {Buchhave}, {Chandrasekaran},
  {Christiansen}, {Ciardi}, {Clarke}, {Dotson}, {Endl}, {Fischer}, {Fressin},
  {Haas}, {Horch}, {Howard}, {Isaacson}, {Kolodziejczak}, {Li}, {MacQueen},
  {Meibom}, {Prsa}, {Quintana}, {Rowe}, {Sherry}, {Tenenbaum}, {Torres},
  {Twicken}, {Van Cleve}, {Walkowicz}, \& {Wu}}]{Borucki11a}
{Borucki}, W.~J., {Koch}, D.~G., {Basri}, G., {et~al.} 2011,
  \href{http://dx.doi.org/10.1088/0004-637X/728/2/117}{\apj, 728, 117}

\bibitem[{Borucki {et~al.}(2011)Borucki, Koch, Basri, Batalha, Brown, Bryson,
  Caldwell, Christensen-Dalsgaard, Cochran, Devore, Dunham, Gautier, Geary,
  Gilliland, Gould, Howell, Jenkins, Latham, Lissauer, Marcy, Rowe, Sasselov,
  Boss, Charbonneau, Ciardi, Doyle, Dupree, Ford, Fortney, Holman, Seager,
  Steffen, Tarter, Welsh, Allen, Buchhave, Christiansen, Clarke, Das, Desert,
  Endl, Fabrycky, Fressin, Haas, Horch, Howard, Isaacson, Kjeldsen,
  Kolodziejczak, Kulesa, Li, Lucas, Machalek, McCarthy, MacQueen, Meibom,
  Miquel, Prsa, Quinn, Quintana, Ragozzine, Sherry, Shporer, Tenenbaum, Torres,
  Twicken, Van~Cleve, Walkowicz, Witteborn, \& Still}]{Borucki11}
Borucki, W.~J., Koch, D.~G., Basri, G., {et~al.} 2011,
  \href{http://dx.doi.org/10.1088/0004-637X/736/1/19}{ApJ, 736, 19}

\bibitem[{{Boyajian} {et~al.}(2012){Boyajian}, {von Braun}, {van Belle},
  {McAlister}, {ten Brummelaar}, {Kane}, {Muirhead}, {Jones}, {White},
  {Schaefer}, {Ciardi}, {Henry}, {L{\'o}pez-Morales}, {Ridgway}, {Gies}, {Jao},
  {Rojas-Ayala}, {Parks}, {Sturmann}, {Sturmann}, {Turner}, {Farrington},
  {Goldfinger}, \& {Berger}}]{Boyajian12}
{Boyajian}, T.~S., {von Braun}, K., {van Belle}, G., {et~al.} 2012,
  \href{http://dx.doi.org/10.1088/0004-637X/757/2/112}{\apj, 757, 112}

\bibitem[{{Boyajian} {et~al.}(2013){Boyajian}, {von Braun}, {van Belle},
  {Farrington}, {Schaefer}, {Jones}, {White}, {McAlister}, {ten Brummelaar},
  {Ridgway}, {Gies}, {Sturmann}, {Sturmann}, {Turner}, {Goldfinger}, \&
  {Vargas}}]{Boyajian13}
---. 2013, \href{http://dx.doi.org/10.1088/0004-637X/771/1/40}{\apj, 771, 40}

\bibitem[{{Brown} {et~al.}(2001){Brown}, {Charbonneau}, {Gilliland}, {Noyes},
  \& {Burrows}}]{Brown01}
{Brown}, T.~M., {Charbonneau}, D., {Gilliland}, R.~L., {Noyes}, R.~W., \&
  {Burrows}, A. 2001, \href{http://dx.doi.org/10.1086/320580}{\apj, 552, 699}

\bibitem[{{Brown} {et~al.}(2011){Brown}, {Latham}, {Everett}, \&
  {Esquerdo}}]{Brown11}
{Brown}, T.~M., {Latham}, D.~W., {Everett}, M.~E., \& {Esquerdo}, G.~A. 2011,
  \href{http://dx.doi.org/10.1088/0004-6256/142/4/112}{\aj, 142, 112}

\bibitem[{{Bryson} {et~al.}(2013){Bryson}, {Jenkins}, {Gilliland}, {Twicken},
  {Clarke}, {Rowe}, {Caldwell}, {Batalha}, {Mullally}, {Haas}, \&
  {Tenenbaum}}]{Bryson13}
{Bryson}, S.~T., {Jenkins}, J.~M., {Gilliland}, R.~L., {et~al.} 2013,
  \href{http://dx.doi.org/10.1086/671767}{\pasp, 125, 889}

\bibitem[{{Buchhave} {et~al.}(2012){Buchhave}, {Latham}, {Johansen},
  {Bizzarro}, {Torres}, {Rowe}, {Batalha}, {Borucki}, {Brugamyer}, {Caldwell},
  {Bryson}, {Ciardi}, {Cochran}, {Endl}, {Esquerdo}, {Ford}, {Geary},
  {Gilliland}, {Hansen}, {Isaacson}, {Laird}, {Lucas}, {Marcy}, {Morse},
  {Robertson}, {Shporer}, {Stefanik}, {Still}, \& {Quinn}}]{Buchhave12nat}
{Buchhave}, L.~A., {Latham}, D.~W., {Johansen}, A., {et~al.} 2012,
  \href{http://dx.doi.org/10.1038/nature11121}{\nat, 486, 375}

\bibitem[{{Burke} {et~al.}(2014){Burke}, {Bryson}, {Mullally}, {Rowe},
  {Christiansen}, {Thompson}, {Coughlin}, {Haas}, {Batalha}, {Caldwell},
  {Jenkins}, {Still}, {Barclay}, {Borucki}, {Chaplin}, {Ciardi}, {Clarke},
  {Cochran}, {Demory}, {Esquerdo}, {Gautier}, {Gilliland}, {Girouard}, {Havel},
  {Henze}, {Howell}, {Huber}, {Latham}, {Li}, {Morehead}, {Morton}, {Pepper},
  {Quintana}, {Ragozzine}, {Seader}, {Shah}, {Shporer}, {Tenenbaum}, {Twicken},
  \& {Wolfgang}}]{Burke14}
{Burke}, C.~J., {Bryson}, S.~T., {Mullally}, F., {et~al.} 2014,
  \href{http://dx.doi.org/10.1088/0067-0049/210/2/19}{\apjs, 210, 19}

\bibitem[{{Butler} \& {Marcy}(1996)}]{Butler96}
{Butler}, R.~P., \& {Marcy}, G.~W. 1996,
  \href{http://dx.doi.org/10.1086/310102}{\apjl, 464, L153}

\bibitem[{{Caffau} {et~al.}(2010){Caffau}, {Ludwig}, {Bonifacio}, {Faraggiana},
  {Steffen}, {Freytag}, {Kamp}, \& {Ayres}}]{Caffau10}
{Caffau}, E., {Ludwig}, H., {Bonifacio}, P., {et~al.} 2010,
  \href{http://dx.doi.org/10.1051/0004-6361/200912227}{\aap, 514, A92}

\bibitem[{{Campbell}(1983)}]{Campbell83}
{Campbell}, B. 1983, \href{http://dx.doi.org/10.1086/131209}{\pasp, 95, 577}

\bibitem[{{Campo} {et~al.}(2011){Campo}, {Harrington}, {Hardy}, {Stevenson},
  {Nymeyer}, {Ragozzine}, {Lust}, {Anderson}, {Collier-Cameron}, {Blecic},
  {Britt}, {Bowman}, {Wheatley}, {Loredo}, {Deming}, {Hebb}, {Hellier},
  {Maxted}, {Pollaco}, \& {West}}]{Campo11}
{Campo}, C.~J., {Harrington}, J., {Hardy}, R.~A., {et~al.} 2011,
  \href{http://dx.doi.org/10.1088/0004-637X/727/2/125}{\apj, 727, 125}

\bibitem[{{Casagrande} {et~al.}(2010){Casagrande}, {Ram{\'{\i}}rez},
  {Mel{\'e}ndez}, {Bessell}, \& {Asplund}}]{Casagrande10}
{Casagrande}, L., {Ram{\'{\i}}rez}, I., {Mel{\'e}ndez}, J., {Bessell}, M., \&
  {Asplund}, M. 2010,
  \href{http://dx.doi.org/10.1051/0004-6361/200913204}{\aap, 512, A54}

\bibitem[{{Castelli} \& {Kurucz}(2003)}]{Castelli03}
{Castelli}, F., \& {Kurucz}, R.~L. 2003, in IAU Symposium, Vol. 210, Modelling
  of Stellar Atmospheres, ed. N.~{Piskunov}, W.~W. {Weiss}, \& D.~F. {Gray},
  20P

\bibitem[{{Cayrel} {et~al.}(1991){Cayrel}, {Perrin}, {Barbuy}, \&
  {Buser}}]{Cayrel91}
{Cayrel}, R., {Perrin}, M.-N., {Barbuy}, B., \& {Buser}, R. 1991, \aap, 247,
  108

\bibitem[{{Chaplin} {et~al.}(2014){Chaplin}, {Basu}, {Huber}, {Serenelli},
  {Casagrande}, {Silva Aguirre}, {Ball}, {Creevey}, {Gizon}, {Handberg},
  {Karoff}, {Lutz}, {Marques}, {Miglio}, {Stello}, {Suran}, {Pricopi},
  {Metcalfe}, {Monteiro}, {Molenda-{\.Z}akowicz}, {Appourchaux},
  {Christensen-Dalsgaard}, {Elsworth}, {Garc{\'{\i}}a}, {Houdek}, {Kjeldsen},
  {Bonanno}, {Campante}, {Corsaro}, {Gaulme}, {Hekker}, {Mathur}, {Mosser},
  {R{\'e}gulo}, \& {Salabert}}]{Chaplin14}
{Chaplin}, W.~J., {Basu}, S., {Huber}, D., {et~al.} 2014,
  \href{http://dx.doi.org/10.1088/0067-0049/210/1/1}{\apjs, 210, 1}

\bibitem[{{Charbonneau} {et~al.}(2000){Charbonneau}, {Brown}, {Latham}, \&
  {Mayor}}]{Charbonneau00}
{Charbonneau}, D., {Brown}, T.~M., {Latham}, D.~W., \& {Mayor}, M. 2000,
  \href{http://dx.doi.org/10.1086/312457}{\apjl, 529, L45}

\bibitem[{{Chiang} \& {Laughlin}(2012)}]{Chiang12}
{Chiang}, E., \& {Laughlin}, G. 2012, ArXiv e-prints,
  \href{http://arxiv.org/abs/1211.1673}{{\sffamily arXiv:1211.1673
  [astro-ph.EP]}}

\bibitem[{{Christiansen} {et~al.}(2012){Christiansen}, {Science Office}, \&
  {Science Operations Center}}]{Christiansen12}
{Christiansen}, J., {Science Office}, K., \& {Science Operations Center}, K.
  2012, in AAS/Division for Planetary Sciences Meeting Abstracts, Vol.~44,
  AAS/Division for Planetary Sciences Meeting Abstracts, 113.01

\bibitem[{Christiansen {et~al.}(2011)Christiansen, Van~Cleve, Jenkins,
  Caldwell, Allen, Barclay, Bryson, Burke, Clarke, Cote, Fanelli, Gilliland,
  Girouard, Haas, Hall, Ibrahim, Kinemuchi, Klaus, Kolodziejczak, Li, Machalek,
  McCauliff, Middour, Morris, Mullally, Quintana, Seader, Smith, Still, Stumpe,
  Tenenbaum, Thompson, Twicken, Uddin, \& Wohler}]{Christiansen:2011}
Christiansen, J.~L., Van~Cleve, J.~E., Jenkins, J.~M., {et~al.} 2011, {Kepler
  Data Characteristics Handbook (KSCI-19040-002)}

\bibitem[{{Coelho} {et~al.}(2005){Coelho}, {Barbuy}, {Mel{\'e}ndez},
  {Schiavon}, \& {Castilho}}]{Coelho05}
{Coelho}, P., {Barbuy}, B., {Mel{\'e}ndez}, J., {Schiavon}, R.~P., \&
  {Castilho}, B.~V. 2005,
  \href{http://dx.doi.org/10.1051/0004-6361:20053511}{\aap, 443, 735}

\bibitem[{Collette(2008)}]{Collette08}
Collette, A. 2008, HDF5 for Python

\bibitem[{{Croll} {et~al.}(2011){Croll}, {Lafreniere}, {Albert},
  {Jayawardhana}, {Fortney}, \& {Murray}}]{Croll11}
{Croll}, B., {Lafreniere}, D., {Albert}, L., {et~al.} 2011,
  \href{http://dx.doi.org/10.1088/0004-6256/141/2/30}{\aj, 141, 30}

\bibitem[{{Cumming} {et~al.}(2008){Cumming}, {Butler}, {Marcy}, {Vogt},
  {Wright}, \& {Fischer}}]{Cumming08}
{Cumming}, A., {Butler}, R.~P., {Marcy}, G.~W., {et~al.} 2008,
  \href{http://dx.doi.org/10.1086/588487}{\pasp, 120, 531}

\bibitem[{{Demarque} {et~al.}(2004){Demarque}, {Woo}, {Kim}, \&
  {Yi}}]{Demarque04}
{Demarque}, P., {Woo}, J.-H., {Kim}, Y.-C., \& {Yi}, S.~K. 2004,
  \href{http://dx.doi.org/10.1086/424966}{\apjs, 155, 667}

\bibitem[{{Dong} \& {Zhu}(2012)}]{Dong12}
{Dong}, S., \& {Zhu}, Z. 2012, ArXiv e-prints,
  \href{http://arxiv.org/abs/1212.4853}{{\sffamily arXiv:1212.4853
  [astro-ph.EP]}}

\bibitem[{{Dotter} {et~al.}(2008){Dotter}, {Chaboyer}, {Jevremovi{\'c}},
  {Kostov}, {Baron}, \& {Ferguson}}]{Dotter08}
{Dotter}, A., {Chaboyer}, B., {Jevremovi{\'c}}, D., {et~al.} 2008,
  \href{http://dx.doi.org/10.1086/589654}{\apjs, 178, 89}

\bibitem[{Doyle {et~al.}(2011)Doyle, Carter, Fabrycky, Slawson, Howell, Winn,
  Orosz, Pr~sa, Welsh, Quinn, Latham, Torres, Buchhave, Marcy, Fortney,
  Shporer, Ford, Lissauer, Ragozzine, Rucker, Batalha, Jenkins, Borucki, Koch,
  Middour, Hall, McCauliff, Fanelli, Quintana, Holman, Caldwell, Still,
  Stefanik, Brown, Esquerdo, Tang, Furesz, Geary, Berlind, Calkins, Short,
  Steffen, Sasselov, Dunham, Cochran, Boss, Haas, Buzasi, \&
  Fischer}]{Doyle:2011ev}
Doyle, L.~R., Carter, J.~A., Fabrycky, D.~C., {et~al.} 2011,
  \href{http://dx.doi.org/10.1126/science.1210923}{Science, 333, 1602}

\bibitem[{{Dressing} \& {Charbonneau}(2013)}]{Dressing2013}
{Dressing}, C.~D., \& {Charbonneau}, D. 2013,
  \href{http://dx.doi.org/10.1088/0004-637X/767/1/95}{\apj, 767, 95}

\bibitem[{{Edvardsson} {et~al.}(1993){Edvardsson}, {Andersen}, {Gustafsson},
  {Lambert}, {Nissen}, \& {Tomkin}}]{Edvardsson93}
{Edvardsson}, B., {Andersen}, J., {Gustafsson}, B., {et~al.} 1993, \aap, 275,
  101

\bibitem[{{ESA}(1997)}]{ESA97}
{ESA}. 1997, VizieR Online Data Catalog, 1239, 0

\bibitem[{{Fang} \& {Margot}(2012)}]{Fang12}
{Fang}, J., \& {Margot}, J.-L. 2012,
  \href{http://dx.doi.org/10.1088/0004-637X/761/2/92}{\apj, 761, 92}

\bibitem[{{Fischer} \& {Valenti}(2005)}]{Fischer05}
{Fischer}, D.~A., \& {Valenti}, J. 2005,
  \href{http://dx.doi.org/10.1086/428383}{\apj, 622, 1102}

\bibitem[{{Fischer} {et~al.}(2005){Fischer}, {Laughlin}, {Butler}, {Marcy},
  {Johnson}, {Henry}, {Valenti}, {Vogt}, {Ammons}, {Robinson}, {Spear},
  {Strader}, {Driscoll}, {Fuller}, {Johnson}, {Manrao}, {McCarthy},
  {Mu{\~n}oz}, {Tah}, {Wright}, {Ida}, {Sato}, {Toyota}, \& {Minniti}}]{N2K}
{Fischer}, D.~A., {Laughlin}, G., {Butler}, P., {et~al.} 2005,
  \href{http://dx.doi.org/10.1086/426810}{\apj, 620, 481}

\bibitem[{Fischer {et~al.}(2011)Fischer, Schwamb, Schawinski, Lintott, Brewer,
  Giguere, Lynn, Parrish, Sartori, Simpson, Smith, Spronck, Batalha, Rowe,
  Jenkins, Bryson, Prsa, Tenenbaum, Crepp, Morton, Howard, Beleu, Kaplan,
  vanNispen, Sharzer, DeFouw, Hajduk, Neal, Nemec, Schuepbach, \&
  Zimmermann}]{Fischer:2011bb}
Fischer, D.~A., Schwamb, M.~E., Schawinski, K., {et~al.} 2011,
  \href{http://dx.doi.org/10.1111/j.1365-2966.2011.19932.x}{Monthly Notices of
  the Royal Astronomical Society, 419, 2900}

\bibitem[{Fraquelli \& Thompson(2012)}]{KeplerArchiveManual}
Fraquelli, D., \& Thompson, S.~E. 2012, Kepler Archive Manual (KDMC-10008-004)

\bibitem[{{Fressin} {et~al.}(2012){Fressin}, {Torres}, {Rowe}, {Charbonneau},
  {Rogers}, {Ballard}, {Batalha}, {Borucki}, {Bryson}, {Buchhave}, {Ciardi},
  {D{\'e}sert}, {Dressing}, {Fabrycky}, {Ford}, {Gautier}, {Henze}, {Holman},
  {Howard}, {Howell}, {Jenkins}, {Koch}, {Latham}, {Lissauer}, {Marcy},
  {Quinn}, {Ragozzine}, {Sasselov}, {Seager}, {Barclay}, {Mullally}, {Seader},
  {Still}, {Twicken}, {Thompson}, \& {Uddin}}]{Fressin12}
{Fressin}, F., {Torres}, G., {Rowe}, J.~F., {et~al.} 2012,
  \href{http://dx.doi.org/10.1038/nature10780}{\nat, 482, 195}

\bibitem[{{Fressin} {et~al.}(2013){Fressin}, {Torres}, {Charbonneau}, {Bryson},
  {Christiansen}, {Dressing}, {Jenkins}, {Walkowicz}, \& {Batalha}}]{Fressin13}
{Fressin}, F., {Torres}, G., {Charbonneau}, D., {et~al.} 2013,
  \href{http://dx.doi.org/10.1088/0004-637X/766/2/81}{\apj, 766, 81}

\bibitem[{{Gonzalez}(1997)}]{Gonzalez97}
{Gonzalez}, G. 1997, \mnras, 285, 403

\bibitem[{{Gray}(1992)}]{Gray92}
{Gray}, D.~F. 1992, {The observation and analysis of stellar photospheres.}

\bibitem[{{Gray}(2005)}]{Gray05}
---. 2005, {The Observation and Analysis of Stellar Photospheres}

\bibitem[{{Grevesse} \& {Sauval}(1998)}]{Grevesse98}
{Grevesse}, N., \& {Sauval}, A.~J. 1998,
  \href{http://dx.doi.org/10.1023/A:1005161325181}{Space Science Reviews, 85,
  161}

\bibitem[{{Gustafsson} {et~al.}(1999){Gustafsson}, {Karlsson}, {Olsson},
  {Edvardsson}, \& {Ryde}}]{Gustafsson99}
{Gustafsson}, B., {Karlsson}, T., {Olsson}, E., {Edvardsson}, B., \& {Ryde}, N.
  1999, \aap, 342, 426

\bibitem[{{Haisch} {et~al.}(2001){Haisch}, {Lada}, \& {Lada}}]{Haisch01}
{Haisch}, Jr., K.~E., {Lada}, E.~A., \& {Lada}, C.~J. 2001,
  \href{http://dx.doi.org/10.1086/320685}{\apjl, 553, L153}

\bibitem[{{Han} {et~al.}(2014){Han}, {Wang}, {Wright}, {Feng}, {Zhao},
  {Fakhouri}, {Brown}, \& {Hancock}}]{Han14}
{Han}, E., {Wang}, S.~X., {Wright}, J.~T., {et~al.} 2014,
  \href{http://dx.doi.org/10.1086/678447}{\pasp, 126, 827}

\bibitem[{{Hansen} \& {Murray}(2012)}]{Hansen12}
{Hansen}, B.~M.~S., \& {Murray}, N. 2012,
  \href{http://dx.doi.org/10.1088/0004-637X/751/2/158}{\apj, 751, 158}

\bibitem[{{Hebb} {et~al.}(2009){Hebb}, {Collier-Cameron}, {Loeillet},
  {Pollacco}, {H{\'e}brard}, {Street}, {Bouchy}, {Stempels}, {Moutou},
  {Simpson}, {Udry}, {Joshi}, {West}, {Skillen}, {Wilson}, {McDonald},
  {Gibson}, {Aigrain}, {Anderson}, {Benn}, {Christian}, {Enoch}, {Haswell},
  {Hellier}, {Horne}, {Irwin}, {Lister}, {Maxted}, {Mayor}, {Norton}, {Parley},
  {Pont}, {Queloz}, {Smalley}, \& {Wheatley}}]{Hebb09}
{Hebb}, L., {Collier-Cameron}, A., {Loeillet}, B., {et~al.} 2009,
  \href{http://dx.doi.org/10.1088/0004-637X/693/2/1920}{\apj, 693, 1920}

\bibitem[{{Henry} {et~al.}(2000){Henry}, {Marcy}, {Butler}, \&
  {Vogt}}]{Henry00}
{Henry}, G.~W., {Marcy}, G.~W., {Butler}, R.~P., \& {Vogt}, S.~S. 2000,
  \href{http://dx.doi.org/10.1086/312458}{\apjl, 529, L41}

\bibitem[{{Hirano} {et~al.}(2011){Hirano}, {Suto}, {Winn}, {Taruya}, {Narita},
  {Albrecht}, \& {Sato}}]{Hirano11}
{Hirano}, T., {Suto}, Y., {Winn}, J.~N., {et~al.} 2011,
  \href{http://dx.doi.org/10.1088/0004-637X/742/2/69}{\apj, 742, 69}

\bibitem[{{Howard} {et~al.}(2010){Howard}, {Marcy}, {Johnson}, {Fischer},
  {Wright}, {Isaacson}, {Valenti}, {Anderson}, {Lin}, \& {Ida}}]{Howard10}
{Howard}, A.~W., {Marcy}, G.~W., {Johnson}, J.~A., {et~al.} 2010,
  \href{http://dx.doi.org/10.1126/science.1194854}{Science, 330, 653}

\bibitem[{{Howard} {et~al.}(2012){Howard}, {Marcy}, {Bryson}, {Jenkins},
  {Rowe}, {Batalha}, {Borucki}, {Koch}, {Dunham}, {Gautier}, {Van Cleve},
  {Cochran}, {Latham}, {Lissauer}, {Torres}, {Brown}, {Gilliland}, {Buchhave},
  {Caldwell}, {Christensen-Dalsgaard}, {Ciardi}, {Fressin}, {Haas}, {Howell},
  {Kjeldsen}, {Seager}, {Rogers}, {Sasselov}, {Steffen}, {Basri},
  {Charbonneau}, {Christiansen}, {Clarke}, {Dupree}, {Fabrycky}, {Fischer},
  {Ford}, {Fortney}, {Tarter}, {Girouard}, {Holman}, {Johnson}, {Klaus},
  {Machalek}, {Moorhead}, {Morehead}, {Ragozzine}, {Tenenbaum}, {Twicken},
  {Quinn}, {Isaacson}, {Shporer}, {Lucas}, {Walkowicz}, {Welsh}, {Boss},
  {Devore}, {Gould}, {Smith}, {Morris}, {Prsa}, {Morton}, {Still}, {Thompson},
  {Mullally}, {Endl}, \& {MacQueen}}]{Howard12}
{Howard}, A.~W., {Marcy}, G.~W., {Bryson}, S.~T., {et~al.} 2012,
  \href{http://dx.doi.org/10.1088/0067-0049/201/2/15}{\apjs, 201, 15}

\bibitem[{Huang {et~al.}(2012)Huang, Bakos, \& Hartman}]{Huang:2012uj}
Huang, X., Bakos, G.~{\'A}., \& Hartman, J.~D. 2012,
  \href{http://arxiv.org/abs/1205.6492v1}{arXiv, astro-ph.EP}

\bibitem[{{Huber} {et~al.}(2013){Huber}, {Chaplin}, {Christensen-Dalsgaard},
  {Gilliland}, {Kjeldsen}, {Buchhave}, {Fischer}, {Lissauer}, {Rowe},
  {Sanchis-Ojeda}, {Basu}, {Handberg}, {Hekker}, {Howard}, {Isaacson},
  {Karoff}, {Latham}, {Lund}, {Lundkvist}, {Marcy}, {Miglio}, {Silva Aguirre},
  {Stello}, {Arentoft}, {Barclay}, {Bedding}, {Burke}, {Christiansen},
  {Elsworth}, {Haas}, {Kawaler}, {Metcalfe}, {Mullally}, \&
  {Thompson}}]{Huber13}
{Huber}, D., {Chaplin}, W.~J., {Christensen-Dalsgaard}, J., {et~al.} 2013,
  \href{http://dx.doi.org/10.1088/0004-637X/767/2/127}{\apj, 767, 127}

\bibitem[{Hunter(2007)}]{matplotlib}
Hunter, J.~D. 2007, Computing In Science \& Engineering, 9, 90

\bibitem[{{Ida} \& {Lin}(2008)}]{Ida08}
{Ida}, S., \& {Lin}, D.~N.~C. 2008,
  \href{http://dx.doi.org/10.1086/590401}{\apj, 685, 584}

\bibitem[{{Ida} \& {Lin}(2010)}]{Ida10}
---. 2010, \href{http://dx.doi.org/10.1088/0004-637X/719/1/810}{\apj, 719, 810}

\bibitem[{{Ida} {et~al.}(2013){Ida}, {Lin}, \& {Nagasawa}}]{Ida13}
{Ida}, S., {Lin}, D.~N.~C., \& {Nagasawa}, M. 2013,
  \href{http://dx.doi.org/10.1088/0004-637X/775/1/42}{\apj, 775, 42}

\bibitem[{{Jenkins} {et~al.}(2010{\natexlab{a}}){Jenkins}, {Caldwell},
  {Chandrasekaran}, {Twicken}, {Bryson}, {Quintana}, {Clarke}, {Li}, {Allen},
  {Tenenbaum}, {Wu}, {Klaus}, {Middour}, {Cote}, {McCauliff}, {Girouard},
  {Gunter}, {Wohler}, {Sommers}, {Hall}, {Uddin}, {Wu}, {Bhavsar}, {Van Cleve},
  {Pletcher}, {Dotson}, {Haas}, {Gilliland}, {Koch}, \& {Borucki}}]{Jenkins10}
{Jenkins}, J.~M., {Caldwell}, D.~A., {Chandrasekaran}, H., {et~al.}
  2010{\natexlab{a}},
  \href{http://dx.doi.org/10.1088/2041-8205/713/2/L87}{\apjl, 713, L87}

\bibitem[{{Jenkins} {et~al.}(2010{\natexlab{b}}){Jenkins}, {Chandrasekaran},
  {McCauliff}, {Caldwell}, {Tenenbaum}, {Li}, {Klaus}, {Cote}, \&
  {Middour}}]{Jenkins10TPS}
{Jenkins}, J.~M., {Chandrasekaran}, H., {McCauliff}, S.~D., {et~al.}
  2010{\natexlab{b}}, \href{http://dx.doi.org/10.1117/12.856764}{in Society of
  Photo-Optical Instrumentation Engineers (SPIE) Conference Series, Vol. 7740}

\bibitem[{{Johansson} {et~al.}(2003){Johansson}, {Litz{\'e}n}, {Lundberg}, \&
  {Zhang}}]{Johansson03}
{Johansson}, S., {Litz{\'e}n}, U., {Lundberg}, H., \& {Zhang}, Z. 2003,
  \href{http://dx.doi.org/10.1086/374037}{\apjl, 584, L107}

\bibitem[{Jones {et~al.}(2001--)Jones, Oliphant, Peterson, {et~al.}}]{scipy}
Jones, E., Oliphant, T., Peterson, P., {et~al.} 2001--, {SciPy}: Open source
  scientific tools for {Python}

\bibitem[{{Kasting} {et~al.}(1993){Kasting}, {Whitmire}, \&
  {Reynolds}}]{Kasting1993}
{Kasting}, J.~F., {Whitmire}, D.~P., \& {Reynolds}, R.~T. 1993,
  \href{http://dx.doi.org/10.1006/icar.1993.1010}{\icarus, 101, 108}

\bibitem[{{Koch} {et~al.}(2010){Koch}, {Borucki}, {Basri}, {Batalha}, {Brown},
  {Caldwell}, {Christensen-Dalsgaard}, {Cochran}, {DeVore}, {Dunham},
  {Gautier}, {Geary}, {Gilliland}, {Gould}, {Jenkins}, {Kondo}, {Latham},
  {Lissauer}, {Marcy}, {Monet}, {Sasselov}, {Boss}, {Brownlee}, {Caldwell},
  {Dupree}, {Howell}, {Kjeldsen}, {Meibom}, {Morrison}, {Owen}, {Reitsema},
  {Tarter}, {Bryson}, {Dotson}, {Gazis}, {Haas}, {Kolodziejczak}, {Rowe}, {Van
  Cleve}, {Allen}, {Chandrasekaran}, {Clarke}, {Li}, {Quintana}, {Tenenbaum},
  {Twicken}, \& {Wu}}]{Koch10}
{Koch}, D.~G., {Borucki}, W.~J., {Basri}, G., {et~al.} 2010,
  \href{http://dx.doi.org/10.1088/2041-8205/713/2/L79}{\apjl, 713, L79}

\bibitem[{{Kopparapu}(2013)}]{Kopparapu2013b}
{Kopparapu}, R.~K. 2013,
  \href{http://dx.doi.org/10.1088/2041-8205/767/1/L8}{\apjl, 767, L8}

\bibitem[{{Kopparapu} {et~al.}(2013){Kopparapu}, {Ramirez}, {Kasting}, {Eymet},
  {Robinson}, {Mahadevan}, {Terrien}, {Domagal-Goldman}, {Meadows}, \&
  {Deshpande}}]{Kopparapu2013}
{Kopparapu}, R.~K., {Ramirez}, R., {Kasting}, J.~F., {et~al.} 2013,
  \href{http://dx.doi.org/10.1088/0004-637X/765/2/131}{\apj, 765, 131}

\bibitem[{Kov{\'a}cs {et~al.}(2005)Kov{\'a}cs, Bakos, \& Noyes}]{Kovacs:2005}
Kov{\'a}cs, G., Bakos, G., \& Noyes, R.~W. 2005,
  \href{http://dx.doi.org/10.1111/j.1365-2966.2004.08479.x}{Monthly Notices of
  the Royal Astronomical Society, 356, 557}

\bibitem[{{Kov{\'a}cs} {et~al.}(2002){Kov{\'a}cs}, {Zucker}, \&
  {Mazeh}}]{Kovacs02}
{Kov{\'a}cs}, G., {Zucker}, S., \& {Mazeh}, T. 2002,
  \href{http://dx.doi.org/10.1051/0004-6361:20020802}{\aap, 391, 369}

\bibitem[{{Kuchner} \& {Seager}(2005)}]{Kuchner05}
{Kuchner}, M.~J., \& {Seager}, S. 2005, ArXiv Astrophysics e-prints,
  \href{http://arxiv.org/abs/arXiv:astro-ph/0504214}{{\sffamily
  arXiv:astro-ph/0504214}}

\bibitem[{{Kurucz}(1992)}]{Kurucz92}
{Kurucz}, R.~L. 1992, in IAU Symposium, Vol. 149, The Stellar Populations of
  Galaxies, ed. {B.~Barbuy \& A.~Renzini}, 225

\bibitem[{{Kurucz} {et~al.}(1984){Kurucz}, {Furenlid}, {Brault}, \&
  {Testerman}}]{Kurucz84}
{Kurucz}, R.~L., {Furenlid}, I., {Brault}, J., \& {Testerman}, L. 1984, {Solar
  flux atlas from 296 to 1300 nm}, ed. {Kurucz, R.~L., Furenlid, I., Brault,
  J., \& Testerman, L.}

\bibitem[{{Lambert}(1978)}]{Lambert78}
{Lambert}, D.~L. 1978, \mnras, 182, 249

\bibitem[{{Latham}(1985)}]{Latham85}
{Latham}, D.~W. 1985, in Stellar Radial Velocities, ed. A.~G.~D. {Philip} \&
  D.~W. {Latham}, 21

\bibitem[{{Latham} {et~al.}(1989){Latham}, {Stefanik}, {Mazeh}, {Mayor}, \&
  {Burki}}]{Latham89}
{Latham}, D.~W., {Stefanik}, R.~P., {Mazeh}, T., {Mayor}, M., \& {Burki}, G.
  1989, \href{http://dx.doi.org/10.1038/339038a0}{\nat, 339, 38}

\bibitem[{{Levy} \& {Lunine}(1993)}]{Lissauer93}
{Levy}, E.~H., \& {Lunine}, J.~I., eds. 1993, {Growth of planets from
  planetesimals}, ed. E.~H. {Levy} \& J.~I. {Lunine}, 1061

\bibitem[{Lintott {et~al.}(2012)Lintott, Schwamb, Sharzer, Fischer, Barclay,
  Parrish, Batalha, Bryson, Jenkins, Ragozzine, Rowe, Schawinski, Gagliano,
  Gilardi, Jek, P{\"a}{\"a}kk{\"o}nen, \& Smits}]{Lintott:2012ut}
Lintott, C., Schwamb, M.~E., Sharzer, C., {et~al.} 2012,
  \href{http://adsabs.harvard.edu/abs/2012arXiv1202.6007L}{eprint
  arXiv:1202.6007}

\bibitem[{Lissauer {et~al.}(2011)Lissauer, Fabrycky, Ford, Borucki, Fressin,
  Marcy, Orosz, Rowe, Torres, Welsh, Batalha, Bryson, Buchhave, Caldwell,
  Carter, Charbonneau, Christiansen, Cochran, Desert, Dunham, Fanelli, Fortney,
  Gautier~III, Geary, Gilliland, Haas, Hall, Holman, Koch, Latham, Lopez,
  McCauliff, Miller, Morehead, Quintana, Ragozzine, Sasselov, Short, \&
  Steffen}]{Lissauer:2011el}
Lissauer, J.~J., Fabrycky, D.~C., Ford, E.~B., {et~al.} 2011,
  \href{http://dx.doi.org/10.1038/nature09760}{Nature, 469, 53}

\bibitem[{{Lissauer} {et~al.}(2011){Lissauer}, {Ragozzine}, {Fabrycky},
  {Steffen}, {Ford}, {Jenkins}, {Shporer}, {Holman}, {Rowe}, {Quintana},
  {Batalha}, {Borucki}, {Bryson}, {Caldwell}, {Carter}, {Ciardi}, {Dunham},
  {Fortney}, {Gautier}, {Howell}, {Koch}, {Latham}, {Marcy}, {Morehead}, \&
  {Sasselov}}]{Lissauer11}
{Lissauer}, J.~J., {Ragozzine}, D., {Fabrycky}, D.~C., {et~al.} 2011,
  \href{http://dx.doi.org/10.1088/0067-0049/197/1/8}{\apjs, 197, 8}

\bibitem[{{Lissauer} {et~al.}(2012){Lissauer}, {Marcy}, {Rowe}, {Bryson},
  {Adams}, {Buchhave}, {Ciardi}, {Cochran}, {Fabrycky}, {Ford}, {Fressin},
  {Geary}, {Gilliland}, {Holman}, {Howell}, {Jenkins}, {Kinemuchi}, {Koch},
  {Morehead}, {Ragozzine}, {Seader}, {Tanenbaum}, {Torres}, \&
  {Twicken}}]{Lissauer:2012}
{Lissauer}, J.~J., {Marcy}, G.~W., {Rowe}, J.~F., {et~al.} 2012,
  \href{http://dx.doi.org/10.1088/0004-637X/750/2/112}{\apj, 750, 112}

\bibitem[{{Luck} \& {Heiter}(2006)}]{Luck06}
{Luck}, R.~E., \& {Heiter}, U. 2006,
  \href{http://dx.doi.org/10.1086/504080}{\aj, 131, 3069}

\bibitem[{{Madhusudhan} {et~al.}(2011){Madhusudhan}, {Harrington}, {Stevenson},
  {Nymeyer}, {Campo}, {Wheatley}, {Deming}, {Blecic}, {Hardy}, {Lust},
  {Anderson}, {Collier-Cameron}, {Britt}, {Bowman}, {Hebb}, {Hellier},
  {Maxted}, {Pollacco}, \& {West}}]{Madhu11}
{Madhusudhan}, N., {Harrington}, J., {Stevenson}, K.~B., {et~al.} 2011,
  \href{http://dx.doi.org/10.1038/nature09602}{\nat, 469, 64}

\bibitem[{{Mandel} \& {Agol}(2002)}]{Mandel02}
{Mandel}, K., \& {Agol}, E. 2002,
  \href{http://dx.doi.org/10.1086/345520}{\apjl, 580, L171}

\bibitem[{{Mann} {et~al.}(2013){Mann}, {Gaidos}, \& {Ansdell}}]{Mann13}
{Mann}, A.~W., {Gaidos}, E., \& {Ansdell}, M. 2013,
  \href{http://dx.doi.org/10.1088/0004-637X/779/2/188}{\apj, 779, 188}

\bibitem[{{Marcy} {et~al.}(2005){Marcy}, {Butler}, {Fischer}, {Vogt}, {Wright},
  {Tinney}, \& {Jones}}]{Marcy05}
{Marcy}, G., {Butler}, R.~P., {Fischer}, D., {et~al.} 2005,
  \href{http://dx.doi.org/10.1143/PTPS.158.24}{Progress of Theoretical Physics
  Supplement, 158, 24}

\bibitem[{{Marcy}(1983)}]{Marcy83}
{Marcy}, G.~W. 1983, in Bulletin of the American Astronomical Society, Vol.~15,
  Bulletin of the American Astronomical Society, 947

\bibitem[{{Marcy} \& {Butler}(1992)}]{Marcy92}
{Marcy}, G.~W., \& {Butler}, R.~P. 1992,
  \href{http://dx.doi.org/10.1086/132989}{\pasp, 104, 270}

\bibitem[{{Marcy} \& {Butler}(1996)}]{Marcy96}
---. 1996, \href{http://dx.doi.org/10.1086/310096}{\apjl, 464, L147}

\bibitem[{{Marcy} {et~al.}(2008){Marcy}, {Butler}, {Vogt}, {Fischer}, {Wright},
  {Johnson}, {Tinney}, {Jones}, {Carter}, {Bailey}, {O'Toole}, \&
  {Upadhyay}}]{Marcy08}
{Marcy}, G.~W., {Butler}, R.~P., {Vogt}, S.~S., {et~al.} 2008, {Exoplanet
  properties from Lick, Keck and AAT}

\bibitem[{{Mayor} \& {Maurice}(1985)}]{Mayor85}
{Mayor}, M., \& {Maurice}, E. 1985, in Stellar Radial Velocities, ed. A.~G.~D.
  {Philip} \& D.~W. {Latham}, 299

\bibitem[{{Mayor} \& {Queloz}(1995)}]{Mayor95}
{Mayor}, M., \& {Queloz}, D. 1995,
  \href{http://dx.doi.org/10.1038/378355a0}{\nat, 378, 355}

\bibitem[{{Mayor} {et~al.}(2011){Mayor}, {Marmier}, {Lovis}, {Udry},
  {S{\'e}gransan}, {Pepe}, {Benz}, {Bertaux}, {Bouchy}, {Dumusque}, {Lo Curto},
  {Mordasini}, {Queloz}, \& {Santos}}]{Mayor11}
{Mayor}, M., {Marmier}, M., {Lovis}, C., {et~al.} 2011, ArXiv e-prints,
  \href{http://arxiv.org/abs/1109.2497}{{\sffamily arXiv:1109.2497
  [astro-ph.EP]}}

\bibitem[{{McCullough} {et~al.}(2005){McCullough}, {Stys}, {Valenti},
  {Fleming}, {Janes}, \& {Heasley}}]{McCullough05}
{McCullough}, P.~R., {Stys}, J.~E., {Valenti}, J.~A., {et~al.} 2005,
  \href{http://dx.doi.org/10.1086/432024}{\pasp, 117, 783}

\bibitem[{{Mel{\'e}ndez} \& {Barbuy}(1999)}]{Melendez99}
{Mel{\'e}ndez}, J., \& {Barbuy}, B. 1999,
  \href{http://dx.doi.org/10.1086/313261}{\apjs, 124, 527}

\bibitem[{{Mishenina} {et~al.}(2004){Mishenina}, {Soubiran}, {Kovtyukh}, \&
  {Korotin}}]{Mishenina04}
{Mishenina}, T.~V., {Soubiran}, C., {Kovtyukh}, V.~V., \& {Korotin}, S.~A.
  2004, \href{http://dx.doi.org/10.1051/0004-6361:20034454}{\aap, 418, 551}

\bibitem[{{Mordasini} {et~al.}(2012){Mordasini}, {Alibert}, {Georgy},
  {Dittkrist}, {Klahr}, \& {Henning}}]{Mordasini12}
{Mordasini}, C., {Alibert}, Y., {Georgy}, C., {et~al.} 2012,
  \href{http://dx.doi.org/10.1051/0004-6361/201118464}{\aap, 547, A112}

\bibitem[{{Morton}(2012)}]{Morton12}
{Morton}, T.~D. 2012, \href{http://dx.doi.org/10.1088/0004-637X/761/1/6}{\apj,
  761, 6}

\bibitem[{{Morton} \& {Johnson}(2011)}]{Morton11}
{Morton}, T.~D., \& {Johnson}, J.~A. 2011,
  \href{http://dx.doi.org/10.1088/0004-637X/738/2/170}{\apj, 738, 170}

\bibitem[{{Muirhead} {et~al.}(2012){Muirhead}, {Johnson}, {Apps}, {Carter},
  {Morton}, {Fabrycky}, {Pineda}, {Bottom}, {Rojas-Ayala}, {Schlawin},
  {Hamren}, {Covey}, {Crepp}, {Stassun}, {Pepper}, {Hebb}, {Kirby}, {Howard},
  {Isaacson}, {Marcy}, {Levitan}, {Diaz-Santos}, {Armus}, \&
  {Lloyd}}]{Muirhead12}
{Muirhead}, P.~S., {Johnson}, J.~A., {Apps}, K., {et~al.} 2012,
  \href{http://dx.doi.org/10.1088/0004-637X/747/2/144}{\apj, 747, 144}

\bibitem[{{Nidever} {et~al.}(2002){Nidever}, {Marcy}, {Butler}, {Fischer}, \&
  {Vogt}}]{Nidever02}
{Nidever}, D.~L., {Marcy}, G.~W., {Butler}, R.~P., {Fischer}, D.~A., \& {Vogt},
  S.~S. 2002, \href{http://dx.doi.org/10.1086/340570}{\apjs, 141, 503}

\bibitem[{{Nordstr{\"o}m} {et~al.}(2004){Nordstr{\"o}m}, {Mayor}, {Andersen},
  {Holmberg}, {J{\o}rgensen}, {Pont}, {Olsen}, {Udry}, \&
  {Mowlavi}}]{Nordstrom04}
{Nordstr{\"o}m}, B., {Mayor}, M., {Andersen}, J., {et~al.} 2004, The Messenger,
  118, 61

\bibitem[{Ofir \& Dreizler(2012)}]{Ofir:2012va}
Ofir, A., \& Dreizler, S. 2012, \href{http://arxiv.org/abs/1206.5347v1}{arXiv,
  astro-ph.EP}

\bibitem[{Oliphant(2007)}]{Oliphant07}
Oliphant, T.~E. 2007, \href{http://dx.doi.org/10.1109/MCSE.2007.58}{Computing
  in Science Engineering, 9, 10}

\bibitem[{{Peek}(2009)}]{PeekThesis}
{Peek}, K. 2009, PhD thesis, University of California, Berkeley

\bibitem[{P\'erez \& Granger(2007)}]{ipython}
P\'erez, F., \& Granger, B.~E. 2007, \href{http://ipython.org}{{C}omput. {S}ci.
  {E}ng., 9, 21}

\bibitem[{{Petigura} {et~al.}(2013{\natexlab{a}}){Petigura}, {Howard}, \&
  {Marcy}}]{Petigura13b}
{Petigura}, E.~A., {Howard}, A.~W., \& {Marcy}, G.~W. 2013{\natexlab{a}},
  Proceedings of the National Academy of Science, 110, 19273

\bibitem[{{Petigura} \& {Marcy}(2012)}]{Petigura12}
{Petigura}, E.~A., \& {Marcy}, G.~W. 2012,
  \href{http://dx.doi.org/10.1086/668291}{\pasp, 124, 1073}

\bibitem[{{Petigura} {et~al.}(2013{\natexlab{b}}){Petigura}, {Marcy}, \&
  {Howard}}]{Petigura13}
{Petigura}, E.~A., {Marcy}, G.~W., \& {Howard}, A.~W. 2013{\natexlab{b}},
  \href{http://dx.doi.org/10.1088/0004-637X/770/1/69}{\apj, 770, 69}

\bibitem[{{Pierrehumbert} \& {Gaidos}(2011)}]{Pierrehumbert2011}
{Pierrehumbert}, R., \& {Gaidos}, E. 2011,
  \href{http://dx.doi.org/10.1088/2041-8205/734/1/L13}{\apjl, 734, L13}

\bibitem[{{Pinsonneault} {et~al.}(2012){Pinsonneault}, {An},
  {Molenda-{\.Z}akowicz}, {Chaplin}, {Metcalfe}, \& {Bruntt}}]{Pinsonneault12}
{Pinsonneault}, M.~H., {An}, D., {Molenda-{\.Z}akowicz}, J., {et~al.} 2012,
  \href{http://dx.doi.org/10.1088/0067-0049/199/2/30}{\apjs, 199, 30}

\bibitem[{{Piskunov} {et~al.}(1995){Piskunov}, {Kupka}, {Ryabchikova}, {Weiss},
  \& {Jeffery}}]{VALD}
{Piskunov}, N.~E., {Kupka}, F., {Ryabchikova}, T.~A., {Weiss}, W.~W., \&
  {Jeffery}, C.~S. 1995, \aaps, 112, 525

\bibitem[{{Pollacco} {et~al.}(2006){Pollacco}, {Skillen}, {Collier Cameron},
  {Christian}, {Hellier}, {Irwin}, {Lister}, {Street}, {West}, {Anderson},
  {Clarkson}, {Deeg}, {Enoch}, {Evans}, {Fitzsimmons}, {Haswell}, {Hodgkin},
  {Horne}, {Kane}, {Keenan}, {Maxted}, {Norton}, {Osborne}, {Parley}, {Ryans},
  {Smalley}, {Wheatley}, \& {Wilson}}]{Pollacco06}
{Pollacco}, D.~L., {Skillen}, I., {Collier Cameron}, A., {et~al.} 2006,
  \href{http://dx.doi.org/10.1086/508556}{\pasp, 118, 1407}

\bibitem[{{Pollack} {et~al.}(1996){Pollack}, {Hubickyj}, {Bodenheimer},
  {Lissauer}, {Podolak}, \& {Greenzweig}}]{Pollack96}
{Pollack}, J.~B., {Hubickyj}, O., {Bodenheimer}, P., {et~al.} 1996,
  \href{http://dx.doi.org/10.1006/icar.1996.0190}{\icarus, 124, 62}

\bibitem[{{Raghavan} {et~al.}(2010){Raghavan}, {McAlister}, {Henry}, {Latham},
  {Marcy}, {Mason}, {Gies}, {White}, \& {ten Brummelaar}}]{Raghavan10}
{Raghavan}, D., {McAlister}, H.~A., {Henry}, T.~J., {et~al.} 2010,
  \href{http://dx.doi.org/10.1088/0067-0049/190/1/1}{\apjs, 190, 1}

\bibitem[{{Ram{\'{\i}}rez} {et~al.}(2007){Ram{\'{\i}}rez}, {Allende Prieto}, \&
  {Lambert}}]{Ramirez07}
{Ram{\'{\i}}rez}, I., {Allende Prieto}, C., \& {Lambert}, D.~L. 2007,
  \href{http://dx.doi.org/10.1051/0004-6361:20066619}{\aap, 465, 271}

\bibitem[{Rasmussen \& Williams(2006)}]{Rasmussen06}
Rasmussen, C.~E., \& Williams, C. K.~I. 2006, {Gaussian Processes for Machine
  Learning} (MIT Press)

\bibitem[{{Reader} {et~al.}(2002){Reader}, {Wiese}, {Martin}, {Musgrove}, \&
  {Fuhr}}]{Reader02}
{Reader}, J., {Wiese}, W.~L., {Martin}, W.~C., {Musgrove}, A., \& {Fuhr}, J.~R.
  2002, 80

\bibitem[{{Reddy} {et~al.}(2006){Reddy}, {Lambert}, \& {Allende
  Prieto}}]{Reddy06}
{Reddy}, B.~E., {Lambert}, D.~L., \& {Allende Prieto}, C. 2006,
  \href{http://dx.doi.org/10.1111/j.1365-2966.2006.10148.x}{\mnras, 367, 1329}

\bibitem[{{Santos} {et~al.}(2004){Santos}, {Israelian}, \& {Mayor}}]{Santos04}
{Santos}, N.~C., {Israelian}, G., \& {Mayor}, M. 2004,
  \href{http://dx.doi.org/10.1051/0004-6361:20034469}{\aap, 415, 1153}

\bibitem[{{Scott} {et~al.}(2009){Scott}, {Asplund}, {Grevesse}, \&
  {Sauval}}]{Scott09}
{Scott}, P., {Asplund}, M., {Grevesse}, N., \& {Sauval}, A.~J. 2009,
  \href{http://dx.doi.org/10.1088/0004-637X/691/2/L119}{\apjl, 691, L119}

\bibitem[{{Seager}(2013)}]{Seager13}
{Seager}, S. 2013, \href{http://dx.doi.org/10.1126/science.1232226}{Science,
  340, 577}

\bibitem[{{Seager} \& {Lissauer}(2010)}]{Seager10}
{Seager}, S., \& {Lissauer}, J.~J. 2010, {Introduction to Exoplanets}, ed.
  S.~{Seager}, 3

\bibitem[{{Sliski} \& {Kipping}(2014)}]{Sliski14}
{Sliski}, D.~H., \& {Kipping}, D.~M. 2014, ArXiv e-prints,
  \href{http://arxiv.org/abs/1401.1207}{{\sffamily arXiv:1401.1207
  [astro-ph.EP]}}

\bibitem[{{Smith} {et~al.}(2012){Smith}, {Stumpe}, {Van Cleve}, {Jenkins},
  {Barclay}, {Fanelli}, {Girouard}, {Kolodziejczak}, {McCauliff}, {Morris}, \&
  {Twicken}}]{Smith12}
{Smith}, J.~C., {Stumpe}, M.~C., {Van Cleve}, J.~E., {et~al.} 2012,
  \href{http://dx.doi.org/10.1086/667697}{\pasp, 124, 1000}

\bibitem[{{Sneden}(1973)}]{Sneden73}
{Sneden}, C.~A. 1973, PhD thesis, The University Of Texas At Austin

\bibitem[{{Spite}(1967)}]{Spite67}
{Spite}, M. 1967, Annales d'Astrophysique, 30, 211

\bibitem[{Staelin(1969)}]{Staelin69}
Staelin, D.~H. 1969,
  \href{http://adsabs.harvard.edu/cgi-bin/nph-data_query?bibcode=1969IEEEP..57..724S&link_type=ABSTRACT}{in
  Proceedings of the IEEE}, National Radio Astronomy Observatory,
  Charlottesville, VA, USA, 724

\bibitem[{{Struve}(1952)}]{Struve52}
{Struve}, O. 1952, The Observatory, 72, 199

\bibitem[{{Stumpe} {et~al.}(2012){Stumpe}, {Smith}, {Van Cleve}, {Twicken},
  {Barclay}, {Fanelli}, {Girouard}, {Jenkins}, {Kolodziejczak}, {McCauliff}, \&
  {Morris}}]{Stumpe12}
{Stumpe}, M.~C., {Smith}, J.~C., {Van Cleve}, J.~E., {et~al.} 2012,
  \href{http://dx.doi.org/10.1086/667698}{\pasp, 124, 985}

\bibitem[{Tamuz {et~al.}(2005)Tamuz, Mazeh, \& Zucker}]{Tamuz:2005}
Tamuz, O., Mazeh, T., \& Zucker, S. 2005,
  \href{http://dx.doi.org/10.1111/j.1365-2966.2004.08585.x}{Monthly Notices of
  the Royal Astronomical Society, 356, 1466}

\bibitem[{Tenenbaum {et~al.}(2010)Tenenbaum, Bryson, Chandrasekaran, Li,
  Quintana, Twicken, \& Jenkins}]{Tenenbaum:2010}
Tenenbaum, P., Bryson, S.~T., Chandrasekaran, H., {et~al.} 2010,
  \href{http://dx.doi.org/10.1117/12.856705}{Software and Cyberinfrastructure
  for Astronomy. Edited by Radziwill, 7740, 16}

\bibitem[{{Torres} {et~al.}(2012){Torres}, {Fischer}, {Sozzetti}, {Buchhave},
  {Winn}, {Holman}, \& {Carter}}]{Torres12}
{Torres}, G., {Fischer}, D.~A., {Sozzetti}, A., {et~al.} 2012,
  \href{http://dx.doi.org/10.1088/0004-637X/757/2/161}{\apj, 757, 161}

\bibitem[{{Torres} {et~al.}(2011){Torres}, {Fressin}, {Batalha}, {Borucki},
  {Brown}, {Bryson}, {Buchhave}, {Charbonneau}, {Ciardi}, {Dunham}, {Fabrycky},
  {Ford}, {Gautier}, {Gilliland}, {Holman}, {Howell}, {Isaacson}, {Jenkins},
  {Koch}, {Latham}, {Lissauer}, {Marcy}, {Monet}, {Prsa}, {Quinn}, {Ragozzine},
  {Rowe}, {Sasselov}, {Steffen}, \& {Welsh}}]{Torres11}
{Torres}, G., {Fressin}, F., {Batalha}, N.~M., {et~al.} 2011,
  \href{http://dx.doi.org/10.1088/0004-637X/727/1/24}{\apj, 727, 24}

\bibitem[{{Traub}(2012)}]{Traub12}
{Traub}, W.~A. 2012, \href{http://dx.doi.org/10.1088/0004-637X/745/1/20}{\apj,
  745, 20}

\bibitem[{{Twicken} {et~al.}(2010){Twicken}, {Chandrasekaran}, {Jenkins},
  {Gunter}, {Girouard}, \& {Klaus}}]{Twicken10}
{Twicken}, J.~D., {Chandrasekaran}, H., {Jenkins}, J.~M., {et~al.} 2010,
  \href{http://dx.doi.org/10.1117/12.856798}{in Society of Photo-Optical
  Instrumentation Engineers (SPIE) Conference Series, Vol. 7740}

\bibitem[{{Udalski} {et~al.}(2002){Udalski}, {Paczynski}, {Zebrun},
  {Szymanski}, {Kubiak}, {Soszynski}, {Szewczyk}, {Wyrzykowski}, \&
  {Pietrzynski}}]{Udalski02}
{Udalski}, A., {Paczynski}, B., {Zebrun}, K., {et~al.} 2002, \actaa, 52, 1

\bibitem[{{Valenti} \& {Fischer}(2005)}]{Valenti05}
{Valenti}, J.~A., \& {Fischer}, D.~A. 2005,
  \href{http://dx.doi.org/10.1086/430500}{\apjs, 159, 141}

\bibitem[{{Valenti} \& {Piskunov}(1996)}]{Valenti96}
{Valenti}, J.~A., \& {Piskunov}, N. 1996, \aaps, 118, 595

\bibitem[{Van~Cleve \& Caldwell(2009)}]{VanCleve:2009}
Van~Cleve, J.~E., \& Caldwell, D.~A. 2009, {Kepler Instrument Handbook
  (KSCI-19033-001)}

\bibitem[{{Vogt} {et~al.}(1994){Vogt}, {Allen}, {Bigelow}, {Bresee}, {Brown},
  {Cantrall}, {Conrad}, {Couture}, {Delaney}, {Epps}, {Hilyard}, {Hilyard},
  {Horn}, {Jern}, {Kanto}, {Keane}, {Kibrick}, {Lewis}, {Osborne},
  {Pardeilhan}, {Pfister}, {Ricketts}, {Robinson}, {Stover}, {Tucker}, {Ward},
  \& {Wei}}]{Vogt94}
{Vogt}, S.~S., {Allen}, S.~L., {Bigelow}, B.~C., {et~al.} 1994, in Society of
  Photo-Optical Instrumentation Engineers (SPIE) Conference Series, Vol. 2198,
  Instrumentation in Astronomy VIII, ed. D.~L. {Crawford} \& E.~R. {Craine},
  362

\bibitem[{{von Braun} {et~al.}(2014){von Braun}, {Boyajian}, {van Belle},
  {Kane}, {Jones}, {Farrington}, {Schaefer}, {Vargas}, {Scott}, {ten
  Brummelaar}, {Kephart}, {Gies}, {Ciardi}, {L{\'o}pez-Morales}, {Mazingue},
  {McAlister}, {Ridgway}, {Goldfinger}, {Turner}, \& {Sturmann}}]{von-Braun14}
{von Braun}, K., {Boyajian}, T.~S., {van Belle}, G.~T., {et~al.} 2014,
  \href{http://dx.doi.org/10.1093/mnras/stt2360}{\mnras, 438, 2413}

\bibitem[{{Weiss} {et~al.}(2013){Weiss}, {Marcy}, {Rowe}, {Howard}, {Isaacson},
  {Fortney}, {Miller}, {Demory}, {Fischer}, {Adams}, {Dupree}, {Howell},
  {Kolbl}, {Johnson}, {Horch}, {Everett}, {Fabrycky}, \& {Seager}}]{Weiss13}
{Weiss}, L.~M., {Marcy}, G.~W., {Rowe}, J.~F., {et~al.} 2013,
  \href{http://dx.doi.org/10.1088/0004-637X/768/1/14}{\apj, 768, 14}

\bibitem[{{Winn}(2010)}]{Winn10}
{Winn}, J.~N. 2010, {Exoplanet Transits and Occultations}, ed. S.~{Seager}, 55

\bibitem[{{Wolszczan} \& {Frail}(1992)}]{Wolszczan92}
{Wolszczan}, A., \& {Frail}, D.~A. 1992,
  \href{http://dx.doi.org/10.1038/355145a0}{\nat, 355, 145}

\bibitem[{{Yi} {et~al.}(2001){Yi}, {Demarque}, {Kim}, {Lee}, {Ree}, {Lejeune},
  \& {Barnes}}]{Yi01}
{Yi}, S., {Demarque}, P., {Kim}, Y.-C., {et~al.} 2001,
  \href{http://dx.doi.org/10.1086/321795}{\apjs, 136, 417}

\bibitem[{{Youdin}(2011)}]{Youdin11}
{Youdin}, A.~N. 2011, \href{http://dx.doi.org/10.1088/0004-637X/742/1/38}{\apj,
  742, 38}

\bibitem[{{Zsom} {et~al.}(2013){Zsom}, {Seager}, {de Wit}, \&
  {Stamenkovic}}]{Zsom2013}
{Zsom}, A., {Seager}, S., {de Wit}, J., \& {Stamenkovic}, V. 2013, ArXiv
  e-prints, \href{http://arxiv.org/abs/1304.3714}{{\sffamily arXiv:1304.3714
  [astro-ph.EP]}}

\end{thebibliography}
